\documentclass[onecolumn,notitlepage,aps,superscriptaddress]{revtex4-1}
\usepackage{tikz}
\usetikzlibrary{shadings}
\usepackage{cmap}
\usepackage{color}
\usepackage[english]{babel}
\usepackage[utf8]{inputenc}
\usepackage{hyperref}
\usepackage{graphicx}
\usepackage{amsmath}
\usepackage{amssymb}
\usepackage{amsfonts}
\usepackage{amsthm}
\usepackage{textcomp}
\usepackage{scrpage2}
\usepackage[braket]{qcircuit}
\usepackage{verbatim}
\usepackage{latexsym}
\usepackage{pifont}
\begin{document}
\newcommand{\hilite}[1]{\color{red} \textit{#1} \color{black}}
\newcommand{\data}{{\mathrm{dat}}}
\newcommand{\aux}{{\mathrm{aux}}}
\newcommand{\auxdata}{{\mathrm{aux}\,\mathrm{dat}}}
\newcommand{\zetto}[1]{\mathbb{Z}_2^{\otimes #1}}
\newcommand{\obar}[1]{\overline{#1}}
\newcommand{\bbs}[1]{\boldsymbol{#1}}
\newcommand{\moto}{\;\mathrm{mod}\;2}
\newcommand{\clra}{black} 
\newcommand{\clrb}{black!80} 
\newcommand{\clrc}{white} 
\newcommand{\clrd}{black!80} 
\newcommand{\clre}{black!40} 
\newcommand{\checker}{black!15} 
\newcommand{\fadedblack}{black!80}
\newcommand{\BKsmaller}[1]{\,\underset{#1}{<}\, }
\newcommand{\inter}{\mathcal{I}}
 \newcommand{\cca}{black!80}
 \newcommand{\ccb}{black!15}
 \newcommand{\ccc}{white}
\newcommand{\ex}{\ding{55}}
\newcommand{\tick}{\ding{51}}

\begin{abstract}

Quantum simulation of fermionic systems is a promising application of quantum computers, but in order to program them, we  need to map fermionic states and operators to  qubit states and quantum gates. While quantum processors may  be built as two-dimensional qubit networks with couplings between  nearest neighbors,  standard Fermion-to-qubit mappings do not account for that kind of connectivity. In this work we concatenate the (one-dimensional) Jordan-Wigner transform with specific quantum codes  defined under the addition of a certain number of auxiliary qubits. This yields a novel class of mappings  with which any fermionic system can be embedded in a two-dimensional qubit setup, fostering scalable quantum simulation. Our technique is demonstrated on the two-dimensional Fermi-Hubbard model, that we transform into a local Hamiltonian. What is more, we adapt  the Verstraete-Cirac transform and Bravyi-Kitaev Superfast simulation to the square lattice connectivity and compare them to our mappings.  An advantage of our approach in this comparison is that it allows us to encode and decode a  logical state  with a simple unitary quantum circuit.

\end{abstract}

\title{Quantum codes for quantum simulation of Fermions on a square lattice of qubits}
\author{Mark Steudtner}\affiliation{Instituut-Lorentz, Universiteit Leiden, P.O. Box 9506, 2300 RA Leiden, The Netherlands}
\affiliation{QuTech, Delft University of Technology, Lorentzweg 1, 2628 CJ Delft, The Netherlands}
\author{Stephanie Wehner}
\affiliation{QuTech, Delft University of Technology, Lorentzweg 1, 2628 CJ Delft, The Netherlands}
\date{\today} \maketitle
\section{Introduction}

It is believed that quantum computers will help to increase our understanding of large molecules and strongly-correlated materials. Simulating these systems with classical computers is difficult, as they are populated by Fermions. The Hilbert space that these particles span,  has a dimension that scales exponentially with the system size, and thus, if no further efforts are undertaken, the same scaling applies for the amount of computational resources required to simulate it. Building on the works of Feynman, Lloyd and Abrams  \cite{feynman1982simulating,lloyd1996universal,abrams1997simulation}, one can however hope to simulate these problems with other quantum systems of similar size. In digital quantum simulation, we not only absorb the fermionic Hilbert space in a system of qubits, but also use a gate-based quantum computer to solve the problem with quantum algorithms  \cite{kitaev1995quantum, cleve1998quantum, aspuru2005simulated, mcclean2016theory}. However, until quantum computers  can outperform their classical counterparts and some day even tackle  real-world problems, many challenges must be overcome. While small quantum simulations have been performed on few-qubit devices across all platforms \cite{lanyon2010towards,du2010nmr,peruzzo2014variational,barends2015digital,wang2015quantum,o2016scalable,hempel2018quantum}, and efforts are undertaken to scale devices up, the simulation of larger systems is still a challenge.  Critical factors that determine the feasibility of an algorithm would be its qubit requirements, its gate cost (in terms of magic states when error-corrected, and in terms of two-qubit gates when noisy) \cite{jones2012faster,wecker2014gate} and circuit depth (a measure of the algorithm run time, where each time step is the duration of one quantum gate).  Quantum algorithms are generally to be kept shallow  to ensure that they can be run before the qubit system has decohered.  It is thus in our interest to decompose the algorithms into many parts that can be run in parallel, i.e.~at the same time. Obviously, one can hope for parallelization if the algorithm is comprised of gate sequences that act on subsets of as few qubits as possible and these subsets do not overlap much. Another factor is that actual quantum devices can have geometric limitations which negatively influence the circuit depth. In a practical setting not every qubit can reach every other qubit, i.e. they cannot be entangled with a single two-qubit gate. To entangle distant qubits, it takes additional efforts in gates and time. Thus another criterion for the reduction of the circuit depth  is that gate sequences only  act on qubits  adjacent on a certain connectivity graph. Although this graph depends on the actual quantum device, we can make an educated guess:
devices on which surface code can be run, require a square lattice connectivity graph.  \\
Unfortunately, it is non-trivial to embed fermionic problems in those lattices, which opposes shallow-depth quantum simulation. Let us illustrate the exact issue. In order to bring the problem into a form  the quantum computer can process, the  fermionic modes need to be embedded into a (two-dimensional) lattice structure  related to the qubit connectivity graph. After that, a \textit{Fermion-to-qubit mapping} translates the interactions of those system to a qubit Hamiltonian fit to be simulated. It is this last step in which the problem lies, as simulating the interaction between as little as two fermionic modes usually requires gates acting on large subsets of qubits.  This is a consequence of the fermionic wave functions being antisymmetric under particle permutations, which causes the interaction of two fermionic modes to also be sensitive to the occupation of seemingly uninvolved modes, turning into gates on the qubits representing them. This is the same issue that prohibits us from describing  Fermions on (two-dimensional) lattices in terms of Bosons, which could be simulated more easily. In fact, the problems are somewhat intertwined  considering that those bosonic descriptions can double as Fermion-to-qubit mappings.
 The \textit{Jordan-Wigner transform} for instance is widely used as a Fermion-to-qubit mapping \cite{lanyon2010towards,peruzzo2014variational,barends2015digital,hempel2018quantum} today, but its appearance in 1928 \cite{wigner1928uber} predates the work of Feynman by half a century. The original work of Jordan and Wigner was rather meant to compare fermionic operators to the operators of (hard-core) Bosons, which on the other hand are easily  mapped to $(1/2)$-spins. For our purposes, the spins are immediately  identified as qubits, rendering the transform a default for Fermion-to-qubit mappings.  However, the Jordan-Wigner transform is effectively one-dimensional and exhibits large deficits in the treatment of two-dimensional systems. In particular it fails to map a fermionic lattice model with local interactions (meaning their interaction range is bounded by a constant) to a model of locally-interacting spins. In contrast to that,  locally-interacting spins on a lattice can be mapped to a locally-interacting Boson lattice, due to the bosonic wave function not being antisymmetric \cite{holstein1940field}.  While there are tricks and generalizations to circumvent the deficits of the Jordan-Wigner transform \cite{fradkin1989jordan,wang1991ground,ball2005fermions,chen2018exact}, not all of them are useful for its role in quantum simulation: there is no ultimate choice for a two-dimensional Fermion-to-qubit mapping.  However, there is a mapping with which  locally-interacting Fermion and qubit lattices can be related:
 the \textit{Verstraete-Cirac transform} (VCT) \cite{verstraete2005mapping} also known as \textit{Auxiliary Fermion Mapping} \cite{whitfield2016local,havlivcek2017operator,zohar2018eliminating}, can be regarded as a manipulation of the Jordan-Wigner transform, in which additional \textit{auxiliary} particles are added, hence the name.
Other works on Fermion-to-qubit mappings \cite{seeley2012bravyi,tranter2015bravyi,havlivcek2017operator,setia2017bravyi}  are based on two transforms proposed by Bravyi and Kitaev in \cite{bravyi2002fermionic}. Firstly, there is what is commonly known as the \textit{Bravyi-Kitaev transform}, that, compared to the Jordan-Wigner transform, exhibits an up to exponential improvement on the number of qubits that each fermionic interaction term acts on. The Bravyi-Kitaev transformation however demands a qubit connectivity that is higher than what a square lattice can offer. Secondly, the mapping referred to as \textit{`Superfast simulation of Fermions on a graph'} (BKSF)  has the power to map local Fermion lattices to local qubit lattices, but the square lattice connectivity is generally only sufficient when the underlying model is an interacting square lattice as well: to make interactions local, the mapping requires a qubit connectivity graph set by the Hamiltonian.  When the given connectivity turns into a limitation, classical tools like sorting networks might be applied \cite{beals2013efficient}. Most notably, there are recent attempts to incorporate swapping networks into the Fermion-to-qubit mapping. With so-called fermionic swaps \cite{bravyi2002fermionic}, not only qubits are swapped but also fermionic modes, in the sense that swapping operations can change the locality of their interactions in the Jordan-Wigner transform. This effectively eliminates the contribution of the Fermion-to-qubit mapping to the gate cost and algorithmic depth which is then dominated by the swapping network alone \cite{babbush2017low,babbush2018low,kivlichan2018quantum}.

In this work, we want to abstain from swapping and sorting networks in order to make use of the (two-dimensional) geometric proximity of qubits inside the quantum device. In this way, the gate cost is determined by the range of interactions  on the fermionic lattice and  distant interactions can be simulated in parallel.
For this purpose, we define two-dimensional (non-perturbative) Fermion-to-qubit mappings that generalize the Jordan-Wigner transform on the square lattice. We here not only demand that local  Hamiltonians of Fermions are mapped to local qubit Hamiltonians but want to go beyond nearest neighbor interactions. The exchange interaction between two (distant) modes should involve only the  two qubits that  these modes correspond to, and some chain of qubits that  connects them geometrically. This means that when we imagine the system as a Fermion lattice with dimension $(\ell_1 \times \ell_2)$, we want an interaction term of any two modes to transform into a term acting on $O(m)$ qubits, when the modes have a Manhattan distance of $m$. As a consequence, we can bound  the weight of  the largest terms by $O(\ell_1+\ell_2)$, rather than $O(\ell_1 \times \ell_2)$ as in the case of the  Jordan-Wigner transform. In this way the entire simulation  only considers operators acting on the shortest possible strings along adjacent qubits, fostering parallelization.
\section{Results}
\begin{figure}
\begin{tikzpicture}[scale=0.85]

\node[] at (0,0) {
\begin{tikzpicture}[scale=.45]
\draw[dotted] (1 + 1 * 0.500000 , 1 * 0.500000)--(6 + 1 * 0.500000 , 1 * 0.500000);
\draw[dotted] (1 + 2 * 0.500000 , 2 * 0.500000)--(6 + 2 * 0.500000 , 2 * 0.500000);
\draw[dotted] (1 + 3 * 0.500000 , 3 * 0.500000)--(6 + 3 * 0.500000 , 3 * 0.500000);
\draw[dotted] (1 + 4 * 0.500000 , 4 * 0.500000)--(6 + 4 * 0.500000 , 4 * 0.500000);
\draw[dotted] (1 + 5 * 0.500000 , 5 * 0.500000)--(6 + 5 * 0.500000 , 5 * 0.500000);
\draw[dotted] (1 + 6 * 0.500000 , 6 * 0.500000)--(6 + 6 * 0.500000 , 6 * 0.500000);

\draw[dotted] (1 + 1 * 0.500000 , 1 * 0.500000)--(1 + 6 * 0.500000 , 6 * 0.500000);
\draw[dotted] (2 + 1 * 0.500000 , 1 * 0.500000)--(2 + 6 * 0.500000 , 6 * 0.500000);
\draw[dotted] (3 + 1 * 0.500000 , 1 * 0.500000)--(3 + 6 * 0.500000 , 6 * 0.500000);
\draw[dotted] (4 + 1 * 0.500000 , 1 * 0.500000)--(4 + 6 * 0.500000 , 6 * 0.500000);
\draw[dotted] (5 + 1 * 0.500000 , 1 * 0.500000)--(5 + 6 * 0.500000 , 6 * 0.500000);
\draw[dotted] (6 + 1 * 0.500000 , 1 * 0.500000)--(6 + 6 * 0.500000 , 6 * 0.500000);

\draw[thick] (2 + 2 * 0.500000 , 2 * 0.500000)--(2 + 1 * 0.500000 , 1 * 0.500000);

\draw[thick] (3 + 3 * 0.500000 , 3 * 0.500000)--(2 + 5 * 0.500000 , 5 * 0.500000);

\draw[thick] (4 + 6 * 0.500000 , 6 * 0.500000).. controls (4.6 + 6.7 * 0.500000 , 6.7 * 0.500000) and (5.4 + 6.7 * 0.500000 , 6.7 * 0.500000) .. (6 + 6 * 0.500000 , 6 * 0.500000);

\draw[fill=white, rotate around = { 90 : ( 1 + 1 * 0.500000 , 1 * 0.500000) } ] (1 + 1 * 0.500000 , 1 * 0.500000) ellipse ( 0.1 * 0.500000 and 0.1 );
\draw[white, fill=black, rotate around = { 90 : ( 2 + 1 * 0.500000 , 1 * 0.500000) } ] (2 + 1 * 0.500000 , 1 * 0.500000) ellipse ( 0.16 * 0.500000 and 0.16 );
\draw[fill=white, rotate around = { 90 : ( 3 + 1 * 0.500000 , 1 * 0.500000) } ] (3 + 1 * 0.500000 , 1 * 0.500000) ellipse ( 0.1 * 0.500000 and 0.1 );
\draw[fill=white, rotate around = { 90 : ( 4 + 1 * 0.500000 , 1 * 0.500000) } ] (4 + 1 * 0.500000 , 1 * 0.500000) ellipse ( 0.1 * 0.500000 and 0.1 );
\draw[fill=white, rotate around = { 90 : ( 5 + 1 * 0.500000 , 1 * 0.500000) } ] (5 + 1 * 0.500000 , 1 * 0.500000) ellipse ( 0.1 * 0.500000 and 0.1 );
\draw[fill=white, rotate around = { 90 : ( 6 + 1 * 0.500000 , 1 * 0.500000) } ] (6 + 1 * 0.500000 , 1 * 0.500000) ellipse ( 0.1 * 0.500000 and 0.1 );

\draw[fill=white, rotate around = { 90 : ( 1 + 2 * 0.500000 , 2 * 0.500000) } ] (1 + 2 * 0.500000 , 2 * 0.500000) ellipse ( 0.1 * 0.500000 and 0.1 );
\draw[white, fill=black, rotate around = { 90 : ( 2 + 2 * 0.500000 , 2 * 0.500000) } ] (2 + 2 * 0.500000 , 2 * 0.500000) ellipse ( 0.16 * 0.500000 and 0.16 );
\draw[fill=white, rotate around = { 90 : ( 3 + 2 * 0.500000 , 2 * 0.500000) } ] (3 + 2 * 0.500000 , 2 * 0.500000) ellipse ( 0.1 * 0.500000 and 0.1 );
\draw[fill=white, rotate around = { 90 : ( 4 + 2 * 0.500000 , 2 * 0.500000) } ] (4 + 2 * 0.500000 , 2 * 0.500000) ellipse ( 0.1 * 0.500000 and 0.1 );
\draw[fill=white, rotate around = { 90 : ( 5 + 2 * 0.500000 , 2 * 0.500000) } ] (5 + 2 * 0.500000 , 2 * 0.500000) ellipse ( 0.1 * 0.500000 and 0.1 );
\draw[fill=white, rotate around = { 90 : ( 6 + 2 * 0.500000 , 2 * 0.500000) } ] (6 + 2 * 0.500000 , 2 * 0.500000) ellipse ( 0.1 * 0.500000 and 0.1 );

\draw[fill=white, rotate around = { 90 : ( 1 + 3 * 0.500000 , 3 * 0.500000) } ] (1 + 3 * 0.500000 , 3 * 0.500000) ellipse ( 0.1 * 0.500000 and 0.1 );
\draw[fill=white, rotate around = { 90 : ( 2 + 3 * 0.500000 , 3 * 0.500000) } ] (2 + 3 * 0.500000 , 3 * 0.500000) ellipse ( 0.1 * 0.500000 and 0.1 );
\draw[white, fill=black, rotate around = { 90 : ( 3 + 3 * 0.500000 , 3 * 0.500000) } ] (3 + 3 * 0.500000 , 3 * 0.500000) ellipse ( 0.16 * 0.500000 and 0.16 );
\draw[fill=white, rotate around = { 90 : ( 4 + 3 * 0.500000 , 3 * 0.500000) } ] (4 + 3 * 0.500000 , 3 * 0.500000) ellipse ( 0.1 * 0.500000 and 0.1 );
\draw[fill=white, rotate around = { 90 : ( 5 + 3 * 0.500000 , 3 * 0.500000) } ] (5 + 3 * 0.500000 , 3 * 0.500000) ellipse ( 0.1 * 0.500000 and 0.1 );
\draw[fill=white, rotate around = { 90 : ( 6 + 3 * 0.500000 , 3 * 0.500000) } ] (6 + 3 * 0.500000 , 3 * 0.500000) ellipse ( 0.1 * 0.500000 and 0.1 );

\draw[fill=white, rotate around = { 90 : ( 1 + 4 * 0.500000 , 4 * 0.500000) } ] (1 + 4 * 0.500000 , 4 * 0.500000) ellipse ( 0.1 * 0.500000 and 0.1 );
\draw[fill=white, rotate around = { 90 : ( 2 + 4 * 0.500000 , 4 * 0.500000) } ] (2 + 4 * 0.500000 , 4 * 0.500000) ellipse ( 0.1 * 0.500000 and 0.1 );
\draw[fill=white, rotate around = { 90 : ( 3 + 4 * 0.500000 , 4 * 0.500000) } ] (3 + 4 * 0.500000 , 4 * 0.500000) ellipse ( 0.1 * 0.500000 and 0.1 );
\draw[fill=white, rotate around = { 90 : ( 4 + 4 * 0.500000 , 4 * 0.500000) } ] (4 + 4 * 0.500000 , 4 * 0.500000) ellipse ( 0.1 * 0.500000 and 0.1 );
\draw[fill=white, rotate around = { 90 : ( 5 + 4 * 0.500000 , 4 * 0.500000) } ] (5 + 4 * 0.500000 , 4 * 0.500000) ellipse ( 0.1 * 0.500000 and 0.1 );
\draw[fill=white, rotate around = { 90 : ( 6 + 4 * 0.500000 , 4 * 0.500000) } ] (6 + 4 * 0.500000 , 4 * 0.500000) ellipse ( 0.1 * 0.500000 and 0.1 );

\draw[fill=white, rotate around = { 90 : ( 1 + 5 * 0.500000 , 5 * 0.500000) } ] (1 + 5 * 0.500000 , 5 * 0.500000) ellipse ( 0.1 * 0.500000 and 0.1 );
\draw[white, fill=black, rotate around = { 90 : ( 2 + 5 * 0.500000 , 5 * 0.500000) } ] (2 + 5 * 0.500000 , 5 * 0.500000) ellipse ( 0.16 * 0.500000 and 0.16 );
\draw[fill=white, rotate around = { 90 : ( 3 + 5 * 0.500000 , 5 * 0.500000) } ] (3 + 5 * 0.500000 , 5 * 0.500000) ellipse ( 0.1 * 0.500000 and 0.1 );
\draw[fill=white, rotate around = { 90 : ( 4 + 5 * 0.500000 , 5 * 0.500000) } ] (4 + 5 * 0.500000 , 5 * 0.500000) ellipse ( 0.1 * 0.500000 and 0.1 );
\draw[fill=white, rotate around = { 90 : ( 5 + 5 * 0.500000 , 5 * 0.500000) } ] (5 + 5 * 0.500000 , 5 * 0.500000) ellipse ( 0.1 * 0.500000 and 0.1 );
\draw[fill=white, rotate around = { 90 : ( 6 + 5 * 0.500000 , 5 * 0.500000) } ] (6 + 5 * 0.500000 , 5 * 0.500000) ellipse ( 0.1 * 0.500000 and 0.1 );

\draw[fill=white, rotate around = { 90 : ( 1 + 6 * 0.500000 , 6 * 0.500000) } ] (1 + 6 * 0.500000 , 6 * 0.500000) ellipse ( 0.1 * 0.500000 and 0.1 );
\draw[fill=white, rotate around = { 90 : ( 2 + 6 * 0.500000 , 6 * 0.500000) } ] (2 + 6 * 0.500000 , 6 * 0.500000) ellipse ( 0.1 * 0.500000 and 0.1 );
\draw[fill=white, rotate around = { 90 : ( 3 + 6 * 0.500000 , 6 * 0.500000) } ] (3 + 6 * 0.500000 , 6 * 0.500000) ellipse ( 0.1 * 0.500000 and 0.1 );
\draw[white, fill=black, rotate around = { 90 : ( 4 + 6 * 0.500000 , 6 * 0.500000) } ] (4 + 6 * 0.500000 , 6 * 0.500000) ellipse ( 0.16 * 0.500000 and 0.16 );
\draw[fill=white, rotate around = { 90 : ( 5 + 6 * 0.500000 , 6 * 0.500000) } ] (5 + 6 * 0.500000 , 6 * 0.500000) ellipse ( 0.1 * 0.500000 and 0.1 );
\draw[white, fill=black, rotate around = { 90 : ( 6 + 6 * 0.500000 , 6 * 0.500000) } ] (6 + 6 * 0.500000 , 6 * 0.500000) ellipse ( 0.16 * 0.500000 and 0.16 );
\end{tikzpicture}
};

\node[] at (0,-2) {
\begin{tikzpicture}[scale=.45]

\draw[thick] ( 6 + 3 * 0.500000 , 3 * 0.500000) .. controls ( 6.6 + 3.2 * 0.500000 , 3.2 * 0.500000) and ( 6.6 + 3.8 * 0.500000 , 3.8 * 0.500000) .. ( 6 + 4 * 0.500000 , 4 * 0.500000);

\draw[thick] ( 6 + 1 * 0.500000 , 1 * 0.500000) .. controls ( 6.6 + 1.2 * 0.500000 , 1.2 * 0.500000) and ( 6.6 + 1.8 * 0.500000 , 1.8 * 0.500000) .. ( 6 + 2 * 0.500000 , 2 * 0.500000);

\draw[thick] ( 1 + 4 * 0.500000 , 4 * 0.500000) .. controls ( .4 + 4.2 * 0.500000 , 4.2 * 0.500000) and ( .4 + 4.8 * 0.500000 , 4.8 * 0.500000) .. ( 1 + 5 * 0.500000 , 5 * 0.500000);

\draw[thick] ( 6 + 6 * 0.500000 , 6 * 0.500000) -- ( 4 + 6 * 0.500000 , 6 * 0.500000);

\draw[thick] ( 2 + 1 * 0.500000 , 1 * 0.500000) -- ( 6 + 1 * 0.500000 , 1 * 0.500000);

\draw[thick] ( 2 + 2 * 0.500000 , 2 * 0.500000) -- ( 6 + 2 * 0.500000 , 2 * 0.500000);

\draw[thick] ( 1 + 4 * 0.500000 , 4 * 0.500000) -- ( 6 + 4 * 0.500000 , 4 * 0.500000);

\draw[thick] ( 3 + 3 * 0.500000 , 3 * 0.500000) -- ( 6 + 3 * 0.500000 , 3 * 0.500000);

\draw[thick] ( 2 + 5 * 0.500000 , 5 * 0.500000) -- ( 1 + 5 * 0.500000 , 5 * 0.500000);

\draw[rotate around = { 90 : ( 1 + 1 * 0.500000 , 1 * 0.500000) } ] (1 + 1 * 0.500000 , 1 * 0.500000) ellipse ( 0.2 * 0.500000 and 0.2 );
\draw[white, fill=black, rotate around = { 90 : ( 2 + 1 * 0.500000 , 1 * 0.500000) } ] (2 + 1 * 0.500000 , 1 * 0.500000) ellipse ( 0.2 * 0.500000 and 0.2 );
\draw[white, fill=black, rotate around = { 90 : ( 3 + 1 * 0.500000 , 1 * 0.500000) } ] (3 + 1 * 0.500000 , 1 * 0.500000) ellipse ( 0.2 * 0.500000 and 0.2 );
\draw[white, fill=black, rotate around = { 90 : ( 4 + 1 * 0.500000 , 1 * 0.500000) } ] (4 + 1 * 0.500000 , 1 * 0.500000) ellipse ( 0.2 * 0.500000 and 0.2 );
\draw[white, fill=black, rotate around = { 90 : ( 5 + 1 * 0.500000 , 1 * 0.500000) } ] (5 + 1 * 0.500000 , 1 * 0.500000) ellipse ( 0.2 * 0.500000 and 0.2 );
\draw[white, fill=black, rotate around = { 90 : ( 6 + 1 * 0.500000 , 1 * 0.500000) } ] (6 + 1 * 0.500000 , 1 * 0.500000) ellipse ( 0.2 * 0.500000 and 0.2 );

\draw[rotate around = { 90 : ( 1 + 2 * 0.500000 , 2 * 0.500000) } ] (1 + 2 * 0.500000 , 2 * 0.500000) ellipse ( 0.2 * 0.500000 and 0.2 );
\draw[white, fill=black, rotate around = { 90 : ( 2 + 2 * 0.500000 , 2 * 0.500000) } ] (2 + 2 * 0.500000 , 2 * 0.500000) ellipse ( 0.2 * 0.500000 and 0.2 );
\draw[white, fill=black, rotate around = { 90 : ( 3 + 2 * 0.500000 , 2 * 0.500000) } ] (3 + 2 * 0.500000 , 2 * 0.500000) ellipse ( 0.2 * 0.500000 and 0.2 );
\draw[white, fill=black, rotate around = { 90 : ( 4 + 2 * 0.500000 , 2 * 0.500000) } ] (4 + 2 * 0.500000 , 2 * 0.500000) ellipse ( 0.2 * 0.500000 and 0.2 );
\draw[white, fill=black, rotate around = { 90 : ( 5 + 2 * 0.500000 , 2 * 0.500000) } ] (5 + 2 * 0.500000 , 2 * 0.500000) ellipse ( 0.2 * 0.500000 and 0.2 );
\draw[white, fill=black, rotate around = { 90 : ( 6 + 2 * 0.500000 , 2 * 0.500000) } ] (6 + 2 * 0.500000 , 2 * 0.500000) ellipse ( 0.2 * 0.500000 and 0.2 );

\draw[rotate around = { 90 : ( 1 + 3 * 0.500000 , 3 * 0.500000) } ] (1 + 3 * 0.500000 , 3 * 0.500000) ellipse ( 0.2 * 0.500000 and 0.2 );
\draw[rotate around = { 90 : ( 2 + 3 * 0.500000 , 3 * 0.500000) } ] (2 + 3 * 0.500000 , 3 * 0.500000) ellipse ( 0.2 * 0.500000 and 0.2 );
\draw[white, fill=black, rotate around = { 90 : ( 3 + 3 * 0.500000 , 3 * 0.500000) } ] (3 + 3 * 0.500000 , 3 * 0.500000) ellipse ( 0.2 * 0.500000 and 0.2 );
\draw[white, fill=black, rotate around = { 90 : ( 4 + 3 * 0.500000 , 3 * 0.500000) } ] (4 + 3 * 0.500000 , 3 * 0.500000) ellipse ( 0.2 * 0.500000 and 0.2 );
\draw[white, fill=black, rotate around = { 90 : ( 5 + 3 * 0.500000 , 3 * 0.500000) } ] (5 + 3 * 0.500000 , 3 * 0.500000) ellipse ( 0.2 * 0.500000 and 0.2 );
\draw[white, fill=black, rotate around = { 90 : ( 6 + 3 * 0.500000 , 3 * 0.500000) } ] (6 + 3 * 0.500000 , 3 * 0.500000) ellipse ( 0.2 * 0.500000 and 0.2 );

\draw[white, fill=black, rotate around = { 90 : ( 1 + 4 * 0.500000 , 4 * 0.500000) } ] (1 + 4 * 0.500000 , 4 * 0.500000) ellipse ( 0.2 * 0.500000 and 0.2 );
\draw[white, fill=black, rotate around = { 90 : ( 2 + 4 * 0.500000 , 4 * 0.500000) } ] (2 + 4 * 0.500000 , 4 * 0.500000) ellipse ( 0.2 * 0.500000 and 0.2 );
\draw[white, fill=black, rotate around = { 90 : ( 3 + 4 * 0.500000 , 4 * 0.500000) } ] (3 + 4 * 0.500000 , 4 * 0.500000) ellipse ( 0.2 * 0.500000 and 0.2 );
\draw[white, fill=black, rotate around = { 90 : ( 4 + 4 * 0.500000 , 4 * 0.500000) } ] (4 + 4 * 0.500000 , 4 * 0.500000) ellipse ( 0.2 * 0.500000 and 0.2 );
\draw[white, fill=black, rotate around = { 90 : ( 5 + 4 * 0.500000 , 4 * 0.500000) } ] (5 + 4 * 0.500000 , 4 * 0.500000) ellipse ( 0.2 * 0.500000 and 0.2 );
\draw[white, fill=black, rotate around = { 90 : ( 6 + 4 * 0.500000 , 4 * 0.500000) } ] (6 + 4 * 0.500000 , 4 * 0.500000) ellipse ( 0.2 * 0.500000 and 0.2 );

\draw[white, fill=black, rotate around = { 90 : ( 1 + 5 * 0.500000 , 5 * 0.500000) } ] (1 + 5 * 0.500000 , 5 * 0.500000) ellipse ( 0.2 * 0.500000 and 0.2 );
\draw[white, fill=black, rotate around = { 90 : ( 2 + 5 * 0.500000 , 5 * 0.500000) } ] (2 + 5 * 0.500000 , 5 * 0.500000) ellipse ( 0.2 * 0.500000 and 0.2 );
\draw[rotate around = { 90 : ( 3 + 5 * 0.500000 , 5 * 0.500000) } ] (3 + 5 * 0.500000 , 5 * 0.500000) ellipse ( 0.2 * 0.500000 and 0.2 );
\draw[rotate around = { 90 : ( 4 + 5 * 0.500000 , 5 * 0.500000) } ] (4 + 5 * 0.500000 , 5 * 0.500000) ellipse ( 0.2 * 0.500000 and 0.2 );
\draw[rotate around = { 90 : ( 5 + 5 * 0.500000 , 5 * 0.500000) } ] (5 + 5 * 0.500000 , 5 * 0.500000) ellipse ( 0.2 * 0.500000 and 0.2 );
\draw[rotate around = { 90 : ( 6 + 5 * 0.500000 , 5 * 0.500000) } ] (6 + 5 * 0.500000 , 5 * 0.500000) ellipse ( 0.2 * 0.500000 and 0.2 );

\draw[rotate around = { 90 : ( 1 + 6 * 0.500000 , 6 * 0.500000) } ] (1 + 6 * 0.500000 , 6 * 0.500000) ellipse ( 0.2 * 0.500000 and 0.2 );
\draw[rotate around = { 90 : ( 2 + 6 * 0.500000 , 6 * 0.500000) } ] (2 + 6 * 0.500000 , 6 * 0.500000) ellipse ( 0.2 * 0.500000 and 0.2 );
\draw[rotate around = { 90 : ( 3 + 6 * 0.500000 , 6 * 0.500000) } ] (3 + 6 * 0.500000 , 6 * 0.500000) ellipse ( 0.2 * 0.500000 and 0.2 );
\draw[white, fill=black, rotate around = { 90 : ( 4 + 6 * 0.500000 , 6 * 0.500000) } ] (4 + 6 * 0.500000 , 6 * 0.500000) ellipse ( 0.2 * 0.500000 and 0.2 );
\draw[white, fill=black, rotate around = { 90 : ( 5 + 6 * 0.500000 , 6 * 0.500000) } ] (5 + 6 * 0.500000 , 6 * 0.500000) ellipse ( 0.2 * 0.500000 and 0.2 );
\draw[white, fill=black, rotate around = { 90 : ( 6 + 6 * 0.500000 , 6 * 0.500000) } ] (6 + 6 * 0.500000 , 6 * 0.500000) ellipse ( 0.2 * 0.500000 and 0.2 );
\end{tikzpicture}
};

\node[] at (0,-4) {
\begin{tikzpicture}[scale=.45]

\draw[thick] ( 2 + 1 * 0.500000 , 1 * 0.500000)--( 2 + 1.5 * 0.500000 , 1.5 * 0.500000);

\draw[thick] ( 6 + 6 * 0.500000 , 6 * 0.500000)--( 4 + 6 * 0.500000 , 6 * 0.500000)--( 4 + 5.5 * 0.500000 , 5.5 * 0.500000);

\draw[thick] ( 3 + 2.5 * 0.500000 , 2.5 * 0.500000)--( 3 + 4 * 0.500000 , 4 * 0.500000)-- ( 2 + 4 * 0.500000 , 4 * 0.500000)--( 2 + 4.5 * 0.500000 , 4.5 * 0.500000);

\draw[rotate around = { 90 : ( 1 + 1 * 0.500000 , 1 * 0.500000) } ] (1 + 1 * 0.500000 , 1 * 0.500000) ellipse ( 0.2 * 0.500000 and 0.2 );
\draw[white, fill=black, rotate around = { 90 : ( 2 + 1 * 0.500000 , 1 * 0.500000) } ] (2 + 1 * 0.500000 , 1 * 0.500000) ellipse ( 0.2 * 0.500000 and 0.2 );
\draw[rotate around = { 90 : ( 3 + 1 * 0.500000 , 1 * 0.500000) } ] (3 + 1 * 0.500000 , 1 * 0.500000) ellipse ( 0.2 * 0.500000 and 0.2 );
\draw[rotate around = { 90 : ( 4 + 1 * 0.500000 , 1 * 0.500000) } ] (4 + 1 * 0.500000 , 1 * 0.500000) ellipse ( 0.2 * 0.500000 and 0.2 );
\draw[rotate around = { 90 : ( 5 + 1 * 0.500000 , 1 * 0.500000) } ] (5 + 1 * 0.500000 , 1 * 0.500000) ellipse ( 0.2 * 0.500000 and 0.2 );
\draw[rotate around = { 90 : ( 6 + 1 * 0.500000 , 1 * 0.500000) } ] (6 + 1 * 0.500000 , 1 * 0.500000) ellipse ( 0.2 * 0.500000 and 0.2 );

\draw[rotate around = { 90 : ( 1 + 2 * 0.500000 , 2 * 0.500000) } ] (1 + 2 * 0.500000 , 2 * 0.500000) ellipse ( 0.2 * 0.500000 and 0.2 );
\draw[rotate around = { 90 : ( 2 + 2 * 0.500000 , 2 * 0.500000) } ] (2 + 2 * 0.500000 , 2 * 0.500000) ellipse ( 0.2 * 0.500000 and 0.2 );
\draw[rotate around = { 90 : ( 3 + 2 * 0.500000 , 2 * 0.500000) } ] (3 + 2 * 0.500000 , 2 * 0.500000) ellipse ( 0.2 * 0.500000 and 0.2 );
\draw[rotate around = { 90 : ( 4 + 2 * 0.500000 , 2 * 0.500000) } ] (4 + 2 * 0.500000 , 2 * 0.500000) ellipse ( 0.2 * 0.500000 and 0.2 );
\draw[rotate around = { 90 : ( 5 + 2 * 0.500000 , 2 * 0.500000) } ] (5 + 2 * 0.500000 , 2 * 0.500000) ellipse ( 0.2 * 0.500000 and 0.2 );
\draw[rotate around = { 90 : ( 6 + 2 * 0.500000 , 2 * 0.500000) } ] (6 + 2 * 0.500000 , 2 * 0.500000) ellipse ( 0.2 * 0.500000 and 0.2 );

\draw[rotate around = { 90 : ( 1 + 3 * 0.500000 , 3 * 0.500000) } ] (1 + 3 * 0.500000 , 3 * 0.500000) ellipse ( 0.2 * 0.500000 and 0.2 );
\draw[rotate around = { 90 : ( 2 + 3 * 0.500000 , 3 * 0.500000) } ] (2 + 3 * 0.500000 , 3 * 0.500000) ellipse ( 0.2 * 0.500000 and 0.2 );
\draw[white, fill=black, rotate around = { 90 : ( 3 + 3 * 0.500000 , 3 * 0.500000) } ] (3 + 3 * 0.500000 , 3 * 0.500000) ellipse ( 0.2 * 0.500000 and 0.2 );
\draw[rotate around = { 90 : ( 4 + 3 * 0.500000 , 3 * 0.500000) } ] (4 + 3 * 0.500000 , 3 * 0.500000) ellipse ( 0.2 * 0.500000 and 0.2 );
\draw[rotate around = { 90 : ( 5 + 3 * 0.500000 , 3 * 0.500000) } ] (5 + 3 * 0.500000 , 3 * 0.500000) ellipse ( 0.2 * 0.500000 and 0.2 );
\draw[rotate around = { 90 : ( 6 + 3 * 0.500000 , 3 * 0.500000) } ] (6 + 3 * 0.500000 , 3 * 0.500000) ellipse ( 0.2 * 0.500000 and 0.2 );

\draw[rotate around = { 90 : ( 1 + 4 * 0.500000 , 4 * 0.500000) } ] (1 + 4 * 0.500000 , 4 * 0.500000) ellipse ( 0.2 * 0.500000 and 0.2 );
\draw[white, fill=black, rotate around = { 90 : ( 2 + 4 * 0.500000 , 4 * 0.500000) } ] (2 + 4 * 0.500000 , 4 * 0.500000) ellipse ( 0.2 * 0.500000 and 0.2 );
\draw[white, fill=black, rotate around = { 90 : ( 3 + 4 * 0.500000 , 4 * 0.500000) } ] (3 + 4 * 0.500000 , 4 * 0.500000) ellipse ( 0.2 * 0.500000 and 0.2 );
\draw[rotate around = { 90 : ( 4 + 4 * 0.500000 , 4 * 0.500000) } ] (4 + 4 * 0.500000 , 4 * 0.500000) ellipse ( 0.2 * 0.500000 and 0.2 );
\draw[rotate around = { 90 : ( 5 + 4 * 0.500000 , 4 * 0.500000) } ] (5 + 4 * 0.500000 , 4 * 0.500000) ellipse ( 0.2 * 0.500000 and 0.2 );
\draw[rotate around = { 90 : ( 6 + 4 * 0.500000 , 4 * 0.500000) } ] (6 + 4 * 0.500000 , 4 * 0.500000) ellipse ( 0.2 * 0.500000 and 0.2 );

\draw[rotate around = { 90 : ( 1 + 5 * 0.500000 , 5 * 0.500000) } ] (1 + 5 * 0.500000 , 5 * 0.500000) ellipse ( 0.2 * 0.500000 and 0.2 );
\draw[rotate around = { 90 : ( 2 + 5 * 0.500000 , 5 * 0.500000) } ] (2 + 5 * 0.500000 , 5 * 0.500000) ellipse ( 0.2 * 0.500000 and 0.2 );
\draw[rotate around = { 90 : ( 3 + 5 * 0.500000 , 5 * 0.500000) } ] (3 + 5 * 0.500000 , 5 * 0.500000) ellipse ( 0.2 * 0.500000 and 0.2 );
\draw[rotate around = { 90 : ( 4 + 5 * 0.500000 , 5 * 0.500000) } ] (4 + 5 * 0.500000 , 5 * 0.500000) ellipse ( 0.2 * 0.500000 and 0.2 );
\draw[rotate around = { 90 : ( 5 + 5 * 0.500000 , 5 * 0.500000) } ] (5 + 5 * 0.500000 , 5 * 0.500000) ellipse ( 0.2 * 0.500000 and 0.2 );
\draw[rotate around = { 90 : ( 6 + 5 * 0.500000 , 5 * 0.500000) } ] (6 + 5 * 0.500000 , 5 * 0.500000) ellipse ( 0.2 * 0.500000 and 0.2 );

\draw[rotate around = { 90 : ( 1 + 6 * 0.500000 , 6 * 0.500000) } ] (1 + 6 * 0.500000 , 6 * 0.500000) ellipse ( 0.2 * 0.500000 and 0.2 );
\draw[rotate around = { 90 : ( 2 + 6 * 0.500000 , 6 * 0.500000) } ] (2 + 6 * 0.500000 , 6 * 0.500000) ellipse ( 0.2 * 0.500000 and 0.2 );
\draw[rotate around = { 90 : ( 3 + 6 * 0.500000 , 6 * 0.500000) } ] (3 + 6 * 0.500000 , 6 * 0.500000) ellipse ( 0.2 * 0.500000 and 0.2 );
\draw[white, fill=black, rotate around = { 90 : ( 4 + 6 * 0.500000 , 6 * 0.500000) } ] (4 + 6 * 0.500000 , 6 * 0.500000) ellipse ( 0.2 * 0.500000 and 0.2 );
\draw[white, fill=black, rotate around = { 90 : ( 5 + 6 * 0.500000 , 6 * 0.500000) } ] (5 + 6 * 0.500000 , 6 * 0.500000) ellipse ( 0.2 * 0.500000 and 0.2 );
\draw[white, fill=black, rotate around = { 90 : ( 6 + 6 * 0.500000 , 6 * 0.500000) } ] (6 + 6 * 0.500000 , 6 * 0.500000) ellipse ( 0.2 * 0.500000 and 0.2 );

\draw[rotate around = { 90 : ( 1 + 1.5 * 0.500000 , 1.5 * 0.500000) } ] (1 + 1.5 * 0.500000 , 1.5 * 0.500000) ellipse ( 0.2 * 0.500000 and 0.2 );
\draw[white, fill=black, rotate around = { 90 : ( 2 + 1.5 * 0.500000 , 1.5 * 0.500000) } ] (2 + 1.5 * 0.500000 , 1.5 * 0.500000) ellipse ( 0.2 * 0.500000 and 0.2 );
\draw[rotate around = { 90 : ( 3 + 1.5 * 0.500000 , 1.5 * 0.500000) } ] (3 + 1.5 * 0.500000 , 1.5 * 0.500000) ellipse ( 0.2 * 0.500000 and 0.2 );
\draw[rotate around = { 90 : ( 4 + 1.5 * 0.500000 , 1.5 * 0.500000) } ] (4 + 1.5 * 0.500000 , 1.5 * 0.500000) ellipse ( 0.2 * 0.500000 and 0.2 );
\draw[rotate around = { 90 : ( 5 + 1.5 * 0.500000 , 1.5 * 0.500000) } ] (5 + 1.5 * 0.500000 , 1.5 * 0.500000) ellipse ( 0.2 * 0.500000 and 0.2 );
\draw[rotate around = { 90 : ( 6 + 1.5 * 0.500000 , 1.5 * 0.500000) } ] (6 + 1.5 * 0.500000 , 1.5 * 0.500000) ellipse ( 0.2 * 0.500000 and 0.2 );

\draw[rotate around = { 90 : ( 1 + 2.5 * 0.500000 , 2.5 * 0.500000) } ] (1 + 2.5 * 0.500000 , 2.5 * 0.500000) ellipse ( 0.2 * 0.500000 and 0.2 );
\draw[rotate around = { 90 : ( 2 + 2.5 * 0.500000 , 2.5 * 0.500000) } ] (2 + 2.5 * 0.500000 , 2.5 * 0.500000) ellipse ( 0.2 * 0.500000 and 0.2 );
\draw[white, fill=black, rotate around = { 90 : ( 3 + 2.5 * 0.500000 , 2.5 * 0.500000) } ] (3 + 2.5 * 0.500000 , 2.5 * 0.500000) ellipse ( 0.2 * 0.500000 and 0.2 );
\draw[rotate around = { 90 : ( 4 + 2.5 * 0.500000 , 2.5 * 0.500000) } ] (4 + 2.5 * 0.500000 , 2.5 * 0.500000) ellipse ( 0.2 * 0.500000 and 0.2 );
\draw[rotate around = { 90 : ( 5 + 2.5 * 0.500000 , 2.5 * 0.500000) } ] (5 + 2.5 * 0.500000 , 2.5 * 0.500000) ellipse ( 0.2 * 0.500000 and 0.2 );
\draw[rotate around = { 90 : ( 6 + 2.5 * 0.500000 , 2.5 * 0.500000) } ] (6 + 2.5 * 0.500000 , 2.5 * 0.500000) ellipse ( 0.2 * 0.500000 and 0.2 );

\draw[rotate around = { 90 : ( 1 + 3.5 * 0.500000 , 3.5 * 0.500000) } ] (1 + 3.5 * 0.500000 , 3.5 * 0.500000) ellipse ( 0.2 * 0.500000 and 0.2 );
\draw[rotate around = { 90 : ( 2 + 3.5 * 0.500000 , 3.5 * 0.500000) } ] (2 + 3.5 * 0.500000 , 3.5 * 0.500000) ellipse ( 0.2 * 0.500000 and 0.2 );
\draw[white, fill=black, rotate around = { 90 : ( 3 + 3.5 * 0.500000 , 3.5 * 0.500000) } ] (3 + 3.5 * 0.500000 , 3.5 * 0.500000) ellipse ( 0.2 * 0.500000 and 0.2 );
\draw[rotate around = { 90 : ( 4 + 3.5 * 0.500000 , 3.5 * 0.500000) } ] (4 + 3.5 * 0.500000 , 3.5 * 0.500000) ellipse ( 0.2 * 0.500000 and 0.2 );
\draw[rotate around = { 90 : ( 5 + 3.5 * 0.500000 , 3.5 * 0.500000) } ] (5 + 3.5 * 0.500000 , 3.5 * 0.500000) ellipse ( 0.2 * 0.500000 and 0.2 );
\draw[rotate around = { 90 : ( 6 + 3.5 * 0.500000 , 3.5 * 0.500000) } ] (6 + 3.5 * 0.500000 , 3.5 * 0.500000) ellipse ( 0.2 * 0.500000 and 0.2 );

\draw[rotate around = { 90 : ( 1 + 4.5 * 0.500000 , 4.5 * 0.500000) } ] (1 + 4.5 * 0.500000 , 4.5 * 0.500000) ellipse ( 0.2 * 0.500000 and 0.2 );
\draw[white, fill=black, rotate around = { 90 : ( 2 + 4.5 * 0.500000 , 4.5 * 0.500000) } ] (2 + 4.5 * 0.500000 , 4.5 * 0.500000) ellipse ( 0.2 * 0.500000 and 0.2 );
\draw[rotate around = { 90 : ( 3 + 4.5 * 0.500000 , 4.5 * 0.500000) } ] (3 + 4.5 * 0.500000 , 4.5 * 0.500000) ellipse ( 0.2 * 0.500000 and 0.2 );
\draw[rotate around = { 90 : ( 4 + 4.5 * 0.500000 , 4.5 * 0.500000) } ] (4 + 4.5 * 0.500000 , 4.5 * 0.500000) ellipse ( 0.2 * 0.500000 and 0.2 );
\draw[rotate around = { 90 : ( 5 + 4.5 * 0.500000 , 4.5 * 0.500000) } ] (5 + 4.5 * 0.500000 , 4.5 * 0.500000) ellipse ( 0.2 * 0.500000 and 0.2 );
\draw[rotate around = { 90 : ( 6 + 4.5 * 0.500000 , 4.5 * 0.500000) } ] (6 + 4.5 * 0.500000 , 4.5 * 0.500000) ellipse ( 0.2 * 0.500000 and 0.2 );

\draw[rotate around = { 90 : ( 1 + 5.5 * 0.500000 , 5.5 * 0.500000) } ] (1 + 5.5 * 0.500000 , 5.5 * 0.500000) ellipse ( 0.2 * 0.500000 and 0.2 );
\draw[rotate around = { 90 : ( 2 + 5.5 * 0.500000 , 5.5 * 0.500000) } ] (2 + 5.5 * 0.500000 , 5.5 * 0.500000) ellipse ( 0.2 * 0.500000 and 0.2 );
\draw[rotate around = { 90 : ( 3 + 5.5 * 0.500000 , 5.5 * 0.500000) } ] (3 + 5.5 * 0.500000 , 5.5 * 0.500000) ellipse ( 0.2 * 0.500000 and 0.2 );
\draw[white, fill=black, rotate around = { 90 : ( 4 + 5.5 * 0.500000 , 5.5 * 0.500000) } ] (4 + 5.5 * 0.500000 , 5.5 * 0.500000) ellipse ( 0.2 * 0.500000 and 0.2 );
\draw[rotate around = { 90 : ( 5 + 5.5 * 0.500000 , 5.5 * 0.500000) } ] (5 + 5.5 * 0.500000 , 5.5 * 0.500000) ellipse ( 0.2 * 0.500000 and 0.2 );
\draw[rotate around = { 90 : ( 6 + 5.5 * 0.500000 , 5.5 * 0.500000) } ] (6 + 5.5 * 0.500000 , 5.5 * 0.500000) ellipse ( 0.2 * 0.500000 and 0.2 );

\end{tikzpicture}
};

\node[] at (3.5,0) {$N$ Fermionic sites};
\node[right] at (3.7,-1) {Jordan-Wigner transform};
\node[] at (3.5,-2) {$N$ Qubits};
\node[right] at (3.7,-3) {Auxiliary qubits + Quantum code };
\node[] at (3.5,-4) {$n > N $ Qubits};
 \draw[-latex] (3.5, -.4)--(3.5,-1.6);
  \draw[-latex] (3.5, -2.4)--(3.5,-3.6);
\node[] at (-3.5,0.5) {\textbf{(a)}};
\node[] at (-3.5,-1.5) {\textbf{(b)}};
\node[] at (-3.5,-3.5) {\textbf{(c)}};
\draw[-latex] (5.5,0)--(10.5,0)--(10.5,-4)--(5.5,-4);
\node[above, rotate=90] at (10.5,-2) {\textbf{AQM}};
\end{tikzpicture}
\caption{Visualizing an Auxiliary Qubit Mapping (\textbf{AQM}) as a concatenation of the Jordan-Wigner transform and a particular quantum code. The three layers represent the lattices of Fermions and qubits. We have highlighted the same three exchange terms on each lattice, so there transformation can be observed.  \textbf{(a)} The starting point: a fermionic lattice or two-dimensional embedding  of a Fermion system with $\ell_1 \times \ell_2$ modes. The three (local) interactions highlighted are brought via the Jordan-Wigner transform onto the (data) qubit layer. \textbf{(b)} The data qubit layer, in which two of the formally local interactions now assume a non-local form. To restore locality, we need to define a quantum code on the data qubits register and some auxiliary qubits, added to the next layer. \textbf{(c)} The final layer:  a composite system of $n$ qubits, where we have placed $n-N$ auxiliary qubits in between the data qubits. By the Auxiliary Qubit code, interactions that were non-local in the top layer can now be made local again. Note also that the interaction in the center of the lattice, which has involved many qubits in the middle layer, is now reduced to act on few qubits again by the Manhattan-distance property.   } \label{fig:layers}
\end{figure}
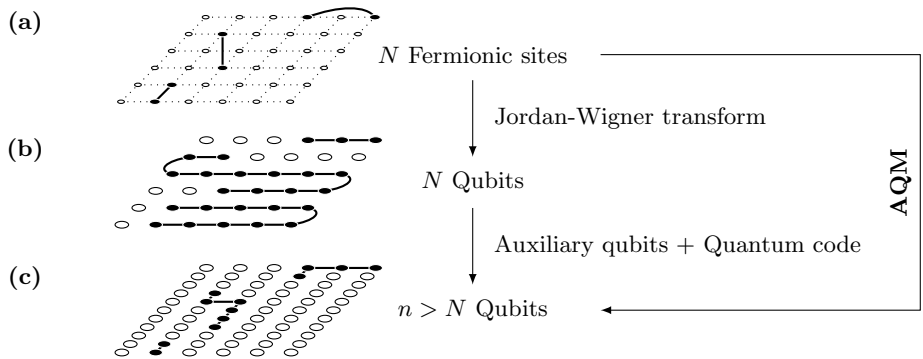

In this work, we introduce a new class of Fermion-to-qubit mappings, that are  two-dimensional generalizations of the Jordan-Wigner transform on a $\ell_1 \times \ell_2$ lattice of fermionic sites. The \textit{Auxiliary Qubit Mappings} (AQMs) are based on the (one-dimensional) Jordan-Wigner transform, concatenated with specific quantum (stabilizer) codes.  Stabilizer codes, which play an important role in quantum error correction, encode a logical basis of $2^N$ degrees  of freedom (here $N=\ell_1 \times \ell_2$) in a subspace of a larger system with $n>N$ qubits. The degrees of freedom left are constrained with so-called stabilizer conditions, which means there are $n-N$ (independent) qubit operators $\lbrace S_i \rbrace_i$ that stabilize this basis, i.e. in the logical subspace the expectation value of all stabilizers is one, $\langle S_i\rangle=1$. In our case, the logical basis encoded is the one of the Jordan Wigner transform, to which $r=n-N$ auxiliary qubits have been added and constrained. The entire procedure is illustrated in Figure \ref{fig:layers}, where the AQM  performs the transition from layer (a) to (c), effectively avoiding the non-local interactions on layer (b).   The codes used for AQMs are planar on the square lattice, and we devise a unitary quantum circuit that switches in between the layers (b) and (c).  This circuit has an algorithmic depth that scales with $\ell_1$, the length of one of the lattice sides.  There is no such operation  for mappings found in prior works, the Verstraete-Cirac transform and Superfast simulation. To compare them with the AQMs, we modify the VCT and BKSF, rendering them planar codes  with the Manhattan-distance property. The contributions of this paper are:
\begin{itemize}
\item   We introduce three types of Auxiliary Qubit Mappings, each requiring a different amount of auxiliary qubits. Our main result of this paper is the \textit{square lattice AQM}, which uses $2N-\ell_1$ qubits in total.  Note that in general, mappings with more auxiliary qubits will in some sense deal better with the second dimension, but  none of the mappings generalizing the Jordan-Wigner transform  has a total qubit number exceeding $2N$. However, one might be interested in using fewer auxiliary qubits: this can be the case for instance when simulating lattice models, where we would like to make the physical lattice as large as possible and `being on a fixed qubit budget' accept a trade-off between circuit depth and the number of auxiliary qubits. A qubit-economic version of this mapping would be the \textit{sparse AQM}, which introduces the parameter $\inter$ to regulate the trade-off. Furthermore, with adding only a few qubits we can already obtain a modified version of this mapping which has easy-to-prepare logical states and is called \textit{E-type AQM}.    A comprehensive list of all considered Fermion-to-qubit mappings, that allows us to compare their properties, is compiled into Table \ref{tab:results}.  For all Auxiliary Qubit Mappings, we provide the initialization circuits of $O(\ell_1)$ depth.

\item We demonstrate the Auxiliary Qubit Mappings on the Fermi-Hubbard model, decreasing its algorithmic depth from  being linear with the number of data qubits, $O(N)$, to  being constant, $O(1)$. This is an important step towards making its simulation scalable (at the expense of more qubits). Lattice models are in general not just interesting by themselves, but also test on how a Fermion-to-qubit mapping deals with the second dimension, i.e. the criteria mentioned in the introduction, in a minimal fashion. We explicitly show how the mappings transform the Fermi-Hubbard model into a model of local qubit interactions on the  lattice.

\item We compare our work, the Auxiliary Qubit Mappings, to  the Verstraete-Cirac transform  \cite{verstraete2005mapping} and the Superfast simulation \cite{bravyi2002fermionic} from the literature. As indicated above, we adjust the latter two slightly to make all three mappings comparable. Advantages and disadvantages of each mapping eventually lead us to conclude which of them to recommend for different situations.
\end{itemize}
While these contributions are covered in  Sections \ref{sec:three}, \ref{sec:four} and \ref{sec:five}, the rest of the paper is organized as follows: in Section \ref{sec:one}, we provide a more structured introduction to the layout of the quantum device and the established Fermion-to-qubit mappings. We discuss  criteria for a `good'  mapping in detail and that the Jordan-Wigner transform has deficits in those regards.  In Section \ref{sec:two}, we illustrate the effect of quantum codes, such as the ones that are the blueprint for the AQMs, on a given Hamiltonian. While the AQMs are an original idea, we cannot claim the same about their theoretical backbone: the foundations for Auxiliary Qubit codes are basically used in \cite{subacsi2016nonperturbative}, although there the stabilizer formalism was not employed. As a consequence, one auxiliary qubit would have to be added for each term in the Hamiltonian, which is a large overhead that can be avoided by using the underlying principle to define quantum codes.  We derive these codes from scratch in Appendix \ref{sec:B}. Some minor contributions are provided outside the main text of this work. In Appendix \ref{sec:A}, we study the class of tree-based mappings, to which the Bravyi-Kitaev transform belongs. The Bravyi-Kitaev transform itself does not do well with the square lattice, but we provide a general method to tailor and embed similar mappings to arbitrary two-dimensional setups. Appendix \ref{sec:mappings} is mostly providing details on the Verstraete-Cirac transform and Superfast simulation, but we also tackle some side issues by deriving the logical basis of both mappings.

\newcommand{\kopf}{.17\textwidth}
\begin{table}[h!]
\begin{tabular}{r|c c c c c c c}
 & \begin{tabular}{c} Jordan-Wigner \\ (S-pattern)\end{tabular}  & \begin{tabular}{c} Verstraete-Cirac \\ transform\end{tabular} & \begin{tabular}{c} Superfast \\ simulation \end{tabular}& \begin{tabular}{c}  Square \\  lattice AQM \end{tabular} & \begin{tabular}{c} E-type \\ AQM \end{tabular} &  \begin{tabular}{c} Sparse \\ AQM \end{tabular} & \\ \hline
Origin &\cite{wigner1928uber}&\cite{verstraete2005mapping}&\cite{bravyi2002fermionic} & [here] & [here] &  [here] \\ \\
  Aux. qubits  & $0$ &$\ell_1 \ell_2$&$\ell_1 \ell_2-\ell_1-\ell_2$&  $\ell_1 \ell_2-\ell_1 $ & $\ell_2$ & $(\ell_2-1) (\frac{\ell_1-1}{\inter}+1)  $ \\ \\
\begin{minipage}{\kopf} \raggedleft  String length (general) \\ $\,$ \end{minipage} & $O(\ell_1 \ell_2)$ &$O(2\ell_1+\ell_2)$&$O(2\ell_1+2\ell_2)$& $O(\ell_1+2\ell_2)$ & $O(2\ell_1+\ell_2)$ & $O(\ell_1+2\ell_2)$ \\  \begin{minipage}{\kopf} \raggedleft  Manhattan-distance property?  \\  $\,$ \end{minipage} & \ex & \tick & \tick & \tick & \ex & approximately \\ \begin{minipage}{\kopf} \raggedleft  String length (lattice) \\ $\,$ \end{minipage} & $O(\ell_1)$ &$O(1)$&$O(1)$& $O(1)$& $O(\ell_1)$  & $O(\inter)$ \\
 \begin{minipage}{\kopf} \raggedleft  Simulation time (lattice) \\ $\,$ \end{minipage} & $O(\ell_1  \ell_2)$ &$O(1)$&$O(1)$ & $O(1)$ & $O(\ell_1  \ell_2)$ & $O(\inter^2)$ \\
Restores locality? & \ex & \tick & \tick & \tick & \ex & approximately \\
 \end{tabular}
\caption{All Fermion-to-qubit mappings discussed in this work. We consider a $N=(\ell_1 \times \ell_2)$ square lattice block of fermionic modes, and compare the number of auxiliary qubits, or more generally the total number of qubits minus $N$. We also compare the scaling of the number of qubits involved in two types of Hamiltonians:  generic ones, in which we expect interactions between  every mode, and lattice models, with only nearest-neighbor interactions. For the former, we  also ask whether long-range interactions can be mapped to operators involving qubits along a direct path (Manhattan-distance property).  For the lattice models, we specify the expected algorithmic depth for simulating the entire Hamiltonian by e.g.~Trotterization and whether their locality is restored after the transformation.  Note that  $\inter$ is a parameter of the last mapping that can be chosen as some integer number: $1 \leq \inter \leq \ell_1-1$.  This parameter determines how well the Manhattan-distance property and locality is approximated.}\label{tab:results}
  \end{table}
\section{Preliminaries}
\label{sec:one}
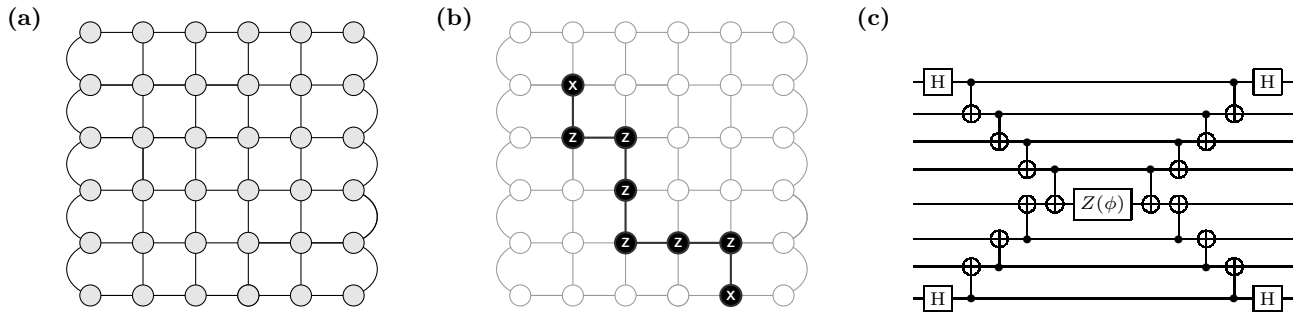
\begin{figure}
\begin{tikzpicture}[scale=0.7,baseline=0]
\node[] at (-.25, 6.25) {\textbf{(a)}};
\draw[](6,2).. controls(6.6,2.2)and (6.6,2.8).. (6,3);
\draw[](1,2).. controls(.4,2.2)and (.4,2.8).. (1,3);
\draw[](1,1).. controls(.4,1.2)and (.4,1.8).. (1,2);
\draw[](6,4).. controls(6.6,4.2)and (6.6,4.8).. (6,5);
\draw[](6,2).. controls(6.6,2.2)and (6.6,2.8).. (6,3);
\draw[](1,3).. controls(.4,3.2)and (.4,3.8).. (1,4);
\draw[](6,3).. controls(6.6,3.2)and (6.6,3.8).. (6,4);
\draw[](6,5).. controls(6.6,5.2)and (6.6,5.8).. (6,6);
\draw[](1,5).. controls(.4,5.2)and (.4,5.8).. (1,6);
\draw[](1,4).. controls(.4,4.2)and (.4,4.8).. (1,5);
\draw[](6,1).. controls(6.6,1.2)and (6.6,1.8).. (6,2);
\draw[](1,1)--(6,1);
\draw[](1,2)--(6,2);
\draw[](1,3)--(6,3);
\draw[](1,4)--(6,4);
\draw[](1,5)--(6,5);
\draw[](1,6)--(6,6);
\draw[](2,1)--(2,6);
\draw[](3,1)--(3,6);
\draw[](4,1)--(4,6);
\draw[](5,1)--(5,6);
\draw[](1,1)--(4,1);
\draw[] (4,4)--(1,4);
\draw[](1,5)--(4,5);
\draw[](4,2)--(6,2);
\draw[] (2,4)--(2,3);

\draw[fill=black!10] (1,6) circle[radius=0.2];
\draw[fill=black!10] (2,6) circle[radius=0.2];
\draw[fill=black!10] (3,6) circle[radius=0.2];
\draw[fill=black!10] (4,6) circle[radius=0.2];
\draw[fill=black!10] (5,6) circle[radius=0.2];
\draw[fill=black!10](6,6) circle[radius=0.2];

\draw[fill=black!10] (6,5) circle[radius=0.2];
\draw[fill=black!10] (6,4) circle[radius=0.2];
\draw[fill=black!10] (6,3) circle[radius=0.2];
\draw[fill=black!10] (2,2) circle[radius=0.2];
\draw[fill=black!10] (6,1) circle[radius=0.2];
\draw[fill=black!10] (1,3) circle[radius=0.2];
\draw[fill=black!10] (2,3) circle[radius=0.2];
\draw[fill=black!10] (3,3) circle[radius=0.2];
\draw[fill=black!10] (4,3) circle[radius=0.2];
\draw[fill=black!10] (1,1) circle[radius=0.2];
\draw[fill=black!10] (2,1) circle[radius=0.2];
\draw[fill=black!10](3,1) circle[radius=0.2];
\draw[fill=black!10] (4,1) circle[radius=0.2];
\draw[fill=black!10] (1,2) circle[radius=0.2];
\draw[fill=black!10](4,2) circle[radius=0.2];
\draw[fill=black!10] (5,2) circle[radius=0.2];
\draw[fill=black!10] (6,2) circle[radius=0.2];

\draw[fill=black!10] (1,4) circle[radius=0.2];
\draw[fill=black!10] (2,4) circle[radius=0.2];
\draw[fill=black!10] (3,4) circle[radius=0.2];
\draw[fill=black!10] (4,4) circle[radius=0.2];

\draw[fill=black!10](4,1) circle[radius=0.2];

\draw[fill=black!10] (1,5) circle[radius=0.2];

\draw[fill=black!10] (2,5) circle[radius=0.2];

\draw[fill=black!10] (3,5) circle[radius=0.2];

\draw[fill=black!10](4,5) circle[radius=0.2];
\draw[fill=black!10] (5,5) circle[radius=0.2];
\draw[fill=black!10] (5,4) circle[radius=0.2];
\draw[fill=black!10] (5,3) circle[radius=0.2];
\draw[fill=black!10] (3,2) circle[radius=0.2];
\draw[fill=black!10] (5,1) circle[radius=0.2];

\end{tikzpicture}  $\quad $ \begin{tikzpicture}[scale=0.7, baseline=0]
\textbf{\node[] at (-.25, 6.25) {\textbf{(b)}};}
\draw[\clre](6,2).. controls(6.6,2.2)and (6.6,2.8).. (6,3);
\draw[\clre](1,2).. controls(.4,2.2)and (.4,2.8).. (1,3);
\draw[\clre](1,1).. controls(.4,1.2)and (.4,1.8).. (1,2);
\draw[\clre](6,4).. controls(6.6,4.2)and (6.6,4.8).. (6,5);
\draw[\clre](6,2).. controls(6.6,2.2)and (6.6,2.8).. (6,3);
\draw[\clre](1,3).. controls(.4,3.2)and (.4,3.8).. (1,4);
\draw[\clre](6,3).. controls(6.6,3.2)and (6.6,3.8).. (6,4);
\draw[\clre](6,5).. controls(6.6,5.2)and (6.6,5.8).. (6,6);
\draw[\clre](1,5).. controls(.4,5.2)and (.4,5.8).. (1,6);
\draw[\clre](1,4).. controls(.4,4.2)and (.4,4.8).. (1,5);
\draw[\clre](6,1).. controls(6.6,1.2)and (6.6,1.8).. (6,2);
\draw[\clre](1,1)--(6,1);
\draw[\clre](1,2)--(6,2);
\draw[\clre](1,3)--(6,3);
\draw[\clre](1,4)--(6,4);
\draw[\clre](1,5)--(6,5);
\draw[\clre](1,6)--(6,6);
\draw[\clre](2,1)--(2,6);
\draw[\clre](3,1)--(3,6);
\draw[\clre](4,1)--(4,6);
\draw[\clre](5,1)--(5,6);
\draw[\clre](1,1)--(4,1);
\draw[\clre] (4,4)--(1,4);
\draw[\clre](1,5)--(4,5);
\draw[\clre](4,2)--(6,2);
\draw[\clre] (2,4)--(2,3);
\draw[\clrd,thick] (2,5)--(2,4)--(3,4)--(3,2)--(5,2)--(5,1);
\draw[\clre,fill=white] (1,6) circle[radius=0.2];
\draw[\clre,fill=white] (2,6) circle[radius=0.2];
\draw[\clre,fill=white] (3,6) circle[radius=0.2];
\draw[\clre,fill=white] (4,6) circle[radius=0.2];
\draw[\clre,fill=white] (5,6) circle[radius=0.2];
\draw[\clre,fill=white] (6,6) circle[radius=0.2];

\draw[\clre,fill=white] (6,5) circle[radius=0.2];
\draw[\clre,fill=white] (6,4) circle[radius=0.2];
\draw[\clre,fill=white] (6,3) circle[radius=0.2];
\draw[\clre,fill=white] (2,2) circle[radius=0.2];
\draw[\clre,fill=white] (6,1) circle[radius=0.2];
\draw[\clre,fill=white] (1,3) circle[radius=0.2];
\draw[\clre,fill=white] (2,3) circle[radius=0.2];
\draw[color=\clrb, fill=\clra, thick] (3,3) circle[radius=0.2];
\draw[\clre,fill=white] (4,3) circle[radius=0.2];
\draw[\clre,fill=white] (1,1) circle[radius=0.2];
\draw[\clre,fill=white] (2,1) circle[radius=0.2];
\draw[\clre,fill=white] (3,1) circle[radius=0.2];
\draw[\clre,fill=white] (4,1) circle[radius=0.2];
\draw[\clre,fill=white] (1,2) circle[radius=0.2];
\draw[color=\clrb, fill=\clra, thick] (4,2) circle[radius=0.2];
\draw[color=\clrb, fill=\clra, thick] (5,2) circle[radius=0.2];
\draw[\clre,fill=white] (6,2) circle[radius=0.2];

\draw[\clre,fill=white] (1,4) circle[radius=0.2];
\draw[color=\clrb, fill=\clra, thick] (2,4) circle[radius=0.2];
\draw[color=\clrb, fill=\clra, thick] (3,4) circle[radius=0.2];
\draw[\clre,fill=white] (4,4) circle[radius=0.2];

\draw[\clre,fill=white] (4,1) circle[radius=0.2];

\draw[\clre,fill=white] (1,5) circle[radius=0.2];

\draw[color=\clrb, fill=\clra, thick] (2,5) circle[radius=0.2];

\draw[\clre,fill=white] (3,5) circle[radius=0.2];

\draw[\clre,fill=white] (4,5) circle[radius=0.2];

\draw[\clre,fill=white] (5,5) circle[radius=0.2];
\draw[\clre,fill=white] (5,4) circle[radius=0.2];
\draw[\clre,fill=white] (5,3) circle[radius=0.2];
\draw[color=\clrb, fill=\clra, thick] (3,2) circle[radius=0.2];
\draw[color=\clrb, fill=\clra, thick] (5,1) circle[radius=0.2];
\node[\clrc] at (2,5) {\tiny $\boldsymbol{\mathsf{X}}$};
\node[\clrc] at (4,2) {\tiny $\boldsymbol{\mathsf{Z}}$};
\node[\clrc] at (3,2) {\tiny $\boldsymbol{\mathsf{Z}}$};
\node[\clrc] at (3,3) {\tiny $\boldsymbol{\mathsf{Z}}$};
\node[\clrc] at (3,4) {\tiny $\boldsymbol{\mathsf{Z}}$};
\node[\clrc] at (2,4) {\tiny $\boldsymbol{\mathsf{Z}}$};
\node[\clrc] at (5,2) {\tiny $\boldsymbol{\mathsf{Z}}$};
\node[\clrc] at (5,1) {\tiny $\boldsymbol{\mathsf{X}}$};
\end{tikzpicture} $\quad $\begin{tikzpicture}[scale=0.7, baseline=0]\node[] at (-.25, 6.25) {\textbf{(c)}}; \node[] at (4,3) {\scriptsize\Qcircuit @C=0.5em @R=0.5em {&		&\gate{\mathrm{H}}	&\ctrl{1}	&\qw	&\qw	&\qw		&\qw	&\qw	&\qw	&\qw	&\ctrl{1}	&\gate{\mathrm{H}}	&\qw	\\
&		&\qw	&\targ	&\ctrl{1}	&\qw	&\qw		&\qw	&\qw	&\qw	&\ctrl{1}	&\targ	&\qw	&\qw	\\
&		&\qw	&\qw	&\targ	&\ctrl{1}	&\qw		&\qw	&\qw	&\ctrl{1}	&\targ	&\qw	&\qw	&\qw	\\
&		&\qw	&\qw	&\qw	&\targ	&\ctrl{1}		&\qw	&\ctrl{1}	&\targ	&\qw	&\qw	&\qw	&\qw	\\
&		&\qw	&\qw	&\qw	&\targ	&\targ		&\gate{Z(\phi)}	&\targ	&\targ	&\qw	&\qw	&\qw	&\qw	\\
&		&\qw	&\qw	&\targ	&\ctrl{-1}	&\qw		&\qw	&\qw	&\ctrl{-1}	&\targ	&\qw	&\qw	&\qw	\\
&		&\qw	&\targ	&\ctrl{-1}	&\qw	&\qw		&\qw	&\qw	&\qw	&\ctrl{-1}	&\targ	&\qw	&\qw	\\
&		&\gate{\mathrm{H}}	&\ctrl{-1}	&\qw	&\qw	&\qw		&\qw	&\qw	&\qw	&\qw	&\ctrl{-1}	&\gate{\mathrm{H}}	&\qw	\\} };
	\end{tikzpicture}
\caption{Simulation of Pauli strings in a system with limited connectivity.  \textbf{(a)} Qubit connectivity graph: the vertices are qubits. Two-qubit gates can be performed only between qubits coupled by an edge. \textbf{(b)}   Simulating a Pauli string  on the quantum device: the qubits involved, and the edges along which entangling gates are performed,  are highlighted. Inscriptions $\mathsf{X}$, $\mathsf{Y}$ and $\mathsf{Z}$ indicate which Pauli operator acts on each qubit. \textbf{(c)} Simulating a Pauli string, here we simulate the propagator $\mathrm{exp}(i\, \phi\, X\otimes Z^{\otimes 6} \otimes X)$, where $\phi$ is an angle. The Pauli string could be the one in (b). In general, this circuit stores the parity information of the involved qubits on one of them, which is done by chains of $\textsc{CNot}$-gates. The inscriptions $\mathsf{X}$, $\mathsf{Z}$ and $\mathsf{Y}$ determine for each individual qubit whether it is in the Hadamard, computational or Y-basis in the process. Note that it does not play a role on which of the qubits the parity of the others is collected, but to optimize the simulation time, a qubit in the middle of the chain is chosen. On that qubit the phase rotation $Z(\phi)=\mathrm{exp}(i\, \phi \,Z )$ is performed, after which the chains are uncomputed.   }\label{fig:simulation}
\end{figure}
In this section, we describe the influence of Fermion-to-qubit mappings on the algorithmic depth of quantum simulation in a setup of square-lattice qubit-connectivity. In particular, we will discuss criteria which render mappings `good' in the sense that they allow for parallelization and low gate costs. For that purpose, we will  give a theoretical description of the qubit layout and sketch the simulation algorithms. Let us start however by stating the role of Fermion-to-qubit mappings for quantum simulation in general. We generally advise the reader familiar with the subject to skip ahead to Section \ref{sec:two}, and if necessary use the table of notations offered in Appendix \ref{sec:notations}.\\

The goal of quantum simulation is to approximate the ground state and the ground-state energy of a given Hamiltonian. When the Hamiltonian acts on a space of Fermions, a Fermion-to-qubit mapping serves as translator between the quantum system to be simulated and the qubit system inside the quantum computer. That not only entails a correspondence of basis states, but also a transformation of the Hamiltonian. The Hamiltonian after its transformation with the mapping, is henceforward acting on the qubits inside the quantum computer. We here consider the case where the qubit system underlies architectural constraints, that we want to abstract with the following model.

Our setup is a two-dimensional quantum device that we describe with a planar graph, where each of the $n$ vertices  is a qubit.   In this model, it is assumed that we can individually and simultaneously  perform Pauli-rotations on every single qubit. However,  entangling gates  can only be applied between two qubits that share an edge in the graph. We assume that we can perform two-qubit gates individually per edge, but qubits involved in one gate cannot be part in another  at the same time. Although we do not want to specify which kind of two-qubit gate is native to the quantum device, we want to assume that we can do $\textsc{CNot}$-gates in $O(1)$ time using only a few native gates. The full qubit connectivity graph  will furthermore be assumed to be a square lattice, so we can only perform entangling gates between qubits that are nearest neighbors,  see Figure \ref{fig:simulation}(a). Note that the individual connectivity graphs, that every Fermion-to-qubit mapping in this work comes with,  are  subgraphs of Figure \ref{fig:simulation}(a), such that every mapping can be embedded in the considered qubit system.
\subsection{Simulating a qubit Hamiltonian}
In order to elucidate the connection between the mapping and the depth and cost of the simulation algorithms, we need to understand these algorithms better.
Let us assume the Fermion-to-qubit mapping transforms a Hamiltonian into  the form  of Pauli strings, i.e.  the sum ${H}=\sum_{h}\, \Gamma^{h}\cdot\, h $, where $\lbrace \Gamma^{h} \rbrace$ are real coefficients associated to a Pauli string on $n$ qubits,
$h \in \lbrace X,\, Y,\, Z,\, \mathbb{I}  \rbrace^{\otimes n} $. Note that we will refer to the number of qubits, that a string $h$ acts on non-trivially, as (operator) weight and (string) length, interchangeably.  \\ Quantum simulation algorithms have  different ways to search for the ground state of $H$.
 Depending on which algorithm is used, the Pauli strings $h$ have to be either measured, or their propagator simulated (conditionally) \cite{kitaev1995quantum,cleve1998quantum}. With a propagator we mean the operator $\mathrm{exp}(i\, \phi \, h)$, where $\phi$ is an angle that typically is some function of $\Gamma^h$. Using $\textsc{CNot}$-gates, we simulate such a propagator with the gadget like in Figure \ref{fig:simulation}(c), where chains of these gates copy parity information across the lattice onto a single qubit, on which then a $Z$-rotation around the angle $\phi$ is performed and afterwards the \textsc{CNot}-chain is uncomputed. For \textit{quantum eigensolvers} \cite{mcclean2016theory}, this qubit will be measured instead. Often we need the rotation to be conditional on the state of another qubit, so conventionally the $Z$-rotation, $Z(\phi)=\mathrm{exp}(i\,\phi\,Z)$, is to be replaced  with  a controlled rotation, $ \mathbb{I}\otimes \ket{0}\!\!\bra{0} + Z(\phi)\otimes \ket{1}\!\!\bra{1}$ where the first qubit is the one that holds the parity information, and the second is the control, typically an auxiliary qubit of a phase estimation procedure. Alternatively, the quantum phase estimation algorithm can be adapted to include control qubits in the string, namely to simulate the propagator $\mathrm{exp}(-i\,\frac{\phi}{2} \, h\otimes Z) = \mathrm{exp}(-i\,\frac{\phi}{2} \, h) \otimes \ket{0}\!\!\bra{0} +\mathrm{exp}(i\,\frac{\phi}{2} \, h)\otimes \ket{1}\!\!\bra{1}$ instead.

For phase estimation-based algorithms, the propagator of the entire Hamiltonian, $\mathrm{exp}(iH \phi)$  needs to be simulated, which invokes the propagator of each string at least once (e.g. \cite{poulin2018quantum,babbush2018encoding}). Other algorithms invoke each string multiple times:  Trotterization \cite{suzuki1990fractal,suzuki1991general}  approximates the Hamiltonian propagator as repeating sequences of all string propagators $\mathrm{exp}(i\,\phi\, h)$, and in \textit{iterative phase estimation} \cite{aspuru2005simulated}, a repeated application of $\mathrm{exp}(i H \phi)$ increases the accuracy of the computed energy. In general $H$ does not even have to be a Hamiltonian: it could also be an operator that prepares a trial state with \textit{Givens rotations} \cite{kivlichan2018quantum} or implements a \textit{unitary coupled-cluster} operator \cite{bartlett1989alternative}. In any case, we will expect there to be a large number of strings in $H$ so we would like to apply the gadgets \ref{fig:simulation}(c) in parallel to keep the simulation shallow whenever possible. Let us coordinate the simulation of all those propagators by switching to layout diagrams like the one in  Figure \ref{fig:simulation}(b), instead of using circuit diagrams like in panel (c).  This gives us an idea of all the qubits involved and how they are coupled, but leaves out certain details about  for instance the specific simulation algorithm. Our ability to parallelize the simulation is determined by the Fermion-to-qubit mapping, in particular in the shape of the  strings that it outputs.  In regard of our connectivity setup \ref{fig:simulation}(c), we consider a Fermion-to-qubit mapping as good, if it outputs Hamiltonians ${H}$ with Pauli strings that are \textit{short}, \textit{continuous} and \textit{non-overlapping}. We will now explain these criteria: \\

\textit{short} - The length of a Pauli string is  the number of qubits that it acts on non-trivially. It was  recently  pointed out by Motzoi et al., that the gadget in Figure \ref{fig:simulation}(c) can be replaced with one that performs the same operation in a number of time steps scaling with the logarithm of the number of qubits involved, so at most $O(\log n)$  \cite{motzoi2017linear}. However, taking into account the limited qubit connectivity,  we have to stick to the gadget of Figure \ref{fig:simulation}(c), and expect a time scaling linear in the string length. As the number of time steps is interchangeably connected to the circuit depth, we have an interest in keeping the Pauli strings as short as possible. \\

\begin{figure}
\begin{align*}
\tiny \Qcircuit @C=0.5em @R=0.5em {	 \lstick{\omega_1} &\ctrl{5} 	&\qw 	 & \push{\omega_1} &\push{\qquad\qquad}&& \lstick{\omega_1}		&\targ	&\qw	&\qw	&\qw	&\ctrl{1}	&\qw	&\qw	&\qw	&\qw	&\qw	& \push{\omega_1+\omega_2} &\push{\qquad}&& \lstick{\omega_1}	&\qw	&\qw	&\qw	&\qw	&\ctrl{1}	&\qw	&\qw	&\qw	&\qw	&\qw	&\rstick{\omega_1}\\
\lstick{\omega_2}&\qw	&\qw	& \push{\omega_2} & \push{\qquad\qquad}&& \lstick{\omega_2}		&\ctrl{-1}	&\targ	&\qw	&\qw	&\targ	&\ctrl{1}	&\qw	&\qw	&\qw	&\qw	& \push{\omega_1+\omega_3} &\push{\qquad\qquad}&& \lstick{\omega_2}	&\qw	&\qw	&\qw	&\ctrl{1}	&\targ	&\ctrl{1}	&\qw	&\qw	&\qw	&\qw	&\rstick{\omega_1+\omega_2}\\
\lstick{\omega_3}&\qw	&\qw	& \push{\omega_3} &\push{\qquad}&& \lstick{\omega_3}		&\qw	&\ctrl{-1}	&\targ	&\qw	&\qw	&\targ	&\ctrl{1}	&\qw	&\qw	&\qw	& \push{\omega_1+\omega_4} &\push{\qquad}&& \lstick{\omega_3}	&\qw	&\qw	&\ctrl{1}	&\targ	&\qw	&\targ	&\ctrl{1}	&\qw	&\qw	&\qw	&\rstick{\omega_1+\omega_3}\\
\lstick{\omega_4}&\qw	&\qw	& \push{\omega_4} &\push{\qquad}&& \lstick{\omega_4}		&\qw	&\qw	&\ctrl{-1}	&\targ	&\qw	&\qw	&\targ	&\ctrl{1}	&\qw	&\qw	& \push{\omega_1+\omega_5} &\push{\qquad}&& \lstick{\omega_4}	&\qw	&\ctrl{1}	&\targ	&\qw	&\qw	&\qw	&\targ	&\ctrl{1}	&\qw	&\qw	&\rstick{\omega_1+\omega_4}\\
\lstick{\omega_5}&\qw	&\qw	& \push{\omega_5} &\push{\qquad}&& \lstick{\omega_5}		&\qw	&\qw	&\qw	&\ctrl{-1}	&\qw	&\qw	&\qw	&\targ	&\ctrl{1}	&\qw	& \push{\omega_1} &\push{\qquad}&& \lstick{\omega_5}	&\ctrl{1}	&\targ	&\qw	&\qw	&\qw	&\qw	&\qw	&\targ	&\ctrl{1}	&\qw	&\rstick{\omega_1+\omega_5}\\
\lstick{\omega_6}&\targ	&\qw	& \push{\omega_1+\omega_6} &\push{\qquad}&& \lstick{\omega_6}		&\qw	&\qw	&\qw	&\qw	&\qw	&\qw	&\qw	&\qw	&\targ	&\qw	& \push{\omega_1+\omega_6} &\push{\qquad}&& \lstick{\omega_6}	&\targ	&\qw	&\qw	&\qw	&\qw	&\qw	&\qw	&\qw	&\targ	&\qw &\rstick{\omega_1+\omega_6}	\\} \normalsize
\end{align*}
\caption{ Skipping several qubits in a \textsc{CNot}-chain. Here we consider the effect of the circuits on a computational basis state $(\bigotimes_i \ket{\omega_i})$, mapping it to a state $(\bigotimes_i \ket{\omega^\prime_i})$. We denote the qubit values $\omega_i$ and $\omega^\prime_i$ (mod 2) on the left and right side of each circuit. \textbf{Left:} The desired circuit, a \textsc{CNot}-gate that adds the parity from the first qubit to the last. For connectivity reasons, this gate is not possible: we can only connect adjacent qubits. \textbf{Center/Right:} Two circuits in which the  middle qubits are compensated for in order to entangle the first and last qubit.  To get rid of the effect on qubits 2 - 5, the gadgets have to be partially uncomputed, but in propagators like in Figure \ref{fig:simulation}(c), this is not necessary.}\label{fig:bridge}
\end{figure}
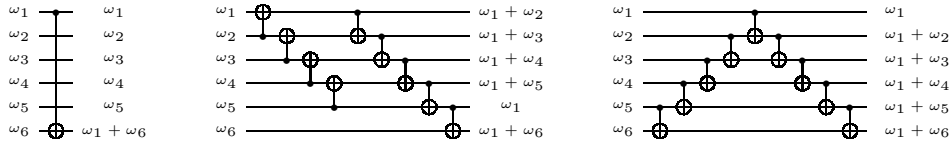

\textit{continuous} - In general, Pauli strings in  ${H}$ will not only act on nearest neighbors, this means we cannot connect the qubits involved along shared edges as it is done in Figure \ref{fig:simulation}(b). Connectivity problems are symptomatic for layouts like this, in which only nearest-neighbors are coupled.  Let us assume that two qubits need to be connected in  a gadget  like \ref{fig:simulation}(c), but they do not share an edge and the shortest path along edges encompasses a number of $m$ uninvolved qubits. In order to skip these qubits, $O(m)$ additional two-qubit gates and time steps are required.  In case the native two-qubit gates are either $i\textsc{Swap}$ or $\sqrt{\textsc{Swap}}$, the outer qubits can be connected by a  chain of  $\textsc{Swap}$  gates, which  costs  $2m$ native gates in the former case and $4m$ in the latter. For systems with native $\textsc{CNot}$-gates the formation  $\textsc{Swap}$  gates with three $\textsc{CNot}$s is unnecessarily expensive, so instead we amend gadgets like in Figure \ref{fig:simulation}(c) with a construction that includes the $m$ inner qubits in the $\textsc{CNot}$-chains, but compensates for their contribution.  We present two  versions of such a compensation circuit  in Figure \ref{fig:bridge}, where the left panel  shows us the gate that we would like to perform but cannot: we would like the configuration of the first qubit to be added to the last qubit by a non-local $\textsc{CNot}$-gate.   In the end, the circuits in the center and on the right achieve that task but render the  $m$ uninvolved qubits  useless until the circuit is uncomputed.  The additional cost in time and gates is $4m$, which means that it is cheaper to include a qubit in a string than to skip it. In conclusion, compensating or swapping of qubits is possible, but we would prefer  to avoid the additional cost and rather deal with continuous strings. \\

\textit{non-overlapping} - The overlap of two (or more)  Pauli strings is the number of qubits  in the intersection the sets of qubits each string acts on.   Two Pauli strings that are both acting non-trivially on a common subset of qubits are hard to simulate in parallel, as these qubits get parity information attached to them like in  Figure \ref{fig:simulation}(c). Thus if the qubits are not located as the first  in a chain, this parity would have to be corrected for. Avoiding any of the additional cost, we ideally would like our mapping to transform every commuting pairs of fermionic operators  into non-overlapping Pauli strings.

\subsection{Fermion-to-qubit mappings based on linear transforms}
\label{subsec:conventional}
Here we will review  Fermion-to-qubit mappings that are based on linear transforms of binary vectors, as these serve as foundation for the Auxiliary Qubit Mappings later.  Let us start this section at the fermionic side of the problem, that we seek to map onto qubits.

In general, we search for the ground state of a system of Fermions, that live on $N$  modes, governed by a Hamiltonian. It is convenient for us to formulate this problem in the language of second quantization, where we consider Fermion creation $c_j^\dagger$ and annihilation operators $c_j$ of modes $j \in [N]$, with $[N]$  just being a shorthand for the set of integers $\lbrace 1, \, 2, \, ... , \, N     \rbrace$ and $[0]=\emptyset$, a notation that will be used continuously throughout this work. The fermionic operators $c_j^\dagger, c_j^{\phantom{\dagger}}$ create and annihilate particles on the  $j$-th fermionic mode, and the antisymmetricity of the wave functions is built into these operators by their anticommutation relations
\begin{align}
\label{eq:aannttii}
[c^{\phantom{\dagger}}_i, \, c^{\phantom{\dagger}}_j]_+ = 0 \, , \qquad  [c^{{\dagger}}_i, \, c^{{\dagger}}_j]_+ = 0 \,, \qquad  [c^{\phantom{\dagger}}_i, \, c^{{\dagger}}_j]_+ = \delta_{ij} \, .
\end{align}
A Hamiltonian can now be formulated by means of these terms, where they will typically appear in pairs. A form typical for a molecular Hamiltonian is
\begin{align}
\label{eq:secquant}
\sum_{i,\, j \,\in\, [N]}\, \mathrm{h}_{ij}\,c^{\dagger}_ic^{\phantom{\dagger}}_j \;+\;  \sum_{i, \,j,\, k, \,l \,\in \, [N]} \,\mathrm{h}_{ijkl} \,c^{\dagger}_ic^{{\dagger}}_jc^{\phantom{\dagger}}_kc^{\phantom{\dagger}}_l\, ,
\end{align}
where $\lbrace \mathrm{h}_{ijkl} \rbrace$ and $ \lbrace \mathrm{h}_{ij} \rbrace$ are complex coefficients dictated by the problem, but they always take values such that the Hamiltonian is Hermitian. Hamiltonians in second quantization have notoriously no regard for the particle number of the system, but rather map the entire fermionic Fock space to itself, where physical Hamiltonians like \eqref{eq:secquant} conserve subspaces with a fixed particle number. A basis encoding the ($N$-mode) Fock space of Fermions can be parametrized by $N$-fold binary vectors $\bbs{\nu}=(\nu_1, \, \nu_2, \, ... ,\, \nu_N)^\top \in \zetto{N}$, where each component $\nu_i$ is an element of $\mathbb{Z}_2=\lbrace 0, 1\rbrace$. Conventionally, the correspondence is
\begin{align}
\label{eq:modes}
\bbs{\nu}=\left( \nu_1, \, \nu_2, \, \dots,\, \nu_N \right)^{\top} \quad\longleftrightarrow \quad \left[\prod_{j\in [N]} (c_j^\dagger)^{\nu_j}\right] \ket{\Theta}\,  ,
\end{align}
where $\ket{\Theta}$ is the fermionic vacuum state and we define $(c_j^\dagger)^0=1$, such that the component $\nu_j$ indicates  the occupation of the $j$-th fermionic mode.  At this point we need to raise awareness of a subtle but important point of that parametrization: we have imposed a certain labeling of Fermion modes that implies an order in the basis states. This order, called canonical order,  needs to be chosen carefully since it is as crucial for the transformed operators as the Fermion-to-qubit mapping.  \\

To encode all $2^N$ basis elements \eqref{eq:modes} into quantum states, one needs a minimum of $N$ qubits. We will now describe how basis states and operators are mapped to the qubit system, but first set-up some further notation. \\
A single qubit will be assigned a label $j$, which will appear as a subscript on its states, e.g. on its basis configurations  $\ket{0}_j , \ket{1}_j$. The label will also be carried by single qubit operators to indicate on which qubit they act on, e.g. $Z_j$ acts on qubit $j$. In Pauli strings, identities will be omitted, so e.g.~$X\otimes \mathbb{I} \otimes X = X_1\otimes X_3 = (\bigotimes_{i\in \lbrace 1, \, 3 \rbrace} X_i)$,  but an identity over all qubits will generally be  denoted by $\mathbb{I}$. Multi-qubit states and operators on the other hand will be branded with a subset of $[n]$ as subscript, for the qubits they have support on.  Let us consider an example: as mentioned before, we only need $N$ qubits to encode the entire Fock space, but will usually have more, i.e. $n\geq N$. In the mappings we consider the first $N$ qubits already describe the system and the other qubits are just there for auxiliary purposes. Hence we will group qubits $1$ to $N$  into a  set referred to as the data register. $N$-qubit states  $\ket{\varphi}$ in that register as well as operators $U$ acting on it will be denoted by the index $\data = [N]$, e.g.  $\ket{\varphi}_{\data}$ and $U_{\data}$.  This notation will be used throughout this work, as $n>N$, but for the moment we stick to the case of $n=N$. First, the bases of Fermions and qubits need to be matched. As a counterpart to \eqref{eq:modes}, the qubit basis can be parametrized by binary vectors $\bbs{\omega} \in \zetto{N}$, where the components $\omega_j$ indicate the quantum state of the $j$-th qubit in a product state. The correspondence is
\begin{align}
\label{eq:qubits}
\bbs{\omega}=\left( \omega_1, \, \omega_2, \, \dots,\, \omega_N \right)^{\top} \quad\longleftrightarrow \quad\ket{\bbs{\omega}}_{\data}\;=\;\bigotimes_{j\in[N]} \ket{\omega_j}_j \, .
\end{align}
The set of states $\lbrace \ket{\bbs{\omega}} \rbrace_{\bbs{\omega}}$ constitutes the computational basis on $N$ qubits, and an arbitrary state in that basis can be defined as: \begin{align}
 \label{eq:genericstate}
 \ket{\varphi}_{\data} \;=\;\sum_{\bbs{\omega} \in \zetto{N}} a_{\bbs{\omega}} \ket{\bbs{\omega}}_{\data} \, ,
 \end{align}
 where $a_{\bbs{\omega}} $ are complex coefficients that normalize the state $\sum_{\bbs{\omega}} \left| a_{\bbs{\omega}}  \right|^2 =1$.\\ A linear Fermion-to-qubit mapping now implies a one-to-one correspondence between all possible basis-defining vectors $\bbs{\nu} \leftrightarrow \bbs{\omega}$, that is done by multiplication with the  invertible binary $(N\times N)$-matrix $A$, such that \cite{seeley2012bravyi,tranter2015bravyi}: \\

\begin{align}
\label{eq:lineartransform}
\bbs{\omega}\;=\;A\, \bbs{\nu} \moto,\qquad \bbs{\nu}\;=\;A^{-1}\bbs{\omega} \moto \qquad \text{with} \qquad A^{-1}A \moto \;=\; \mathbb{I} \, ,
\end{align} where $\mathbb{I}$ is the $(N \times N)$ identity matrix.
Thus, by the transform \eqref{eq:lineartransform}, we have related the basis of Fermions \eqref{eq:modes} to the qubit basis \eqref{eq:qubits}. We will now show how this transform translates the Hamiltonian \eqref{eq:secquant}. Mimicking the effect of the Fermion operators $c^{\dagger}_j, \, c^{\phantom{\dagger}}_j$ on the basis states, we find \cite{steudtner2018fermion}

\begin{align}
c^{\dagger}_j\;\hat{=}\;\frac{1}{2}\left(\bigotimes_{k\in {U}(j)} X_k\right)\left(\mathbb{I}+\bigotimes_{l\in {F}(j)} Z_l\right)  \left(\bigotimes_{m\in {P}(j)} Z_m \right) \, ,  \notag \\ \label{eq:singleop}
c^{\;}_j\;\hat{=}\;\frac{1}{2}\left(\bigotimes_{k\in {U}(j)} X_k\right)\left(\mathbb{I}-\bigotimes_{l\in {F}(j)} Z_l\right)\left(  \bigotimes_{m\in {P}(j)} Z_m \right) \, ,
\end{align}
where the hatted equal signs $\hat{=}$ denote the correspondence between operators on the Fermion and qubit space. The relations \eqref{eq:singleop} feature the generalized update, flip and parity sets of modes $j \in [N]$: $U(j)$, $F(j)$ and $P(j)$ (in a notation slightly different from \cite{seeley2012bravyi}). These are sets of integers, subsets of $[n]$ to be exact. The sets $F(j)$ and $P(j)$ are made up by the column indices of all `1'-entries, in the $j$-th row of the matrices $A$ and $R A \moto$, where $R$ is the lower triangular matrix,
\begin{align}
\label{eq:triangular}
 R\;=\; \scriptsize \left[\begin{matrix} 0 \\
1 & 0 \\
1& 1 & 0\\
\vdots & \vdots	& &  \ddots	\\
 1&1 & \cdots & 1 & 0
	 \end{matrix} \right] \, .
\end{align}
The update sets $U(j)$ are comprised of all row numbers of `1'-entries in the $j$-th column of  $A^{-1}$. Operators in \eqref{eq:secquant} will  have to be transformed into Pauli strings according to \eqref{eq:singleop}.

\subsection{S-pattern Jordan-Wigner transform}
Based on the insights of the previous sections, we will now review what is probably the standard Fermion-to-qubit mapping \cite{wigner1928uber}.
In case of the Jordan-Wigner transform, the transformation matrix  $A$ can be regarded as the identity: $A=A^{-1}=\mathbb{I}$.
From \eqref{eq:singleop}, we derive the number operators
\begin{align}
\label{eq:number}
c^{\dagger}_jc^{\phantom{\dagger}}_j\; \hat{=}\; \frac{1}{2} \left(\mathbb{I}-Z_j \right) \,
\end{align}
 and hopping terms (for $i<j$)
\begin{align}
\mathrm{h}_{ij} \,c^{\dagger}_ic^{\phantom{\dagger}}_j +  (\mathrm{h}_{ij})^* \,c^{\dagger}_jc^{\phantom{\dagger}}_i \quad \hat{=}\quad & \frac{1}{2}\, \mathrm{Re}(\mathrm{h}_{ij}) \,\left(\bigotimes_{k=i+1}^{j-1} Z_k \right) \left(  X_i\otimes X_j +   Y_i\otimes Y_j  \right)\notag \\  +\, & \frac{1}{2} \, \mathrm{Im}(\mathrm{h}_{ij}) \, \left(\bigotimes_{k=i+1}^{j-1} Z_k \right) \left( Y_i\otimes X_j  - X_i\otimes Y_j  \right) \, . \label{eq:hopping}
\end{align}

While the number operator is transformed into just a constant term and a term that acts on one qubit only, the hopping terms are transformed into a string that exhibits long substrings of $Z$-operators, $(\bigotimes_{k=i+1}^{j-1} Z_k)$, sometimes called parity (sub-)strings.
  The right-hand side of \eqref{eq:hopping}, which describes an interaction of the fermionic modes $i$ and $j$, translates into several strings with $X$- and $Y$-operators on the corresponding qubits of $i$ and $j$, and all qubits of indices $k$, with $i<k<j$, are part of the parity substring. Although the parity string does us the service of connecting the qubits $i$ and $j$ in that way, it is also the reason that Pauli strings produced by the Jordan-Wigner transform are of length $O(N)$.\\
While the nature of our problem determines the Hamiltonian coefficients (such as $\mathrm{h}_{ij}$) with respect to the fermionic wave functions, it is up to us to  label each fermionic mode such that we minimize the appearance of long Pauli strings in $H$. While problems that are intrinsically one-dimensional can be mapped to local Hamiltonians, long strings can generally not be avoided for systems in higher spatial dimensions.

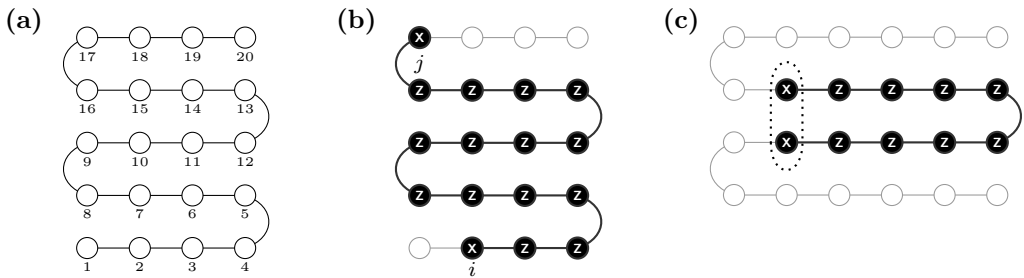
\begin{figure}
 \tiny
 \begin{tabular}{ccc}
\begin{tikzpicture}[scale=0.7,baseline=0]
\node[] at (-0.2,4.3) {\normalsize\textbf{(a)}};
\draw(4,0).. controls(4.6,.2)and (4.6,.8).. (4,1);
\draw(1,1).. controls(.4,1.2)and (.4,1.8).. (1,2);
\draw(4,2).. controls(4.6,2.2)and (4.6,2.8).. (4,3);
\draw(1,3).. controls(.4,3.2)and (.4,3.8).. (1,4);
\draw(1,0) -- (4,0);
\draw(1,1) -- (4,1);
\draw(1,2) -- (4,2);
\draw(1,3) -- (4,3);
\draw(1,4) -- (4,4);
\draw[fill=white] (1,0) circle[radius=0.2];
\draw[fill=white] (2,0) circle[radius=0.2];
\draw[fill=white](3,0) circle[radius=0.2];
\draw[fill=white] (4,0) circle[radius=0.2];
\draw[fill=white] (1,1) circle[radius=0.2];
\draw[fill=white] (2,1) circle[radius=0.2];
\draw[fill=white] (3,1) circle[radius=0.2];
\draw[fill=white] (4,1) circle[radius=0.2];
\draw[fill=white] (1,2) circle[radius=0.2];
\draw[fill=white] (2,2) circle[radius=0.2];
\draw[fill=white] (3,2) circle[radius=0.2];
\draw[fill=white] (4,2) circle[radius=0.2];
\draw[fill=white] (1,3) circle[radius=0.2];
\draw[fill=white] (2,3) circle[radius=0.2];
\draw[fill=white] (3,3) circle[radius=0.2];
\draw[fill=white] (4,3) circle[radius=0.2];
\draw[fill=white] (1,4) circle[radius=0.2];
\draw[fill=white] (2,4) circle[radius=0.2];
\draw[fill=white] (3,4) circle[radius=0.2];
\draw[fill=white] (4,4) circle[radius=0.2];
\node[below] at (1,-.15) { 1} ;
\node[below] at (2,-.15) {2} ;
\node[below] at (3,-.15) {3};
\node[below] at (4,-.15) {4} ;
\node[below] at (4,.85) { 5} ;
\node[below] at (3,.85) { 6} ;
\node[below] at (2,.85) {7} ;
\node[below] at (1,.85) {8};
\node[below] at (1,1.85) { 9} ;
\node[below] at (2,1.85) {10} ;
\node[below] at (3,1.85) {11};
\node[below] at (4,1.85) {12} ;
\node[below] at (4,2.85) { 13} ;
\node[below] at (3,2.85) { 14} ;
\node[below] at (2,2.85) {15} ;
\node[below] at (1,2.85) {16};
\node[below] at (1,3.85) { 17} ;
\node[below] at (2,3.85) {18} ;
\node[below] at (3,3.85) {19};
\node[below] at (4,3.85) {20} ;
\end{tikzpicture} $\qquad$& \tiny\begin{tikzpicture}[scale=0.7,baseline=0]
\node[] at (-0.2,4.3) {\normalsize\textbf{(b)}};
\draw[color=\clre](1,0) -- (4,0);
\draw[color=\clre](1,1) -- (4,1);
\draw[color=\clre](1,2) -- (4,2);
\draw[color=\clre](1,3) -- (4,3);
\draw[color=\clre](1,4) -- (4,4);
\draw[\clrd,thick] (2,0)--(4,0);
\draw[\clrd,thick](1,1) -- (4,1);
\draw[\clrd,thick](1,2) -- (4,2);
\draw[\clrd,thick](1,3) -- (4,3);
\draw[\clrd,thick](4,0).. controls(4.6,.2)and (4.6,.8).. (4,1);
\draw[\clrd,thick](1,1).. controls(.4,1.2)and (.4,1.8).. (1,2);
\draw[\clrd,thick](4,2).. controls(4.6,2.2)and (4.6,2.8).. (4,3);
\draw[\clrd,thick](1,3).. controls(.4,3.2)and (.4,3.8).. (1,4);

\draw[color=\clre,fill=white] (1,0) circle[radius=0.2];
\draw[color=\clrb, fill=\clra, thick] (2,0) circle[radius=0.2];
\draw[color=\clrb, fill=\clra, thick](3,0) circle[radius=0.2];
\draw[color=\clrb, fill=\clra, thick] (4,0) circle[radius=0.2];
\draw[color=\clrb, fill=\clra, thick] (1,1) circle[radius=0.2];
\draw[color=\clrb, fill=\clra, thick] (2,1) circle[radius=0.2];
\draw[color=\clrb, fill=\clra, thick] (3,1) circle[radius=0.2];
\draw[color=\clrb, fill=\clra, thick] (4,1) circle[radius=0.2];
\draw[color=\clrb, fill=\clra, thick] (1,2) circle[radius=0.2];
\draw[color=\clrb, fill=\clra, thick] (2,2) circle[radius=0.2];
\draw[color=\clrb, fill=\clra, thick] (3,2) circle[radius=0.2];
\draw[color=\clrb, fill=\clra, thick] (4,2) circle[radius=0.2];
\draw[color=\clrb, fill=\clra, thick] (1,3) circle[radius=0.2];
\draw[color=\clrb, fill=\clra, thick] (2,3) circle[radius=0.2];
\draw[color=\clrb, fill=\clra, thick] (3,3) circle[radius=0.2];
\draw[color=\clrb, fill=\clra, thick] (4,3) circle[radius=0.2];
\draw[color=\clrb, fill=\clra, thick] (1,4) circle[radius=0.2];
\draw[color=\clre,  fill=white] (2,4) circle[radius=0.2];
\draw[color=\clre,  fill=white] (3,4) circle[radius=0.2];
\draw[color=\clre,  fill=white] (4,4) circle[radius=0.2];
\node[\clrc] at (2,0) {\tiny $\boldsymbol{\mathsf{X}}$} ;
\node[\clrc] at (3,0) {\tiny $\boldsymbol{\mathsf{Z}}$};
\node[\clrc] at (4,0) {\tiny $\boldsymbol{\mathsf{Z}}$} ;
\node[\clrc] at (4,1) {\tiny $\boldsymbol{\mathsf{Z}}$} ;
\node[\clrc] at (3,1) {\tiny $\boldsymbol{\mathsf{Z}}$} ;
\node[\clrc] at (2,1) {\tiny $\boldsymbol{\mathsf{Z}}$} ;
\node[\clrc] at (1,1) {\tiny $\boldsymbol{\mathsf{Z}}$};
\node[\clrc] at (1,2) {\tiny $\boldsymbol{\mathsf{Z}}$} ;
\node[\clrc] at (2,2) {\tiny $\boldsymbol{\mathsf{Z}}$} ;
\node[\clrc] at (3,2) {\tiny $\boldsymbol{\mathsf{Z}}$};
\node[\clrc] at (4,2) {\tiny $\boldsymbol{\mathsf{Z}}$} ;
\node[\clrc] at (4,3) {\tiny $\boldsymbol{\mathsf{Z}}$} ;
\node[\clrc] at (3,3) {\tiny $\boldsymbol{\mathsf{Z}}$} ;
\node[\clrc] at (2,3) {\tiny $\boldsymbol{\mathsf{Z}}$} ;
\node[\clrc] at (1,3) {\tiny $\boldsymbol{\mathsf{Z}}$};
\node[\clrc] at (1,4) {\tiny $\boldsymbol{\mathsf{X}}$} ;
\node[below] at (2,-.15) {\footnotesize $i$} ;
\node[below] at (1,3.80) {\footnotesize $j$} ;
\end{tikzpicture}$\qquad$ & \begin{tikzpicture}[scale=0.7,baseline=-40]
\draw[\clrd, thick] (4,0).. controls(4.6,0.2)and (4.6,0.8).. (4,1);
\draw[\clre](-1,1).. controls(-1.6,1.2)and (-1.6,1.8).. (-1,2);
\draw[\clre](-1,0).. controls(-1.6,-0.2)and (-1.6,-.8).. (-1,-1);
\node[] at (-2,2.3) {\normalsize\textbf{(c)}};
\draw[color=\clre] (4 ,1)--(-1.,1);
\draw[color=\clre] (4 ,0)--(-1.,0);
\draw[color=\clre] (4 ,2)--(-1.,2);
\draw[color=\clre] (4 ,-1)--(-1.,-1);

\draw[\clrd,thick, ](0,1)--(4,1);
\draw[color=\clrd, thick](4,0)--(0,0);
\draw[color=\clrb, fill=\clra, thick] (0,0) circle[radius=0.2];
\node[\clrc] at (0,0) {\tiny $\boldsymbol{\mathsf{X}}$};
\draw[color=\clrb, fill=\clra, thick] (1,0) circle[radius=0.2];
\node[\clrc] at (1,0) {\tiny $\boldsymbol{\mathsf{Z}}$};
\draw[color=\clrb, fill=\clra, thick] (2,0) circle[radius=0.2];
\node[\clrc] at (2,0) {\tiny $\boldsymbol{\mathsf{Z}}$};
\draw[color=\clrb, fill=\clra, thick] (3,0) circle[radius=0.2];
\node[\clrc] at (3,0) {\tiny $\boldsymbol{\mathsf{Z}}$};
\draw[color=\clrb, fill=\clra, thick] (4,0) circle[radius=0.2];
\node[\clrc] at (4,0) {\tiny $\boldsymbol{\mathsf{Z}}$};
\draw[color=\clrb, fill=\clra, thick] (0,1) circle[radius=0.2];
\node[\clrc] at (0,1) {\tiny $\boldsymbol{\mathsf{X}}$};
\draw[color=\clrb, fill=\clra, thick] (1,1) circle[radius=0.2];
\node[\clrc] at (1,1) {\tiny $\boldsymbol{\mathsf{Z}}$};
\draw[color=\clrb, fill=\clra, thick] (2,1) circle[radius=0.2];
\node[\clrc] at (2,1) {\tiny $\boldsymbol{\mathsf{Z}}$};
\draw[color=\clrb, fill=\clra, thick] (3,1) circle[radius=0.2];
\node[\clrc] at (3,1) {\tiny $\boldsymbol{\mathsf{Z}}$};
\draw[color=\clrb, fill=\clra, thick] (4,1) circle[radius=0.2];
\node[\clrc] at (4,1) {\tiny $\boldsymbol{\mathsf{Z}}$};

\draw[color=\clre,fill=white] (-1,-1) circle[radius=0.2];
\draw[color=\clre,fill=white] (-1,0) circle[radius=0.2];
\draw[color=\clre,fill=white] (-1,1) circle[radius=0.2];
\draw[color=\clre,fill=white] (-1,2) circle[radius=0.2];

\draw[color=\clre,fill=white] (0,2) circle[radius=0.2];

\draw[color=\clre,fill=white] (1,2) circle[radius=0.2];

\draw[color=\clre,fill=white] (2,2) circle[radius=0.2];

\draw[color=\clre,fill=white] (3,2) circle[radius=0.2];

\draw[color=\clre,fill=white] (4,2) circle[radius=0.2];

\draw[color=\clre,fill=white] (0,-1) circle[radius=0.2];

\draw[color=\clre,fill=white] (1,-1) circle[radius=0.2];

\draw[color=\clre,fill=white] (2,-1) circle[radius=0.2];

\draw[color=\clre,fill=white] (3,-1) circle[radius=0.2];

\draw[color=\clre,fill=white] (4,-1) circle[radius=0.2];

\draw[thick, dotted] (-0.3,0)..controls(-.3,-0.7)and(.3,-0.7)..(0.3,0)--(0.3,1)..controls(.3,1.7)and(-.3,1.7)..(-0.3,1)--cycle;
\end{tikzpicture} $\qquad$
\end{tabular}
\caption{\textbf{(a)} The connectivity graph for the S-pattern Jordan-Wigner transform.  \textbf{(b)} Simulating a Pauli string $(X_i\otimes Z_{i+1} \otimes \dots \otimes Z_{j-1} \otimes X_j )$, that can be considered half of a hopping term. \textbf{(c)} Simulation of a Pauli string associated with a fermionic hopping between the two encircled qubits (dotted line). The hopping is in the vertical direction (diagonal to the S-pattern) which unfortunately involves gates on all qubits on the S-pattern between the two qubits.    }
\label{fig:Stype}
\end{figure}

The question is  how to incorporate the Jordan-Wigner transform into the square lattice layout. There is a natural solution: given a $N=(\ell_1 \times \ell_2)$-matrix of qubits, we need  to use only $N-1$ edges to connect them in canonical order like beads on a string, see Figure \ref{fig:Stype}(a). Due to the windings of the pattern on the block boundaries, we will refer to this particular way of using the Jordan-Wigner transform on a square lattice as  \textit{S-pattern Jordan-Wigner transform}. Let us now describe its properties in order to assert how good a mapping it is.  The mapping produces strings that are  \textit{continuous}: although arbitrary terms (like $c^{\dagger}_ic^{{\dagger}}_jc^{\phantom{\dagger}}_kc^{\phantom{\dagger}}_l$) will in general not be transformed into continuous Pauli strings,  creation/annihilation operator pairs  $c^{\dagger}_ic^{\phantom{\dagger}}_j$  will. Unfortunately the resulting Pauli-strings are neither \textit{short} nor  \textit{non-overlapping}. As the  parity strings encompass all the qubits in between $i$ and $j$, the string can  even span several rows, see Figure \ref{fig:Stype}(b). This leads not just to a high gate count and algorithmic depth, but also occupies  a large portion  of qubits at once, effectively hindering parallelization.

Let us consider an illustrative example: if we want our quantum device to simulate a two-dimensional lattice of sites with fermionic occupation and nearest-neighbor hopping, we encounter two kinds of terms. Short ones, where the exchange between nearest-neighbors $c^{\dagger}_ic^{\phantom{\dagger}}_{i+1}+\text{h.c.}$ yields the Pauli strings $(X_i\otimes X_{i+1}+Y_{i}\otimes Y_{i+1})/2$, and long ones, as the nearest-neighbor hoppings in the vertical direction will result in strings that can be seen in Figure \ref{fig:Stype}(c). Although these are nearest-neighbor interactions, they use all qubits around the winding linking the two rows, so all vertical hopping terms between two sites in the same two rows will overlap.  The S-pattern Jordan-Wigner transform thus has the property to transform operators, that are geometrically local in second quantization into non-local Pauli strings on the lattice.  In Section \ref{sec:four}, we will learn that it is those vertical hopping terms, that prevent us from simulating lattice models efficiently.

The verdict for the S-pattern Jordan-Wigner transform is that it is not good in the sense of our criteria, but good enough to serve as a foundation for better mappings.
 In the following, we will introduce mappings modifying the Jordan-Winger transform in using quantum codes to cancel  non-local parity strings, which will make the resulting strings short and non-overlapping. This will lead to a certain overhead in auxiliary qubits, placed along with the  original $(\ell_1\times \ell_2)$-block of data qubits on a square lattice. In contrast to the S-pattern Jordan-Wigner transform, the mappings to follow embrace the second dimension as a useful tool.\\

 Note that there are other alternatives to the Jordan-Wigner transform. The Bravyi-Kitaev transform \cite{bravyi2002fermionic,seeley2012bravyi,tranter2015bravyi, havlivcek2017operator} is known to produce Pauli strings of weight $O(\log N)$ instead of $O(N)$. For $N>16$ it can however be rather difficult to embed the mapping into a square lattice such that it outputs continuous strings. For a geometric interpretation of the Bravyi-Kitaev transform and related mappings we would like to refer the reader to Appendix \ref{sec:A}.

\section{Techniques}
\label{sec:two}
\subsection{Motivation}
Here we motivate the general concept of Auxiliary Qubit Mappings. The starting point will be a non-local Hamiltonian obtained by transformation with some  linear mapping from Section \ref{subsec:conventional}. We then define  quantum codes in order to restore operator locality. These codes will act on the original system extended  by several `auxiliary' qubits. The effect of such codes on the Hamiltonian will be studied.

Consider that we have an $N$-qubit Hamiltonian $H_{\data}$,
\begin{align} \label{eq:hamil}
H_{\data}\; = \; \sum_{h\in \mathcal{S}} \Gamma^{h} \cdot \,  h_{\data} \, ,
\end{align}
where $\mathcal{S}$ is the set of all Pauli strings occurring in the Hamiltonian, $\mathcal{S}\subseteq \lbrace X, Y, Z, \mathbb{I} \rbrace^{\otimes N}$  with all $\Gamma^h$ being real, non-zero coefficients. Let us omit the qubit subscripts for now.  Although we want to remain fairly general at this point, the reader can already think of \eqref{eq:hamil} as the result of a Jordan-Wigner-transformed Hamiltonian \eqref{eq:secquant}. In general, the problem with this Hamiltonian is that $\mathcal{S}$ contains variations of Pauli strings that are either too long, discontinuous or otherwise inconvenient to us. Thus we would like to somehow replace these strings inside the Hamiltonian, even if it means that we need to add qubits to the system. Let us first consider a na\"ive approach which indicates the challenges of the method. We then tackle these challenges with a more sophisticated proposal.
 For the moment, let there be for exactly one inconvenient string $p \in\lbrace X, Y, Z, \mathbb{I}  \rbrace^{\otimes N}$, that either appears in the Hamiltonian directly, or is the non-local substring of some Hamiltonian strings $\lbrace h^\prime \rbrace  \subset \mathcal{S}$.
To bring the Hamiltonian in a convenient form, we would like to multiply  every such string $h^\prime$ with  $p$.
 Now we entangle an additional qubit to the system. Ideally, we would like to find the Pauli operator $\sigma \in \pm\lbrace X, Y, Z \rbrace$, acting on the added qubit, such that for every state $\ket{\varphi}$ on the original system of $N$ qubits, there exists a state  $\ket{\widetilde{\varphi}}$ on the system extended by the $(N+1)$-th qubit, on which $H$ has the same effect as on $\ket{\varphi}$, but  $(p\otimes \sigma)$ is a stabilizer:
\begin{align}\label{eq:stabcondition}
\left(p\otimes \sigma\right) \ket{\widetilde{\varphi}} =  \ket{\widetilde{\varphi}}  \qquad \text{implying} \quad \left(p \otimes \mathbb{I} \right)\ket{\widetilde{\varphi}}\;= \; \left( \mathbb{I}^{\otimes N}\otimes \, \sigma \right)  \ket{\widetilde{\varphi}} \, .
\end{align}
If this was true, then  every time $p$ appears as a string in the Hamiltonian we could just replace it with $\sigma$, or multiply inconvenient strings $(h^\prime \otimes \mathbb{I})$ by $(p\otimes \sigma)$ to cancel the non-local substrings. However, this is generally not possible: when there are  terms in $\mathcal{S}$ that anticommute with $p$, then $H$ will destroy the stabilizer state $\ket{\widetilde{\varphi}} $. This means that the state is altered in a way that \eqref{eq:stabcondition} is no longer valid. The simulation of the adjusted Hamiltonian on such a broken stabilizer state subsequently no longer describes the correct time evolution of the underlying $N$-qubit system. We thus need to adjust the Hamiltonian $H\to {H}^{(\kappa)}$, where ${H}^{(\kappa)}$ generally acts on $N+1$ qubits even without having its terms multiplied by stabilizers yet. This has to be done in a way as to ensure that the time evolution of $\ket{\widetilde{\varphi}}$ according to  ${H}^{(\kappa)}$ can be mapped back to the time evolution of  $\ket{\varphi}$ according to  $H$. At the same time we need to demand $[{H}^{(\kappa)},\, p \otimes \sigma]=0$ and that $(p \otimes \sigma)$ is a stabilizer like in \eqref{eq:stabcondition}. Only then we can use $(p\otimes \sigma)$ to cancel $p$ inside the terms of  ${H}^{(\kappa)}$,  and so obtain a convenient Hamiltonian $\widetilde{H}$.   \\

We now refine our approach accordingly, considering also the appearance of multiple strings  $p$ (and  picking up qubit subscripts as well).
 In $H_{\data}$, we identify $r$ Pauli strings  $p^i_{\data}$  (for $i\in [r]$) that we would like to cancel as we have done with a single string $p$ above. Furthermore, we would like to have the option for every Hamiltonian  term $h_{\data}$ to multiply it with either several, one or none of the strings $\lbrace p^i_{\data}\rbrace$. This is done by repeating the above procedure for each of the $r$ strings. To that end, we add $r$ qubits to the system: grouping them together we introduce the $r$-qubit auxiliary register $\aux=\lbrace N+1,\, N+2, \, \dots , \, N+r \rbrace $.  We assume that at the beginning, the  $\aux$-register is initialized in the state $\ket{0^r} = \ket{0}^{\otimes r}$. Our goal is to cancel the $i$-th string $p^i_{\data}$ with a single Pauli operator on the $(N+i)$-th qubit: $\sigma^i_{N+i}$. Thus we need to find a unitary quantum circuit which entangles the $\aux$-register with the data qubits in a certain way:  it has to implement a unitary $V_{\auxdata}$, such that for every state $\ket{\varphi}_{\data}$ \eqref{eq:genericstate}, we have a state in the composite system, $\ket{\widetilde{\varphi}}_{\auxdata}$ with
 \begin{align}
 \label{eq:stabsystem}
 V_{\auxdata} \, \ket{\varphi}_{\data} \otimes \ket{0^r}_{\aux} \,= \, \ket{\widetilde{\varphi}}_{\auxdata}\qquad \text{and} \qquad  (p^i_{\data} \otimes \sigma^i_{N+i}) \ket{\widetilde{\varphi}}_{\auxdata} \, = \, \ket{\widetilde{\varphi}}_{\auxdata} \, ,
\end{align}
for all $i \in [r]$.
 To make this work even on a conceptual level, we need to demand that all $p^i_{\data}$  commute pairwise,  otherwise there cannot be a common stabilizer state of all $(p^i_{\data} \otimes \sigma^i_{N+i})$. Once the stabilizer state is obtained, we maintain it by adjusting every term of  Hamiltonian \eqref{eq:hamil} with a Pauli string on the auxiliary register. This is done in a way such that the action of the adjusted term on the enlarged system is the same as the action of the original term on the original system. The adjustments are:
 \begin{align}
 \label{eq:stabhamil}
 \, h_{\data}\; \to \;  (h_{\data} \otimes \kappa^h_{\aux})  \quad \text{with} \quad  V^{\dagger}_{\auxdata}\,(h_{\data} \otimes \kappa^h_{\aux})\, \ket{\widetilde{\varphi}}_{\auxdata} \;=\; h_{\data} \, \ket{\varphi}_{\data}\otimes \ket{0^r}_{\aux}\, ,
 \end{align}
where $\kappa^h_{\aux}$ is the Pauli substring on the auxiliary register that is  correcting  $h_{\data}$.  Note that in case $h_{\data}$ already commutes with all the stabilizers, $\kappa^h_{\aux}$ is the identity. Of course we would like the above relation to hold for every string in the Hamiltonian, $h_{\data} \in \mathcal{S}$, but as we have effectively defined a quantum code encoding the entire Hilbert space of the $N$ data qubits, $h_\data$ can be an arbitrary $N$-qubit Pauli string.  Now by virtue of the stabilizer conditions \eqref{eq:stabsystem}, we can multiply the adjusted terms  $(h_{\data} \otimes \kappa^h_{\aux})$ by any of the operators  $(p^i_{\data} \otimes \sigma^i_{N+i})$, and thus get rid of their detrimental parts.  The resulting logical operators $\widetilde{h}_{\auxdata}$  define a convenient (logical) Hamiltonian
\begin{align}
\label{eq:stabhamil2}
\widetilde{H}_{\auxdata}=\sum_{h\in \mathcal{S}} \, \Gamma^h \cdot \, \widetilde{h}_{\auxdata} \, .
\end{align}

 \subsection{Definitions}

  Generally, the auxiliary qubits can be added in the computational basis to cancel strings $p^i_\data \in \lbrace \mathbb{I}, \, Z \rbrace^{\otimes N}$ with $Z$-operators $\sigma^i_{N+i} = Z_{N+i}$. As an enhancement of  the Jordan-Wigner transform, codes like this can be used to cancel non-local parity strings. The adjustment strings (of a term $h_\data$) $\kappa_\aux^h$ would then for all $k\in [r]$ contain $X_{N+k}$ if $h_\data$ anticommutes with $p^k_\data$. Note that the codes defined in this way (with only $Z$-stabilizers) have the property to map $N$-qubit computational basis states to  states in the computational basis on $n$ qubits, a trait that is useful for state preparation.  These codes however have their limitations, as  they can easily demand adjustment strings $\kappa^h_\aux$ of weight $O(r)$. \\

Other schemes specifically minimize the weight of $\kappa^h_\aux$ .  The methods of  Suba{\c{s}}{\i} and Jarzynski \cite{subacsi2016nonperturbative}
  effectively define codes with auxiliary qubits in Hadamard basis  that allow for an arbitrary choice of Pauli strings $p^i_\data$, as long as all $r$ strings commute pairwise.  The $p$-strings are subsequently replaced with $X$-operators, $\sigma^i_{N+i}=X_{N+i}$, and the adjustments $\kappa_{\aux}^h$ contain $Z_{N+k}$ for every string $p^k_{\data}$, that anticommutes with $h_\data$.  In  \cite{subacsi2016nonperturbative} some  concern is expressed  that the operator weight might generally scale with the number of auxiliary qubits added -  a key problem addressed by our work. We will in the following pick a set of strings  $\lbrace p^i_\data \rbrace$ such that every term  $h_\data \in \mathcal{S}$, resulting from any fermionic Hamiltonian, anticommutes with only a small number of stabilizers.  \\

In Appendix \ref{sec:B} we give more details about these Auxiliary Qubit codes, such as their logical basis and the  derivation of their stabilizers, adjustment terms as well as of the initialization unitaries $V_{\auxdata}$.   There are a few ways to extend the Auxiliary Qubit Mappings. In replacing the Pauli operators $\lbrace\sigma^i_{N+i}\rbrace$ with a set of Pauli strings $\lbrace \gamma^i_\aux \rbrace$, we can even stabilize Pauli strings $\lbrace p^i_\data \rbrace$ that anticommute.  In a similar vein, we can express the Verstraete-Cirac transform as a quantum code, which allows us to make modifications and to verify its operator transforms, see Appendix \ref{sec:mappings}.

\section{Auxiliary qubit mappings}
\label{sec:three}
\subsection{E-type AQM}
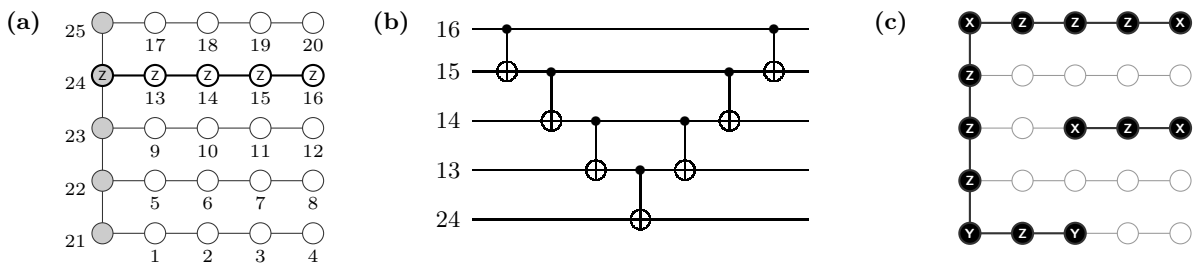
\begin{figure}[h!]
\centering \begin{tikzpicture}[scale=0.7, baseline=0]
\node[] at (-1.5,4){\textbf{(a)}};
\draw[color=\fadedblack](0,0)--(0,4)--(4,4);
\draw[color=black, thick](0,3)--(4,3);
\draw[color=\fadedblack](0,2)--(4,2);
\draw[color=\fadedblack](0,1)--(4,1);
\draw[color=\fadedblack](0,0)--(4,0);

\draw[color=\fadedblack, fill=black!20] (0,0) circle[radius=0.2];
\draw[color=\fadedblack, fill=white] (1,0) circle[radius=0.2];
\draw[color=\fadedblack, fill=white] (2,0) circle[radius=0.2];
\draw[color=\fadedblack, fill=white](3,0) circle[radius=0.2];
\draw[color=\fadedblack, fill=white] (4,0) circle[radius=0.2];
\draw[color=\fadedblack, fill=black!20] (0,1) circle[radius=0.2];
\draw[color=\fadedblack, fill=white] (1,1) circle[radius=0.2];
\draw[color=\fadedblack, fill=white] (2,1) circle[radius=0.2];
\draw[color=\fadedblack, fill=white] (3,1) circle[radius=0.2];
\draw[color=\fadedblack, fill=white] (4,1) circle[radius=0.2];
\draw[color=\fadedblack, fill=black!20] (0,2) circle[radius=0.2];
\draw[color=\fadedblack, fill=white] (1,2) circle[radius=0.2];

\draw[color=\fadedblack, fill=white] (2,2) circle[radius=0.2];
\draw[color=\fadedblack, fill=white] (3,2) circle[radius=0.2];
\draw[color=\fadedblack, fill=white] (4,2) circle[radius=0.2];
\draw[thick, color=black, fill=black!20] (0,3) circle[radius=0.2];
\draw[thick, color=black, fill=white] (1,3) circle[radius=0.2];
\draw[thick, color=black, fill=white] (2,3) circle[radius=0.2];
\draw[thick, color=black, fill=white] (3,3) circle[radius=0.2];
\draw[thick, color=black, fill=white] (4,3) circle[radius=0.2];
\draw[color=\fadedblack, fill=black!20] (0,4) circle[radius=0.2];
\draw[color=\fadedblack, fill=white] (1,4) circle[radius=0.2];
\draw[color=\fadedblack, fill=white] (2,4) circle[radius=0.2];
\draw[color=\fadedblack, fill=white] (3,4) circle[radius=0.2];
\draw[color=\fadedblack, fill=white] (4,4) circle[radius=0.2];
\node[black] at (0,3) { \tiny $\mathsf{Z}$} ;
\node[black] at (1,3) { \tiny $\mathsf{Z}$} ;
\node[black] at (2,3) { \tiny $\mathsf{Z}$} ;
\node[black] at (3,3) { \tiny $\mathsf{Z}$} ;
\node[black] at (4,3) { \tiny $\mathsf{Z}$} ;
\node[below] at (1,-.15) {\scriptsize 1} ;
\node[below] at (2,-.15) {\scriptsize 2} ;
\node[below] at (3,-.15) {\scriptsize 3};
\node[below] at (4,-.15) {\scriptsize 4} ;
\node[below] at (4,.85) { \scriptsize 8} ;
\node[below] at (3,.85) { \scriptsize 7} ;
\node[below] at (2,.85) {\scriptsize 6} ;
\node[below] at (1,.85) {\scriptsize 5};

\node[below] at (1,1.85) {\scriptsize 9} ;
\node[below] at (2,1.85) {\scriptsize 10} ;
\node[below] at (3,1.85) {\scriptsize 11};
\node[below] at (4,1.85) {\scriptsize 12} ;
\node[below] at (4,2.85) { \scriptsize 16} ;
\node[below] at (3,2.85) { \scriptsize 15} ;
\node[below] at (2,2.85) {\scriptsize 14} ;
\node[below] at (1,2.85) {\scriptsize 13};

\node[below] at (4,3.85) { \scriptsize 20} ;
\node[below] at (3,3.85) { \scriptsize 19} ;
\node[below] at (2,3.85) {\scriptsize 18} ;
\node[below] at (1,3.85) {\scriptsize 17};
\node[left] at (-.15,-0.15) {\scriptsize 21} ;
\node[left] at (-.15,0.85) {\scriptsize 22} ;
\node[left] at (-.15,1.85) {\scriptsize 23} ;
\node[left] at (-.15,2.85) {\scriptsize 24} ;
\node[left] at (-.15,3.85) {\scriptsize 25} ;
\end{tikzpicture}  $\quad$\begin{tikzpicture}[scale=0.7, baseline=0]
\node[] at (-1.5,4){\textbf{(b)}};
\node[] at (3,2) {\Qcircuit @C=1em @R=1.2em{
&\lstick{\scriptsize 16}	&\ctrl{1}	&\qw	&\qw	&\qw	&\qw	&\qw	&\ctrl{1}	&\qw	\\
&\lstick{\scriptsize 15}	&\targ	&\ctrl{1}	&\qw	&\qw	&\qw	&\ctrl{1}	&\targ	&\qw	\\
&\lstick{\scriptsize 14}	&\qw	&\targ	&\ctrl{1}	&\qw	&\ctrl{1}	&\targ	&\qw	&\qw	\\
&\lstick{\scriptsize 13}	&\qw	&\qw	&\targ	&\ctrl{1}	&\targ	&\qw	&\qw	&\qw	\\
&\lstick{\scriptsize 24}	&\qw	&\qw	&\qw	&\targ	&\qw	&\qw	&\qw	&\qw	\\
}}; \end{tikzpicture}$\qquad$\begin{tikzpicture}[scale=0.7, baseline=0]
\node[] at (-1.5,4){\textbf{(c)}};
\draw[color=\clre,](0,0)--(0,4)--(4,4);
\draw[color=\clre,](0,3)--(4,3);
\draw[color=\clre,](0,2)--(4,2);
\draw[color=\clre,](0,1)--(4,1);
\draw[color=\clre,](0,0)--(4,0);
\draw[\clrd,thick] (2,0)--(0,0)--(0,4)--(4,4);
\draw[\clrd,thick](2,2)--(4,2);

\draw[color=\clrb, fill=\clra, thick] (0,0) circle[radius=0.2];
\draw[color=\clrb, fill=\clra, thick] (1,0) circle[radius=0.2];
\draw[color=\clrb, fill=\clra, thick] (2,0) circle[radius=0.2];
\draw[color=\clre,fill=white](3,0) circle[radius=0.2];
\draw[color=\clre,fill=white] (4,0) circle[radius=0.2];
\draw[color=\clrb, fill=\clra, thick] (0,1) circle[radius=0.2];
\draw[color=\clre,fill=white] (1,1) circle[radius=0.2];
\draw[color=\clre,fill=white] (2,1) circle[radius=0.2];
\draw[color=\clre,fill=white] (3,1) circle[radius=0.2];
\draw[color=\clre,fill=white] (4,1) circle[radius=0.2];
\draw[color=\clrb, fill=\clra, thick] (0,2) circle[radius=0.2];
\draw[color=\clre,fill=white] (1,2) circle[radius=0.2];

\draw[color=\clrb, fill=\clra, thick] (2,2) circle[radius=0.2];
\draw[color=\clrb, fill=\clra, thick] (3,2) circle[radius=0.2];
\draw[color=\clrb, fill=\clra, thick] (4,2) circle[radius=0.2];
\draw[color=\clrb, fill=\clra, thick] (0,3) circle[radius=0.2];
\draw[color=\clre,fill=white] (1,3) circle[radius=0.2];
\draw[color=\clre,fill=white] (2,3) circle[radius=0.2];
\draw[color=\clre,fill=white] (3,3) circle[radius=0.2];
\draw[color=\clre,fill=white] (4,3) circle[radius=0.2];
\draw[color=\clrb, fill=\clra, thick] (0,4) circle[radius=0.2];
\draw[color=\clrb, fill=\clra, thick] (1,4) circle[radius=0.2];
\draw[color=\clrb, fill=\clra, thick] (2,4) circle[radius=0.2];
\draw[color=\clrb, fill=\clra, thick] (3,4) circle[radius=0.2];
\draw[color=\clrb, fill=\clra, thick] (4,4) circle[radius=0.2];
\node[\clrc] at (0,0) { \tiny $\boldsymbol{\mathsf{Y}}$} ;
\node[\clrc] at (1,0) { \tiny $\boldsymbol{\mathsf{Z}}$} ;
\node[\clrc] at (2,0) {\tiny $\boldsymbol{\mathsf{Y}}$} ;
\node[\clrc] at (0,1) {\tiny $\boldsymbol{\mathsf{Z}}$} ;
\node[\clrc] at (0,2) { \tiny $\boldsymbol{\mathsf{Z}}$} ;
\node[\clrc] at (0,3) {\tiny $\boldsymbol{\mathsf{Z}}$} ;
\node[\clrc] at (0,4) {\tiny $\boldsymbol{\mathsf{X}}$} ;
\node[\clrc] at (1,4) {\tiny $\boldsymbol{\mathsf{Z}}$} ;
\node[\clrc] at (2,4) {\tiny $\boldsymbol{\mathsf{Z}}$} ;
\node[\clrc] at (3,4) {\tiny $\boldsymbol{\mathsf{Z}}$};
\node[\clrc] at (4,4) {\tiny $\boldsymbol{\mathsf{X}}$} ;
\node[\clrc] at (4,2) {\tiny $\boldsymbol{\mathsf{X}}$};
\node[\clrc] at (3,2) {\tiny $\boldsymbol{\mathsf{Z}}$};
\node[\clrc] at (2,2) {\tiny $\boldsymbol{\mathsf{X}}$};
\end{tikzpicture}
\caption{ E-type AQM. \textbf{(a)} A block of ($4\times 5$) data qubits (white) enhanced with 5 auxiliary qubits (gray).  A single stabilizer is highlighted in the graph.  \textbf{(b)} Initializing one of the stabilizers $(\bigotimes_{i = 13}^{16}Z_i)\otimes Z_{24}$. \textbf{(c)} Simulating Pauli strings $h_\data=(X \otimes Z \otimes \cdots \otimes Z \otimes X)$ on the E-type AQM. While long strings are rerouted to skip rows, extending along the corresponding auxiliary qubits instead, shorter strings that do not switch rows can be simulated in parallel. }\label{fig:Etype}
\end{figure}
Here we present a mapping that remedies the biggest drawback of the S-pattern Jordan-Wigner transform  under a moderate overhead of qubits. Given a $(\ell_1 \times \ell_2)$ block of data qubits, we are going to add $\ell_2$ qubits as auxiliaries in computational basis. With this overhead, we will not manage to achieve any advantage for lattice models, but the scaling of  long-range interactions (on the fermionic lattice) is improved.  The following mapping will be referred to as E-type AQM.  We will first illustrate its graph, along with instructions on how to initialize the stabilizer state from $\ket{\varphi}_{\data}\otimes \ket{0^r}_{\aux}$. Afterwards, a discussion of the resulting Pauli strings will elucidate the advantages of the E-type AQM. \\
The idea of the E-type AQM is to store the parity of distinct data-qubit subsets  permanently  on auxiliary qubits. As we will see shortly, choosing to attach an auxiliary qubit to each of the  $\ell_2$ data-qubit rows is providing us with a geometric interpretation of the resulting strings. The result is shown in Figure \ref{fig:Etype}(a). Note that two things are different between the S-pattern Jordan-Wigner transform and the E-type AQM: firstly, the connectivity graph has changed. A row of qubits is now coupled to one auxiliary qubit, and only those auxiliary qubits are coupled together, data qubits in different rows are not coupled anymore. Although such connections between data qubits might be useful for simulating many-body terms, they are not necessarily required.  Secondly,  we have also changed the labeling of the qubits: the indices $i\in [\ell_1  \ell_2]$ still correspond to the indices attached to Fermion operators in \eqref{eq:modes}, but their order in the graph does no longer  resemble an S-pattern of  the canonical indices. \\ From $\ket{\varphi}_\data \otimes \ket{0^r}_\aux$ the logical state $\ket{\widetilde{\varphi}}_\auxdata$ can be initialized in $O(\ell_1)$-time and a total of  $O(\ell_1\ell_2)$ gates.  Here a chain of $\textsc{CNot}$s is used to mirror the collective parity information of an entire row of qubits on the attached auxiliary. The scaling in time is due to the fact that the preparation  circuit in Figure \ref{fig:Etype}(b),   can theoretically be implemented on every row in parallel.  The stabilizers of the system are
\begin{align}
\label{eq:Etypestab}
\left(\bigotimes_{i \,\in\, \text{row $k$}} Z_{i}\right)\otimes Z_{N+k}\, ,
\end{align}
for all rows $k \in [\ell_2]$ in the data qubit block.
We now turn to describe the  resulting Pauli strings, for which we  need to discuss the adjustments $\kappa^h_\aux$. Diagonal terms \eqref{eq:hopping} in the Hamiltonian do not influence the stabilizer state, as well as hopping terms \eqref{eq:number} between qubits in the same row. Our attention is thus focused on Pauli strings of the form $h_\data=(X_i\otimes Z_{i+1} \otimes\cdots \otimes Z_{j-1}\otimes X_{j})$, where qubits $i$ and $j$ are situated in different rows $k$ and $l$, where $k<l$. Those  Pauli strings are subsequently adjusted by  $\kappa_\aux^h=(X_{N+k}\otimes X_{N+l})$. \\

In order to make these terms more convenient, we multiply the adjusted strings with the corresponding stabilizers \eqref{eq:Etypestab} of rows $k^\prime$, for all $k\leq k^\prime < l$. Here we discover the benefit of this mapping: wherever Pauli strings act as $Z$-strings on entire rows, the parity is inferred instead from the auxiliary qubits attached. This limits the length of parity substrings and so  Pauli strings (originating from hopping terms) have a maximal length  $2\ell_1+\ell_2$, instead of $\ell_1\ell_2$. This is not just a benefit in time and gates, but also allows us to simulate single-row strings at the same time as long strings spanning these rows, see Figure \ref{fig:Etype}(c).   \\
 Although we expect the  E-type AQM to be useful for problems long-range interactions, it has no advantage compared to the S-pattern Jordan-Wigner transform if one considers locally-interacting lattice Hamiltonians. With only single-row Pauli strings or strings between adjacent rows, no savings in gates and algorithmic depth can be anticipated. In the following, we will define a mapping that can transform those models into local qubit-Hamiltonians.
 \subsection{Square lattice AQM}
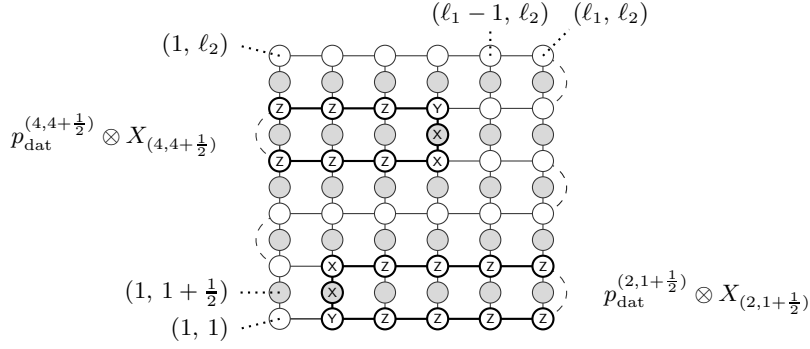
\begin{figure}[h!]
\begin{tikzpicture}[scale=0.7]
\draw[color=\fadedblack,dashed](1,2).. controls(.4,2.2)and (.4,2.8).. (1,3);
\draw[color=black!90,dashed](6,3).. controls(6.6,3.2)and (6.6,3.8).. (6,4);
\draw[color=\fadedblack,dashed](6,5).. controls(6.6,5.2)and (6.6,5.8).. (6,6);
\draw[color=\fadedblack,dashed](1,4).. controls(.4,4.2)and (.4,4.8).. (1,5);
\draw[color=\fadedblack,dashed](6,1).. controls(6.6,1.2)and (6.6,1.8).. (6,2);

\draw[\fadedblack](1,1)--(6,1);
\draw[\fadedblack](1,2)--(6,2);
\draw[\fadedblack](1,3)--(6,3);
\draw[\fadedblack](1,4)--(6,4);
\draw[\fadedblack](1,5)--(6,5);
\draw[\fadedblack](1,6)--(6,6);

\draw[\fadedblack](1,1)--(1,6);
\draw[\fadedblack](2,1)--(2,6);
\draw[\fadedblack](3,1)--(3,6);
\draw[\fadedblack](4,1)--(4,6);
\draw[\fadedblack](5,1)--(5,6);
\draw[\fadedblack](6,1)--(6,6);

\draw[\fadedblack] (1,1)--(2,1);
\draw[\fadedblack] (4,4)--(1,4)--(1,5)--(4,5);
\draw[\fadedblack] (2,2)--(4,2);
\draw[black,thick](6,2)--(2,2)--(2,1)--(6,1);
\draw[black,thick](1,4)--(4,4)--(4,5)--(1,5);
\draw[\fadedblack, fill=white, ] (1,1) circle[radius=0.2];

\draw[black, fill=white,thick ] (2,1) circle[radius=0.2];

\draw[black,fill=white,thick] (3,1) circle[radius=0.2];
\draw[black,fill=white,thick] (4,1) circle[radius=0.2];
\draw[\fadedblack,fill=white] (1,2) circle[radius=0.2];
\draw[black,fill=white,thick] (2,2) circle[radius=0.2];

\draw[black,fill=white,thick] (3,2) circle[radius=0.2];

\draw[black,fill=white,thick] (4,2) circle[radius=0.2];

\draw[\fadedblack,fill=white] (1,3) circle[radius=0.2];
\draw[\fadedblack,fill=white] (2,3) circle[radius=0.2];
\draw[\fadedblack,fill=white] (3,3) circle[radius=0.2];
\draw[\fadedblack,fill=white] (4,3) circle[radius=0.2];
\draw[black, fill=white ,thick] (1,4) circle[radius=0.2];

\draw[fill=white,thick] (2,4) circle[radius=0.2];

\draw[fill=white,thick] (3,4) circle[radius=0.2];

\draw[fill=white,thick] (4,4) circle[radius=0.2];

\draw[fill=white,thick] (1,5) circle[radius=0.2];

\draw[fill=white,thick] (2,5) circle[radius=0.2];

\draw[fill=white,thick] (3,5) circle[radius=0.2];

\draw[fill=white,thick] (4,5) circle[radius=0.2];

\draw[\fadedblack,fill=white] (5,5) circle[radius=0.2];
\draw[\fadedblack,fill=white] (5,4) circle[radius=0.2];
\draw[\fadedblack,fill=white] (5,3) circle[radius=0.2];
\draw[black,fill=white,thick] (5,2) circle[radius=0.2];
\draw[black,fill=white,thick] (5,1) circle[radius=0.2];

\draw[\fadedblack,fill=\checker] (1,1.5) circle[radius=0.2];
\draw[\fadedblack,fill=\checker] (1,2.5) circle[radius=0.2];
\draw[\fadedblack,fill=\checker] (1,3.5) circle[radius=0.2];
\draw[\fadedblack,fill=\checker] (1,4.5) circle[radius=0.2];

\draw[black, fill=\checker,thick] (2,1.5) circle[radius=0.2];
\draw[\fadedblack,fill=\checker] (2,2.5) circle[radius=0.2];
\draw[\fadedblack,fill=\checker] (2,3.5) circle[radius=0.2];
\draw[\fadedblack,fill=\checker] (2,4.5) circle[radius=0.2];

\draw[\fadedblack,fill=\checker] (3,1.5) circle[radius=0.2];
\draw[\fadedblack,fill=\checker] (3,2.5) circle[radius=0.2];
\draw[\fadedblack,fill=\checker] (3,3.5) circle[radius=0.2];
\draw[\fadedblack,fill=\checker] (3,4.5) circle[radius=0.2];

\draw[\fadedblack,fill=\checker] (4,1.5) circle[radius=0.2];
\draw[\fadedblack,fill=\checker] (4,2.5) circle[radius=0.2];
\draw[\fadedblack,fill=\checker] (4,3.5) circle[radius=0.2];
\draw[black, fill=\checker,thick] (4,4.5) circle[radius=0.2];

\draw[\fadedblack,fill=\checker] (5,1.5) circle[radius=0.2];
\draw[\fadedblack,fill=\checker] (5,2.5) circle[radius=0.2];
\draw[\fadedblack,fill=\checker] (5,3.5) circle[radius=0.2];
\draw[\fadedblack,fill=\checker] (5,4.5) circle[radius=0.2];

\draw[\fadedblack,fill=\checker] (6,5.5) circle[radius=0.2];
\draw[\fadedblack,fill=\checker] (5,5.5) circle[radius=0.2];
\draw[\fadedblack,fill=\checker] (4,5.5) circle[radius=0.2];
\draw[\fadedblack,fill=\checker] (3,5.5) circle[radius=0.2];
\draw[\fadedblack,fill=\checker] (2,5.5) circle[radius=0.2];
\draw[\fadedblack,fill=\checker] (1,5.5) circle[radius=0.2];
\draw[\fadedblack,fill=white] (1,6) circle[radius=0.2];
\draw[\fadedblack,fill=white] (2,6) circle[radius=0.2];
\draw[\fadedblack,fill=white] (3,6) circle[radius=0.2];
\draw[\fadedblack,fill=white] (4,6) circle[radius=0.2];
\draw[\fadedblack,fill=white] (5,6) circle[radius=0.2];
\draw[\fadedblack,fill=white] (6,6) circle[radius=0.2];

\draw[\fadedblack,fill=\checker] (6,1.5) circle[radius=0.2];
\draw[\fadedblack,fill=\checker] (6,2.5) circle[radius=0.2];
\draw[\fadedblack,fill=\checker] (6,3.5) circle[radius=0.2];
\draw[\fadedblack,fill=\checker] (6,4.5) circle[radius=0.2];
\draw[\fadedblack,fill=white] (6,5) circle[radius=0.2];
\draw[\fadedblack,fill=white] (6,4) circle[radius=0.2];
\draw[\fadedblack,fill=white] (6,3) circle[radius=0.2];
\draw[black,fill=white,thick] (6,2) circle[radius=0.2];
\draw[black,fill=white,thick] (6,1) circle[radius=0.2];

\node[] at (6,1) {\tiny $\mathsf{Z}$};
\node[] at (6,2) {\tiny $\mathsf{Z}$};

\node[] at (2,1) {\tiny $\mathsf{Y}$};
\node[] at (3,1) {\tiny $\mathsf{Z}$};
\node[] at (4,1) {\tiny $\mathsf{Z}$};
\node[] at (5,1) {\tiny $\mathsf{Z}$};

\node[] at (2,2) {\tiny $\mathsf{X}$};
\node[] at (3,2) {\tiny $\mathsf{Z}$};
\node[] at (4,2) {\tiny $\mathsf{Z}$};
\node[] at (5,2) {\tiny $\mathsf{Z}$};
\node[] at (2,1.5) {\tiny $\mathsf{X}$};
\node[] at (1,4) {\tiny $\mathsf{Z}$};
\node[] at (2,4) {\tiny $\mathsf{Z}$};
\node[] at (3,4) {\tiny $\mathsf{Z}$};
\node[] at (4,4) {\tiny $\mathsf{X}$};
\node[] at (1,5) {\tiny $\mathsf{Z}$};
\node[] at (2,5) {\tiny $\mathsf{Z}$};
\node[] at (3,5) {\tiny $\mathsf{Z}$};
\node[] at (4,5) {\tiny $\mathsf{Y}$};
\node[] at (4,4.5) {\tiny $\mathsf{X}$};
\draw[thick, dotted] (1,1)--(0.2,0.8);
\draw[thick,dotted] (1,1.5)--(0.2,1.5);
\draw[thick,dotted] (1,6)--(0.2,6.2);
\draw[thick,dotted] (6,6)--(6.5,6.5) ;
\draw[thick,dotted] (5,6)--(5,6.5) ;
\node[left] at (0.2,0.8) {$\tiny(1, \, 1)$};
\node[left] at (0.2,1.5) {$\tiny(1, \, 1+\frac{1}{2})$};
\node[left] at (0.2,6.2) {$\tiny(1, \, \ell_2)$};
\node[above] at (5,6.4) {$\tiny(\ell_1-1, \, \ell_2)$};
\node[above right] at (6.4,6.4) {$\tiny(\ell_1, \, \ell_2)$};
\node[left] at (0,4.5) {$p^{(4,4+\frac{1}{2})}_{\data}\otimes X_{(4,4+\frac{1}{2})}$};
\node[right] at (7,1.5) {$p^{(2,1+\frac{1}{2})}_{\data}\otimes X_{(2,1+\frac{1}{2})}$};
\end{tikzpicture}
\caption{Square lattice AQM, defined on a $\ell_1 \times( 2\ell_2-1)$ square lattice of qubits, here $\ell_1=\ell_2=6$. The gray qubits form the $\aux$-register. Some qubits are labeled with their coordinates (dotted lines), where the auxiliary qubits generally sit on half-integer positions.  The dashed lines do not couple qubits, but only indicate  the windings of the S-pattern of the underlying Jordan-Wigner transform. The highlighted qubits and edges are two examples of stabilizers for odd and  even rows, respectively.}
\label{fig:stabsquare}
\end{figure}

\begin{figure}

 \begin{tikzpicture}[scale=0.6]
 \node[] at (0.5,6.5) {\textbf{(a)}};
\draw[color=\clre,dashed](1,2).. controls(.4,2.2)and (.4,2.8).. (1,3);
\draw[color=black, thick, fill=white,dashed](6,3).. controls(6.6,3.2)and (6.6,3.8).. (6,4);
\draw[color=\clre,dashed](6,5).. controls(6.6,5.2)and (6.6,5.8).. (6,6);
\draw[color=black, thick, fill=white,dashed](1,4).. controls(.4,4.2)and (.4,4.8).. (1,5);
\draw[color=\clre,dashed](6,1).. controls(6.6,1.2)and (6.6,1.8).. (6,2);
\draw[\clre](1,1)--(6,1);
\draw[\clre](1,2)--(6,2);
\draw[\clre](1,3)--(6,3);
\draw[\clre](1,4)--(6,4);
\draw[\clre](1,5)--(6,5);
\draw[\clre](1,6)--(6,6);

\draw[\clre](1,1)--(1,6);
\draw[\clre](2,1)--(2,6);
\draw[\clre](3,1)--(3,6);
\draw[\clre](4,1)--(4,6);
\draw[\clre](5,1)--(5,6);
\draw[\clre](6,1)--(6,6);

\draw[black, thick] (1,2)--(1,3)--(6,3);
\draw[black, thick] (6,4)--(1,4);
\draw[black, thick] (1,5)--(5,5)--(5,5.5);
\draw[color=\clre,fill=white] (1,1) circle[radius=0.2];

\draw[color=\clre,fill=white] (2,1) circle[radius=0.2];

\draw[color=\clre,fill=white] (3,1) circle[radius=0.2];
\draw[color=\clre,fill=white] (4,1) circle[radius=0.2];
\draw[color=\clre,fill=white] (1,2) circle[radius=0.2];
\draw[color=\clre,fill=white] (2,2) circle[radius=0.2];

\draw[color=\clre,fill=white] (3,2) circle[radius=0.2];

\draw[color=\clre,fill=white] (4,2) circle[radius=0.2];

\draw[color=\clre,fill=white] (1,3) circle[radius=0.2];
\draw[color=\clre,fill=white] (2,3) circle[radius=0.2];
\draw[color=\clre,fill=white] (3,3) circle[radius=0.2];
\draw[color=\clre,fill=white] (4,3) circle[radius=0.2];
\draw[color=\clre,fill=white] (1,4) circle[radius=0.2];

\draw[color=\clre,fill=white] (2,4) circle[radius=0.2];

\draw[color=\clre,fill=white] (3,4) circle[radius=0.2];

\draw[color=\clre,fill=white] (4,4) circle[radius=0.2];

\draw[color=\clre,fill=white] (1,5) circle[radius=0.2];

\draw[color=\clre,fill=white] (2,5) circle[radius=0.2];

\draw[color=\clre,fill=white] (3,5) circle[radius=0.2];

\draw[color=\clre,fill=white] (4,5) circle[radius=0.2];

\draw[color=\clre,fill=white] (5,5) circle[radius=0.2];
\draw[color=\clre,fill=white] (5,4) circle[radius=0.2];
\draw[color=\clre,fill=white] (5,3) circle[radius=0.2];
\draw[color=\clre,fill=white] (5,2) circle[radius=0.2];
\draw[color=\clre,fill=white] (5,1) circle[radius=0.2];

\draw[color=\clre,fill=white] (1,1.5) circle[radius=0.2];
\draw[color=\clre,fill=white] (1,2.5) circle[radius=0.2];
\draw[color=\clre,fill=white] (1,3.5) circle[radius=0.2];
\draw[color=\clre,fill=white] (1,4.5) circle[radius=0.2];

\draw[color=\clre,fill=white] (2,1.5) circle[radius=0.2];
\draw[color=\clre,fill=white] (2,2.5) circle[radius=0.2];
\draw[color=\clre,fill=white] (2,3.5) circle[radius=0.2];
\draw[color=\clre,fill=white] (2,4.5) circle[radius=0.2];

\draw[color=\clre,fill=white] (3,1.5) circle[radius=0.2];
\draw[color=\clre,fill=white] (3,2.5) circle[radius=0.2];
\draw[color=\clre,fill=white] (3,3.5) circle[radius=0.2];
\draw[color=\clre,fill=white] (3,4.5) circle[radius=0.2];

\draw[color=\clre,fill=white] (4,1.5) circle[radius=0.2];
\draw[color=\clre,fill=white] (4,2.5) circle[radius=0.2];
\draw[color=\clre,fill=white] (4,3.5) circle[radius=0.2];
\draw[color=\clre,fill=white] (4,4.5) circle[radius=0.2];

\draw[color=\clre,fill=white] (5,1.5) circle[radius=0.2];
\draw[color=\clre,fill=white] (5,2.5) circle[radius=0.2];
\draw[color=\clre,fill=white] (5,3.5) circle[radius=0.2];
\draw[color=\clre,fill=white] (5,4.5) circle[radius=0.2];

\draw[color=\clre,fill=white] (6,5.5) circle[radius=0.2];
\draw[color=\clre,fill=white] (5,5.5) circle[radius=0.2];
\draw[color=\clre,fill=white] (4,5.5) circle[radius=0.2];
\draw[color=\clre,fill=white] (3,5.5) circle[radius=0.2];
\draw[color=\clre,fill=white] (2,5.5) circle[radius=0.2];
\draw[color=\clre,fill=white] (1,5.5) circle[radius=0.2];
\draw[color=\clre,fill=white] (1,6) circle[radius=0.2];
\draw[color=\clre,fill=white] (2,6) circle[radius=0.2];
\draw[color=\clre,fill=white] (3,6) circle[radius=0.2];
\draw[color=\clre,fill=white] (4,6) circle[radius=0.2];
\draw[color=\clre,fill=white] (5,6) circle[radius=0.2];
\draw[color=\clre,fill=white] (6,6) circle[radius=0.2];

\draw[color=\clre,fill=white] (6,1.5) circle[radius=0.2];
\draw[color=\clre,fill=white] (6,2.5) circle[radius=0.2];
\draw[color=\clre,fill=white] (6,3.5) circle[radius=0.2];
\draw[color=\clre,fill=white] (6,4.5) circle[radius=0.2];
\draw[color=\clre,fill=white] (6,5) circle[radius=0.2];
\draw[color=\clre,fill=white] (6,4) circle[radius=0.2];
\draw[color=\clre,fill=white] (6,3) circle[radius=0.2];
\draw[color=\clre,fill=white] (6,2) circle[radius=0.2];
\draw[color=\clre,fill=white] (6,1) circle[radius=0.2];

\draw[color=black, thick, fill=white] (1,2) circle[radius=0.2];
\draw[color=black, thick, fill=white]  (1,2.5) circle[radius=0.2];
\draw[color=black, thick, fill=white]  (1,3) circle[radius=0.2];
\draw[color=black, thick, fill=white]  (2,3) circle[radius=0.2];
\draw[color=black, thick, fill=white]  (3,3) circle[radius=0.2];
\draw[color=black, thick, fill=white]  (4,3) circle[radius=0.2];
\draw[color=black, thick, fill=white]  (5,3) circle[radius=0.2];
\draw[color=black, thick, fill=white]  (6,3) circle[radius=0.2];
\draw[color=black, thick, fill=white]  (1,4) circle[radius=0.2];
\draw[color=black, thick, fill=white]  (2,4) circle[radius=0.2];
\draw[color=black, thick, fill=white]  (3,4) circle[radius=0.2];
\draw[color=black, thick, fill=white]  (4,4) circle[radius=0.2];
\draw[color=black, thick, fill=white]  (5,4) circle[radius=0.2];
\draw[color=black, thick, fill=white]  (6,4) circle[radius=0.2];
\draw[color=black, thick, fill=white]  (1,5) circle[radius=0.2];
\draw[color=black, thick, fill=white]  (2,5) circle[radius=0.2];
\draw[color=black, thick, fill=white]  (3,5) circle[radius=0.2];
\draw[color=black, thick, fill=white]  (4,5) circle[radius=0.2];
\draw[color=black, thick, fill=white]  (5,5) circle[radius=0.2];
\draw[color=black, thick, fill=white]  (5,5.5) circle[radius=0.2];
\node[black] at (1,2) {\tiny $\boldsymbol{\mathsf{X}}$};
\node[black] at (1,2.5) {\tiny $\boldsymbol{\mathsf{Z}}$};
\node[black] at (1,3) {\tiny $\boldsymbol{\mathsf{Z}}$};
\node[black] at (2,3) {\tiny $\boldsymbol{\mathsf{Z}}$};
\node[black] at (3,3) {\tiny $\boldsymbol{\mathsf{Z}}$};
\node[black] at (4,3) {\tiny $\boldsymbol{\mathsf{Z}}$};
\node[black] at (5,3) {\tiny $\boldsymbol{\mathsf{Z}}$};
\node[black] at (6,3) {\tiny $\boldsymbol{\mathsf{Z}}$};
\node[black] at (1,4) {\tiny $\boldsymbol{\mathsf{Z}}$};
\node[black] at (2,4) {\tiny $\boldsymbol{\mathsf{Z}}$};
\node[black] at (3,4) {\tiny $\boldsymbol{\mathsf{Z}}$};
\node[black] at (4,4) {\tiny $\boldsymbol{\mathsf{Z}}$};
\node[black] at (5,4) {\tiny $\boldsymbol{\mathsf{Z}}$};
\node[black] at (6,4) {\tiny $\boldsymbol{\mathsf{Z}}$};
\node[black] at (1,5) {\tiny $\boldsymbol{\mathsf{Z}}$};
\node[black] at (2,5) {\tiny $\boldsymbol{\mathsf{Z}}$};
\node[black] at (3,5) {\tiny $\boldsymbol{\mathsf{Z}}$};
\node[black] at (4,5) {\tiny $\boldsymbol{\mathsf{Z}}$};
\node[black] at (5,5) {\tiny $\boldsymbol{\mathsf{X}}$};
\node[black] at (5,5.5) {\tiny $\boldsymbol{\mathsf{Z}}$};

\end{tikzpicture}  \begin{tikzpicture}[scale=0.6]
\node[] at (0.5,6.5) {\textbf{(b)}};
\draw[color=\clre,dashed](1,2).. controls(.4,2.2)and (.4,2.8).. (1,3);
\draw[color=\clre,dashed](6,3).. controls(6.6,3.2)and (6.6,3.8).. (6,4);
\draw[color=\clre,dashed](6,5).. controls(6.6,5.2)and (6.6,5.8).. (6,6);
\draw[color=\clre,dashed](1,4).. controls(.4,4.2)and (.4,4.8).. (1,5);
\draw[color=\clre,dashed](6,1).. controls(6.6,1.2)and (6.6,1.8).. (6,2);
\draw[\clre](1,1)--(6,1);
\draw[\clre](1,2)--(6,2);
\draw[\clre](1,3)--(6,3);
\draw[\clre](1,4)--(6,4);
\draw[\clre](1,5)--(6,5);
\draw[\clre](1,6)--(6,6);

\draw[\clre](1,1)--(1,6);
\draw[\clre](2,1)--(2,6);
\draw[\clre](3,1)--(3,6);
\draw[\clre](4,1)--(4,6);
\draw[\clre](5,1)--(5,6);
\draw[\clre](6,1)--(6,6);

\draw[color=\clre,fill=white] (1,1) circle[radius=0.2];

\draw[color=\clre,fill=white] (2,1) circle[radius=0.2];

\draw[color=\clre,fill=white] (3,1) circle[radius=0.2];
\draw[color=\clre,fill=white] (4,1) circle[radius=0.2];
\draw[color=\clre,fill=white] (1,2) circle[radius=0.2];
\draw[color=\clre,fill=white] (2,2) circle[radius=0.2];

\draw[color=\clre,fill=white] (3,2) circle[radius=0.2];

\draw[color=\clre,fill=white] (4,2) circle[radius=0.2];

\draw[color=\clre,fill=white] (1,3) circle[radius=0.2];
\draw[color=\clre,fill=white] (2,3) circle[radius=0.2];
\draw[color=\clre,fill=white] (3,3) circle[radius=0.2];
\draw[color=\clre,fill=white] (4,3) circle[radius=0.2];
\draw[color=\clre,fill=white] (1,4) circle[radius=0.2];

\draw[color=\clre,fill=white] (2,4) circle[radius=0.2];

\draw[color=\clre,fill=white] (3,4) circle[radius=0.2];

\draw[color=\clre,fill=white] (4,4) circle[radius=0.2];

\draw[color=\clre,fill=white] (1,5) circle[radius=0.2];

\draw[color=\clre,fill=white] (2,5) circle[radius=0.2];

\draw[color=\clre,fill=white] (3,5) circle[radius=0.2];

\draw[color=\clre,fill=white] (4,5) circle[radius=0.2];

\draw[color=\clre,fill=white] (5,5) circle[radius=0.2];
\draw[color=\clre,fill=white] (5,4) circle[radius=0.2];
\draw[color=\clre,fill=white] (5,3) circle[radius=0.2];
\draw[color=\clre,fill=white] (5,2) circle[radius=0.2];
\draw[color=\clre,fill=white] (5,1) circle[radius=0.2];

\draw[color=\clre,fill=white] (1,1.5) circle[radius=0.2];
\draw[color=\clre,fill=white] (1,2.5) circle[radius=0.2];
\draw[color=\clre,fill=white] (1,3.5) circle[radius=0.2];
\draw[color=\clre,fill=white] (1,4.5) circle[radius=0.2];

\draw[color=\clre,fill=white] (2,1.5) circle[radius=0.2];
\draw[color=\clre,fill=white] (2,2.5) circle[radius=0.2];
\draw[color=\clre,fill=white] (2,3.5) circle[radius=0.2];
\draw[color=\clre,fill=white] (2,4.5) circle[radius=0.2];

\draw[color=\clre,fill=white] (3,1.5) circle[radius=0.2];
\draw[color=\clre,fill=white] (3,2.5) circle[radius=0.2];
\draw[color=\clre,fill=white] (3,3.5) circle[radius=0.2];
\draw[color=\clre,fill=white] (3,4.5) circle[radius=0.2];

\draw[color=\clre,fill=white] (4,1.5) circle[radius=0.2];
\draw[color=\clre,fill=white] (4,2.5) circle[radius=0.2];
\draw[color=\clre,fill=white] (4,3.5) circle[radius=0.2];
\draw[color=\clre,fill=white] (4,4.5) circle[radius=0.2];

\draw[color=\clre,fill=white] (5,1.5) circle[radius=0.2];
\draw[color=\clre,fill=white] (5,2.5) circle[radius=0.2];
\draw[color=\clre,fill=white] (5,3.5) circle[radius=0.2];
\draw[color=\clre,fill=white] (5,4.5) circle[radius=0.2];

\draw[color=\clre,fill=white] (6,5.5) circle[radius=0.2];
\draw[color=\clre,fill=white] (5,5.5) circle[radius=0.2];
\draw[color=\clre,fill=white] (4,5.5) circle[radius=0.2];
\draw[color=\clre,fill=white] (3,5.5) circle[radius=0.2];
\draw[color=\clre,fill=white] (2,5.5) circle[radius=0.2];
\draw[color=\clre,fill=white] (1,5.5) circle[radius=0.2];
\draw[color=\clre,fill=white] (1,6) circle[radius=0.2];
\draw[color=\clre,fill=white] (2,6) circle[radius=0.2];
\draw[color=\clre,fill=white] (3,6) circle[radius=0.2];
\draw[color=\clre,fill=white] (4,6) circle[radius=0.2];
\draw[color=\clre,fill=white] (5,6) circle[radius=0.2];
\draw[color=\clre,fill=white] (6,6) circle[radius=0.2];

\draw[color=\clre,fill=white] (6,1.5) circle[radius=0.2];
\draw[color=\clre,fill=white] (6,2.5) circle[radius=0.2];
\draw[color=\clre,fill=white] (6,3.5) circle[radius=0.2];
\draw[color=\clre,fill=white] (6,4.5) circle[radius=0.2];
\draw[color=\clre,fill=white] (6,5) circle[radius=0.2];
\draw[color=\clre,fill=white] (6,4) circle[radius=0.2];
\draw[color=\clre,fill=white] (6,3) circle[radius=0.2];
\draw[color=\clre,fill=white] (6,2) circle[radius=0.2];
\draw[color=\clre,fill=white] (6,1) circle[radius=0.2];

\draw[color=\clrd,  thick] (1,2.5)--(1,5)--(5,5)--(5,5.5);
\draw[color=\clrb, fill=\clra, thick](1,2.5) circle[radius=0.21];
\draw[color=\clrb, fill=\clra, thick] (1,3) circle[radius=0.21];
\draw[color=\clrb, fill=\clra, thick] (1,3.5) circle[radius=0.21];
\draw[color=\clrb, fill=\clra, thick] (1,4) circle[radius=0.21];
\draw[color=\clrb, fill=\clra, thick] (1,4.5) circle[radius=0.21];
\draw[color=\clrb, fill=\clra, thick] (1,5) circle[radius=0.21];
\draw[color=\clrb, fill=\clra, thick] (2,5) circle[radius=0.21];
\draw[color=\clrb, fill=\clra, thick] (3,5) circle[radius=0.21];
\draw[color=\clrb, fill=\clra, thick] (4,5) circle[radius=0.21];
\draw[color=\clrb, fill=\clra, thick](5,5) circle[radius=0.21];
\draw[color=\clrb, fill=\clra, thick] (5,5.5) circle[radius=0.21];

\node[\clrc] at (1,2.5) {\tiny $\boldsymbol{\mathsf{Y}}$};
\node[\clrc] at (1,3) {\tiny $\boldsymbol{\mathsf{Z}}$};
\node[\clrc] at (1,3.5) {\tiny $\boldsymbol{\mathsf{X}}$};
\node[\clrc] at (1,4) {\tiny $\boldsymbol{\mathsf{Z}}$};
\node[\clrc] at (1,4.5) {\tiny $\boldsymbol{\mathsf{X}}$};
\node[\clrc] at (1,5) {\tiny $\boldsymbol{\mathsf{X}}$};
\node[\clrc] at (2,5) {\tiny $\boldsymbol{\mathsf{Z}}$};
\node[\clrc] at (3,5) {\tiny $\boldsymbol{\mathsf{Z}}$};
\node[\clrc] at (4,5) {\tiny $\boldsymbol{\mathsf{Z}}$};
\node[\clrc] at (5,5) {\tiny $\boldsymbol{\mathsf{X}}$};
\node[\clrc] at (5,5.5) {\tiny $\boldsymbol{\mathsf{Z}}$};

\end{tikzpicture} \begin{tikzpicture}[scale=0.6]
\node[] at (0.5,6.5) {\textbf{(c)}};
\draw[color=\clre,dashed](1,2).. controls(.4,2.2)and (.4,2.8).. (1,3);
\draw[color=\clre,dashed](6,3).. controls(6.6,3.2)and (6.6,3.8).. (6,4);
\draw[color=\clre,dashed](6,5).. controls(6.6,5.2)and (6.6,5.8).. (6,6);
\draw[color=\clre,dashed](1,4).. controls(.4,4.2)and (.4,4.8).. (1,5);
\draw[color=\clre,dashed](6,1).. controls(6.6,1.2)and (6.6,1.8).. (6,2);
\draw[\clre](1,1)--(6,1);
\draw[\clre](1,2)--(6,2);
\draw[\clre](1,3)--(6,3);
\draw[\clre](1,4)--(6,4);
\draw[\clre](1,5)--(6,5);
\draw[\clre](1,6)--(6,6);

\draw[\clre](1,1)--(1,6);
\draw[\clre](2,1)--(2,6);
\draw[\clre](3,1)--(3,6);
\draw[\clre](4,1)--(4,6);
\draw[\clre](5,1)--(5,6);
\draw[\clre](6,1)--(6,6);
\draw[\clrd,  thick] (1,2.5)--(1,2)--(5,2)--(5,5.5);
\draw[color=\clre,fill=white] (1,1) circle[radius=0.2];

\draw[color=\clre,fill=white] (2,1) circle[radius=0.2];

\draw[color=\clre,fill=white] (3,1) circle[radius=0.2];
\draw[color=\clre,fill=white] (4,1) circle[radius=0.2];
\draw[color=\clre,fill=white] (1,2) circle[radius=0.2];
\draw[color=\clre,fill=white] (2,2) circle[radius=0.2];

\draw[color=\clre,fill=white] (3,2) circle[radius=0.2];

\draw[color=\clre,fill=white] (4,2) circle[radius=0.2];

\draw[color=\clre,fill=white] (1,3) circle[radius=0.2];
\draw[color=\clre,fill=white] (2,3) circle[radius=0.2];
\draw[color=\clre,fill=white] (3,3) circle[radius=0.2];
\draw[color=\clre,fill=white] (4,3) circle[radius=0.2];
\draw[color=\clre,fill=white] (1,4) circle[radius=0.2];

\draw[color=\clre,fill=white] (2,4) circle[radius=0.2];

\draw[color=\clre,fill=white] (3,4) circle[radius=0.2];

\draw[color=\clre,fill=white] (4,4) circle[radius=0.2];

\draw[color=\clre,fill=white] (1,5) circle[radius=0.2];

\draw[color=\clre,fill=white] (2,5) circle[radius=0.2];

\draw[color=\clre,fill=white] (3,5) circle[radius=0.2];

\draw[color=\clre,fill=white] (4,5) circle[radius=0.2];

\draw[color=\clre,fill=white] (5,5) circle[radius=0.2];
\draw[color=\clre,fill=white] (5,4) circle[radius=0.2];
\draw[color=\clre,fill=white] (5,3) circle[radius=0.2];
\draw[color=\clre,fill=white] (5,2) circle[radius=0.2];
\draw[color=\clre,fill=white] (5,1) circle[radius=0.2];

\draw[color=\clre,fill=white] (1,1.5) circle[radius=0.2];
\draw[color=\clre,fill=white] (1,2.5) circle[radius=0.2];
\draw[color=\clre,fill=white] (1,3.5) circle[radius=0.2];
\draw[color=\clre,fill=white] (1,4.5) circle[radius=0.2];

\draw[color=\clre,fill=white] (2,1.5) circle[radius=0.2];
\draw[color=\clre,fill=white] (2,2.5) circle[radius=0.2];
\draw[color=\clre,fill=white] (2,3.5) circle[radius=0.2];
\draw[color=\clre,fill=white] (2,4.5) circle[radius=0.2];

\draw[color=\clre,fill=white] (3,1.5) circle[radius=0.2];
\draw[color=\clre,fill=white] (3,2.5) circle[radius=0.2];
\draw[color=\clre,fill=white] (3,3.5) circle[radius=0.2];
\draw[color=\clre,fill=white] (3,4.5) circle[radius=0.2];

\draw[color=\clre,fill=white] (4,1.5) circle[radius=0.2];
\draw[color=\clre,fill=white] (4,2.5) circle[radius=0.2];
\draw[color=\clre,fill=white] (4,3.5) circle[radius=0.2];
\draw[color=\clre,fill=white] (4,4.5) circle[radius=0.2];

\draw[color=\clre,fill=white] (5,1.5) circle[radius=0.2];
\draw[color=\clre,fill=white] (5,2.5) circle[radius=0.2];
\draw[color=\clre,fill=white] (5,3.5) circle[radius=0.2];
\draw[color=\clre,fill=white] (5,4.5) circle[radius=0.2];

\draw[color=\clre,fill=white] (6,5.5) circle[radius=0.2];
\draw[color=\clre,fill=white] (5,5.5) circle[radius=0.2];
\draw[color=\clre,fill=white] (4,5.5) circle[radius=0.2];
\draw[color=\clre,fill=white] (3,5.5) circle[radius=0.2];
\draw[color=\clre,fill=white] (2,5.5) circle[radius=0.2];
\draw[color=\clre,fill=white] (1,5.5) circle[radius=0.2];
\draw[color=\clre,fill=white] (1,6) circle[radius=0.2];
\draw[color=\clre,fill=white] (2,6) circle[radius=0.2];
\draw[color=\clre,fill=white] (3,6) circle[radius=0.2];
\draw[color=\clre,fill=white] (4,6) circle[radius=0.2];
\draw[color=\clre,fill=white] (5,6) circle[radius=0.2];
\draw[color=\clre,fill=white] (6,6) circle[radius=0.2];

\draw[color=\clre,fill=white] (6,1.5) circle[radius=0.2];
\draw[color=\clre,fill=white] (6,2.5) circle[radius=0.2];
\draw[color=\clre,fill=white] (6,3.5) circle[radius=0.2];
\draw[color=\clre,fill=white] (6,4.5) circle[radius=0.2];
\draw[color=\clre,fill=white] (6,5) circle[radius=0.2];
\draw[color=\clre,fill=white] (6,4) circle[radius=0.2];
\draw[color=\clre,fill=white] (6,3) circle[radius=0.2];
\draw[color=\clre,fill=white] (6,2) circle[radius=0.2];
\draw[color=\clre,fill=white] (6,1) circle[radius=0.2];

\draw[color=\clrb, fill=\clra, thick]  (1,2.5) circle[radius=0.21];
\draw[color=\clrb, fill=\clra, thick]   (1,2) circle[radius=0.21];
\draw[color=\clrb, fill=\clra, thick]   (2,2) circle[radius=0.21];
\draw[color=\clrb, fill=\clra, thick]   (3,2) circle[radius=0.21];
\draw[color=\clrb, fill=\clra, thick]   (4,2) circle[radius=0.21];
\draw[color=\clrb, fill=\clra, thick]   (5,2) circle[radius=0.21];
\draw[color=\clrb, fill=\clra, thick]  (5,2.5) circle[radius=0.21];
\draw[color=\clrb, fill=\clra, thick]   (5,3) circle[radius=0.21];
\draw[color=\clrb, fill=\clra, thick]   (5,3.5) circle[radius=0.21];
\draw[color=\clrb, fill=\clra, thick]   (5,4) circle[radius=0.21];
\draw[color=\clrb, fill=\clra, thick]   (5,4.5) circle[radius=0.21];
\draw[color=\clrb, fill=\clra, thick]   (5,5) circle[radius=0.21];
\draw[color=\clrb, fill=\clra, thick]   (5,5.5) circle[radius=0.21];

\node[\clrc] at (1,2.5) {\tiny $\boldsymbol{\mathsf{Z}}$};
\node[\clrc] at (1,2) {\tiny $\boldsymbol{\mathsf{Y}}$};
\node[\clrc] at (2,2) {\tiny $\boldsymbol{\mathsf{Z}}$};
\node[\clrc] at (3,2) {\tiny $\boldsymbol{\mathsf{Z}}$};
\node[\clrc] at (4,2) {\tiny $\boldsymbol{\mathsf{Z}}$};
\node[\clrc] at (5,2) {\tiny $\boldsymbol{\mathsf{X}}$};
\node[\clrc] at (5,2.5) {\tiny $\boldsymbol{\mathsf{X}}$};
\node[\clrc] at (5,3) {\tiny $\boldsymbol{\mathsf{Z}}$};
\node[\clrc] at (5,3.5) {\tiny $\boldsymbol{\mathsf{X}}$};
\node[\clrc] at (5,4) {\tiny $\boldsymbol{\mathsf{Z}}$};
\node[\clrc] at (5,4.5) {\tiny $\boldsymbol{\mathsf{X}}$};
\node[\clrc] at (5,5) {\tiny $\boldsymbol{\mathsf{Z}}$};
\node[\clrc] at (5,5.5) {\tiny $\boldsymbol{\mathsf{Z}}$};

\end{tikzpicture}
\begin{tikzpicture}[scale=0.6]
\node[] at (0.5,6.5) {\textbf{(d)}};
\draw[color=\clre,dashed](1,2).. controls(.4,2.2)and (.4,2.8).. (1,3);
\draw[color=\clre,dashed](6,3).. controls(6.6,3.2)and (6.6,3.8).. (6,4);
\draw[color=\clre,dashed](6,5).. controls(6.6,5.2)and (6.6,5.8).. (6,6);
\draw[color=\clre,dashed](1,4).. controls(.4,4.2)and (.4,4.8).. (1,5);
\draw[color=\clre,dashed](6,1).. controls(6.6,1.2)and (6.6,1.8).. (6,2);
\draw[\clre](1,1)--(6,1);
\draw[\clre](1,2)--(6,2);
\draw[\clre](1,3)--(6,3);
\draw[\clre](1,4)--(6,4);
\draw[\clre](1,5)--(6,5);
\draw[\clre](1,6)--(6,6);

\draw[\clre](1,1)--(1,6);
\draw[\clre](2,1)--(2,6);
\draw[\clre](3,1)--(3,6);
\draw[\clre](4,1)--(4,6);
\draw[\clre](5,1)--(5,6);
\draw[\clre](6,1)--(6,6);

\draw[color=\clre,fill=white] (1,1) circle[radius=0.2];

\draw[color=\clre,fill=white] (2,1) circle[radius=0.2];

\draw[color=\clre,fill=white] (3,1) circle[radius=0.2];
\draw[color=\clre,fill=white] (4,1) circle[radius=0.2];
\draw[color=\clre,fill=white] (1,2) circle[radius=0.2];
\draw[color=\clre,fill=white] (2,2) circle[radius=0.2];

\draw[color=\clre,fill=white] (3,2) circle[radius=0.2];

\draw[color=\clre,fill=white] (4,2) circle[radius=0.2];

\draw[color=\clre,fill=white] (1,3) circle[radius=0.2];
\draw[color=\clre,fill=white] (2,3) circle[radius=0.2];
\draw[color=\clre,fill=white] (3,3) circle[radius=0.2];
\draw[color=\clre,fill=white] (4,3) circle[radius=0.2];
\draw[color=\clre,fill=white] (1,4) circle[radius=0.2];

\draw[color=\clre,fill=white] (2,4) circle[radius=0.2];

\draw[color=\clre,fill=white] (3,4) circle[radius=0.2];

\draw[color=\clre,fill=white] (4,4) circle[radius=0.2];

\draw[color=\clre,fill=white] (1,5) circle[radius=0.2];

\draw[color=\clre,fill=white] (2,5) circle[radius=0.2];

\draw[color=\clre,fill=white] (3,5) circle[radius=0.2];

\draw[color=\clre,fill=white] (4,5) circle[radius=0.2];

\draw[color=\clre,fill=white] (5,5) circle[radius=0.2];
\draw[color=\clre,fill=white] (5,4) circle[radius=0.2];
\draw[color=\clre,fill=white] (5,3) circle[radius=0.2];
\draw[color=\clre,fill=white] (5,2) circle[radius=0.2];
\draw[color=\clre,fill=white] (5,1) circle[radius=0.2];

\draw[color=\clre,fill=white] (1,1.5) circle[radius=0.2];
\draw[color=\clre,fill=white] (1,2.5) circle[radius=0.2];
\draw[color=\clre,fill=white] (1,3.5) circle[radius=0.2];
\draw[color=\clre,fill=white] (1,4.5) circle[radius=0.2];

\draw[color=\clre,fill=white] (2,1.5) circle[radius=0.2];
\draw[color=\clre,fill=white] (2,2.5) circle[radius=0.2];
\draw[color=\clre,fill=white] (2,3.5) circle[radius=0.2];
\draw[color=\clre,fill=white] (2,4.5) circle[radius=0.2];

\draw[color=\clre,fill=white] (3,1.5) circle[radius=0.2];
\draw[color=\clre,fill=white] (3,2.5) circle[radius=0.2];
\draw[color=\clre,fill=white] (3,3.5) circle[radius=0.2];
\draw[color=\clre,fill=white] (3,4.5) circle[radius=0.2];

\draw[color=\clre,fill=white] (4,1.5) circle[radius=0.2];
\draw[color=\clre,fill=white] (4,2.5) circle[radius=0.2];
\draw[color=\clre,fill=white] (4,3.5) circle[radius=0.2];
\draw[color=\clre,fill=white] (4,4.5) circle[radius=0.2];

\draw[color=\clre,fill=white] (5,1.5) circle[radius=0.2];
\draw[color=\clre,fill=white] (5,2.5) circle[radius=0.2];
\draw[color=\clre,fill=white] (5,3.5) circle[radius=0.2];
\draw[color=\clre,fill=white] (5,4.5) circle[radius=0.2];

\draw[color=\clre,fill=white] (6,5.5) circle[radius=0.2];
\draw[color=\clre,fill=white] (5,5.5) circle[radius=0.2];
\draw[color=\clre,fill=white] (4,5.5) circle[radius=0.2];
\draw[color=\clre,fill=white] (3,5.5) circle[radius=0.2];
\draw[color=\clre,fill=white] (2,5.5) circle[radius=0.2];
\draw[color=\clre,fill=white] (1,5.5) circle[radius=0.2];
\draw[color=\clre,fill=white] (1,6) circle[radius=0.2];
\draw[color=\clre,fill=white] (2,6) circle[radius=0.2];
\draw[color=\clre,fill=white] (3,6) circle[radius=0.2];
\draw[color=\clre,fill=white] (4,6) circle[radius=0.2];
\draw[color=\clre,fill=white] (5,6) circle[radius=0.2];
\draw[color=\clre,fill=white] (6,6) circle[radius=0.2];

\draw[color=\clre,fill=white] (6,1.5) circle[radius=0.2];
\draw[color=\clre,fill=white] (6,2.5) circle[radius=0.2];
\draw[color=\clre,fill=white] (6,3.5) circle[radius=0.2];
\draw[color=\clre,fill=white] (6,4.5) circle[radius=0.2];
\draw[color=\clre,fill=white] (6,5) circle[radius=0.2];
\draw[color=\clre,fill=white] (6,4) circle[radius=0.2];
\draw[color=\clre,fill=white] (6,3) circle[radius=0.2];
\draw[color=\clre,fill=white] (6,2) circle[radius=0.2];
\draw[color=\clre,fill=white] (6,1) circle[radius=0.2];
\draw[ thick, \clrd] (1,2.5)--(1,3)--(2,3)--(2,4)--(4,4)--(4,5)--(5,5)--(5,5.5);
\draw[color=\clrb, fill=\clra, thick] (1,2.5) circle[radius=0.215];
\draw[color=\clrb, fill=\clra, thick](1,3) circle[radius=0.215];
\draw[color=\clrb, fill=\clra, thick] (2,3) circle[radius=0.215];
\draw[color=\clrb, fill=\clra, thick]  (2,3.5) circle[radius=0.215];
\draw[color=\clrb, fill=\clra, thick]  (2,4) circle[radius=0.215];
\draw[color=\clrb, fill=\clra, thick]  (3,4) circle[radius=0.215];
\draw[color=\clrb, fill=\clra, thick]  (4,4) circle[radius=0.215];
\draw[color=\clrb, fill=\clra, thick]  (4,4.5) circle[radius=0.215];
\draw[color=\clrb, fill=\clra, thick]  (4,5) circle[radius=0.215];
\draw[color=\clrb, fill=\clra, thick] (5,5) circle[radius=0.215];
\draw[color=\clrb, fill=\clra, thick] (5,5.5) circle[radius=0.215];

\node[\clrc] at (1,2.5) {\tiny $\boldsymbol{\mathsf{Y}}$};
\node[\clrc] at (1,3) {\tiny $\boldsymbol{\mathsf{X}}$};
\node[\clrc] at (2,3) {\tiny $\boldsymbol{\mathsf{X}}$};
\node[\clrc] at (2,3.5) {\tiny $\boldsymbol{\mathsf{X}}$};
\node[\clrc] at (2,4) {\tiny $\boldsymbol{\mathsf{X}}$};
\node[\clrc] at (3,4) {\tiny $\boldsymbol{\mathsf{Z}}$};
\node[\clrc] at (4,4) {\tiny $\boldsymbol{\mathsf{X}}$};
\node[\clrc] at (4,4.5) {\tiny $\boldsymbol{\mathsf{X}}$};
\node[\clrc] at (4,5) {\tiny $\boldsymbol{\mathsf{X}}$};
\node[\clrc] at (5,5) {\tiny $\boldsymbol{\mathsf{X}}$};
\node[\clrc] at (5,5.5) {\tiny $\boldsymbol{\mathsf{Z}}$};

\end{tikzpicture}
\caption{A hopping term $h_\data=(X \otimes Z \otimes \cdots \otimes Z \otimes X)$ spanning several rows and columns in the  square lattice. \textbf{(a)} Adjusted term $(h_\data \otimes \kappa_\aux^h)$, not yet multiplied with any stabilizer. Note that this string is not connected on the lattice, and the windings on which the string is disconnected are highlighted. \textbf{(b)-(d)} Pauli strings $\widetilde{h}_{\auxdata}$ that are equivalent to $(h_\data \otimes \kappa_\aux^h)$ by multiplication with stabilizers. All those strings are continuous on the connectivity graph. The strings in (b) and (d) have the same weight  (and the string in (c) is just slightly longer) which is determined by the Manhattan distance of the string endpoints.} \label{fig:manhattan}
\end{figure}
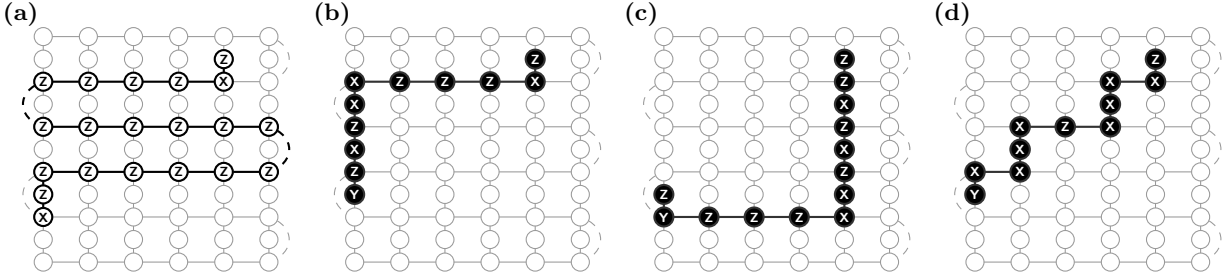

 Our main result, the square lattice AQM, is a mapping that requires a square lattice connectivity graph of $\ell_1 \times (2\ell_2-1)$ qubits for a $(\ell_1 \times \ell_2)$ fermionic lattice. With the large amount of $\ell_1 (\ell_2-1)$ qubits added, we make sure that the code space  can be initialized in $O(\ell_1)$ time steps; a time frame that is better than linear in the total number of data qubits. In the resulting mapping, we will be able to reroute and deform Pauli strings, such that  strings originating from hopping terms have an operator weight of the order of the  Manhattan distance between the two qubits on the lattice. The implication of this mapping for lattice Hamiltonians is that vertical hopping terms have a constant weight, and the algorithmic depth required to simulate such  a model (after the stabilizer state is prepared) is constant, i.e. independent of the lattice dimension.   \\
Before we start describing the mapping, we want to introduce  some helpful notation concerning qubit labeling. For the sake of a geometric interpretation, we will migrate to a geometric labeling, where each qubit index denotes its coordinate  on a grid. In the following, qubits in the data register will bear labels $(i, \, j) \in [\ell_1]\otimes [\ell_2]$, so each data qubit sits on integer positions of a grid and the qubit in the south-west corner of the block has coordinate $(1,\, 1)$. Beginning from that very qubit, the index of each qubit  is given  according to the canonical order of the S-pattern in Figure \ref{fig:Stype}. \\
We will now describe the placement of the auxiliary qubits on the lattice.
The idea of the square lattice AQM is to insert auxiliary qubits in between data qubits of different rows, so in between $(i,\,j)$ and $(i,\,j+1)$ into  half-integer positions $(i, \, j + \frac{1}{2})$, in order to cancel the parity strings in between those qubits. However, we also want the $p$-strings  to have (anti-)commutation relations like Majorana-pair operators.  This is an integral ingredient to avoid long  adjustments substrings $\kappa_{\aux}^h$. To that end, we use a Hadamard-basis Auxiliary Qubit code with stabilizers
\begin{align}
\label{eq:stabsquare0}
 p^{(i,\, j+\frac{1}{2})}_{\data}\otimes X_{(i,\, j+\frac{1}{2})} \, ,
\end{align}
 which act on the data qubits at  $(i,\, j)$ and $(i, \,j+1)$ as $X$- or $Y$-operators and as $Z$-operators on all other data qubits  along the S-pattern in between them.  The position of the auxiliary qubits and the choice of stabilizers can be seen in Figure \ref{fig:stabsquare}.
Note that it is unnecessary for the auxiliary qubits to be connected to each other in the horizontal direction, although it might come in handy in the process of initializing the code space.
 As indicated in the figure, the Pauli terms on  $(i, \, j)$ and $(i, \, j+1)$ in  the stabilizers of qubits $(i,\,j+\frac{1}{2})$ are different for even and odd rows numbers $j$. The sole reason for this decision is to render both terms of the vertical hopping terms  with real coefficients \eqref{eq:hopping} of the same weight. For every vertical connection $(i, \, j + \frac{1}{2})$, the $p$-substrings of the stabilizers \eqref{eq:stabsquare0} are defined as:

 \begin{align}
 \label{eq:stabsqarestabs1}
p^{(i, \, j+\frac{1}{2})}_\data \;&= \; \left(\bigotimes_{k=i+1}^{\ell_1} Z_{(k,\, j)} \right) \left(\bigotimes_{l=\ell_1}^{i+1}  Z_{(l,\,j+1)}\right) \otimes Y_{(i, \,j)} \otimes X_{(i, \, j+1)} \, , \quad \text{for odd $j$, } \\  \label{eq:stabsqarestabs2}
 &= \; \left(\bigotimes_{k=i-1}^{1} Z_{(k, \, j)} \right) \left(\bigotimes_{l=1}^{i-1}  Z_{(l,\,j+1)}\right)\otimes   X_{(i, \, j)} \otimes Y_{(i, \, j+1)} \, , \quad \text{for even $j$. }
\end{align}
Now we are going to give instructions on how to initialize the state $\ket{\widetilde{\varphi}}$ within $O(\ell_1)$ depth, starting from a disentangled state $\ket{\varphi}_{\data}\otimes \ket{0^r}_{\aux}$. First we apply Hadamard gates on all auxiliary qubits.
In all rows with odd [even] row numbers $j$, we then simultaneously apply the strings $(Y_{(\ell_1,\, j)} \otimes X_{(\ell_1,\, j+1)})$ $\left[(X_{(1,\, j)} \otimes Y_{(1,\, j+1)}) \right]$ conditional on the qubit at $(\ell_1, \, j+\frac{1}{2})$ $\left[(1, \, j+\frac{1}{2})\right]$.
Entangling these auxiliaries is easy as the stabilizers are at the windings and therefore local, the operation can be performed in $O(1)$ time steps.
We  then proceed by applying the strings
\begin{align} \notag
X_{(\ell_1-s+1,\, j)} \otimes Y_{(\ell_1-s+1,\, j+1)} \otimes  Y_{(\ell_1-s,\, j)} \otimes  X_{(\ell_1-s+1, \, j+\frac{1}{2})}\otimes X_{(\ell_1-s,\, j+1)} \\ \left[  Y_{(s,\, j)} \otimes X_{(s,\, j+1)} \otimes  X_{(s+1,\, j)} \otimes  X_{(s, \, j+\frac{1}{2})}\otimes Y_{(s+1,\, j+1)} \right] \label{eq:sequence}
\end{align}
conditionally on the qubits $(\ell_1-s,\, j+\frac{1}{2})$ $[(s+1, \, j+\frac{1}{2})]$. We do this  sequentially from  $s=1$ to $s=(\ell_1-1)$, which means we require $O(\ell_1)$ time steps in total. This concludes the definition $V_\auxdata$, as can be verified considering its formal definition in Appendix \ref{sec:B}, and  where we use that \eqref{eq:sequence} is obtained from the multiplication of a $p$-string with the closest stabilizer.  A  measurement-based approach for state preparation is discussed in Section \ref{sec:five}.\\
We are now going to describe the logical operators of the code space defined.  In Figure  \ref{fig:manhattan}(a), the adjusted term $\widetilde{h}_\auxdata$ to a string $h_\data=(X \otimes Z \otimes \cdots \otimes Z \otimes X)$ is presented. One can show, either directly from definitions of strings like $h_\data$ with $p^{(i,  \, j + \frac{1}{2})}_\data$ or by relations between Majorana-pair operators,  that for Pauli strings originating from hopping terms \eqref{eq:hopping} between two sites $(i,j)$ and  $(k,\,l)$, it is sufficient to check  for adjustments on only the auxiliary qubits at $(i,\,j\pm \frac{1}{2})$ and $(k,\,l\pm \frac{1}{2})$. If $j$ and $l$ are different rows, it follows that the string is not continuous, see Figure \ref{fig:manhattan}(a). We then choose  to multiply the adjusted term with the stabilizers  involving the auxiliary qubits on which we wish the string to cross rows. For vertical hoppings of lattice Hamiltonians, this choice is trivial. For arbitrary hoppings however it is not. Considering that we likely have several such terms inside one Hamiltonian,  we want commuting strings not to overlap so we would deform them (by multiplying other stabilizers) to go around each other. This allows us to simulate them in parallel. In Figure \ref{fig:manhattan}, panels (b)-(d), different paths have been chosen for the logical operator $\widetilde{h}_\auxdata$ to run along. Only deformed by the multiplication of stabilizers, all of those choices are in fact  equivalent.  Note that taking a direct path, the resulting strings will always be of  roughly the same length, as every direct path connecting two nodes on a square lattice has the same distance: the Manhattan distance.  \\ In the following, we will generalize this mapping to yield an AQM-version that requires fewer auxiliary qubits.

\subsection{Sparse AQM}
The sparse AQM is  a modification of the  square lattice AQM that allows us to make a trade-off between the number of auxiliary qubits required and the locality in the resulting strings.  The latter directly influences the performance of any quantum simulation algorithm.

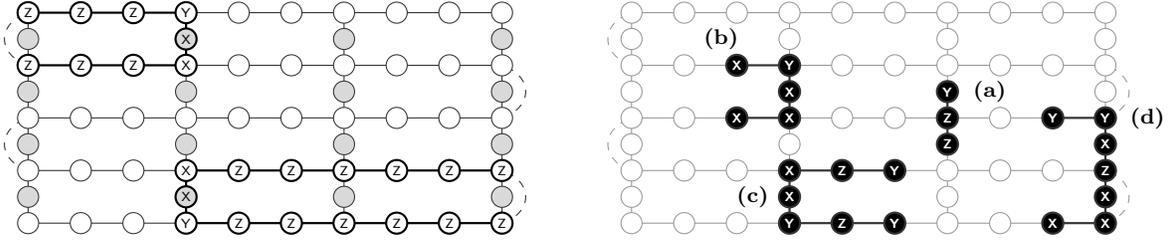
\begin{figure}[h!]

\begin{tikzpicture}[scale=0.7,baseline=0]

\draw[\fadedblack] (0,0)--(0,4);
\draw[\fadedblack] (9,0)--(9,4);
\draw[\fadedblack,dashed](9,0).. controls(9.6,.2)and (9.6,.8).. (9,1);
\draw[\fadedblack,dashed](0,1).. controls(-.6,1.2)and (-.6,1.8).. (0,2);
\draw[\fadedblack,dashed](9,2).. controls(9.6,2.2)and (9.6,2.8).. (9,3);
\draw[\fadedblack,dashed](0,3).. controls(-.6,3.2)and (-.6,3.8).. (0,4);
\draw[\fadedblack] (0,0)--(9,0);
\draw[\fadedblack] (0,1)--(9,1);
\draw[\fadedblack] (0,2)--(9,2);
\draw[\fadedblack] (0,3)--(9,3);
\draw[\fadedblack] (0,4)--(9,4);
\draw[\fadedblack] (3,0)--(3,4);
\draw[\fadedblack] (6,0)--(6,4);
\draw[\fadedblack,fill=\checker] (0,1.5) circle[radius=0.2];
\draw[\fadedblack,fill=\checker] (9,0.5) circle[radius=0.2];
\draw[\fadedblack,fill=\checker] (0,3.5) circle[radius=0.2];
\draw[\fadedblack,fill=\checker] (9,2.5) circle[radius=0.2];
\draw[\fadedblack,fill=white] (1,0) circle[radius=0.2];
\draw[\fadedblack,fill=white] (2,0) circle[radius=0.2];
\draw[thick] (9,0)--(3,0)--(3,1)--(9,1);
\draw[thick] (0,3)--(3,3)--(3,4)--(0,4);
\draw[thick, fill=white] (3,0) circle[radius=0.2];
\draw[thick, fill=white] (4,0) circle[radius=0.2];
\draw[thick, fill=white] (5,0) circle[radius=0.2];
\draw[thick, fill=white] (6,0) circle[radius=0.2];
\draw[thick, fill=white] (7,0) circle[radius=0.2];
\draw[thick, fill=white] (8,0) circle[radius=0.2];

\draw[\fadedblack,fill=white] (1,1) circle[radius=0.2];
\draw[\fadedblack,fill=white] (2,1) circle[radius=0.2];
\draw[thick, fill=white] (3,1) circle[radius=0.2];
\draw[thick, fill=white] (4,1) circle[radius=0.2];
\draw[thick, fill=white] (5,1) circle[radius=0.2];
\draw[thick, fill=white] (6,1) circle[radius=0.2];
\draw[thick, fill=white] (7,1) circle[radius=0.2];
\draw[thick, fill=white] (8,1) circle[radius=0.2];

\draw[\fadedblack,fill=white] (1,2) circle[radius=0.2];
\draw[\fadedblack,fill=white] (2,2) circle[radius=0.2];
\draw[\fadedblack,fill=white] (3,2) circle[radius=0.2];
\draw[\fadedblack,fill=white] (4,2) circle[radius=0.2];
\draw[\fadedblack,fill=white] (5,2) circle[radius=0.2];
\draw[\fadedblack,fill=white] (6,2) circle[radius=0.2];
\draw[\fadedblack,fill=white] (7,2) circle[radius=0.2];
\draw[\fadedblack,fill=white] (8,2) circle[radius=0.2];

\draw[thick, fill=white] (1,3) circle[radius=0.2];
\draw[thick, fill=white] (2,3) circle[radius=0.2];
\draw[thick, fill=white] (3,3) circle[radius=0.2];
\draw[\fadedblack,fill=white] (4,3) circle[radius=0.2];
\draw[\fadedblack,fill=white] (5,3) circle[radius=0.2];
\draw[\fadedblack,fill=white] (6,3) circle[radius=0.2];
\draw[\fadedblack,fill=white] (7,3) circle[radius=0.2];
\draw[\fadedblack,fill=white] (8,3) circle[radius=0.2];

\draw[thick, fill=white] (1,4) circle[radius=0.2];
\draw[thick, fill=white] (2,4) circle[radius=0.2];
\draw[thick, fill=white] (3,4) circle[radius=0.2];
\draw[\fadedblack,fill=white] (4,4) circle[radius=0.2];
\draw[\fadedblack,fill=white] (5,4) circle[radius=0.2];
\draw[\fadedblack,fill=white] (6,4) circle[radius=0.2];
\draw[\fadedblack,fill=white] (7,4) circle[radius=0.2];
\draw[\fadedblack,fill=white] (8,4) circle[radius=0.2];

\draw[thick, fill=\checker] (3,0.5) circle[radius=0.2];
\draw[\fadedblack,fill=\checker] (3,1.5) circle[radius=0.2];
\draw[\fadedblack,fill=\checker] (3,2.5) circle[radius=0.2];
\draw[thick, fill=\checker] (3,3.5) circle[radius=0.2];

\draw[\fadedblack,fill=\checker] (6,0.5) circle[radius=0.2];
\draw[\fadedblack,fill=\checker] (6,1.5) circle[radius=0.2];
\draw[\fadedblack,fill=\checker] (6,2.5) circle[radius=0.2];
\draw[\fadedblack,fill=\checker] (6,3.5) circle[radius=0.2];

\draw[\fadedblack,fill=\checker] (0,0.5) circle[radius=0.2];
\draw[\fadedblack,fill=\checker] (0,2.5) circle[radius=0.2];
\draw[\fadedblack,fill=\checker] (9,1.5) circle[radius=0.2];
\draw[\fadedblack,fill=\checker] (9,3.5) circle[radius=0.2];

\draw[thick,fill=white] (9,0) circle[radius=0.2];
\draw[thick,fill=white] (9,1) circle[radius=0.2];
\draw[\fadedblack,fill=white] (9,2) circle[radius=0.2];
\draw[\fadedblack,fill=white] (9,3) circle[radius=0.2];
\draw[\fadedblack,fill=white] (9,4) circle[radius=0.2];
\draw[\fadedblack,fill=white] (0,0) circle[radius=0.2];
\draw[\fadedblack,fill=white] (0,1) circle[radius=0.2];
\draw[\fadedblack,fill=white] (0,2) circle[radius=0.2];
\draw[thick,fill=white] (0,3) circle[radius=0.2];
\draw[thick,fill=white] (0,4) circle[radius=0.2];
\node[] at (3,1) {\tiny $\mathsf{X}$};
\node[] at (4,1) {\tiny $\mathsf{Z}$};
\node[] at (5,1) {\tiny $\mathsf{Z}$};
\node[] at (6,1) {\tiny $\mathsf{Z}$};
\node[] at (7,1) {\tiny $\mathsf{Z}$};
\node[] at (8,1) {\tiny $\mathsf{Z}$};
\node[] at (3,0) {\tiny $\mathsf{Y}$};
\node[] at (4,0) {\tiny $\mathsf{Z}$};
\node[] at (5,0) {\tiny $\mathsf{Z}$};
\node[] at (6,0) {\tiny $\mathsf{Z}$};
\node[] at (7,0) {\tiny $\mathsf{Z}$};
\node[] at (8,0) {\tiny $\mathsf{Z}$};
\node[] at (3,0.5) {\tiny $\mathsf{X}$};

\node[] at (1,4) {\tiny $\mathsf{Z}$};
\node[] at (2,4) {\tiny $\mathsf{Z}$};
\node[] at (3,4) {\tiny $\mathsf{Y}$};
\node[] at (1,3) {\tiny $\mathsf{Z}$};
\node[] at (2,3) {\tiny $\mathsf{Z}$};
\node[] at (3,3) {\tiny $\mathsf{X}$};
\node[] at (3,3.5) {\tiny $\mathsf{X}$};
\node[] at (9,0) {\tiny $\mathsf{Z}$};
\node[] at (9,1) {\tiny $\mathsf{Z}$};
\node[] at (0,4) {\tiny $\mathsf{Z}$};
\node[] at (0,3) {\tiny $\mathsf{Z}$};
\end{tikzpicture} $\qquad$ \begin{tikzpicture}[scale=0.7,baseline=0]
\draw[\clre, dashed](9,0).. controls(9.6,.2)and (9.6,.8).. (9,1);
\draw[color=\clre, dashed](0,1).. controls(-.6,1.2)and (-.6,1.8).. (0,2);
\draw[color=\clre, dashed](9,2).. controls(9.6,2.2)and (9.6,2.8).. (9,3);
\draw[color=\clre, dashed](0,3).. controls(-.6,3.2)and (-.6,3.8).. (0,4);
\draw[color=\clre] (0,0)--(9,0);
\draw[color=\clre] (0,1)--(9,1);
\draw[color=\clre] (0,2)--(9,2);
\draw[color=\clre] (0,3)--(9,3);
\draw[color=\clre] (0,4)--(9,4);
\draw[color=\clre] (3,0)--(3,4);
\draw[color=\clre] (6,0)--(6,4);
\draw[\clre] (0,0)--(0,4);
\draw[\clre] (9,0)--(9,4);
\draw[\clrd,thick, ](2,3)--(3,3)--(3,2)--(2,2);
\draw[\clrd,thick, ](6,2.5)--(6,1.5);
\draw[\clrd,thick, ](8,0)--(9,0)--(9,2)--(8,2);
\draw[\clrd,thick, ](5,1)--(3,1)--(3,0)--(5,0);
\draw[color=\clre,fill=white] (1,0) circle[radius=0.2];
\draw[color=\clre,fill=white] (2,0) circle[radius=0.2];
\draw[\clre,fill=white] (0,1.5) circle[radius=0.2];
\draw[color=\clrb, fill=\clra, thick] (9,0.5) circle[radius=0.2];
\draw[\clre,fill=white] (0,3.5) circle[radius=0.2];
\draw[\clre,fill=white] (9,2.5) circle[radius=0.2];
\draw[color=\clrb, fill=\clra, thick] (3,0) circle[radius=0.2];
\node[\clrc] at (3,0) {\tiny $\boldsymbol{\mathsf{Y}}$};
\draw[color=\clrb, fill=\clra, thick] (4,0) circle[radius=0.2];
\node[\clrc] at (4,0) {\tiny $\boldsymbol{\mathsf{Z}}$};
\draw[color=\clrb, fill=\clra, thick] (5,0) circle[radius=0.2];
\node[\clrc] at (5,0) {\tiny $\boldsymbol{\mathsf{Y}}$};
\draw[color=\clre,fill=white] (6,0) circle[radius=0.2];
\draw[color=\clre,fill=white] (7,0) circle[radius=0.2];
\draw[color=\clrb, fill=\clra, thick] (9,0) circle[radius=0.2];
\node[\clrc] at (9,0) {\tiny $\boldsymbol{\mathsf{X}}$};

\draw[color=\clre,fill=white] (1,1) circle[radius=0.2];
\draw[color=\clre,fill=white] (2,1) circle[radius=0.2];
\draw[color=\clrb, fill=\clra, thick] (3,1) circle[radius=0.2];
\node[\clrc] at (3,1) {\tiny $\boldsymbol{\mathsf{X}}$};
\draw[color=\clrb, fill=\clra, thick] (4,1) circle[radius=0.2];
\node[\clrc] at (4,1) {\tiny $\boldsymbol{\mathsf{Z}}$};
\draw[color=\clrb, fill=\clra, thick] (5,1) circle[radius=0.2];
\node[\clrc] at (5,1) {\tiny $\boldsymbol{\mathsf{Y}}$};
\draw[color=\clre,fill=white] (6,1) circle[radius=0.2];
\draw[color=\clre,fill=white] (7,1) circle[radius=0.2];
\draw[color=\clrb, fill=\clra, thick] (9,1) circle[radius=0.2];
\node[\clrc] at (9,1) {\tiny $\boldsymbol{\mathsf{Z}}$};

\draw[color=\clre,fill=white] (1,2) circle[radius=0.2];
\draw[color=\clrb, fill=\clra, thick] (2,2) circle[radius=0.2];
\node[\clrc] at (2,2) {\tiny $\boldsymbol{\mathsf{X}}$};
\draw[color=\clrb, fill=\clra, thick] (3,2) circle[radius=0.2];
\node[\clrc] at (3,2) {\tiny $\boldsymbol{\mathsf{X}}$};
\draw[color=\clre,fill=white] (4,2) circle[radius=0.2];
\draw[color=\clre,fill=white] (5,2) circle[radius=0.2];
\draw[color=\clrb, fill=\clra, thick] (6,2) circle[radius=0.2];
\node[\clrc] at (6,2) {\tiny $\boldsymbol{\mathsf{Z}}$};
\draw[color=\clre,fill=white] (7,2) circle[radius=0.2];

\draw[color=\clrb, fill=\clra, thick]  (8,2) circle[radius=0.2];
\node[\clrc] at (8,2) {\tiny $\boldsymbol{\mathsf{Y}}$};

\draw[color=\clre,fill=white] (1,3) circle[radius=0.2];
\draw[color=\clrb, fill=\clra, thick] (2,3) circle[radius=0.2];
\node[\clrc] at (2,3) {\tiny $\boldsymbol{\mathsf{X}}$};
\draw[color=\clrb, fill=\clra, thick] (3,3) circle[radius=0.2];
\node[\clrc] at (3,3) {\tiny $\boldsymbol{\mathsf{Y}}$};
\draw[color=\clre,fill=white] (4,3) circle[radius=0.2];
\draw[color=\clre,fill=white] (5,3) circle[radius=0.2];
\draw[color=\clre,fill=white] (6,3) circle[radius=0.2];

\draw[color=\clre,fill=white] (7,3) circle[radius=0.2];

\draw[color=\clre,fill=white] (8,3) circle[radius=0.2];

\draw[color=\clre,fill=white] (1,4) circle[radius=0.2];
\draw[color=\clre,fill=white] (2,4) circle[radius=0.2];
\draw[color=\clre,fill=white] (3,4) circle[radius=0.2];
\draw[color=\clre,fill=white] (4,4) circle[radius=0.2];
\draw[color=\clre,fill=white] (5,4) circle[radius=0.2];
\draw[color=\clre,fill=white] (6,4) circle[radius=0.2];
\draw[color=\clre,fill=white] (7,4) circle[radius=0.2];
\draw[color=\clre,fill=white] (8,4) circle[radius=0.2];

\draw[color=\clrb, fill=\clra, thick] (3,0.5) circle[radius=0.2];
\node[\clrc] at (3,0.5) {\tiny $\boldsymbol{\mathsf{X}}$};
\draw[color=\clre,fill=white] (3,1.5) circle[radius=0.2];
\draw[color=\clrb, fill=\clra, thick] (3,2.5) circle[radius=0.2];
\node[\clrc] at (3,2.5) {\tiny $\boldsymbol{\mathsf{X}}$};
\draw[color=\clre,fill=white] (3,3.5) circle[radius=0.2];

\draw[color=\clre,fill=white] (6,0.5) circle[radius=0.2];
\draw[color=\clrb, fill=\clra, thick] (6,1.5) circle[radius=0.2];
\node[\clrc] at (6,1.5) {\tiny $\boldsymbol{\mathsf{Z}}$};
\draw[color=\clrb, fill=\clra, thick] (6,2.5) circle[radius=0.2];
\node[\clrc] at (6,2.5) {\tiny $\boldsymbol{\mathsf{Y}}$};
\draw[color=\clre, fill=white] (6,3.5) circle[radius=0.2];

\draw[color=\clre,fill=white] (0,0) circle[radius=0.2];
\draw[color=\clre,fill=white] (0,1) circle[radius=0.2];
\draw[color=\clre,fill=white] (0,2) circle[radius=0.2];
\draw[color=\clre,fill=white] (0,3) circle[radius=0.2];
\draw[color=\clre,fill=white] (0,4) circle[radius=0.2];

\draw[color=\clrb, fill=\clra, thick] (8,0) circle[radius=0.2];
\draw[color=\clre,fill=white] (8,1) circle[radius=0.2];
\draw[color=\clrb, fill=\clra, thick]  (9,2) circle[radius=0.2];
\draw[color=\clre,fill=white] (9,3) circle[radius=0.2];
\draw[color=\clre,fill=white] (9,4) circle[radius=0.2];

\draw[\clre,fill=white] (0,0.5) circle[radius=0.2];
\draw[\clre,fill=white](0,2.5) circle[radius=0.2];
\draw[color=\clrb, fill=\clra, thick](9,1.5) circle[radius=0.2];
\draw[\clre,fill=white] (9,3.5) circle[radius=0.2];

\node[] at (2.3,0.5){\footnotesize \textbf{(c)}};
\node[] at (1.7,3.5){\footnotesize \textbf{(b)}};
\node[] at (6.8,2.5){\footnotesize \textbf{(a)}};
\node[] at (9.8,2){\footnotesize \textbf{(d)}};
\node[\clrc] at (9,0.5) {\tiny $\boldsymbol{\mathsf{X}}$};
\node[\clrc] at (9,1.5) {\tiny $\boldsymbol{\mathsf{X}}$};
\node[\clrc] at (9,2) {\tiny $\boldsymbol{\mathsf{Y}}$};
\node[\clrc] at (8,0) {\tiny $\boldsymbol{\mathsf{X}}$};
\end{tikzpicture}
\caption{Sparse AQM with a periodicity of three $(\inter=3)$. \textbf{Left:} Structure and stabilizers. The gray qubits are auxiliaries, placed sparsely on half-integer positions, connecting different rows. We depict one of the stabilizers in an odd and an even row, respectively. \textbf{Right:} Mappings of different $h_\data=(X\otimes Z\otimes \dots\otimes Z\otimes X)$ strings originating from vertical hoppings. \textbf{(a)} A vertical hopping along a vertical connection. The mapping yields the same $(Z \otimes Z \otimes Y)$-string as we would expect from the square lattice AQM.  \textbf{(b)}  The string is connecting $(3, \, 3)$ and $(3, \,4)$. This example shows the virtue of the sparse AQM: the parity string takes a shortcut along the closest vertical connection. \textbf{(c)} Here we connect the qubits on $(6,\, 1)$ and $(6,\, 2)$ from the other direction: over the vertical connection between $(4, \,1)$ and $(4, \, 2)$.  \textbf{(d)} A next-nearest-neighbor vertical hopping term between $(9,\, 1)$ and $(9, \, 3)$. }\label{fig:dollar}
\end{figure}

In the square lattice AQM, each data qubit (of the interior) has two non-local connections in the vertical direction. This can be regarded as quite wasteful, as a mapping with fewer vertical connections would work in the same way while effectively reducing the number of auxiliary qubits. Here we introduce the sparse AQM, in which vertical connections have a certain distance  from each other. Let us say vertical connections are always placed  $\inter$ qubits apart. The periodicity $\inter$ thus becomes a parameter of the mapping and is generally an integer number $\inter \in [\ell_1-1]$, where the case $\inter=1$  reproduces the square lattice AQM. We have excluded the case in which we have only one vertical connection between every pair of rows, as it is covered  by the E-type AQM already. For convenience let us say that $(\ell_1 - 1)/\inter$ is an integer such that we can place vertical connections at the right and left boundary of the grid without spacing unequally.  The connectivity graph that puts auxiliary qubits on half integer positions along $\inter$-spaced columns can be seen in Figure \ref{fig:dollar}(a), along with the typical stabilizers. In this mapping the auxiliary register holds $r=(\ell_2-1)\cdot(\frac{\ell_1-1}{\inter}+1)$ qubits, which is somewhere in between the square lattice and E-type AQM. For the initialization circuit,   $V_\auxdata$, the sequence \eqref{eq:sequence} has to be changed into applying the strings
\begin{align} \notag
\left(X_{(\ell_1-s+\inter,\, j+ \frac{1}{2})}  \otimes p^{(\ell_1-s+\inter,\, j+ \frac{1}{2})}_\data \right) \cdot p^{(\ell_1-s,\, j+ \frac{1}{2})}_\data  \\ \left[  \left(X_{(s+1-\inter,\, j+ \frac{1}{2})}  \otimes p^{(s+1-\inter,\, j+ \frac{1}{2})}_\data \right) \cdot p^{(s+1,\, j+ \frac{1}{2})}_\data \right]
\end{align}
conditionally on qubits $(\ell_1-s,\, j+ \frac{1}{2})$ $[(s+1,\, j+ \frac{1}{2})]$ for $s= \inter, \, 2 \inter, \, 3\inter, \, \dots \, , \,  \ell_1-1$. All those strings in the sequence are of weight $O(\inter)$, but there are just $(\ell_1-1)/\inter$ of them, which brings the depth of the entire circuit to $O(\ell_1)$. \\
Figure \ref{fig:dollar}(b) shows some output strings of this mapping. While crossing rows works like in the square lattice AQM, the sparsity of vertical connections makes for a more limited choice on where the strings can run along.  As a consequence, hopping terms between modes with a horizontal distance smaller than $\inter$ will transform into strings like in the E-type mapping. The effect of  sparsity on simulations of a lattice model is discussed in the following section. \\

 Note that we have made two arbitrary design choices for the connectivity graph of this mapping: firstly, we have chosen  for the auxiliary qubits to be situated in between rows of data qubits. In order to fit this mapping to a compact square lattice, we can take the auxiliary qubits from in between the rows and insert them into the rows, so e.g.~take them from $(i,\, j+\frac{1}{2})$ and  insert them at $(i+\frac{1}{2}, \, j)$. Then, the auxiliaries have to be connected to the data qubits $(i, \,j)$ and $(i+1,\, j)$, as well as the auxiliary qubits at $(i+\frac{1}{2}, \, j \pm 1)$. In the end, no qubits will be in the spaces between rows - this makes the array more dense and we can map it to a square lattice, but also requires us to skip auxiliary qubits in some horizontal hopping strings. Secondly, we have decided to place auxiliary qubits inside the same column of every other vertical connection. Alternatively, the vertical connections could be arranged in a  brickwork pattern in order to minimize the weight of the adjustments $\kappa_\aux^h$, but then vertical connections along a straight line are no longer possible.

\section{Example: Fermi-Hubbard lattice model}
\label{sec:four}

\subsection{Second quantization and Jordan-Wigner transform}
Here we demonstrate the use of  AQMs on the Fermi-Hubbard model. In this model, we describe spin-$\frac{1}{2}$ Fermions hopping on a square lattice, with a repulsion term whenever spin-up and -down particles are present on the same site. In the following, we will describe the Hamiltonian in both, second quantization and in terms of Pauli strings after Jordan-Wigner transform. Investigating the shortcomings of this mapping with respect to circuit depth will be the motivation for the application of AQMs in the next step.
Let us consider an $(L\times L)$-site square lattice of spatial sites populated by spin-$(1/2)$ Fermions: as every such site hosts a spin-up and -down mode, a total of $N=2L^2$ qubits are minimally required. For convenience, the spin-up and -down modes of the fermionic site with the physical location $(x, \, y)$ shall be placed at the coordinates $(2x, \, y)$ and $(2x-1, \, y)$ in the two-dimensional embedding. This means the spin-partners are horizontal neighbors, which is advantageous for the
Jordan-Wigner transform (and square lattice AQM). The Fermi-Hubbard Hamiltonian is defined as
\begin{align}
\label{eq:fermihubbard0}
&\overbrace{\sum_{(i,j)}\left(  t^{\leftrightarrow}_{ij}\; c^{\dagger}_{(i, \, j)}c^{\phantom{\dagger}}_{(i+2,\,j)}  + \text{h.c.} \right)}^{\text{horizontal hoppings} }
+  \overbrace{\sum_{(i, \, j)}\left( t^{\updownarrow}_{ij}\; c^{\dagger}_{(i, \, j)}c^{\phantom{\dagger}}_{(i, \, j+1)}  + \text{h.c.} \right)}^{\text{vertical hoppings}}\notag\\ +\;&
\underbrace{\sum_{(i, \, j)}\;\epsilon_{ij} \;c^{\dagger}_{(i, \, j)}c^{\phantom{\dagger}}_{(i, \, j)}}_{\text{on-site detunings}}
+\underbrace{\sum_{(2i, \, j)}\;U_{ij} \;c^{\dagger}_{(2i,\,j)}c^{\phantom{\dagger}}_{(2i,\,j)}c^{\dagger}_{(2i-1,\,j)}c^{\phantom{\dagger}}_{(2i-1,\,j)}}_{\text{Hubbard interactions}} \, ,
\end{align}
where $ t^{\leftrightarrow}_{ij}$,  $t^{\updownarrow}_{ij}$,  $\epsilon_{ij}$ and $U_{ij}$ are real parameters. In this particular example sums run over all possible coordinates $(i, \, j)$, $(2i,\,j)$ respectively, but implement open boundary conditions. With an S-pattern Jordan-Wigner transform, the Hamiltonian can now be mapped onto an $(2L\times L)$ square lattice of qubits:
 \begin{align}
 \label{eq:fermihubbard}
 H=&\sum_{(i,\, j) }\frac{t^{\leftrightarrow}_{ij}}{2} \left( X_{(i, \,j)} \otimes Z_{(i+1, \,j)}\otimes X_{(i+2, \, j)}+ Y_{(i, \, j)}\otimes Z_{(i+1, \, j)}\otimes Y_{(i+2, \, j)} \right) \notag \\
 +&\sum_{(i, \, j), \,\text{odd} \,j} \frac{t^{\updownarrow}_{ij}}{2} \left(\bigotimes_{k=i+1}^{2L} Z_{(k, \, j)} \right) \left(\bigotimes_{l=2L}^{i+1}  Z_{(l, \, j+1)}\right) \left( X_{(i, \, j)} \otimes X_{(i, \, j+1)}+ Y_{(i, \, j)}\otimes  Y_{(i, \, j+1)} \right) \notag \\
  +&\sum_{(i, \, j), \,\text{even} \,j} \frac{t^{\updownarrow}_{ij}}{2} \left(\bigotimes_{k=i-1}^{1} Z_{(k, \, j)} \right) \left(\bigotimes_{l=1}^{i-1}  Z_{(l, \, j+1)}\right) \left( X_{(i, \, j)} \otimes X_{(i, \, j+1)}+ Y_{(i, \, j)}\otimes  Y_{(i, \, j+1)} \right) \notag \\
 +&\sum_{(i, \, j)} \frac{\epsilon_{ij}}{2} \left( \mathbb{I}-Z_{(i, \, j)}\right)
 +\sum_{(2i, \, j)} \frac{U_{ij}}{4} \left( \mathbb{I}-Z_{(2i, \, j)}\right)\left( \mathbb{I}-Z_{(2i-1, \, j)}\right)\, .
 \end{align}
 Let us discuss the terms of this Hamiltonian, and finally arrive at the shortcomings of the mapping applied. We note that the vertical hopping terms are different with respect to even and odd columns, due to different directions of the S-pattern. All  terms but the vertical hoppings have a constant weight and can be simulated in $O(1)$ time: only the latter  can assume a length of up to $4L$. Unfortunately, we  have $O(L)$ terms of weight $O(L)$ per row pair. Although these strings commute, they do overlap, which means we cannot simulate them in parallel: if no cancellations are possible, the entire algorithm has an algorithmic depth of  $O(L^2)$, so it scales with the lattice area. In this case the simulation time and the gate count cannot be better than being proportional to the total number of qubits, which renders increasing lattice size expensive. If the simulation algorithm allows us to cancel substrings of consecutively simulated Pauli strings (see for instance \cite{hastings2014improving}), the algorithmic depth can improve to up to $O(L)$. To achieve even better scalings, we will employ the square lattice  AQM and sparse AQM on \eqref{eq:fermihubbard0}. A detailed consideration of the E-type AQM is omitted, as it does not improve upon the scaling in case of lattice models.
 \subsection{Square lattice  and sparse AQM}
With the square lattice AQM, the Fermi-Hubbard Hamiltonian can be simulated in constant time, neglecting the algorithmic depth necessary to initialize the code space, which is $O(L)$ or $O(1)$ depending on the exact method used. We will now describe how the square lattice AQM  modifies the terms of the Hamiltonian \eqref{eq:fermihubbard}, after which we will discuss the sparse AQM in that regard. \\

 We now use the square lattice AQM  to render the vertical hopping terms local: after adjusting each  term  of \eqref{eq:fermihubbard} by $h_\data \to h_\data\otimes \kappa^h_\aux$, the multiplication of adjusted hopping terms between   $(i, \,j)$ and $(i,\,j+1)$ with stabilizers $(p^{(i,\, j+\frac{1}{2})}_{\data}\otimes X_{(i,\, j+\frac{1}{2})})$ is resulting in local operators of weight 3.  While the hopping terms in \eqref{eq:fermihubbard0} only have real coefficients,  the operator weight of more general vertical hopping terms varies, but remains 3 on average. For complex hopping amplitudes $t^{\updownarrow}_{ij}$, we find
 \begin{align}
t^{\updownarrow}_{ij} \;c^{\dagger}_{(i,\,j)}c^{\phantom{\dagger}}_{(i,\,j+1)} + {(t^{\updownarrow}_{ij})}^*\,c^{\dagger}_{(i,\,j+1)}c^{\phantom{\dagger}}_{(i,\,j)} \quad \hat{=}\quad & \frac{(-1)^j}{2}\, \mathrm{Re}(t^{\updownarrow}_{ij}) \, \left(  Z_{(i,\,j-\frac{1}{2})} \otimes Z_{(i,\,j)} \otimes Y_{(i,\,j+\frac{1}{2})}  \right) \notag\\\notag
- \, &\frac{(-1)^j}{2}\, \mathrm{Re}(t^{\updownarrow}_{ij}) \, \left(   Y_{(i,\,j+\frac{1}{2})} \otimes Z_{(i,\,j+1)} \otimes Z_{(i,\,j+\frac{3}{2})}  \right) \\\notag   +\, & \frac{(-1)^j}{2} \, \mathrm{Im}(t^{\updownarrow}_{ij}) \,  \left(Z_{(i,\, j-\frac{1}{2})} \otimes Z_{(i, \, j)} \otimes X_{(i,\, j+\frac{1}{2})} \otimes Z_{(i,\, j+1)} \otimes Z_{(i, \, j+\frac{3}{2})} \right)\\ - \, & \frac{(-1)^j}{2} \, \mathrm{Im}(t^{\updownarrow}_{ij}) \; \; X_{(i,\, j+\frac{1}{2})} \, . \label{eq:hopping2}
\end{align}
  The improvements that we  make on vertical terms come at the cost of the adjustments  $\kappa_{\aux}^h$ to other terms in \eqref{eq:fermihubbard}. However, as already mentioned, the structure of the strings $\lbrace p^i_\data \rbrace$ guarantees to keep those other terms local. For horizontal  hopping terms that are (like the vertical strings) of the form $h_\data= (\mathrm{A}_i \otimes Z_{i+1} \otimes ... \otimes Z_{j-1} \otimes \mathrm{B}_{j})$, with  $\mathrm{A}, \mathrm{B} \in \lbrace X, \, Y\rbrace $, the substrings $\kappa^h_\aux$ invoke $Z$-operators at the end of the strings which makes for an additional weight of 2.  On the other hand, if $\mathrm{A}, \mathrm{B} = Z$, $\kappa^h_\aux$  features $Z$-operators along the entire string. This means that while single $Z$-operators are in this way adjusted to   $Z_{(i,\, j)}\to Z_{(i,\,j-\frac{1}{2})}\otimes Z_{(i,\, j)} \otimes Z_{(i,\, j+\frac{1}{2})} $, the two-qubit Hubbard terms gain 4 qubits worth of weight. \\
With the square lattice AQM, we have thus managed to reduce the weight of every term to a constant independent  of the system size.  A list of relevant terms, that compares Jordan-Wigner and square lattice AQM can be found in Table \ref{tab:squareterms}.    Having achieved locality of every Hamiltonian term, we can trotterize  $\widetilde{H}_{\auxdata}$ by for instance applying all horizontal hopping terms in $O(1)$ time, then continue with a time slice in which we simulate all vertical hoppings,  follow-up with all on-site interactions and Hubbard terms, and so on. Alternatively, one may apply Hamiltonian simulation strategies to simulate patches of the lattice more accurately and then  interweave these patches with the HHKL algorithm, \cite{haah2018quantum}.

With the  square lattice AQM, we have made the simulation scalable in terms of algorithmic depth and gate count. The requirement on the qubit number has however almost doubled. In order to be more economic with the number of auxiliary qubits, we consider the sparse AQM, which will help us to maximize the size of the simulated lattice on a fixed qubit budget. Placing vertical connections $\inter$ qubits apart, the required number of auxiliary qubits is $r=\left(\frac{2L^2-2L+1}{\inter}+L-1\right)$.   The weight of vertical hopping strings now largely depend upon their distance to the next vertical connection: let us say there is a vertical connection across $(i, \, j+\frac{1}{2})$, then the vertical hoppings between $(i, \, j)$ and $(i, \, j +1)$ are of (constant) weight 3, like in the square lattice AQM, while the vertical hoppings of modes to their left and right rather resemble  the strings of E-type AQM.  The worst case is certainly met for vertical hoppings in the middle of two vertical connections, so between $(i\pm \frac{1}{2}\inter,\, j)$ and  $(i\pm \frac{1}{2}\inter,\, j+1)$. Thus per vertical connection, there are $O(\inter)$ strings of weight $O(\inter)$ overlapping with one another.  The simulation time is thus $O(\inter)$ if we allow cancellations and $O(\inter^2)$ in the general case.

 \begin{table}[h!]
\begin{tabular}{rcl}
Jordan-Wigner transform  & &  Square lattice AQM \\ \hline \\
\hline
\end{tabular}
\caption{Comparing the Jordan-Wigner transform \eqref{eq:fermihubbard} to square lattice AQM when applied to the Hubbard model \eqref{eq:fermihubbard}. Vertical hopping terms are displayed between odd rows $j$ and even rows $j+1$ only. For $j$ even, the two $\widetilde{h}_{\auxdata}$-terms are exchanged. Not on display are the on-site terms and single-qubit contributions from  Hubbard interactions, $Z_{(i, \, j)}$, which are adjusted into $(Z_{(i, \, j-\frac{1}{2})}\otimes Z_{(i, \, j)} \otimes Z_{(i, \, j+\frac{1}{2})})$. }
\label{tab:squareterms}
\end{table}
\newpage
\subsection{Verstraete-Cirac transform and Superfast simulation}
The Fermi-Hubbard model can also  be made local by the Verstraete-Cirac transform or Superfast simulation.  In this section, we will compare the weights of  Pauli strings appearing in those cases to the strings resulting from transforming the Hubbard model with the square lattice AQM. We have compiled a list of the operator weights in Table \ref{tab:hubbardall}, and the interested reader may find a visual representation of  the strings from BKSF and VCT in Appendix \ref{sec:mappings}. Let us briefly discuss how the weights of the terms come to be. The VCT and AQM are quite similar in the sense that both concatenate  the Jordan-Wigner transform with a quantum code. However, the data-qubit substrings of the VCT stabilizers just consist of  $Z$-strings, which has two consequences: firstly, the stabilizers commute with diagonal terms like on-site detunings and Hubbard interactions, leaving them unadjusted and without any gain of weight. With this feature, the VCT distinguished itself from the other mapping in producing strings of the lowest weight. Secondly, while in the AQM a hopping string would just be adjusted on its end points, adjustments have to be made all along the strings in the VCT: fortunately, the auxiliary-qubit substring of the VCT stabilizers cancel these adjustments, causing this mapping to have shorter strings in the vertical direction (see Section \ref{sec:five}). We thus place spin-up and -down modes of the same spatial site vertically adjacent, like we have placed them horizontally adjacent in the AQM.  This leads to the weights of horizontal and vertical hoppings to be interchanged between VCT and AQM (on average). The stabilizers of both mappings can be made local with a weight of 6 (and weight-3 stabilizers at the boundaries), which is also the weight of  stabilizers in the BKSF. The BKSF, defined on the least amount of qubits, has surprisingly the longest strings. The reason for this is that logical $Z$-operators have weight $4$ - a consequence of the square lattice connectivity.  With this, the BKSF has also the largest variety of weights in hopping strings, while in the VCT, there is no variety at all among strings in the same direction. While the VCT appears to be the favorable option when comparing string lengths (followed by the AQM), it also uses the most qubits,  as becomes apparent in Appendix \ref{sec:mappings}.

 \begin{table}
\begin{tabular}{r|ccc}
&\begin{tabular}{l} Square lattice  \\ AQM \end{tabular}  &  \begin{tabular}{c} Verstraete-Cirac  \\ transform (VCT) \end{tabular}& \begin{tabular}{c}  Superfast  simulation \\  (BKSF) \end{tabular}\\
\hline
Stabilizer (interior)& 6 & 6 &6 \\

\begin{tabular}{r}Vertical hoppings \\ ${XX\,|\,YY\,|\,XY\,|\,YX}$  \end{tabular} & $3|3|5|1$ & $5|5|5|5$ & $2|6|5|4$ \\
 \begin{tabular}{r} Horizontal hoppings \\   ${XX\,|\,YY\,|\,XY\,|\,YX}$  \end{tabular} & $5|5|5|5$ &$3|3|3|3$ &$8|4|5|7$ \\
Two-qubit Hubbard terms & $6$ & $2$ & $6+2$ \\
On-site terms & $3$ & $1$ & $4$ \\
\end{tabular}
\caption{String lengths of the Fermi-Hubbard model transformed by all three mappings. We compare the weight of the Pauli strings, that originate from the  square lattice AQM, the Verstraete-Cirac transform and the Superfast simulation. For hopping terms, we consider the strings $h_\data=(\mathrm{A}_i \otimes Z_{i+1} \otimes \dots \otimes Z_{j-1} \otimes \mathrm{B}_j)$, with all variations of $\mathrm{A},  \mathrm{B}  \in \lbrace X, \,  Y \rbrace$. For vertical hoppings (in the AQM) we fix the case of $j$ being in an even row. Two-qubit Hubbard terms are of the form $h_\data=(Z \otimes Z)$, and on-site terms are singular $Z$-operators. In the BKSF it is  required to skip a qubit, which we penalize with an additional cost of two gates. In conclusion, the Verstraete-Cirac transform seems to exhibit the shortest strings, with the weights of the hopping terms being the same for all $\mathrm{A}_i, \, \mathrm{B}_j$.  Regarding string lengths, the square lattice AQM is in between the Verstraete-Cirac transform and the Superfast simulation, where the latter has the longest strings and largest variations in length.   }\label{tab:hubbardall}
\end{table}
\section{Comparison of AQM, VCT and BKSF}
\label{sec:five}
In this section, we will compare the Auxiliary Qubit Mapping, Superfast simulation and Verstraete-Cirac transform. Not only can the latter two be used to simulate the Hubbard model with local interactions, but we can also give them the Manhattan-distance property to align them with our notions of a good mapping for square lattices of qubits. This is done  in Appendix \ref{sec:mappings}. The reader completely unfamiliar with those mappings may also find an introduction reviewing the original proposals \cite{bravyi2002fermionic,verstraete2005mapping}.  Let us here compare AQM, VCT and BKSF regarding state preparation, qubit requirements,   Manhattan-distance property and the possibility of error mitigation. Afterwards, we can conclude and identify cases in which each mapping is advantageous. \\

\begin{figure}
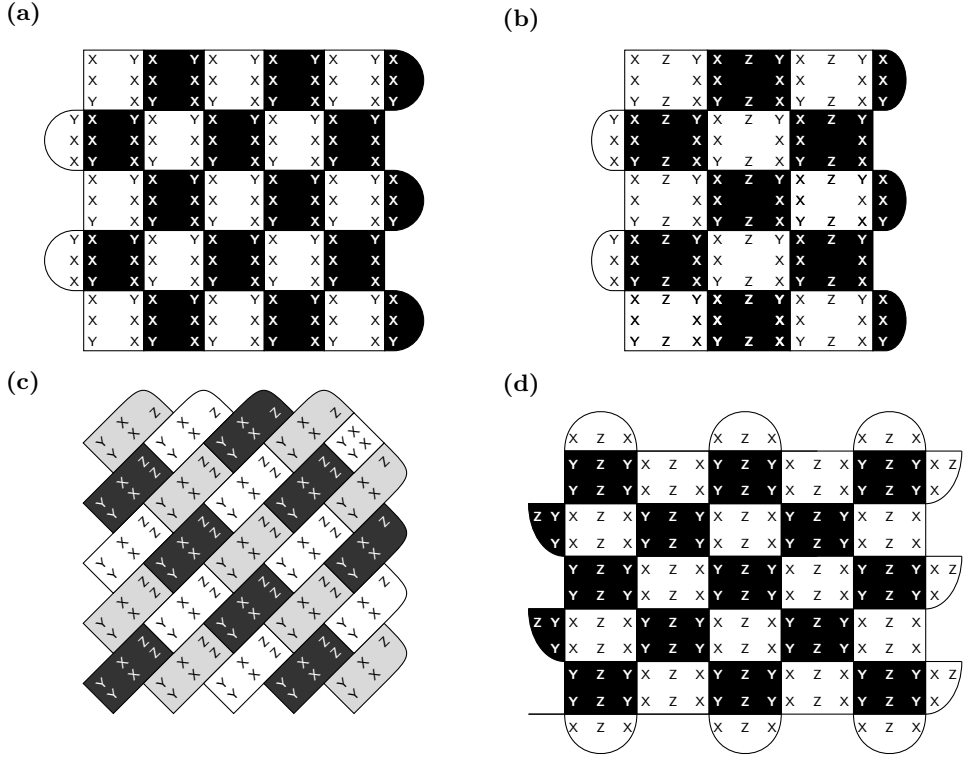


\caption{Tilings of local stabilizers for square lattice and sparse AQMs, BKSF and VCT.  Every tile represents a local stabilizer involving qubits along its perimeter. Inside the tiles,  $\mathsf{X}$, $\mathsf{Y}$ and $\mathsf{Z}$ indicate the Pauli operators that every qubit contributes to the corresponding stabilizer. We have shaded the tiles to as a visual aid for error mitigation. \textbf{(a)} Square lattice AQM with dimensions $\ell_1=\ell_2=6$. The stabilizers of all tiles are the same, except at the windings. \textbf{(b)} Sparse AQM with dimensions $\ell_1=7$, $\, \ell_2=6$ and  $\inter=2$. \textbf{(c)} BKSF of a $\ell_1=\ell_2=6$ fermionic lattice. The tiling is a three-colorable brickwork pattern. \textbf{(d)} VCT with dimensions $\ell_1=\ell_2=6$. The stabilizer tiles are alternating in a checkerboard pattern, that resembles the rotated surface code except for the $Z$-operators on the data qubits.  } \label{fig:measurement}
\end{figure}

\textit{State preparation} -  As we have shown, there is a unitary quantum circuit for the AQM to elevate an $N$-qubit state to its equivalent in the logical basis.
The VCT on the other hand has a logical basis that is entangled in a more complicated way, such that we cannot find a unitary quantum circuit of the same simplicity. Although the BKSF has no clear distinction between data and auxiliary qubits, there is a set of $N-1$ qubits that is only relevant for an S-pattern and one could argue that only  vertical connections  add the remaining qubits and introduce stabilizers. As each connection is implemented by just one entangled qubit, we believe that there might be a unitary circuit as simple as  $V_{\auxdata}$. As of now, we would have to resort to syndrome measurements to initialize the code space of VCT and BKSF. By syndrome measurements, we mean the measurement and readout of a  generating set of  stabilizers and correct for  outcomes inconsistent with the code space. While measurement and readout-times of state-of-the-art quantum devices might make this strategy challenging at present, we can at least arrange for local stabilizers such that the time overhead per measurement cycle is constant.  In Figure \ref{fig:measurement}(c)-(d)  the local stabilizer tilings of VCT and BKSF are shown.  A planar tiling for stabilizers of square lattice and sparse AQM follows  from multiplication of adjacent stabilizer generators

\begin{align}
\left( p^{(i,j+\frac{1}{2})}_{\data} \otimes X_{(i, \,j+\frac{1}{2})}\right) \cdot \left( p^{(i+1, \,j+\frac{1}{2})}_{\data} \otimes X_{(i+1, \, j+\frac{1}{2})} \right) \quad \text{and} \quad \left( p^{(i, \, j+\frac{1}{2})}_{\data} \otimes X_{(i, \, j+\frac{1}{2})}\right) \cdot \left( p^{(i+\inter,\,j+\frac{1}{2})}_{\data} \otimes X_{(i+\inter, \, j+\frac{1}{2})} \right)  \, ,
\end{align}
excluding the stabilizers at the windings, which are local already. The result is a repeating pattern of tiles with \textit{ears} at the windings, shown in Figure \ref{fig:measurement}(a)-(b). Note that we have implicitly used these tilings already in the respective definitions of $V_\auxdata$. While with the unitary quantum circuit we can prepare the state on only the data qubits  before encoding it into the logical basis, the same thing seems impossible with syndrome measurements. Even if the protective operations would not change the data-qubit state, there is still an ambiguity in the logical bases of VCT and AQM, that we now want to discuss. As can be seen in Appendix \ref{sec:B}, the quantum code layer included in these mappings transform any computational basis state $\ket{\bbs{\omega}}_\data$ into a logical basis state $\left[\prod_{i\in [r]} \frac{1}{\sqrt{2}}(\mathbb{I}+S^i_\auxdata)\right] \ket{\bbs{\omega}}_\data \otimes \ket{\bbs{\chi}}_\aux$, where $\lbrace S^i_\auxdata \rbrace_i$ is a generating set of stabilizers and $\bbs{\chi}=(\chi_1, \, \chi_2, \, ... \, , \chi_r)^\top\in \zetto{r}$ is a constant binary vector. While in the VCT, the set of stabilizers limit (not constrain) the choice of $\bbs{\chi}$, (square lattice and sparse) AQMs are properly stabilized for all possible $\bbs{\chi}\in \zetto{r}$. However, for both mappings the (signs of) adjustments made to operators $h_\data$ depend on $\bbs{\chi}$. For  AQMs  we  rely on $\bbs{\chi}=(0)^{\otimes r}$ for the substrings $\kappa^h_\aux$ to be free of signs. Obviously, for any basis with an unintended $\bbs{\chi}$-shift, the logical Hamiltonian $\widetilde{H}_\auxdata$ will not replicate the action of $H_\data$. As we cannot detect this $\bbs{\chi}$-offset,   we have to ignore it, e.g.~pretend that $\ket{\bbs{\chi}}=\ket{0^r}$ in AQMs: this effectively means that the state $\ket{\widetilde{\varphi}}_{\auxdata}$, which is created with an unknown $\bbs{\chi}$-shift in the $\aux$-register, becomes a state $(\prod_i \,[p^i_{\data}]^{\chi_i}) \ket{\widetilde{\varphi}}_{\auxdata}$ without shift, a state we have not intended to prepare.  To combat   ambiguities in all mappings, the system has to be constrained to the correct subspace before any state preparation can happen.  This  means we have to measure not only the stabilizers, but also logical operators until all degrees of freedom are eliminated. Apart form the tiles, we could measure all logical $Z$-operators, i.e. all logical encodings of $(2 c^{\dagger}_{j} c^{\phantom{\dagger}}_j-1)$. When all measurement outcomes yield `$+1$', we have prepared the logical zero state,  $|\widetilde{0^N}\rangle_{\auxdata}$. From there on, we directly prepare  $|\widetilde{\varphi}\rangle_\auxdata$ by e.g. Givens rotations  \cite{wecker2015solving,kivlichan2018quantum} using logical operators. This strategy appears to be the only option for measurement-based  preparation of states in any mapping, although practically one will certainly want to  perform only one cycle of measurements form the outcome of which the logical state and the (signs of the) stabilizers  are defined. For the modest E-type AQM on the other hand, neither syndrome measurements  nor unitary quantum circuits are necessary to prepare a logical state.  Due to the fact that its logical basis is in the computational basis, the product state  $(|0^N\rangle_{\data}\otimes |0^r\rangle_{\aux})$ is in fact the logical zero state, even though the two registers are obviously not entangled. Initializing all qubits in zero at first is thus a sufficient preliminary to prepare the state $\ket{\widetilde{\varphi}}_\auxdata$ with logical operators.
 \\

\textit{Qubit requirements} - For all mappings we find the highest number of qubits they require to be $\leq 2N$, in fact only the VCT demands exactly $2N$ qubits, the square lattice AQM on the other hand requires $\ell_1$ qubits less, and the BKSF requires even $\ell_2$ less than the AQM.  As for the  AQM, we can think about reducing the amount of qubits with sparse AQMs. For the VCT such a modification is discussed in Appendix \ref{sec:mappings}. As the qubits added to the VCT are generally added into the rows, its sparse version can be mapped back to a compact square lattice more easily than the AQM. In the BKSF, we can also make vertical connections more sparse, but as its layout is rotated, mapping the sparse  BKSF to a compact square lattice requires changes in the connectivity graph, which will influence the continuity of resulting strings.  \\

\textit{Manhattan-distance property} - With all mappings we manage to transform long-range hopping terms of a $\ell_1 \times \ell_2$ fermionic lattice to continuous Pauli strings on a qubit lattice, that can be deformed by the multiplication of stabilizers. For all mappings, the shortest version of those strings involve a number of qubits  scaling with the Manhattan distance of the fermionic modes on their lattice, but their exact weight differs from mapping to mapping - and is an interesting figure of merit. Let us say that on the fermionic lattice we  have a hopping term
 \begin{align}
t\; c^{\dagger}_{(i, \,j)} c^{\phantom{\dagger}}_{(i+x, \,j+y)} + t^{*}\; c^{\dagger}_{(i+x, \,j+y)} c^{\phantom{\dagger}}_{(i, \,j)}  \,  ,
\end{align}
where $t$ and $t^*$ is a complex coefficient and its Hermitian conjugate. Here the shortest path connecting those modes is over $x$ modes in horizontal and $y$ in vertical direction, the Manhattan distance is $x+y$.
Transforming a string with such a distance by one of the three mappings, the connecting string  is  supported on roughly $O(x+y)$ qubits, but its operator weight is not going to be $x+y$ exactly.  In case of the AQM, we will have twice the number of qubits per mode in the vertical direction, which means
that overcoming a vertical distance is more difficult, the string has the weight $x+2y$. In the VCT, the situation is exactly opposite and the horizontal distance is more costly to overcome due to the adjustment costs of the auxiliary modes: the operator weight of the connecting string is $2x+y$. For the BKSF, we find that horizontal and vertical paths are of equal weight, unfortunately the cost is doubled, so $2(x+y)$. Note that different versions of the BKSF  exist,  where the one version that yields these results is similar to the mapping in  \cite{chen2018exact} - others produce strings of higher weight, for some they are even disconnected. \\ Note that so far we have omitted the discussion of  constant weight overheads, that can arise at the end points of each string, and as such they are just relevant for small Manhattan distances. Around the modes labeled $(i,\, j)$ and $(i+x, \,j+y)$, BKSF and AQM can yield additional terms that matter predominantly for the local hoppings. As discussed, strings in the AQM can have one additional $Z$-operator around each end-mode, due to costs of the adjustments $\kappa^h_\aux$. In the BKSF, the strings might differ by up to one  logical $Z$-operator on each end, meaning there can be an additional cost of up to three (physical) $Z$-operators per end. Most notably, the VCT does not have such additional costs making it attractive for the simulation of lattice models, where $x+y$ is small.  \\

\textit{Error mitigation} - The reduction of the algorithmic depth, that all three mappings aim at, is the main tool in the reduction of noise. However, as the mappings  can be regarded as stabilizer codes, it is fair to ask if they can be used for mitigating the effect of noise, as has recently been proposed on a small scale \cite{mcardle2018error, bonet2018low}. Intriguingly, the AQM and VCT have local stabilizer tilings that resemble the stabilizers of surface code \cite{bombin2007optimal}. However, in contrast to those error correction codes,  we cannot achieve topological protection against logical errors. For the planar code of the VCT to correct errors, we necessarily  would need the data qubits (the qubits with $\textsf{Z}$ on them  in Figure \ref{fig:measurement}(d)) to be error free, as $X$- and  $Y$- errors would masquerade  syndromes of  errors on the auxiliary qubits. Furthermore, the code cannot detect $Z$-errors on the data qubits, and even increases their $Z$-error rate, as syndromes which are stabilizers in surface code  differ by some $Z$-operators from the stabilizers of the VCT.  A similar statement can be made for the square lattice AQM, where the auxiliary qubits  would have to be perfect, and their $X$-error rate is increased, see Figure \ref{fig:measurement} (a). Using fewer auxiliary qubits, the square lattice AQM  has fewer ears to mitigate errors with (as compared to Figure \ref{fig:measurement}(d)), they could however be added with more auxiliary qubits encoding the corresponding horizontal (local) connections. Unlike the surface code, the BKSF  (Figure \ref{fig:measurement}(c))  has a three-colorable brickwork-pattern in its tiling, that theoretically allows to detect all single-Pauli errors, but  like before some weight-two errors tend to masquerade themselves and go undetected when too close together. Although none of the codes allow for topological error correction, they exhibit a limited potential for error mitigation, in which  one might be able to catch some errors if the rate is low enough. Whether this is feasible is left to be decided.  \\

In conclusion, although the BKSF has the longest operators, it also requires the fewest qubits. As it is defined on a rotated square lattice, its shape might be the perfect fit for actual devices, as a patch of rotated surface code (including measurement qubits) is a rhombus. The BKSF is probably the most feasible candidate for error  mitigation strategies. With its output strings having the lowest weight of all three mappings, the VCT is perhaps the most sophisticated.  However, its theoretical backbone  is also the most complicated   -   when using the VCT one would probably have to adhere to the surface-code-like structure of the original proposal. With the weight of the output strings in between the two mappings, AQMs are a compromise for the cases that demand  more flexibility. The most unique feature of the AQMs is  that we can just use a unitary circuit to promote a data-qubit state into its logical equivalent  and if necessary even  release it from the auxiliary qubits.
The stabilizer state can also be manipulated during the simulation, e.g. accounting for swaps or basis transforms. The state preparation with $V_\auxdata$ might  make this mapping even interesting  for  NISQ devices \cite{preskill2018quantum}, especially for cloud-based quantum computing.

\section{Conclusion and Outlook}

In this work, we have developed a new class of Fermion-to-qubit  mappings that truly generalize the Jordan-Wigner transform to  two dimensions.  Moreover, this class can be regarded as a quantum code layer on top of the mapping provided by the Jordan-Wigner transform, and with the unitary $V_{\auxdata}^{(\dagger)}$ we find  a means to  encode (decode)  quantum states in the code layer. The quantum code is shown to require a certain number of auxiliary qubits that is close to $N$, but this number is not strict. In fact, sparse mappings with a reduced number of auxiliary qubits can achieve similar results, which might be of great practical advantage. More generally, there is a statement that we can make not just about the  Auxiliary Qubit Mapping, but also the Verstraete-Cirac transform and the Bravyi-Kitaev Superfast simulation. Versions of all these transforms can be used as one-dimensional linear Fermion-to-qubit mapping with $N$ (resp. $N-1$) qubits, but at the expense of additional qubits we can pre-compute certain Pauli strings, which allows us to take shortcuts when mapping operators. This pre-computation is done when said strings are stabilized in a quantum code that entangles  data qubits with the qubits added. The usage of these codes allows a quantum computer to do what was not manageable classically: the local treatment of two-dimensional Fermion systems. In this way we can not only simulate fermionic lattices, but embed every Fermion system on a two-dimensional layout. \\

We hope that future work will extend these results: we for instance have not taken into account specific limitations on either the qubit connectivity graph or the ability to perform quantum gates, which can be found in proposals for actual devices \cite{versluis2017scalable,li2018crossbar}. It would also be interesting to incorporate the mappings into specific simulation algorithms, to see for instance how phase estimation  or qubitization could deal with the planar layout.

\section{Acknowledgments}
We would like to thank Jonas Helsen and Ben Criger for the many scientific discussions about this work, as well as Kenneth Goodenough for his help with the manuscript.  We would also like to thank CWJ Beenakker for his support.
MS was supported by the Netherlands Organization for
Scientific Research (NWO/OCW) and an ERC Synergy Grant. SW was supported by  STW Netherlands, an NWO VIDI Grant and an ERC Starting Grant.

\bibliographystyle{unsrt}
\bibliography{addQbib}

\begin{thebibliography}{10}

\bibitem{feynman1982simulating}
Richard~P Feynman.
\newblock Simulating physics with computers.
\newblock {\em International journal of theoretical physics}, 21(6-7):467--488,
  1982.

\bibitem{lloyd1996universal}
Seth Lloyd.
\newblock Universal quantum simulators.
\newblock {\em Science}, pages 1073--1078, 1996.

\bibitem{abrams1997simulation}
Daniel~S Abrams and Seth Lloyd.
\newblock Simulation of many-body fermi systems on a universal quantum
  computer.
\newblock {\em Physical Review Letters}, 79(13):2586, 1997.

\bibitem{kitaev1995quantum}
A~Yu Kitaev.
\newblock Quantum measurements and the {A}belian stabilizer problem.
\newblock {\em arXiv preprint quant-ph/9511026}, 1995.

\bibitem{cleve1998quantum}
Richard Cleve, Artur Ekert, Chiara Macchiavello, and Michele Mosca.
\newblock Quantum algorithms revisited.
\newblock In {\em Proceedings of the Royal Society of London A: Mathematical,
  Physical and Engineering Sciences}, volume 454, pages 339--354. The Royal
  Society, 1998.

\bibitem{aspuru2005simulated}
Al{\'a}n Aspuru-Guzik, Anthony~D Dutoi, Peter~J Love, and Martin Head-Gordon.
\newblock Simulated quantum computation of molecular energies.
\newblock {\em Science}, 309(5741):1704--1707, 2005.

\bibitem{mcclean2016theory}
Jarrod~R McClean, Jonathan Romero, Ryan Babbush, and Al{\'a}n Aspuru-Guzik.
\newblock The theory of variational hybrid quantum-classical algorithms.
\newblock {\em New Journal of Physics}, 18(2):023023, 2016.

\bibitem{lanyon2010towards}
Benjamin~P Lanyon, James~D Whitfield, Geoff~G Gillett, Michael~E Goggin,
  Marcelo~P Almeida, Ivan Kassal, Jacob~D Biamonte, Masoud Mohseni, Ben~J
  Powell, Marco Barbieri, et~al.
\newblock Towards quantum chemistry on a quantum computer.
\newblock {\em Nature chemistry}, 2(2):106--111, 2010.

\bibitem{du2010nmr}
Jiangfeng Du, Nanyang Xu, Xinhua Peng, Pengfei Wang, Sanfeng Wu, and Dawei Lu.
\newblock Nmr implementation of a molecular hydrogen quantum simulation with
  adiabatic state preparation.
\newblock {\em Physical review letters}, 104(3):030502, 2010.

\bibitem{peruzzo2014variational}
Alberto Peruzzo, Jarrod McClean, Peter Shadbolt, Man-Hong Yung, Xiao-Qi Zhou,
  Peter~J Love, Al{\'a}n Aspuru-Guzik, and Jeremy~L O’brien.
\newblock A variational eigenvalue solver on a photonic quantum processor.
\newblock {\em Nature communications}, 5, 2014.

\bibitem{barends2015digital}
R~Barends, L~Lamata, J~Kelly, L~Garc{\'\i}a-{\'A}lvarez, AG~Fowler, A~Megrant,
  E~Jeffrey, TC~White, D~Sank, JY~Mutus, et~al.
\newblock Digital quantum simulation of fermionic models with a superconducting
  circuit.
\newblock {\em Nature communications}, 6:7654, 2015.

\bibitem{wang2015quantum}
Ya~Wang, Florian Dolde, Jacob Biamonte, Ryan Babbush, Ville Bergholm, Sen Yang,
  Ingmar Jakobi, Philipp Neumann, Al{\'a}n Aspuru-Guzik, James~D Whitfield,
  et~al.
\newblock Quantum simulation of helium hydride cation in a solid-state spin
  register.
\newblock {\em ACS nano}, 9(8):7769--7774, 2015.

\bibitem{o2016scalable}
PJJ O’Malley, Ryan Babbush, ID~Kivlichan, Jonathan Romero, JR~McClean, Rami
  Barends, Julian Kelly, Pedram Roushan, Andrew Tranter, Nan Ding, et~al.
\newblock Scalable quantum simulation of molecular energies.
\newblock {\em Physical Review X}, 6(3):031007, 2016.

\bibitem{hempel2018quantum}
Cornelius Hempel, Christine Maier, Jonathan Romero, Jarrod McClean, Thomas
  Monz, Heng Shen, Petar Jurcevic, Ben Lanyon, Peter Love, Ryan Babbush, et~al.
\newblock Quantum chemistry calculations on a trapped-ion quantum simulator.
\newblock {\em arXiv preprint arXiv:1803.10238}, 2018.

\bibitem{jones2012faster}
N~Cody Jones, James~D Whitfield, Peter~L McMahon, Man-Hong Yung, Rodney
  Van~Meter, Al{\'a}n Aspuru-Guzik, and Yoshihisa Yamamoto.
\newblock Faster quantum chemistry simulation on fault-tolerant quantum
  computers.
\newblock {\em New Journal of Physics}, 14(11):115023, 2012.

\bibitem{wecker2014gate}
Dave Wecker, Bela Bauer, Bryan~K Clark, Matthew~B Hastings, and Matthias
  Troyer.
\newblock Gate-count estimates for performing quantum chemistry on small
  quantum computers.
\newblock {\em Physical Review A}, 90(2):022305, 2014.

\bibitem{wigner1928uber}
Eugene~P Wigner and Pascual Jordan.
\newblock \"{U}ber das {P}aulische \"{A}quivalenzverbot.
\newblock {\em Z. Phys}, 47:631, 1928.

\bibitem{holstein1940field}
T~Holstein and Hl~Primakoff.
\newblock Field dependence of the intrinsic domain magnetization of a
  ferromagnet.
\newblock {\em Physical Review}, 58(12):1098, 1940.

\bibitem{fradkin1989jordan}
Eduardo Fradkin.
\newblock {J}ordan-{W}igner transformation for quantum-spin systems in two
  dimensions and fractional statistics.
\newblock {\em Physical review letters}, 63(3):322, 1989.

\bibitem{wang1991ground}
YR~Wang.
\newblock Ground state of the two-dimensional antiferromagnetic {H}eisenberg
  model studied using an extended {W}igner-{J}ordon transformation.
\newblock {\em Physical Review B}, 43(4):3786, 1991.

\bibitem{ball2005fermions}
RC~Ball.
\newblock Fermions without fermion fields.
\newblock {\em Physical review letters}, 95(17):176407, 2005.

\bibitem{chen2018exact}
Yu-An Chen, Anton Kapustin, and {\DJ}or{\dj}e Radi{\v{c}}evi{\'c}.
\newblock Exact bosonization in two spatial dimensions and a new class of
  lattice gauge theories.
\newblock {\em Annals of Physics}, 393:234--253, 2018.

\bibitem{verstraete2005mapping}
Frank Verstraete and J~Ignacio Cirac.
\newblock Mapping local {H}amiltonians of fermions to local {H}amiltonians of
  spins.
\newblock {\em Journal of Statistical Mechanics: Theory and Experiment},
  2005(09):P09012, 2005.

\bibitem{whitfield2016local}
James~D Whitfield, Vojt{\v{e}}ch Havl{\'\i}{\v{c}}ek, and Matthias Troyer.
\newblock Local spin operators for fermion simulations.
\newblock {\em Physical Review A}, 94(3):030301, 2016.

\bibitem{havlivcek2017operator}
Vojt{\v{e}}ch Havl{\'\i}{\v{c}}ek, Matthias Troyer, and James~D Whitfield.
\newblock Operator locality in the quantum simulation of fermionic models.
\newblock {\em Physical Review A}, 95(3):032332, 2017.

\bibitem{zohar2018eliminating}
Erez Zohar and J~Ignacio Cirac.
\newblock Eliminating fermionic matter fields in lattice gauge theories.
\newblock {\em arXiv preprint arXiv:1805.05347}, 2018.

\bibitem{seeley2012bravyi}
Jacob~T Seeley, Martin~J Richard, and Peter~J Love.
\newblock The {B}ravyi-{K}itaev transformation for quantum computation of
  electronic structure.
\newblock {\em The Journal of chemical physics}, 137(22):224109, 2012.

\bibitem{tranter2015bravyi}
Andrew Tranter, Sarah Sofia, Jake Seeley, Michael Kaicher, Jarrod McClean, Ryan
  Babbush, Peter~V Coveney, Florian Mintert, Frank Wilhelm, and Peter~J Love.
\newblock The {B}ravyi--{K}itaev transformation: {P}roperties and applications.
\newblock {\em International Journal of Quantum Chemistry}, 115(19):1431--1441,
  2015.

\bibitem{setia2017bravyi}
Kanav Setia and James~D Whitfield.
\newblock {B}ravyi-{K}itaev {S}uperfast simulation of fermions on a quantum
  computer.
\newblock {\em arXiv preprint arXiv:1712.00446}, 2017.

\bibitem{bravyi2002fermionic}
Sergey~B Bravyi and Alexei~Yu Kitaev.
\newblock Fermionic quantum computation.
\newblock {\em Annals of Physics}, 298(1):210--226, 2002.

\bibitem{beals2013efficient}
Robert Beals, Stephen Brierley, Oliver Gray, Aram~W Harrow, Samuel Kutin, Noah
  Linden, Dan Shepherd, and Mark Stather.
\newblock Efficient distributed quantum computing.
\newblock {\em Proc. R. Soc. A}, 469(2153):20120686, 2013.

\bibitem{babbush2017low}
Ryan Babbush, Nathan Wiebe, Jarrod McClean, James McClain, Hartmut Neven, and
  Garnet~Kin Chan.
\newblock Low depth quantum simulation of electronic structure.
\newblock {\em arXiv preprint arXiv:1706.00023}, 2017.

\bibitem{babbush2018low}
Ryan Babbush, Nathan Wiebe, Jarrod McClean, James McClain, Hartmut Neven, and
  Garnet Kin-Lic Chan.
\newblock Low-depth quantum simulation of materials.
\newblock {\em Physical Review X}, 8(1):011044, 2018.

\bibitem{kivlichan2018quantum}
Ian~D Kivlichan, Jarrod McClean, Nathan Wiebe, Craig Gidney, Al{\'a}n
  Aspuru-Guzik, Garnet Kin-Lic Chan, and Ryan Babbush.
\newblock Quantum simulation of electronic structure with linear depth and
  connectivity.
\newblock {\em Physical review letters}, 120(11):110501, 2018.

\bibitem{subacsi2016nonperturbative}
Yi{\u{g}}it Suba{\c{s}}{\i} and Christopher Jarzynski.
\newblock Nonperturbative embedding for highly nonlocal {H}amiltonians.
\newblock {\em Physical Review A}, 94(1):012342, 2016.

\bibitem{poulin2018quantum}
David Poulin, Alexei Kitaev, Damian~S Steiger, Matthew~B Hastings, and Matthias
  Troyer.
\newblock Quantum algorithm for spectral measurement with a lower gate count.
\newblock {\em Physical review letters}, 121(1):010501, 2018.

\bibitem{babbush2018encoding}
Ryan Babbush, Craig Gidney, Dominic~W Berry, Nathan Wiebe, Jarrod McClean,
  Alexandru Paler, Austin Fowler, and Hartmut Neven.
\newblock Encoding electronic spectra in quantum circuits with linear {T}
  complexity.
\newblock {\em arXiv preprint arXiv:1805.03662}, 2018.

\bibitem{suzuki1990fractal}
Masuo Suzuki.
\newblock Fractal decomposition of exponential operators with applications to
  many-body theories and monte carlo simulations.
\newblock {\em Physics Letters A}, 146(6):319--323, 1990.

\bibitem{suzuki1991general}
Masuo Suzuki.
\newblock General theory of fractal path integrals with applications to
  many-body theories and statistical physics.
\newblock {\em Journal of Mathematical Physics}, 32(2):400--407, 1991.

\bibitem{bartlett1989alternative}
Rodney~J Bartlett, Stanislaw~A Kucharski, and Jozef Noga.
\newblock Alternative coupled-cluster ans{\"a}tze ii. the unitary
  coupled-cluster method.
\newblock {\em Chemical physics letters}, 155(1):133--140, 1989.

\bibitem{motzoi2017linear}
Felix Motzoi, MP~Kaicher, and FK~Wilhelm.
\newblock Linear and logarithmic time compositions of quantum many-body
  operators.
\newblock {\em Physical review letters}, 119(16):160503, 2017.

\bibitem{steudtner2018fermion}
Mark Steudtner and Stephanie Wehner.
\newblock Fermion-to-qubit mappings with varying resource requirements for
  quantum simulation.
\newblock {\em New Journal of Physics}, 2018.

\bibitem{hastings2014improving}
Matthew~B Hastings, Dave Wecker, Bela Bauer, and Matthias Troyer.
\newblock Improving quantum algorithms for quantum chemistry.
\newblock {\em arXiv preprint arXiv:1403.1539}, 2014.

\bibitem{haah2018quantum}
Jeongwan Haah, Matthew~B Hastings, Robin Kothari, and Guang~Hao Low.
\newblock Quantum algorithm for simulating real time evolution of lattice
  {H}amiltonians.
\newblock {\em arXiv preprint arXiv:1801.03922}, 2018.

\bibitem{wecker2015solving}
Dave Wecker, Matthew~B Hastings, Nathan Wiebe, Bryan~K Clark, Chetan Nayak, and
  Matthias Troyer.
\newblock Solving strongly correlated electron models on a quantum computer.
\newblock {\em Physical Review A}, 92(6):062318, 2015.

\bibitem{mcardle2018error}
Sam McArdle, Xiao Yuan, and Simon Benjamin.
\newblock Error mitigated quantum computational chemistry.
\newblock {\em arXiv preprint arXiv:1807.02467}, 2018.

\bibitem{bonet2018low}
X~Bonet-Monroig, R~Sagastizabal, M~Singh, and TE~O'Brien.
\newblock Low-cost error mitigation by symmetry verification.
\newblock {\em arXiv preprint arXiv:1807.10050}, 2018.

\bibitem{bombin2007optimal}
H~Bombin and Miguel~A Martin-Delgado.
\newblock Optimal resources for topological two-dimensional stabilizer codes:
  Comparative study.
\newblock {\em Physical Review A}, 76(1):012305, 2007.

\bibitem{preskill2018quantum}
John Preskill.
\newblock Quantum {C}omputing in the {NISQ} era and beyond.
\newblock {\em {Quantum}}, 2:79, August 2018.

\bibitem{versluis2017scalable}
R~Versluis, S~Poletto, N~Khammassi, B~Tarasinski, N~Haider, DJ~Michalak,
  A~Bruno, K~Bertels, and L~DiCarlo.
\newblock Scalable quantum circuit and control for a superconducting surface
  code.
\newblock {\em Physical Review Applied}, 8(3):034021, 2017.

\bibitem{li2018crossbar}
Ruoyu Li, Luca Petit, David~P Franke, Juan~Pablo Dehollain, Jonas Helsen, Mark
  Steudtner, Nicole~K Thomas, Zachary~R Yoscovits, Kanwal~J Singh, Stephanie
  Wehner, et~al.
\newblock A crossbar network for silicon quantum dot qubits.
\newblock {\em Science advances}, 4(7):eaar3960, 2018.

\bibitem{zhu2018hardware}
Guanyu Zhu, Yi{\u{g}}it Suba{\c{s}}{\i}, James~D Whitfield, and Mohammad
  Hafezi.
\newblock Hardware-efficient fermionic simulation with a cavity--{QED} system.
\newblock {\em npj Quantum Information}, 4(1):16, 2018.

\bibitem{barenco1995elementary}
Adriano Barenco, Charles~H Bennett, Richard Cleve, David~P DiVincenzo, Norman
  Margolus, Peter Shor, Tycho Sleator, John~A Smolin, and Harald Weinfurter.
\newblock Elementary gates for quantum computation.
\newblock {\em Physical review A}, 52(5):3457, 1995.

\end{thebibliography}

\newpage \appendix

\section{Auxiliary Qubit codes}
\label{sec:B}
Here we will  set up the quantum codes used for the AQMs, which includes the review of the methods developed in \cite{subacsi2016nonperturbative}. We adapt those methods for quantum codes and contribute ideas  which can be used to speed up the initialization of the logical basis.  \\

As mentioned before, the stabilizing the Pauli strings $(p^i_\data \otimes \sigma^i_{N+i})$  effectively describes a quantum code: a larger Hilbert space of $n=N+r$ qubits is constrained to the dimension $2^N$ by $r$ stabilizer conditions. In contrast to codes for quantum error correction,  we do not want to encode information non-locally, i.e. obtain non-local logical operators, but want to localize  operators that were non-local to begin with. When characterizing a quantum error correction code, one is usually interested in the  generating set of stabilizers,  the logical basis states, e.g. $\ket{\obar{0}}$, $\ket{\obar{1}}$ and the logical operators, $\obar{X}$, $\obar{Z}$. In the following, we will look at the AQM equivalents of those quantities: while  $\lbrace p^i_\data \otimes \sigma^i_{N+i}\$ $ clearly are the stabilizer generators, the extended computational basis $ V_{\auxdata} \ket{\bbs{\omega}}_{\data}\otimes \ket{0^r}_{\aux}$ spans the logical subspace and the adjusted Pauli strings $\widetilde{h}_{\auxdata}$ are its logical operators. \\

In the initialization of the code space via the unitary $V_\auxdata$, the auxiliary qubits are entangled with data qubits, but not before the former are possibly rotated into some basis other than the computational basis: the basis choice of the auxiliary qubits can have consequences for other methods of state preparation and for sure determines the form of the operators  $\sigma_{N+i}^i$ and $\kappa_{\aux}^h$.
In the following, we will introduce the two logical bases, to which AQMs  resort. For each of these we will outline the following points:
\begin{description}
\item[i. Extended basis]\textit{Relation of  $N$-qubit states $\ket{\varphi}_{\data}$ to $(N+r)$-qubit states $\ket{\widetilde{\varphi}}_{\auxdata}$ with respect to their bases.  }
\item[ii. Entangling operation] \textit{The unitary $V_{\auxdata}$, for initializing the stabilizer state by quantum gates.}
\item[iii. Hamiltonian adjustments] \textit{Adjustments to be made to Pauli strings and operator mappings to obtain $\widetilde{H}_{\auxdata}$.}
\end{description}
We want to deliver the last point in a two-fold way: on the one hand, we present the adjustments to a Hamiltonian in Pauli string form \eqref{eq:hamil}, where we replace every term $h_{\data} \to (h_{\data} \otimes \kappa^h_{\aux})$. The origin of such a Hamiltonian can be arbitrary. On the other hand we want to focus on Hamiltonians that originate from certain many-body problems of Fermions. Therefore, we fuse the Hamiltonian adjustments with the linear transform \eqref{eq:lineartransform}, such that terms  $(h_\data\otimes \kappa^h_\aux)$ can be obtained directly  from  second quantization \eqref{eq:secquant}. The result is a redefinition of relation \eqref{eq:singleop}:
\begin{align}
c^{\dagger}_j\;\hat{=}\;\frac{1}{2}\left(\bigotimes_{k\in \widetilde{U}(j)} X_k\right)\left(\mathbb{I}+\bigotimes_{l\in \widetilde{F}(j)} Z_l\right)  \left(\bigotimes_{m\in \widetilde{P}(j)} Z_m \right) \, ,  \notag \\ \label{eq:stabtrafo}
c^{\;}_j\;\hat{=}\;\frac{1}{2}\left(\bigotimes_{k\in \widetilde{U}(j)} X_k\right)\left(\mathbb{I}-\bigotimes_{l\in \widetilde{F}(j)} Z_l\right)\left(  \bigotimes_{m\in \widetilde{P}(j)} Z_m \right) \, .
\end{align}
The redefined transform stays close to the spirit of the original in the sense that only the flip, parity and update sets are replaced by adjusted versions $\widetilde{F}(j)$, $\widetilde{P}(j)$ and $\widetilde{U}(j)$. \\ Apart from the two bases, we also take a look at an extension of the principle, that allows to build a stabilizer set with strings $\lbrace p^i_\data \rbrace$, that might anticommute. Interestingly, one could in this way encode all terms of a Hamiltonian into a mapping. The resulting code is perhaps most akin to the original method \cite{subacsi2016nonperturbative}, where a new auxiliary qubit is spent for every Hamiltonian term to be multiplied with a stabilizer.
\subsection{Auxiliary qubits in computational basis}
With the parity strings being the detrimental substrings of the Jordan-Wigner-transformed Hamiltonians, our main goal is to cancel long strings of $Z$-operators. In \cite{zhu2018hardware}, this is achieved in collecting the parity information of subsets of qubits with a  circuit QED resonator. In a hardware-unspecific approach, computational basis AQMs store parity information on auxiliary qubits, which can be updated and they have never to be uncomputed. \\

We generally restrict computational-basis Auxiliary Qubit codes to  strings $p^i_{\data}\subseteq \lbrace \mathbb{I}, Z\rbrace^{\otimes N}$. The $p^i_{\data}$-strings are here canceled with auxiliary Pauli-$Z$ operators $\sigma^i_{N+i}=Z_{N+i}$.
Let us say that the stabilizers are characterized by the  $(r\times N)$ binary matrix $B$, such that an entry `$1$' in the $j$-th column on line $i$ of $B$ means that $Z_j$ is part of $p^i_\data$:
\begin{align}
\label{eq:Zstabs}
p^i_{\data} \otimes \sigma^i_{N+i} \;=\; \left( \bigotimes_{j\in [N]}(Z_j)^{B_{ij}} \right) \otimes Z_{N+i} \, .
\end{align}
\begin{description}
\item[i. Extended basis] \textit{ A quantum state $\ket{\widetilde{\varphi}}_\auxdata$ that  is  based on the generic $N$-qubit state $\ket{\varphi}_\data$  \eqref{eq:genericstate} and stabilized by \eqref{eq:Zstabs} takes the form
\begin{align}
\label{eq:Zbasis}
\ket{\widetilde{\varphi}}_{\auxdata}\;=\;\sum_{\bbs{\omega} \in \zetto{N}} a_{\bbs{\omega}} \ket{\bbs{\omega}}_{\data}\otimes \ket{B \bbs{\omega}\moto}_{\aux} \, .
\end{align}
From \eqref{eq:genericstate} to \eqref{eq:Zbasis}, the computational basis has obviously been extended:  $\ket{\bbs{\omega}}_{\data}\to \ket{\bbs{\omega}}_{\data}\otimes \ket{B \bbs{\omega}\moto}_{\aux}$, where $(\mathrm{mod}\; 2)$ acts on every component separately.}
\end{description}
It is easy to verify that this new basis is stabilized by \eqref{eq:Zstabs} considering $Z_j\ket{b}_j\,=\,\left(-1\right)^{b}\ket{b}_j$, where $b\in\mathbb{Z}_2$.

\begin{description}
\item[ii. Entangling operation] \textit{The entangling operation can be described as a (commuting) sequence of  $\textsc{CNot}$-gates that depend on the matrix $B$. If $B_{ij}=1$, then there is a $\textsc{CNot}$-gate in $V_{\auxdata}$, that, controlled on data qubit $j$, targets the auxiliary qubit  labeled  $N+i$:
\begin{align}
\label{eq:Zinit}
V_{\auxdata} \,= \,\prod_{i\in [r]}\prod_{\scriptsize \begin{tabular}{c} $j\in [N]$ \\ $\text{with}\; B_{ij}=1$ \end{tabular}} \textsc{CNot}\left(j\to N+i \right) \, .
\end{align}}
\end{description}
The unitary $V_{\auxdata}$, acting on a basis element $(\ket{\bbs{\omega}}_{\data}\otimes \ket{0^r}_{\aux})$ yields the extended basis of \eqref{eq:Zbasis}, considering that \\$\textsc{CNot}(j\to k) \ket{a}_j \otimes \ket{b}_k \,=\, \ket{a}_j \otimes \ket{a+b\moto}_k$, where $a, \, b \in \mathbb{Z}_2$. The entangling operation basically stores parity information  of subsets of data qubits (as defined by the rows of $B$) on auxiliaries.  For the exact implementation of $V_{\auxdata}$, \eqref{eq:Zinit} needs to be adjusted to the connectivity graph of the qubit layout. For square lattice connectivity, the above formula requires  $O(rN)$ time steps  in the worst case, but there is a way to improve the depth of $V_{\auxdata}$: for the auxiliary qubits $i$ and $k$, we can replace the circuit

\begin{align}
&\left[\prod_{ j: B_{ij}=1} \textsc{CNot}(j\to N+i)\; \right]  \left[ \prod_{ \; l: B_{kl}=1} \textsc{CNot}(l\to N+k) \; \right] \\ \label{eq:Zboost}  & \text{by}\quad \left[\prod_{ j: B_{ij}+B_{kj}=1} \textsc{CNot}(j\to N+i) \;\right]\; \textsc{CNot}(N+k\to N+j)\; \left[ \prod_{\; l: B_{kl}=1} \textsc{CNot}(l\to N+k) \;\right] \, .
\end{align}
In this (non-commuting) sequence of gates, we let the $i$-th auxiliary qubit inherit the parity information of the $k$-th auxiliary qubit by a $\textsc{CNot}$-gate inside the $\aux$-register. This is a useful trick when the parity information that is to be stored on these two auxiliary qubits has a large overlap in data qubits, i.e. when the vectors $\bigoplus_x (B_{ix})$ and $\bigoplus_y (B_{ky})$ have a small Hamming distance.  In that case, the leftmost product contains only few \textsc{CNot}-gates, as the bulk of the parity information has been inherited from the $(N+k)$-th qubit.
\begin{description}
\item[iii. Hamiltonian adjustments] \textit{In order to maintain the stabilizer state \eqref{eq:stabhamil}, we adjust a Pauli string $h_{\data}$ on the data qubits by $h_{\data} \to (h_{\data}\otimes \kappa^h_{\aux})$ with \begin{align}
\kappa^{h}_{\aux} \;=\;\bigotimes_{m \in [r]} (X_{N+m})^{\lambda_m}\, ,
\end{align}
where $\bbs{\lambda}=(\lambda_1, \lambda_2, \dots, \lambda_r   )^{\top} \in \zetto{r}$ is obtained by
\begin{align}
\bbs{\lambda}=\sum_j B \bbs{u_j} \moto \,
\end{align}
with $\bbs{u_j}$ being the $j$-th unit vector of $\zetto{N}$, and the sum extending over all $j\in [N]$, for which $h_{\data}$ acts on the qubit space as $X_j$ or $Y_j$.
Hamiltonian of adjusted terms $(h_\data \otimes \kappa^h_\aux)$ as in \eqref{eq:stabhamil} can be obtained by the redefined transforms \eqref{eq:stabtrafo}, with the same flip and parity sets, $\widetilde{F}(j)=F(j)$ and $\widetilde{P}(j)=P(j) $, but the sets $\widetilde{U}(j)$ defined from the columns of the matrix
\begin{align}
\left[ \begin{tabular}{c}
$A$ \\ \hline $B$
\end{tabular} \right]\, .
\end{align}}
\textit{We recall that $A$ is the matrix the underlying linear transform is based on \eqref{eq:lineartransform} and $B$ is defining the stabilizers as in \eqref{eq:Zstabs}.}
\end{description}
In case a Pauli string $h_{\data}$  flips a data qubit, that is entangled with a qubit in the $\aux$-register, we have to flip the latter qubit as well. In fact we need to flip all other auxiliaries to which the data qubit contributes: so if we apply the operator $X_j$ to a basis state $\ket{\bbs{\omega}}_{\data}\otimes\ket{B\bbs{\omega}\moto}_{\aux}$ for  $j\in[N]$, we leave the stabilized basis, unless we update the configuration of the auxiliary qubits by $B\bbs{\omega} \to B(\bbs{\omega}+\bbs{u_j})$.
\subsubsection*{Example}
Let us consider a minimal example, in which the data register holds five qubits, and a sixth,  an auxiliary qubit, is in the configuration $B\bbs{\omega}$, where $B$ is a $(1 \times 5)$ binary matrix.
We consider a Hamiltonian term  $h_{\data}=(X_1\otimes Z_2\otimes  Z_3\otimes  Z_4\otimes X_5)$. After  adjusting $h_{\data}\to (h_\data \otimes \kappa^h_\aux)$, we have the choice to multiply with the stabilizer or not. In Table \ref{tab:Zexample} we present the adjusted Hamiltonian before and after multiplication with the stabilizer, considering different choices of $B$.

\begin{table}[h!]
\begin{tabular}{clr}
$\!\!\!\!B$  &  $h_{\data}\otimes\kappa^h_\aux$   &$(h_{\data}\otimes\kappa^h_\aux) \cdot (p^1_\data  \otimes Z_6)$
\\  \hline
$[\begin{tabular}{ccccc}
0 & 1 & 0 & 0 & 0
\end{tabular}] \quad$ & $(X_1 \otimes Z_2 \otimes  Z_3 \otimes  Z_4 \otimes X_5)$ & $(X_1 \otimes  Z_3 \otimes  Z_4 \otimes X_5 \otimes Z_6)$  \\
$[\begin{tabular}{ccccc}
0 & 1 & 1 & 1 & 0
\end{tabular}]\quad$ & $(X_1 \otimes Z_2 \otimes  Z_3 \otimes  Z_4 \otimes X_5)$ & $(X_1   \otimes X_5 \otimes Z_6)$  \\
 $[\begin{tabular}{ccccc}
1 & 1 & 1 & 0 & 0
\end{tabular}]\quad$&   $(X_1 \otimes Z_2 \otimes  Z_3 \otimes  Z_4 \otimes X_5 \otimes X_6)$ & $-(Y_1 \otimes Z_4 \otimes X_5 \otimes Y_6)$ \\
$[\begin{tabular}{ccccc}
1 & 1 & 1 & 1 & 1
\end{tabular}]\quad$& $(X_1 \otimes Z_2 \otimes  Z_3 \otimes  Z_4 \otimes X_5) $ & $-(Y_1\otimes Y_5 \otimes Z_6)$
\end{tabular}
\caption{Adjusted Hamiltonian terms $\widetilde{h}_\auxdata$ with respect to the original string  $h_{\data}=(X_1\otimes Z_2\otimes  Z_3\otimes  Z_4\otimes X_5)$, depending on the matrix $(1 \times 5)$ matrix $B$. }
\label{tab:Zexample}
\end{table}
\subsection{Auxiliary qubits in Hadamard basis}
\label{subsec:hadamard}
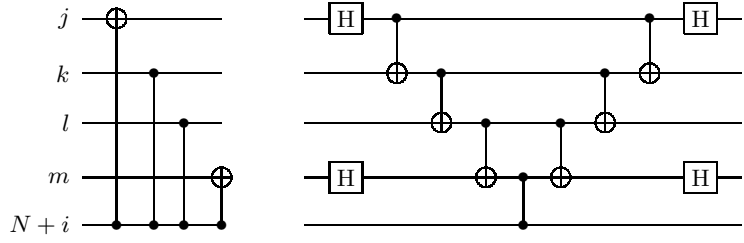
\begin{figure}
  $\qquad\qquad\quad$\Qcircuit @C=1em @R=1.2em {&\lstick{j}	&\targ	&\qw	&\qw	&\qw	&	&	&	&\gate{\mathrm{H}}	&\ctrl{1}	&\qw	&\qw	&\qw	&\qw	&\qw	&\ctrl{1}	&\gate{\mathrm{H}}	 &\qw\\
&\lstick{k}	&\qw	&\control \qw	&\qw	&\qw	&	&	&	&\qw	&\targ	&\ctrl{1}	&\qw	&\qw	&\qw	&\ctrl{1}	&\targ	&\qw	 &\qw\\
&\lstick{l}	&\qw	&\qw	&\control \qw	&\qw	&	&  	&	&\qw	&\qw	&\targ	&\ctrl{1}	&\qw	&\ctrl{1}	&\targ	&\qw	&\qw	 &\qw\\
&\lstick{m}	&\qw	&\qw	&\qw	&\targ	&	&	&	&\gate{\mathrm{H}}	&\qw	&\qw	&\targ	&\ctrl{1}	&\targ	&\qw	&\qw	&\gate{\mathrm{H}}  &\qw	\\
&\lstick{N+i}	&\ctrl{-4}	&\ctrl{-3}	&\ctrl{-2}	&\ctrl{-1}	&	&	&	&\qw	&\qw	&\qw	&\qw	&\control \qw	&\qw	&\qw	&\qw	&\qw   &\qw	\\}
\caption{Two versions of the controlled application of the Pauli string $p_{\data}^i= (X_j\otimes Z_k \otimes Z_l \otimes X_m)$ on the $i$-th qubit in the auxiliary register. }
\label{fig:controlpi}
 \end{figure}

Extending the idea of \cite{subacsi2016nonperturbative}, we can cancel a set of  arbitrary (commuting) strings $\lbrace p^i_\data \rbrace$, where   $p^i_{\data}\in \lbrace X, Y, Z, \mathbb{I}  \rbrace^{\otimes N} $, by $X$-operators: $ \sigma^i_{N+i}  =  X_{N+i} $.  Let us characterize the choice of the strings $p^i_{\data}$ by three $(r\times N)$ binary matrices $C^X$, $C^Y$ and $C^Z$. Here an entry `$1$' in $C^s_{ji}$, with $s\in\lbrace X, Y, Z \rbrace$, indicates that the string $p_{\data}^i$ acts as $s$ on the $j$-th qubit.
\begin{description}
\item[i. Extended basis]  \textit{An extended state $\ket{\widetilde{\varphi}}_{\auxdata}$, stabilized by $\lbrace p^i_{\data} \otimes X_{N+i}\rbrace_{i\in [r]}$ and based on an arbitrary state $N$-qubit state $\ket{\varphi}_{\data}$ \eqref{eq:genericstate}  is given by \begin{align}
\label{eq:Xstates}
\ket{\widetilde{\varphi}}_\data \; = \;\left[\prod_{i \in [r]} \frac{1}{\sqrt{2}} \left( \mathbb{I}+p^i_\data \otimes X_{N+i}\right) \right] \ket{{\varphi}}_\data \otimes \ket{0^r}_\aux \, .
\end{align}
This state can be regarded as being of the form \eqref{eq:genericstate}, where the computational basis is replaced by: \begin{align}
\label{eq:Xbasis}
\ket{\bbs{\omega}}_{\data}\;\to\; \frac{1}{2^{r/2}} \sum_{\bbs{\mu} \in \zetto{r}} \left[\prod_{k\in[r]} \left(p_{\data}^k \right)^{\mu_k} \right]\ket{\bbs{\omega}}_{\data} \otimes \ket{\bbs{\mu}}_{\aux} \, .
\end{align}}
\end{description}
The sums in \eqref{eq:Xbasis}  invoke all the possible qubit configurations $\bbs{\mu}\in \zetto{r}$ with equal weight. This is a result of the auxiliary qubits being in Hadamard basis. This choice of basis becomes plausible by multiplying a basis state \eqref{eq:Xbasis} with one of the stabilizers $(p^i_{\data}\otimes X_{N+i})$:
\begin{align}
\left(p_{\data}^i \otimes X_{N+i} \right) \; &\frac{1}{2^{r/2}} \sum_{\bbs{\mu} \in \mathbb{Z}_2^{\otimes r}} \left[\prod_{k\in[r]} \left(p^k_{\data} \right)^{\mu_k} \right]\ket{\bbs{\omega}}_{\data} \otimes \ket{\bbs{\mu}}_{\aux} \notag \\ \label{eq:very1}
=\;&\frac{1}{2^{r/2}} \sum_{\bbs{\mu} \in \mathbb{Z}_2^{\otimes r}} \left[\prod_{k\in[r]} \left(p^k_{\data} \right)^{\mu_k+\delta_{ik}} \right]\ket{\bbs{\omega}}_{\data} \otimes \ket{\bbs{\mu}+\bbs{u_i}\moto}_{\aux} \, .
\end{align}
If we now shift the binary vector in the sum by the $i$-th unit vector $\bbs{u_i}$ to  $\bbs{\mu} \to \bbs{\mu} + \bbs{u_i} \moto$, the original basis element on the right-hand side of \eqref{eq:Xbasis} is recovered and  thus the set of  Pauli strings $(p_{\data}^i \otimes X_{N+i} )$  stabilizes every state $\ket{\widetilde{\varphi}}_{\auxdata}$ that is in the subspace spanned by  \eqref{eq:Xbasis}.
\begin{description}
\item[ii. Entangling operation] \textit{ Following  \cite{subacsi2016nonperturbative},  the entangling operation can be described as
\begin{align}
\label{eq:Xinit}
V_{\auxdata}\;=\; \prod_{i\in[r]}\left( \ket{0}\!\!\bra{0}_{N+i}+p^i_{\data}\otimes\ket{1}\!\!\bra{1}_{N+i}\right) \mathrm{H}_{N+i}\, ,
\end{align}
where $\mathrm{H}_{N+i}$ is the Hadamard gate on the $(N+i)$-th qubit. In words, $V_{\auxdata}$ can be realized by a unitary quantum circuit that first applies Hadamard gates to every auxiliary qubit, and then applies each string $ p^k_{\data} $ controlled by the $k$-th auxiliary qubit.    }
\end{description}
We notice that the circuit \eqref{eq:Xinit}, when acting on a state $\ket{\varphi}_{\data}\otimes\ket{0^r}_{\aux}$, firstly changes the basis of the auxiliary register into $
\ket{+^r}_{\aux}=(\bigotimes_{i\in [r]} \ket{+}_{N+i}) = r^{-\frac{1}{2}} \sum_{\bbs{\mu}\in \zetto{r}} \ket{\bbs{\mu}}_{\aux}$. Then the controlled application of the strings $p^i_{\data}$ entangles auxiliary and data qubits.  In principle, this can be done by $\textsc{CNot}$, $\textsc{CPhase}$ and controlled-$Y$ gates according to the action of a string $p^i_{\data}$ on each data qubit, see Figure \ref{fig:controlpi} (left).  In practice, the required qubit connectivity might however not be available, such that we may resort to an implementation of the circuit as in Figure \ref{fig:controlpi} (right). Like for the codes with computational-basis auxiliary qubits, we can here apply tricks to make $V_{\auxdata}$ more shallow whenever two  strings $p^i_{\data}$, $p^k_{\data}$  are similar to one another: after the Hadamard-gates are applied to the auxiliary qubits $i$ and $k$, we can replace the circuit
\begin{align}
&\left( \ket{0}\!\!\bra{0}_{N+i}+p^i_{\data}\otimes\ket{1}\!\!\bra{1}_{N+i}\right)\; \left( \ket{0}\!\!\bra{0}_{N+k}+p^k_{\data}\otimes\ket{1}\!\!\bra{1}_{N+k}\right) \\ &  \label{eq:Xboost}\text{by} \quad  \left( \ket{0}\!\!\bra{0}_{N+i}+\left(p^i_{\data} \cdot\, p^k_{\data} \right) \otimes X_{N+k}\otimes\ket{1}\!\!\bra{1}_{N+i}\right)\; \left( \ket{0}\!\!\bra{0}_{N+k}+p^k_{\data}\otimes\ket{1}\!\!\bra{1}_{N+k}\right)\, ,
\end{align}
which means that instead of applying the string $p^i_{\data}$, we conditionally apply the string that results from the operator product of $p^i_{\data}$ with $p^k_{\data}$, and an $X$-operator on the $k$-th auxiliary qubit.  What we use here is the fact that the $(N+k)$-th qubit is already entangled with the data qubits after the right sequence of controlled gates, such that we can use the stabilizer condition in the sequence on the left. For this to work, the order in which the two resulting strings are initialized is now fixed.  A minus sign that might occur in the operator product can be reproduced by adding a $Z_{N+i}$, \cite{barenco1995elementary}.
 \\

Before presenting the Hamiltonian adjustments, it is left for us to verify that the controlled applications of $p^i_{\data}$  on $\ket{\bbs{\omega}}_{\data}\otimes\ket{+^r}_{\aux}$  yield the corresponding element of the extended basis \eqref{eq:Xbasis}. Let us consider the following reformulation of the controlled-($p^i_{\data}$) terms:
\begin{align}
\label{eq:very2}
&\prod_{i\in[r]}\left( \ket{0}\!\!\bra{0}_{N+i}+p^i_{\data}\otimes\ket{1}\!\!\bra{1}_{N+i}\right) \;=\; \prod_{i\in[r]}\left( \sum_{\mu_i^\prime\in \mathbb{Z}_2} \left(p^i_{\data}\right)^{\mu_i^\prime}\otimes\ket{\mu_i^\prime}\!\!\bra{\mu_i^\prime}_{N+i}\right) \;=\; \sum_{\bbs{\mu^\prime}\in \zetto{r}} \left[\prod_{k\in[r]} \left( p^k_{\data}\right)^{\mu_k^\prime}\right] \otimes \ket{\bbs{\mu^\prime}}\!\!\bra{\bbs{\mu^\prime}}_{\aux} \,  .
\end{align}
Considering the expansion of $\ket{+^r}_{\aux}$ in the computational basis, we can proceed to arrive at \eqref{eq:Xbasis} by inspection.
\begin{description}
\item[iii. Hamiltonian adjustments] \textit{ For a Pauli string  $h_{\data}$ to maintain the stabilizer state \eqref{eq:stabhamil}, we adjust it by
\begin{align}
\label{eq:Xadjust}
\kappa_{\aux}^h\;=\;\bigotimes_{j\in T(h)} Z_{N+j}  \,
\end{align}
 where the set $T(h) \subseteq [r]$ contains $k$ if $p_{\data}^k$ anticommutes with $h_{\data}$.  As a consequence, a Hamiltonian of terms $(h_{\data} \otimes \kappa^h_\aux)$ can be obtained from second quantization  using the redefined transformations \eqref{eq:stabtrafo}, where the update sets are defined as before $\widetilde{U}(j)=U(j)$, but flip and parity sets $\widetilde{F}(j)$, $\widetilde{P}(j)$ are redefined by the rows of the matrices
 \begin{align}
 \label{eq:Xtrafo}
  \left[\vphantom{C^X+C^Y}A\right.\left|C^X+C^Y\right] \moto \, , \quad   \left[\vphantom{(C^X+C^Y)} RA\right.\left|R(C^X+C^Y)+C^Y+C^Z\right] \moto \, .
 \end{align}
 \textit{As  explained above, the $C$-matrices define the stabilizers while  $A$ and $R$  stem from the underlying linear mapping, see \eqref{eq:lineartransform} and \eqref{eq:triangular}.  }
}
\end{description}
We will now show that the adjusted Pauli string  $(h_{\data}\otimes \kappa_{A}^{h})$ acts on a state $\ket{\widetilde{\varphi}}_{\auxdata}$ such that after application of $V_{\auxdata}^\dagger$,   we recover $h_{\data} \ket{\varphi}_{\data} \otimes \ket{0^r}_{\aux}$. We start by applying the adjusted term to the extended state. The goal is to use (anti-)commutation relations with the strings $p^k_{\data}$ to let $h_{\data}$ act on the data register first. It turns out that minus signs that we pick up by anticommutations are exactly canceled by sign changes  originating from $\kappa_{A}^{h}$ acting on the $\aux$-register.

In general, we find  if $h_\data$ now anticommutes with a string $p^k_\data$, then $k\in T(h)$ such that $(h_\data\otimes \kappa^h_\aux)$  commutes with $(\mathbb{I}+p^k_{\data}\otimes X_{N+k})$, and we find
 \eqref{eq:stabsystem} satisfied.
 For the transform \eqref{eq:Xtrafo}, we take into account all sorts of Pauli operators that originate from parity, update and flip operators, by which we mean the strings $(\bigotimes_{m \in P(j)} Z_m)$, $(\bigotimes_{k \in U(j)} X_k)$ and $(\bigotimes_{l \in F(j)} Z_l)$ in \eqref{eq:singleop}. If $X$- and $Y$-operators in a string $p^i_\data$ anticommute with the $Z$-operators in the $j$-th flip operator,  we have to counteract by adjusting it with a $Z$-operator on the $i$-th auxiliary:   $(\bigotimes_{l\in F(j)} Z_l )\otimes Z_{N+i}$.  The same argument holds for the parity operators, but we also add $Z$-operators there, stemming from anticommutations of the update operator with $Z$- and $Y$-operators in $p^i_{\data}$.  Considering that the operators $X$, $Y$ and $Z$ appear in the strings $p^i_\data$ according to the $C$-matrices, we can use these matrices to describe the contents of the flip and parity sets, by which we obtain \eqref{eq:Xtrafo}.
 \subsubsection*{Example}
As an example we examine a 5-qubit Hamiltonian term, $h_{\data}=(X_1\otimes Z_2\otimes  Z_3\otimes  Z_4\otimes X_5)$. The sixth qubit is a Hadamard-basis auxiliary, used to cancel various substrings $p^1_{\data}$. In Table \ref{tab:Xexample}, we find the adjusted terms  $(h_\data \otimes \kappa^h_\aux)$ and the deformed terms, $(p^1_\data \otimes X_6) \cdot (h_\data \otimes \kappa^h_\aux)$, for various choices of the stabilizer $(p^1_{\data}\otimes X_6)$.
\begin{table} [h!]
 \begin{tabular}{llr}
$p^1_{\data}$  &  $(h_\data \otimes \kappa^h_\aux)$   & $(p^1_\data \otimes X_6) \cdot (h_\data \otimes \kappa^h_\aux)$
\\  \hline $(Z_2\otimes Z_3 \otimes Z_4)$ & $(X_1\otimes Z_2\otimes  Z_3\otimes  Z_4\otimes X_5)$ & $(X_1\otimes X_5 \otimes X_6)$ \\
$(X_1 \otimes Z_2 \otimes Z_3 \otimes X_4) $ & $(X_1\otimes Z_2\otimes  Z_3\otimes  Z_4\otimes X_5 \otimes Z_6)$  & $ -(Y_4 \otimes X_5 \otimes Y_6)  $ \\
$(X_1\otimes Z_2\otimes  Z_3\otimes  Z_4\otimes X_5)$ & $(X_1\otimes Z_2\otimes  Z_3\otimes  Z_4\otimes X_5)$ & $X_6$
\end{tabular}
\caption{Adjusted Hamiltonians $\widetilde{h}_{\auxdata}$ to $h_{\data}=(X_1\otimes Z_2\otimes  Z_3\otimes  Z_4\otimes X_5)$, depending on the choice of $p^1_{\data}$.}\label{tab:Xexample}
\end{table}

\subsection{Stabilizing anticommuting data-qubit strings}
We present a more general quantum code based on auxiliary qubits in Hadamard basis, but in which the strings $\lbrace p_{\data}^i\rbrace $ do not necessarily have to commute. Using this code, an entire Hamiltonian can in principle be transformed into interactions on only the auxiliary qubits.
   The general idea here is to amend the scheme by the following notion: in order to counter  anticommutations, we replace the (single-qubit) Pauli operators $\sigma^i_{N+i}$ with Pauli strings on the auxiliary register $\gamma^i_{\aux}$, such that $\gamma^i_{\aux}$ contains $X_{N+i}$ as before, but for every other string $p^k_{\data}$ with $k<i$, that anticommutes with $p^i_{\data}$, it contains a $Z$-operator, $Z_{N+k}$. For convenience we define the operation $\star$ as:
\begin{align}
i\star k = \begin{cases}
0 \qquad \text{if} \quad [p^i_{\data}, \, p^k_{\data}]\phantom{_+}=\,0 \\
1 \qquad \text{if} \quad [p^i_{\data}, \, p^k_{\data}]_+=\,0
\end{cases}\, .
\end{align}
Using this notation, we define the stabilizers of our system as
\begin{align}
\label{eq:noncommstabs}
p^i_{\data}\otimes \gamma^i_{\aux} \;=\; p^i_{\data}\otimes\left(\bigotimes_{k\in [i-1]}(Z_{N+k})^{i\star k}\right)\otimes X_{N+i} \,  ,
\end{align}
since all Pauli strings $(p^i_{\data}\otimes \gamma^i_{\aux})$ have to commute pairwise for all $i\in [r]$ as defined above.
We will now turn to describe the mapping in the established way.
 \begin{description}
\item[i. Extended basis]\textit{The  computational basis $\ket{\bbs{\omega}}_{\data}$ is extended to:}
\begin{align}
\label{eq:noncommbasis}
\ket{\bbs{\omega}}_{\data}\;\to\; \frac{1}{2^{r/2}} \sum_{\bbs{\mu} \in \zetto{r}} \left[\left(p_{\data}^1 \right)^{\mu_1} \;\cdots \; \; \left(p_{\data}^r \right)^{\mu_r}\right]\ket{\bbs{\omega}}_{\data} \otimes \ket{\bbs{\mu}}_{\aux} \, .
\end{align}
\end{description}
This basis  resembles \eqref{eq:Xbasis}, with the subtle difference that  the order of the strings $p^i_{\data}$ matters here.  When stabilizer $(p^i_\data\otimes \gamma^i_\aux)$ are multiplied to \eqref{eq:noncommbasis} from the right, the operators $\gamma_\aux^i$ cancel all minus signs from anticommutations, and flip the $i$-th qubit in the auxiliary register. Note that the order of the strings $p^i_\data$ in \eqref{eq:noncommbasis} is to be taken into account when we attempt to encode $\ket{\widetilde{\varphi}}_\auxdata$ from $\ket{\varphi}\otimes \ket{0^r}_\aux$.
\setcounter{enumi}{1}
 \begin{description}
\item[ii. Entangling operation]\textit{We pick a sequence $i_1, \, i_2, \, ..., \, i_r $ that is some permutation of $1, \, 2, \, ..., \, r $, in which we want to perform the entangling operation for the stabilizers $(p^{i_m}_{\data}\otimes\gamma^{i_m}_{\aux})$, where the stabilizer of number $i_r$ is  taken care of first, and the one labeled $i_1$ last. The entangling operation associated with that sequence is
\begin{align}
\label{eq:noncomminit}
V_{\auxdata}\;= \; \prod_{m=1}^{r}\left(\ket{0}\!\!\bra{0}_{N+i_m}+  p^{i_m}_{\data}\otimes \ket{1}\!\!\bra{1}_{N+i_m}\otimes \left[ \bigotimes_{k>m} (Z_{N+i_k})^{(i_m\,\star\, i_k)\;\theta_{i_m\, i_k}} \right]\right)\mathrm{H}_{N+i_m}\, ,
\end{align}
where $\theta_{ij}$ is a binary version of the Heaviside function,
\begin{align}
\label{eq:heavy}
\theta_{ij}=\begin{cases} 1 & i>j \\ 0 & \text{else} \quad . \end{cases}
\end{align}}
\end{description}
Note that if the we pick the original order, $i_m=m$, the circuit almost looks like  \eqref{eq:Xinit}, but, again,  here the exact order matters.
The Hamiltonian adjustments are identical to \eqref{eq:Xadjust} and \eqref{eq:Xtrafo}, as the only difference, the ordering of the strings $p^i_{\data}$, does not matter there: a Hamiltonian term $\widetilde{h}_{\auxdata}$ needs to pass all $p^k_\data$ in \eqref{eq:noncommbasis}, picking up all minus signs possible. \\ We have thus obtained an auxiliary qubit mapping with completely arbitrary set of strings $p^i_{\data}$. If this string  is  a Hamiltonian term  $h_\data=p^i_{\data}$, we can eliminate its action on the data qubits by replacing
\begin{align}
h_\data \; \to \; ({h}_{\data} \otimes \kappa^h_{\aux}) \cdot \left(p^i_\data \otimes  \gamma^i_{\aux}\right) \; =  \; X_{N+i}\otimes \left[ \bigotimes_{k>i} \left(Z_{N+k}\right)^{i \star k} \right]\, .
\end{align}
 The entire Hamiltonian can in this way be pre-computed and reduced to an action on only the auxiliary register.
\newcommand{\trees}{\tau}
\newcommand{\kids}{\Gamma}
\newcommand{\levels}{\Lambda}

\section{Tree-based transforms}
\label{sec:A}

In this section, we consider  Fermion-to-qubit mappings defined on tree structures for a setup with limited connectivity. This particular class of mappings  is part of the mappings considered in Section \ref{subsec:conventional}  (so $n=N$), where the tree structures are inherent in the definition of the transformation matrix $A$. Although this class  technically contains the Jordan-Wigner transform, our motivation is to obtain mappings that are more akin to the Bravyi-Kitaev transform, in order to keep parity strings short. While the Bravyi-Kitaev transform itself does this job perfectly,  we will show that it cannot be reconciled  with a square lattice connectivity graph: in this section, we instead develop a method to tailor mappings to preexisting connectivity graphs, and provide an algorithm with which short parity strings can be guaranteed and the operator weight bounded. Let us start by reviewing the Bravyi-Kitaev transform.

\begin{figure}[h!]
\footnotesize
\begin{tabular}{lr}
\begin{tikzpicture}[baseline=4, scale=1]
\node[] at (-2,4) {\normalsize \textbf{(a)}};

\draw[](0,4)--(0,0);
\draw[](0,4)--(-1,3);
\draw[](0,3)--(-1,2)--(-2,1);
\draw[](0,2)--(-1,1);
\draw[](0,4)--(1,3)--(1,2);
\draw[](1,3)--(2,2)--(2,1);
\draw[](0,4)--(2,3)--(3,2);
\draw[](0,3)--(-2,2);
\draw[fill=white] (0,0) circle[radius=0.15];
\node[below right] at (0,-.1) {1} ;
\draw[fill=white] (0,1) circle[radius=0.15];
\node[below right] at (0,.9) {2} ;
\draw[fill=white] (0,2) circle[radius=0.15];
\node[below right] at (0,1.9) {4} ;
\draw[fill=white] (0,3) circle[radius=0.15];
\node[below right] at (0,2.9) {8} ;
\draw[fill=white] (0,4) circle[radius=0.15];
\node[above right] at (0,4.1) {16} ;
\draw[fill=white] (-1,3) circle[radius=0.15];
\node[below] at (-1.1,2.9) {15} ;
\draw[fill=white] (2,3) circle[radius=0.15];
\node[below] at (2,2.9) {14} ;
\draw[fill=white] (1,3) circle[radius=0.15];
\node[below left] at (1,2.9) {12} ;
\draw[fill=white] (-1,2) circle[radius=0.15];
\node[below right] at (-1,1.9) {6} ;
\draw[fill=white] (1,2) circle[radius=0.15];
\node[below right] at (1,1.9) {11} ;
\draw[fill=white] (2,2) circle[radius=0.15];
\node[below right] at (2,1.9) {10} ;
\draw[fill=white] (3,2) circle[radius=0.15];
\node[below right] at (3,1.9) {13};
\draw[fill=white] (-1,1) circle[radius=0.15];
\node[below right] at (-1,0.9) {3};
\draw[fill=white] (2,1) circle[radius=0.15];
\node[below right] at (2,0.9) {9};
\draw[fill=white] (-2,2) circle[radius=0.15];
\node[below right] at (-2,1.9) {7} ;
\draw[fill=white] (-2,1) circle[radius=0.15];
\node[below right] at (-2,0.9) {5};
\end{tikzpicture} & \begin{tikzpicture}[baseline=3.5, scale=1]
\node[] at (-0.5,3.5) {\normalsize \textbf{(b)}};
\draw[](0,3)--(3,3);
\draw[](0,2)--(3,2);
\draw[](0,1)--(3,1);
\draw[](2,0)--(2,3);
\draw[](0,0)--(0,1);
\draw[](1,0)--(1,1);
\draw[](2,0)--(2,1);
\draw[](3,0)--(3,1);
\draw[fill=white] (0,0) circle[radius=0.15];
\node[below] at (0,-.2) {9} ;
\draw[fill=white] (1,0) circle[radius=0.15];
\node[below ] at (1,-.2) {11} ;
\draw[fill=white] (2,0) circle[radius=0.15];
\node[below  ] at (2,-.2) {15} ;
\draw[fill=white] (3,0) circle[radius=0.15];
\node[below] at (3,-.2) {13} ;
\draw[fill=white] (0,1) circle[radius=0.15];
\node[below right ] at (0,0.9) {10} ;
\draw[fill=white] (1,1) circle[radius=0.15];
\node[below right ] at (1,0.9) {12} ;
\draw[fill=white] (2,1) circle[radius=0.15];
\node[below right] at (2,0.9) {16} ;
\draw[fill=white] (3,1) circle[radius=0.15];
\node[below right] at (3,0.9) {14} ;

\draw[fill=white] (0,2) circle[radius=0.15];
\node[below] at (0,1.8) {5} ;
\draw[fill=white] (1,2) circle[radius=0.15];
\node[below] at (1,1.8) {6} ;
\draw[fill=white] (2,2) circle[radius=0.15];
\node[below right ] at (2,1.8) {8} ;
\draw[fill=white] (3,2) circle[radius=0.15];
\node[below] at (3,1.8) {7} ;

\draw[fill=white] (0,3) circle[radius=0.15];
\node[below] at (0,2.8) {1} ;
\draw[fill=white] (1,3) circle[radius=0.15];
\node[below] at (1,2.8) {2} ;
\draw[fill=white] (2,3) circle[radius=0.15];
\node[below right] at (2,2.8) {4} ;
\draw[fill=white] (3,3) circle[radius=0.15];
\node[below] at (3,2.8) {3} ;
\end{tikzpicture} \\
 \begin{tikzpicture}[baseline=4, scale=1]
 \node[] at (-2,4) {\normalsize \textbf{(c)}};
\draw[\clre](0,4)--(0,0);
\draw[\clre](0,4)--(-1,3);
\draw[\clre](0,3)--(-1,2)--(-2,1);
\draw[\clre](0,2)--(-1,1);
\draw[\clre](0,4)--(1,3)--(1,2);
\draw[\clre](1,3)--(2,2)--(2,1);
\draw[\clre](0,4)--(2,3)--(3,2);
\path (0,3)--(0,2) node [midway,above=-0.05cm,sloped] {$\longrightarrow$};
\path (0,3)--(-1,2) node [midway,below=-0.02cm,sloped] {$\longleftarrow$};
\path (0,3)--(-2,2) node [midway, above=-0.05cm ,sloped] {$\longleftarrow$};
\path (2,2)--(1,3) node [midway, above=-0.05cm ,sloped] {$\longleftarrow$};
\path (1,3)--(0,4) node [midway, above=-0.05cm ,sloped] {$\longleftarrow$};
\draw[\clrd,very thick](0,3)--(-2,2);
\draw[\clrd,very thick](0,3)--(-1,2);
\draw[\clrd,very thick] (0,4)--(2,2);
\draw[\clrd,very thick] (0,3)--(0,2);
\draw[\clre,fill=white] (0,0) circle[radius=0.15];

\draw[\clre,fill=white] (0,1) circle[radius=0.15];

\draw[color=\clrb, fill=\clra, thick] (0,2) circle[radius=0.15];
\node[\clrc] at (0,2) {\scriptsize $\boldsymbol{\mathsf{Z}}$};

\draw[color=\clrb, fill=\clra, thick] (0,3) circle[radius=0.15];
\node[\clrc] at (0,3) {\scriptsize $\boldsymbol{\mathsf{Z}}$};

\draw[color=\clrb, fill=\clra, thick] (0,4) circle[radius=0.15];
\node[\clrc] at (0,4) {\scriptsize $\boldsymbol{\mathsf{X}}$};
\draw[\clre,fill=white] (-1,3) circle[radius=0.15];

\draw[\clre,fill=white] (2,3) circle[radius=0.15];

\draw[color=\clrb, fill=\clra, thick] (1,3) circle[radius=0.15];
\node[\clrc] at (1,3) {\scriptsize $\boldsymbol{\mathsf{X}}$};
\draw[color=\clrb, fill=\clra, thick] (-1,2) circle[radius=0.15];
\node[\clrc] at (-1,2) {\scriptsize $\boldsymbol{\mathsf{Z}}$};
\draw[\clre,fill=white] (1,2) circle[radius=0.15];

\draw[color=\clrb, fill=\clra, thick] (2,2) circle[radius=0.15];
\node[\clrc] at (2,2) {\scriptsize $\boldsymbol{\mathsf{X}}$};

\draw[\clre,fill=white] (3,2) circle[radius=0.15];

\draw[\clre,fill=white] (-1,1) circle[radius=0.15];

\draw[\clre,fill=white] (2,1) circle[radius=0.15];

\draw[color=\clrb, fill=\clra, thick] (-2,2) circle[radius=0.15];
\node[\clrc] at (-2,2) {\scriptsize $\boldsymbol{\mathsf{Z}}$};

\draw[\clre,fill=white] (-2,1) circle[radius=0.15];

\end{tikzpicture} & \begin{tikzpicture}[baseline=3.5, scale=1]
\node[] at (-0.5,3.5) {\normalsize \textbf{(d)}};
\draw[\clre](0,3)--(3,3);
\draw[\clre](0,2)--(3,2);
\draw[\clre](0,1)--(3,1);
\draw[\clre](2,0)--(2,3);
\draw[\clre](0,0)--(0,1);
\draw[\clre](1,0)--(1,1);
\draw[\clre](2,0)--(2,1);
\draw[\clre](3,0)--(3,1);
\node[above=-0.05cm] at (0.5,1) {$\longrightarrow$};
\node[above=-0.05cm] at (1.5,1) {$\longrightarrow$};
\node[above=-0.05cm] at (1.5,2) {$\longleftarrow$};
\node[above=-0.05cm] at (2.5,2) {$\longrightarrow$};
\path (2,2)--(2,3) node [midway,below=-0.02cm,sloped] at (0.5,1) {$\longrightarrow$};
\draw[very thick, \clrd] (0,1)--(2,1);
\draw[very thick, \clrd] (1,2)--(3,2);
\draw[very thick, \clrd] (2,3)--(2,2);
\draw[\clre,fill=white] (0,0) circle[radius=0.15];

\draw[\clre,fill=white] (1,0) circle[radius=0.15];

\draw[\clre,fill=white] (2,0) circle[radius=0.15];

\draw[\clre,fill=white] (3,0) circle[radius=0.15];

\draw[color=\clrb, fill=\clra, thick] (0,1) circle[radius=0.15];
\node[\clrc] at (0,1) {\scriptsize $\boldsymbol{\mathsf{X}}$} ;
\draw[color=\clrb, fill=\clra, thick] (1,1) circle[radius=0.15];
\node[\clrc ] at (1,1) {\scriptsize $\boldsymbol{\mathsf{X}}$} ;
\draw[color=\clrb, fill=\clra, thick] (2,1) circle[radius=0.15];
\node[\clrc] at (2,1) {\scriptsize $\boldsymbol{\mathsf{X}}$} ;
\draw[\clre,fill=white] (3,1) circle[radius=0.15];

\draw[\clre,fill=white] (0,2) circle[radius=0.15];

\draw[color=\clrb, fill=\clra, thick] (1,2) circle[radius=0.15];
\node[\clrc] at (1,2) {\scriptsize $\boldsymbol{\mathsf{Z}}$} ;
\draw[color=\clrb, fill=\clra, thick] (2,2) circle[radius=0.15];
\node[\clrc] at (2,2) {\scriptsize $\boldsymbol{\mathsf{Z}}$} ;
\draw[color=\clrb, fill=\clra, thick] (3,2) circle[radius=0.15];
\node[\clrc] at (3,2) {\scriptsize $\boldsymbol{\mathsf{Z}}$} ;

\draw[\clre,fill=white] (0,3) circle[radius=0.15];

\draw[\clre,fill=white] (1,3) circle[radius=0.15];

\draw[color=\clrb, fill=\clra, thick] (2,3) circle[radius=0.15];
\node[\clrc] at (2,3) {\scriptsize $\boldsymbol{\mathsf{Z}}$} ;
\draw[\clre,fill=white] (3,3) circle[radius=0.15];

\end{tikzpicture} \end{tabular}
\caption{\textbf{(a)} Tree of the Bravyi-Kitaev transform for 16 qubits. Qubits are labeled from 1 to 16 according to the underlying binary tree rule.  \textbf{(b)} Embedding the tree of 16 qubits into a  $(4\times 4)$ square lattice. \textbf{(c) \& (d)} Pauli strings  $(\bigotimes_{i\in U(10)} X_i )$ and $(\bigotimes_{i\in F(8)} Z_i )$ on the tree and the square lattice, where the arrows indicate the rules that determine the update set  $U(10)$, and the flip set $F(8)$ respectively: $F(i)$ would involve node $i$ and all its children, whereas  $U(j)$ would involve involves node $j$ and all its ancestors including the root. }
\label{fig:BKtree}
\end{figure}
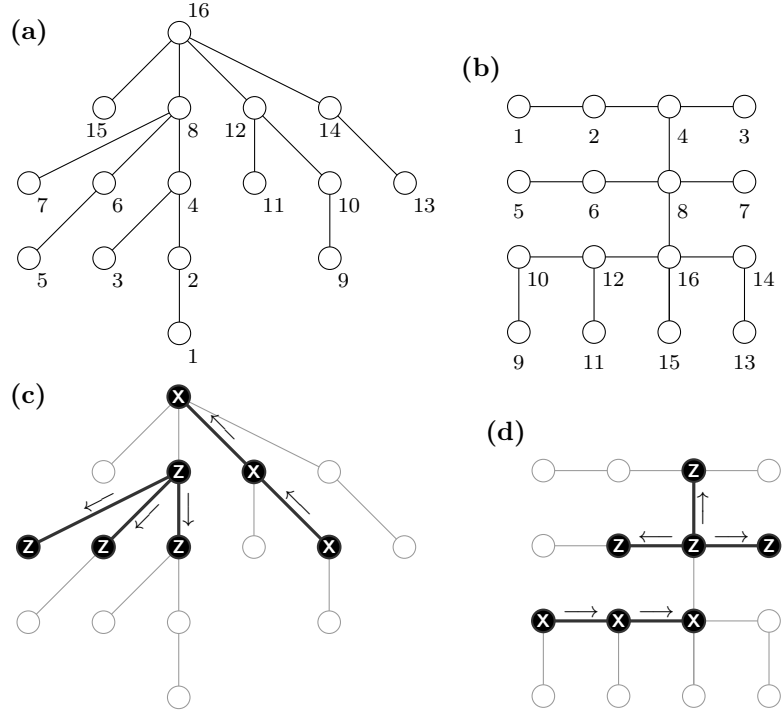

 In \cite{bravyi2002fermionic}, the mapping is introduced in order to reduce the weight of transformed fermionic operators to $O(\log N)$, which is an exponential improvement over the Jordan-Wigner transform. In the original paper, the (classical) encoding and decoding are defined by a partially ordering the mode indices according to some rules defined by their representation as binary numbers. Later works then developed the notion of flip, update and parity sets and provided a  method to construct the binary matrices $A^{-1}$ and $A$  in $\log N$ steps \cite{seeley2012bravyi,tranter2015bravyi}. Instead of being one-dimensional, the partial order can be regarded  placing all mode indices onto nodes inside a tree structure, which is the reason the mapping is sometimes referred to as binary-tree transform (even though the tree is not a binary tree).  As pointed out in \cite{havlivcek2017operator}, the flip and update operators of every mode $j$, $(\bigotimes_{k\in F(j)} Z_k)$  and $(\bigotimes_{l\in U(j)} X_l) $, have a geometric interpretation on that tree (as will be illustrated shortly), so we would naturally like to match  it with the qubit-connectivity graph. While an embedding is possible for small such trees, increasing $N$ will make the tree outgrow the square lattice rather quickly. In fact, the binary rule implies that the node with   index $2^j$ has exactly $j$ children, and all nodes with indices below $2^j$ have fewer than $j$ children.  This  means that trees with $N>16$ modes, cannot be embedded in the square lattice where every site has 4 nearest neighbors. The tree for $N=16$  can be found in Figure \ref{fig:BKtree}(a) and its embedding in the square lattice is presented in panel (b). This particular tree is however not the end of the story. In \cite{havlivcek2017operator}, it was argued that the Bravyi-Kitaev transform can be optimized to produce more local strings, in particular when considering Hamiltonians of locally-interacting Fermions. For that purpose, the `binary' trees are replaced with segmented Fenwick-tree structures. These structures are explicitly allowed to contain multiple trees, and  the number of trees is even a parameter of the mapping. This number  can range from $1$ to $N$ (the number of modes), where at $N$  the mapping is identical to the Jordan-Wigner transform and  at $1$ it corresponds the Bravyi-Kitaev transform (in case $N$ is an integer power of two).  However, we can go even further and define mappings based on an arbitrary number of arbitrary trees. In particular, we can define tree structures that can be embedded on  arbitrary qubit connectivity graphs, like our square lattice, and the associated mappings still yield small  parity operators $(\bigotimes_{m\in P(j)}Z_m)$. Let us consider one specific connectivity graph. \\  We need to pick a forest (a set of trees) which in total has a number of $N$ nodes. As each node will correspond to one qubit, the trees need to be connected to each other, and so we connect their respective roots. It is sufficient here for each root to be connected to two others, such that they are linked like a chain with their order foreshadowing some canonical ordering. We now  choose a set of trees, such that the graph created by connecting them can be embedded in the actual qubit-connectivity graph. Let us now turn to the description of the mapping itself. For that purpose, we firstly need to assign an index to every node, a process for which we later will provide an algorithm, but for now let us assume we have done so in a prudent way. For the definition of the transform, it is sufficient to give a definition of all update and flip sets,  as by  corresponding sets $F(j)$ and $U(j)$  the matrices $A$  and $A^{-1}$ can be inferred column- and row-wise.  For the flip set of index $j$, $F(j)$, we consider the node with index $j$ and all its children in the tree it is on, i.e. all the nodes directly connected to $j$ on edges that lead away from the root. The update set $U(j)$ includes the node $j$ and all its ancestors, i.e. all nodes on the direct line to the root (of the tree it is on), where the root is also included. A visual representation of these operators can be found in Figure \ref{fig:BKtree}(c), where the direction with respect to the root is indicated by arrows. Their embedded version can be found in panel (d) of the figure. Note that this means that by the encoding of this mappings, qubit $j$ stores the parity information of mode $j$ and all other modes whose index is beneath $j$ in the tree.  \\ For anticommutation relations like $[c^{\phantom{\dagger}}_i,c^{{\dagger}}_j]_+=\delta_{ij}$, it is important that
\begin{align}
 \left(\bigotimes_{k\in F(i)} Z_k\right)\left(\bigotimes_{l\in U(j)} X_l\right)= (-1)^{\delta_{ij}}\left(\bigotimes_{l\in U(j)} X_l\right)\left(\bigotimes_{k\in F(i)} Z_k\right) \, ,
 \end{align}
  which we now want to verify  by the definitions of the flip and update sets. If $j$ is any descendant of $i$, then the two operators overlap on two qubits, which means they commute. If it is not an ancestor, then the only case where the operators have overlap is when $i=j$, where they exactly overlap on that very qubit and anticommute.  \\
We so far have suppressed the discussion of the parity operators, that will now lead into an algorithm for the index assigning and a bound for the operator weight. Let us assume that our forest consisted of $\trees$ trees, each of which has at most $\levels$ levels and every node at most $\kids$ children. We know that the operator weight of update and flip operators scales as $ O(\levels+1)$ and $O(\kids+1)$, the structure of the parity set however now depends on the index assigned to the nodes. By a binary rule, the Bravyi-Kitaev transform manages to only involve $O(\log N)$ qubits in the parity operators, and we can devise a labeling that mirrors its principle.  The parity operator of $j$ is only the product of flip operators of $i<j$. On the other hand,  multiplying the flip operator of a parent node $k$ with all flip operators of its descendants will cancel all $Z$-operators but $Z_k$. Thus, in order  for the  parity operator of $j$ to have low weight, as many nodes with labels $i<j$ as possible need to be descendants of $j$. Subsequently, the mapping with the smallest parity sets is characterized by a tree where every node has only one child, i.e. a vertical line. This mapping, described in \cite{seeley2012bravyi} as \textit{parity transform}, has however the problem of $O(N)$-weight update operators, and is thus of the same quality as the Jordan-Wigner transform. Indeed, one being characterized by a vertical line, the other by a horizontal line (connected one-node trees),  makes both mappings  effectively one-dimensional. In order to minimize the weight of update and parity operators altogether, we need to reconcile the cancellation strategy with the tree structure. The idea is to involve only qubits in $P(j)$, that are children of the nodes in $U(j)$. Of course, this is not quite possible. If an entire tree only contains nodes $i<j$, then $P(j)$ will always contain the root of this tree. According to the formulas \eqref{eq:singleop}, transforming $c_j^{(\dagger)}$ thus results in strings of weight $O(\trees+\levels \kids)$. Not only this, but the strings produced will also be continuous for transforms of single operators. Unfortunately, for pairs of operators like $c_i^{\dagger} c_{j}^{\phantom{\dagger}}$, the strings are discontinuous on the first qubit that is both, an ancestor of $i$ and $j$ - a situation we cannot remedy. \\
 The question is now how to assign the labels to the nodes such that this mapping is implemented, or in other words: given an unlabeled  forest with connected roots, how can we obtain a mapping that outputs strings of weight  $O(\trees+\levels \kids)$?  For that purpose, we put labels $1$ to $N$ (in order) on the nodes according to the little program below.

\begin{description}
\item[Line 1] Consider the first tree in line.
\item[Line 2] Choose a leaf and put a label on it.
\item[Line 3] Check whether there are unlabeled siblings.  If there are, choose such a sibling for the consideration in the following step. If not, proceed to Line 5.
\item[Line 4] Check whether the  current node is a leaf, and if it is, label it, otherwise put a label on a leaf chosen from the sub-tree of which the current node is the root. Continue from Line 3 with the last-labeled node.
\item[Line 5] Check whether the last node considered has a parent. If there is a parent, put a label on it and continue from Line 3 with it. In case there is none, the previous node was a root, and we label it and proceed with the next line.
\item[Line 6] If the root is the top of the last tree, the program ends, but if it is not, the next tree in line is considered and the program continues from Line 2.
\end{description}

By the end of the program, all nodes are labeled in a way such that the resulting mapping outputs strings of weight $O(\trees+\levels \kids)$. Note that there might be  variations on how this process can turn out, since in several lines  an element of choice is involved. We can now consider customized trees and root-connected forests. For instance, we can consider a perfect binary tree (a real one this time), which yields a $O(\log N)$ scaling as well. Although with such a tree, every node is only required to have three nearest-neighbors, the embedding of an arbitrarily-sized tree into a square lattice is still not possible. This is due to the children that run into each other as we expand the tree-embedding on the lattice. We hope however that for future work the tools provided in this section will help to tailor tree-based transforms directly to specific device layouts.

\section{Superfast simulation and  Verstraete-Cirac transform}
\label{sec:mappings}
The goal of this section is to review the Superfast simulation \cite{bravyi2002fermionic} and Verstraete-Cirac transform \cite{verstraete2005mapping}, adapt them to the square lattice layout and give them the Manhattan-distance property. Lastly we obtain the strings of the Hubbard model that are referenced in Table \ref{tab:hubbardall}. This is going to be done within four parts: firstly, we will make some general remarks about the Manhattan-distance property that applies to all mappings, even the AQMs. The second and third part will concern the review and adaption of the Verstraete-Cirac transform and Superfast simulation, respectively. Each mapping will  be treated as some linear transform concatenated with a quantum code: we study the logical bases of the codes to learn how the mappings might be practically implemented. Lastly, we turn to the Hubbard model.
\subsection{Manhattan-distance property}
Verstraete-Cirac transform, Superfast simulation and the square lattice AQM   -  all three mappings inherently posses the Manhattan-distance property, which means that when we use them to transform hopping  interaction of two fermionic modes, the weight of the (shortest) resulting Pauli string  can be bounded with the Manhattan distance of the modes on the fermionic lattice. Here we will show that all mappings work in a similar fashion that enables us to use this property and elucidate why  it is necessary to make use of it in a limited qubit layout. However,  before we begin, we need to introduce the tools provided by the properties of Majorana modes.\\

Majorana particles are Fermions as their many-body wave-functions are anti-symmetric under permutation. Majorana operators $m_j^{(\dagger)}$ thus satisfy anticommutation relations like \eqref{eq:aannttii}, but they are also their own antiparticles, making the operators Hermitian: $m_j^{\dagger}=m_j^{\phantom{\dagger}}$. In general, these operators   describe the relations
\begin{align}
\label{eq:anticommix}
\left[ m_i,\, m_j\right]_{+} \;=\; 2\delta_{ij}  \quad \text{and} \quad  m_i m_i = 1 \, .
\end{align}
Per fermionic mode, we need two Majoranas, such that the fermionic operators $c_j^{(\dagger)}$ are described by two Majorana species $m_j$ and $\obar{m}_j$, where $\obar{m}_j$ obey the same relations \eqref{eq:anticommix}, and are indistinguishable to $m_j$, so $m_i \obar{m}_j\, =\, - \obar{m}_j m_i $.  We define

\begin{align}
\label{eq:majoran}
&c^{\dagger}_j=\frac{1}{2}\left( m_j - i \,\obar{m}_j \right) \quad \text{and} \quad c^{\phantom{\dagger}}_j=\frac{1}{2}\left( m_j + i \,\obar{m}_j \right) \, .
\end{align}
We thus can represent the operators $m_j$, $\obar{m}_j$ with the Jordan-Wigner transform as
\begin{align}
\label{eq:majoran2}
m_j \;\hat{=} \;\left( \bigotimes_{k=1}^{j-1} Z_k \right) \otimes X_j \quad \text{and} \quad \obar{m}_j \;\hat{=}\; \left( \bigotimes_{k=1}^{j-1} Z_k \right) \otimes Y_j \, .
\end{align}
Majorana-pair operators  $ (m_j \, m_k)$ are used in the original proposals of VCT and BKSF, and their structure is also an element in the  AQM. This is because these operators can be  transformed into single Pauli strings that describe  the interaction of two fermionic modes $j$ and $k$, making them a useful tool for modeling it. As already established, they also have quite convenient (anti-) commutation relations.
All mappings  introduce extra qubits to encode operators corresponding to Majorana pairs   $   (m_j \, m_k) \,\hat{\propto} \,\mathcal{O}_{jk}$.
In one or the other way, all mappings use these operators to prevent hopping terms, as they occur in  fermionic Hamiltonians, to become non-local Pauli strings in the qubit Hamiltonian. When non-local connections of modes $i$ with $k$, and $k$ with $j$,  as well as $i$ with $j$ appear in a fermionic Hamiltonian,   one might think of encoding three operators $\mathcal{O}_{ik}$,  $\mathcal{O}_{kj}$ and $\mathcal{O}_{ij}$.  However, all  mappings exhibit repercussions for adding qubits to encode these operators, such as a weight increase in the substrings $\kappa^h_\aux$ in case of the AQM. There is also the issue that we need to connect all the modes in a way that would mimic the connectivity graph of the fermionic Hamiltonian - a Hamiltonian that is generally more complicated than a lattice model. In order to be modest with the amount of qubits to be added and to be able to deal with the limited  connectivity of the setup,  we reconsider encoding  operators $\mathcal{O}_{ij}$ of all possible combinations $ij$ by adding qubits. Instead, under the cost of a slightly higher operator weight, we can obtain some non-local $\mathcal{O}_{ij}$ by multiplying operators that are already encoded: $\mathcal{O}_{ij}\propto \mathcal{O}_{ik} \mathcal{O}_{kj}$.

 This is possible since for  Majorana pairs we find   $(m_i \, m_j)=(m_i \, m_k)\cdot(m_k \, m_j)$.  We report only a  `slightly' higher weight as $\mathcal{O}_{ik}$ and $\mathcal{O}_{kj}$ have been introduced to localize their respective links in the first place. With the same argument we can take a walk over an arbitrary sequence of indices $k_1, \, k_2, \, \dots, \, k_l$, where $k_s$ and $k_{s+1}$ are connected by an operator $\mathcal{O}_{k_{s}k_{s+1}}$,   just to obtain the operator that links the first and the last mode $k_1$ and $k_l$
\begin{align}
\label{eq:manhattanops}
\mathcal{O}_{k_1 k_l}\; \propto \;  \prod_{s=1}^l \mathcal{O}_{k_s k_{s+1}} \, .
\end{align}
This is the foundation for the Manhattan-distance property of all three mappings.  \\
\subsection{Verstraete-Cirac transform}
\subsubsection{Review}
Here we will review the Verstraete-Cirac transform starting with the original proposal \cite{verstraete2005mapping}, that, like the AQMs, can be regarded as manipulation of the Jordan-Wigner transform in which non-local strings are canceled with stabilizers. There, the auxiliary degrees of freedom that produce these stabilizers are added on the side of the model, where we find them in the form of  Majorana modes. However, in the investigation of this mapping we found the consideration of the mapping as a quantum code more practical for a rigorous derivation of the stabilizers and outputs. This is why after a short motivation in the original language, we will describe the general concept of this mapping as a quantum code quite similar to the concept of the Auxiliary Qubit codes, which allows for the description of customized mappings such as a mapping with an odd number of rows or a qubit-economic version.

The idea of \cite{verstraete2005mapping} is to extend the fermionic systems by doubling the number of modes, where the  modes added are denoted by primed
numbers from $1^\prime$ to $N^\prime$.
 For all indices $k$, $k^\prime$ does not denote another variable but is the primed version of the value of $k$.  For the Jordan-Wigner transform, we need to impose the canonical order of $2N$ sites, and so we stagger primed and unprimed indices: $1, \, 1^\prime, \, 2 , \,  2^\prime ,\, \dots \, \, N, \, N^\prime  $. \\
 Adding those primed sites, we practically increase the length of Pauli strings, since all hopping terms on the original system hop over primed sites, even turning horizontal nearest-neighbor hoppings  into next-nearest neighbor interactions.

 \begin{align}
 \label{eq:primed}
 (i<j):\qquad c_i^{\dagger} c_j^{\phantom{\dagger}} +  c_j^{\dagger} c_i^{\phantom{\dagger}} \; \; \hat{=}\; \; &\frac{1}{2}\left[ \bigotimes_{k=i+1}^{j-1}  Z_k\right]  \left(X_i \otimes X_j + Y_i\otimes Y_j \right)\notag \\  \to \; \; &\frac{1}{2}\left[ \bigotimes_{k=i+1}^{j-1} \left( Z_k\otimes Z_{k^\prime}\right)\right]  \left(X_i \otimes Z_{i^\prime} \otimes X_j + Y_i\otimes Z_{i^\prime} \otimes Y_j \right) \, .
 \end{align}

The hopping terms are thus made sensitive to the primed subsystem, and the original system is recovered if all primed modes are empty. In their original work, Verstraete and Cirac   define a fermionic quantum code, that constrains the primed subsystem completely by means of majoranic stabilizers $(i \, m_{j^\prime} \,\obar{m}_{k^\prime})$ for certain pairs of modes $j^\prime$ and $k^\prime$.  These are translated to the qubit side by Jordan-Wigner transform  $(i \, m_{j^\prime} \, \obar{m}_{k^\prime}) \,\hat{=}\,\mathcal{P}_{jk} $. While in the original proposal, the majoranic stabilizers  $(i m_{j^\prime} \obar{m}_{k^\prime})$ are fixed as gap terms in the model Hamiltonian, it is suggested in \cite{whitfield2016local} to prepare the entangled state by making syndrome measurements with the transformed stabilizers $\mathcal{P}_{jk}$. \\

Stabilizers like $(i m_{j^\prime} \obar{m}_{k^\prime})$ are useful to cancel non-local connections between $j$ and $k$. Let us here assume that such a stabilizer is present, then the hopping between those modes can be modified by multiplication of the corresponding fermionic terms in the model Hamiltonians:
  \begin{align}
\left(  c_j^{\dagger} c_k^{\phantom{\dagger}} +  c_k^{\dagger} c_j^{\phantom{\dagger}} \right) i\, m_{j^\prime}\, \obar{m}_{k^\prime} \; \hat{=}  \;-\frac{1}{2}\,  X_j\otimes X_{j^\prime} \otimes Y_{k} \otimes Y_{k^\prime}\; + \; \frac{1}{2}\, Y_j \otimes X_{j^\prime} \otimes X_{k} \otimes Y_{k^\prime} \, .
\end{align}
As one can see, the re-sized parity string has been canceled. Although all operators involved satisfy the correct (anti-)commutation relations, it is not possible to attribute the correct sign to all stabilizers and Hamiltonian terms without considering the code space. To do so, we now derive the quantum code version of the VCT, starting by the constructing the logical basis, that has to determine the adjustments to the Jordan-Wigner-transformed Hamiltonian terms.

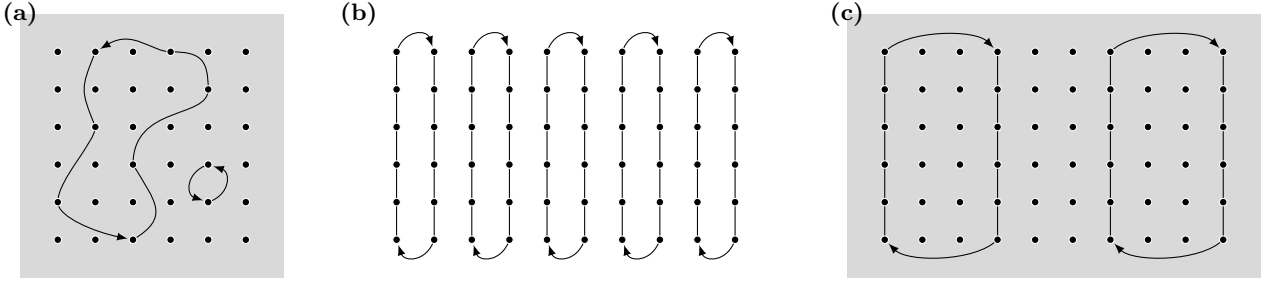
\begin{figure}
\begin{tikzpicture}[scale=0.5]

\fill[\checker] (0,-1)--(7,-1)--(7,6)--(0,6)--cycle;


\draw[](3,2)..controls(3,3.6) and (5,3.2) ..(5,4);
\draw[](5,4)..controls(5,4.8) and (4.8,5) ..(4,5);
\draw[-latex,shorten >= 2](4,5)..controls(3.5,5) and (3,5.6)..(2,5);
\draw[](2,5)..controls(1.6,4)..(2,3);
\draw[](2,3)..controls(2,2.6) and (1,1.6)..(1,1);
\draw[-latex,shorten >= 2](1,1)..controls(1,0.6) and (2,0.2)..(3,0);
\draw[](3,0)..controls(4,0.6) and (3.6,1)..(3,2);

\draw[-latex,shorten >= 2](5,1).. controls(5.6,1.2)and (5.6,1.8).. (5,2);
\draw[-latex,shorten >= 2](5,2).. controls(4.4,1.8)and (4.4,1.2).. (5,1);
\draw[white,fill=black](1,0) circle[radius=0.1];
\draw[white,fill=black](2,0) circle[radius=0.1];
\draw[white,fill=black] (3,0) circle[radius=0.1];
\draw[white,fill=black] (4,0) circle[radius=0.1];
\draw[white,fill=black] (5,0) circle[radius=0.1];
\draw[white,fill=black] (6,0) circle[radius=0.1];

\draw[white,fill=black] (1,1) circle[radius=0.1];
\draw[white,fill=black] (2,1) circle[radius=0.1];
\draw[white,fill=black] (3,1) circle[radius=0.1];
\draw[white,fill=black] (4,1) circle[radius=0.1];
\draw[white,fill=black] (5,1) circle[radius=0.1];
\draw[white,fill=black] (6,1) circle[radius=0.1];

\draw[white,fill=black] (1,2) circle[radius=0.1];
\draw[white,fill=black] (2,2) circle[radius=0.1];
\draw[white,fill=black] (3,2) circle[radius=0.1];
\draw[white,fill=black] (4,2) circle[radius=0.1];
\draw[white,fill=black] (5,2) circle[radius=0.1];
\draw[white,fill=black] (6,2) circle[radius=0.1];

\draw[white,fill=black] (1,3) circle[radius=0.1];
\draw[white,fill=black] (2,3) circle[radius=0.1];
\draw[white,fill=black] (3,3) circle[radius=0.1];
\draw[white,fill=black] (4,3) circle[radius=0.1];
\draw[white,fill=black] (5,3) circle[radius=0.1];
\draw[white,fill=black] (6,3) circle[radius=0.1];

\draw[white,fill=black] (1,4) circle[radius=0.1];
\draw[white,fill=black] (2,4) circle[radius=0.1];
\draw[white,fill=black] (3,4) circle[radius=0.1];
\draw[white,fill=black] (4,4) circle[radius=0.1];
\draw[white,fill=black] (5,4) circle[radius=0.1];
\draw[white,fill=black] (6,4) circle[radius=0.1];

\draw[white,fill=black] (1,5) circle[radius=0.1];
\draw[white,fill=black](2,5) circle[radius=0.1];
\draw[white,fill=black] (3,5) circle[radius=0.1];
\draw[white,fill=black](4,5) circle[radius=0.1];
\draw[white,fill=black] (5,5) circle[radius=0.1];
\draw[white,fill=black] (6,5) circle[radius=0.1];

\node[] at (0,6) {\textbf{(a)}};

\end{tikzpicture}$\qquad$\begin{tikzpicture}[scale=0.5]
\fill[white] (0,-1)--(11,-1)--(11,6)--(0,6)--cycle;
\draw[-latex, shorten >= 2 ](1,5).. controls(1.2,5.6)and (1.8,5.6).. (2,5);
\draw[-latex, shorten >= 2 ](2,0).. controls(1.8,-.6)and (1.2,-.6).. (1,0);
\draw[] (1,5)--(1,0);
\draw[] (2,5)--(2,0);

\draw[-latex, shorten >= 2 ](3,5).. controls(3.2,5.6)and (3.8,5.6).. (4,5);
\draw[-latex, shorten >= 2 ](4,0).. controls(3.8,-.6)and (3.2,-.6).. (3,0);
\draw[] (3,5)--(3,0);
\draw[] (4,5)--(4,0);

\draw[-latex, shorten >= 2 ](5,5).. controls(5.2,5.6)and (5.8,5.6).. (6,5);
\draw[-latex, shorten >= 2 ](6,0).. controls(5.8,-.6)and (5.2,-.6).. (5,0);
\draw[] (5,5)--(5,0);
\draw[] (6,5)--(6,0);

\draw[-latex, shorten >= 2 ](7,5).. controls(7.2,5.6)and (7.8,5.6).. (8,5);
\draw[-latex, shorten >= 2 ](8,0).. controls(7.8,-.6)and (7.2,-.6).. (7,0);
\draw[] (7,5)--(7,0);
\draw[] (8,5)--(8,0);

\draw[-latex, shorten >= 2 ](9,5).. controls(9.2,5.6)and (9.8,5.6).. (10,5);
\draw[-latex, shorten >= 2 ](10,0).. controls(9.8,-.6)and (9.2,-.6).. (9,0);
\draw[] (9,5)--(9,0);
\draw[] (10,5)--(10,0);
\draw[white,fill=black](1,0) circle[radius=0.1];
\draw[white,fill=black](2,0) circle[radius=0.1];
\draw[white,fill=black] (3,0) circle[radius=0.1];
\draw[white,fill=black] (4,0) circle[radius=0.1];
\draw[white,fill=black] (5,0) circle[radius=0.1];
\draw[white,fill=black] (6,0) circle[radius=0.1];
\draw[white,fill=black] (7,0) circle[radius=0.1];

\draw[white,fill=black] (1,1) circle[radius=0.1];
\draw[white,fill=black] (2,1) circle[radius=0.1];
\draw[white,fill=black] (3,1) circle[radius=0.1];
\draw[white,fill=black] (4,1) circle[radius=0.1];
\draw[white,fill=black] (5,1) circle[radius=0.1];
\draw[white,fill=black] (6,1) circle[radius=0.1];
\draw[white,fill=black] (7,1) circle[radius=0.1];

\draw[white,fill=black] (1,2) circle[radius=0.1];
\draw[white,fill=black] (2,2) circle[radius=0.1];
\draw[white,fill=black] (3,2) circle[radius=0.1];
\draw[white,fill=black] (4,2) circle[radius=0.1];
\draw[white,fill=black] (5,2) circle[radius=0.1];
\draw[white,fill=black] (6,2) circle[radius=0.1];
\draw[white,fill=black] (7,2) circle[radius=0.1];

\draw[white,fill=black] (1,3) circle[radius=0.1];
\draw[white,fill=black] (2,3) circle[radius=0.1];
\draw[white,fill=black] (3,3) circle[radius=0.1];
\draw[white,fill=black] (4,3) circle[radius=0.1];
\draw[white,fill=black] (5,3) circle[radius=0.1];
\draw[white,fill=black] (6,3) circle[radius=0.1];
\draw[white,fill=black] (7,3) circle[radius=0.1];

\draw[white,fill=black] (1,4) circle[radius=0.1];
\draw[white,fill=black] (2,4) circle[radius=0.1];
\draw[white,fill=black] (3,4) circle[radius=0.1];
\draw[white,fill=black] (4,4) circle[radius=0.1];
\draw[white,fill=black] (5,4) circle[radius=0.1];
\draw[white,fill=black] (6,4) circle[radius=0.1];
\draw[white,fill=black](7,4) circle[radius=0.1];

\draw[white,fill=black] (1,5) circle[radius=0.1];
\draw[white,fill=black](2,5) circle[radius=0.1];
\draw[white,fill=black] (3,5) circle[radius=0.1];
\draw[white,fill=black](4,5) circle[radius=0.1];
\draw[white,fill=black] (5,5) circle[radius=0.1];
\draw[white,fill=black] (6,5) circle[radius=0.1];
\draw[white,fill=black] (7,5) circle[radius=0.1];

\draw[white,fill=black] (8,1) circle[radius=0.1];
\draw[white,fill=black] (8,2) circle[radius=0.1];
\draw[white,fill=black] (8,3) circle[radius=0.1];
\draw[white,fill=black](8,4) circle[radius=0.1];
\draw[white,fill=black] (8,5) circle[radius=0.1];
\draw[white,fill=black](8,0) circle[radius=0.1];

\draw[white,fill=black] (9,1) circle[radius=0.1];
\draw[white,fill=black] (9,2) circle[radius=0.1];
\draw[white,fill=black] (9,3) circle[radius=0.1];
\draw[white,fill=black] (9,4) circle[radius=0.1];
\draw[white,fill=black] (9,5) circle[radius=0.1];
\draw[white,fill=black] (9,0) circle[radius=0.1];

\draw[white,fill=black] (10,1) circle[radius=0.1];
\draw[white,fill=black] (10,2) circle[radius=0.1];
\draw[white,fill=black] (10,3) circle[radius=0.1];
\draw[white,fill=black] (10,4) circle[radius=0.1];
\draw[white,fill=black] (10,5) circle[radius=0.1];
\draw[white,fill=black] (10,0) circle[radius=0.1];
\node[] at (0,6) {\textbf{(b)}};
\end{tikzpicture}$\qquad$\begin{tikzpicture}[scale=0.5]

\fill[\checker] (0,-1)--(11,-1)--(11,6)--(0,6)--cycle;
\draw[-latex,shorten >= 2](1,5).. controls(1.6,5.6)and (3.4,5.6).. (4,5);
\draw[-latex,shorten >= 2](4,0).. controls(3.4,-.6)and (1.6,-.6).. (1,0);
\draw[-latex,shorten >= 2](7,5).. controls(7.6,5.6)and (9.4,5.6).. (10,5);
\draw[-latex,shorten >= 2](10,0).. controls(9.4,-.6)and (7.6,-.6).. (7,0);


\draw[] (1,0)--(1,5);
\draw[] (4,0)--(4,5);
\draw[] (7,0)--(7,5);
\draw[] (10,0)--(10,5);

\draw[white,fill=black](1,0) circle[radius=0.1];
\draw[white,fill=black](2,0) circle[radius=0.1];
\draw[white,fill=black] (3,0) circle[radius=0.1];
\draw[white,fill=black] (4,0) circle[radius=0.1];
\draw[white,fill=black] (5,0) circle[radius=0.1];
\draw[white,fill=black] (6,0) circle[radius=0.1];
\draw[white,fill=black] (7,0) circle[radius=0.1];

\draw[white,fill=black] (1,1) circle[radius=0.1];
\draw[white,fill=black] (2,1) circle[radius=0.1];
\draw[white,fill=black] (3,1) circle[radius=0.1];
\draw[white,fill=black] (4,1) circle[radius=0.1];
\draw[white,fill=black] (5,1) circle[radius=0.1];
\draw[white,fill=black] (6,1) circle[radius=0.1];
\draw[white,fill=black] (7,1) circle[radius=0.1];

\draw[white,fill=black] (1,2) circle[radius=0.1];
\draw[white,fill=black] (2,2) circle[radius=0.1];
\draw[white,fill=black] (3,2) circle[radius=0.1];
\draw[white,fill=black] (4,2) circle[radius=0.1];
\draw[white,fill=black] (5,2) circle[radius=0.1];
\draw[white,fill=black] (6,2) circle[radius=0.1];
\draw[white,fill=black] (7,2) circle[radius=0.1];

\draw[white,fill=black] (1,3) circle[radius=0.1];
\draw[white,fill=black] (2,3) circle[radius=0.1];
\draw[white,fill=black] (3,3) circle[radius=0.1];
\draw[white,fill=black] (4,3) circle[radius=0.1];
\draw[white,fill=black] (5,3) circle[radius=0.1];
\draw[white,fill=black] (6,3) circle[radius=0.1];
\draw[white,fill=black] (7,3) circle[radius=0.1];

\draw[white,fill=black] (1,4) circle[radius=0.1];
\draw[white,fill=black] (2,4) circle[radius=0.1];
\draw[white,fill=black] (3,4) circle[radius=0.1];
\draw[white,fill=black] (4,4) circle[radius=0.1];
\draw[white,fill=black] (5,4) circle[radius=0.1];
\draw[white,fill=black] (6,4) circle[radius=0.1];
\draw[white,fill=black](7,4) circle[radius=0.1];

\draw[white,fill=black] (1,5) circle[radius=0.1];
\draw[white,fill=black](2,5) circle[radius=0.1];
\draw[white,fill=black] (3,5) circle[radius=0.1];
\draw[white,fill=black](4,5) circle[radius=0.1];
\draw[white,fill=black] (5,5) circle[radius=0.1];
\draw[white,fill=black] (6,5) circle[radius=0.1];
\draw[white,fill=black] (7,5) circle[radius=0.1];

\draw[white,fill=black] (8,1) circle[radius=0.1];
\draw[white,fill=black] (8,2) circle[radius=0.1];
\draw[white,fill=black] (8,3) circle[radius=0.1];
\draw[white,fill=black](8,4) circle[radius=0.1];
\draw[white,fill=black] (8,5) circle[radius=0.1];
\draw[white,fill=black](8,0) circle[radius=0.1];

\draw[white,fill=black] (9,1) circle[radius=0.1];
\draw[white,fill=black] (9,2) circle[radius=0.1];
\draw[white,fill=black] (9,3) circle[radius=0.1];
\draw[white,fill=black] (9,4) circle[radius=0.1];
\draw[white,fill=black] (9,5) circle[radius=0.1];
\draw[white,fill=black] (9,0) circle[radius=0.1];

\draw[white,fill=black] (10,1) circle[radius=0.1];
\draw[white,fill=black] (10,2) circle[radius=0.1];
\draw[white,fill=black] (10,3) circle[radius=0.1];
\draw[white,fill=black] (10,4) circle[radius=0.1];
\draw[white,fill=black] (10,5) circle[radius=0.1];
\draw[white,fill=black] (10,0) circle[radius=0.1];

\node[] at (0,6) {\textbf{(c)}};
\end{tikzpicture}
\caption{Verstraete-Cirac transform. \textbf{(a)} An arbitrary mapping showcasing the constraints on the VCT code space. The black dots correspond to data qubits. Directed loops of operators $\mathcal{P}^b_{jk}$ are drawn into this grid, where the direction of one loop is indicated by arrows. With 9 vertices involved, we entangle 9 auxiliary qubits to that system. \textbf{(b)} Graph of the original proposal  \cite{verstraete2005mapping}. \textbf{(c)} One possibility for a qubit-economic version of the VCT. } \label{fig:vc}
\end{figure}

Although it was recently pointed out in \cite{whitfield2016local}, that keeping the stabilizers majoranic is unnecessary, we will stick to the original concept and merely add the freedom to `flip' the stabilizer by introducing a sign
\begin{align}
\label{eq:VCstabs}
\mathcal{P}^{b_s}_{\alpha_s \beta_s} \; \hat{=} \;(-1)^{b_s}\, i \,m_{\alpha^\prime_s} \obar{m}_{\beta^\prime_{s}}\, ,
\end{align}
where $\bbs{\beta}, \, \bbs{\alpha}=(\alpha_1 , \, \alpha_2, \, \dots \, ,\, \alpha_r)\in[N]^{\otimes r}$ and $\bbs{b}=(b_1 , \, b_2, \, \dots \, ,\, b_r)\in \zetto{r}$ are  sequences that parameterize the mapping. A `flipped' stabilizer would practically be implemented by requiring that syndrome measurements have the outcomes $(-1)$, so a stabilizer $\mathcal{P}^{b_s}_{\alpha_s \beta_s}$ constrains the code space to $\langle \mathcal{P}^{0}_{\alpha_s \beta_s} \rangle = (-1)^{b_s}$. Instead of the primed and unprimed subspace to host indistinguishable Fermions and being interleaved in the canonical order, we separate those modes (qubits) in an attempt to regard the primed subspace as the auxiliary register. The $\aux$-register is not even required to have size $N$, instead a smaller number of auxiliary qubits can be chosen, $r\leq N$. Although separated into different registers, each auxiliary qubit is still affiliated with a data qubit, or rather their corresponding modes are. Our intention is to keep the previous notation and let the auxiliary register contain the primed labels.  For that purpose, we  introduce the set $W$ as a $r$-sized subset of the mode numbers, $W\subseteq [N]$, such that the auxiliary register is comprised of qubits labeled $(\bigcup_{k\in W} k^\prime)$. In this way every data qubit  $k\in W$ has an auxiliary qubit $k^\prime$ associated with it. Let us now characterize a general version of this  mapping. We consider the $(\ell_1 \times \ell_2)$ block of data qubits and for every $s\in[r]$ connect the qubits $\alpha_s$ and $\beta_s$ in a directed graph. For every qubit $k$ that is a vertex of this graph, we add an auxiliary qubit $k^\prime$ somewhere, and the number $k$ becomes a member of $W$. Generalizing \eqref{eq:VCstabs}, the stabilizers of the qubit system are
\begin{align}
\mathcal{P}^{b_s}_{\alpha_s\beta_{s}} \; =  \; &(-1)^{b_s} \left(\bigotimes_{j=\alpha_{s}+1}^{\beta_{s}} Z_{j} \right)  \otimes Y_{\alpha_s^\prime} \otimes   \left( \bigotimes_{\begin{smallmatrix} k \in W \\ \alpha_{s}<k<\beta_{s} \end{smallmatrix}} Z_{k^\prime} \right) \otimes Y_{\beta^\prime_{s}} \qquad \text{if} \quad \alpha_s < \beta_{s} \notag \\ \; =  \; &(-1)^{b_s} \left(\bigotimes_{j=\beta_{s}+1}^{\alpha_{s}} Z_{j} \right)  \otimes X_{\beta_{s}^\prime} \otimes   \left(  \bigotimes_{ \begin{smallmatrix} k \in W \\  \beta_{s}<k<\alpha_s \end{smallmatrix}} Z_{k^\prime}\right) \otimes X_{\alpha^\prime_s} \qquad \text{if} \quad \alpha_s > \beta_{s} \label{eq:VCstabs2} \,.
\end{align}
Note that for the quantum code, that we intent to construct with the set  $\lbrace \mathcal{P}^{b_s}_{\alpha_s \beta_s} \rbrace_{\bbs{\alpha}, \, \bbs{\beta}}$ as  stabilizer generators, certain conditions on
 $\bbs{\alpha}$ and $\bbs{\beta}$  are to be met. While these conditions are intrinsically fulfilled for the mappings in \cite{verstraete2005mapping}, we want to briefly spell them out for the sake of generality.
The following conditions must be met by the directed graph  on which $\bbs{\alpha}$ and $\bbs{\beta}$ are defined: (i) the graph must be composed of closed loops on the $(\ell_1 \times \ell_2)$-grid. (ii) The loops do not overlap in their vertices. (iii) The loops are uniformly directed, which means that within one loop no two edges  point towards the same vertex. \\
Statement (i) is just a consequence of the fact that we need to constrain the auxiliary system completely. As the stabilizers \eqref{eq:VCstabs2} are associated with edges, we need to consider closed loops, otherwise  degrees of freedom remain undetermined.  We also need to make sure that all stabilizes  commute and so, considering \eqref{eq:VCstabs}, we find that every vertex can host one incident and one outbound edge. This, together with statement (i), explains statements (ii) and (iii). An example of such a mapping, for which all three statements hold, is depicted in Figure \ref{fig:vc}(a), where we consider two  loops in counter-clockwise directions. While in (a), we eliminate some arbitrary non-local connections, Figure \ref{fig:vc}(b) exhibits the original proposal, where the stabilizer implement the vertical connections. Of course we need to involve a few horizontal connections in order to comply with statement (i). As loops cannot be closed in it, the original proposal deals with an odd number $\ell_1$ in ignoring the last column.  Alternatively, we suggest that one could just create loops between vertically adjacent modes in that last column,  like it is done in the right-most loop in panel (a). It is of course only possible to stabilize roughly half of all vertical connections in this way, i.e. all even or all odd pairs. Assuming an underlying S-pattern of  the canonical ordering, nothing else would be required, since  half of the links are  local anyways. The original proposal yields a decent mapping already, as we can shorten vertical hoppings along the last column by multiplications  stabilizers of the second-to-last column. In fact, the idea that not every column needs to have their own auxiliary qubits is  the foundation for qubit-conserving versions of the VCT, as is shown in Figure \ref{fig:vc}(c). Note that in order to comply with the three statements, the periodicity $\inter$ has to be chosen such that $(\ell_1-1)/\inter$ is an odd number, the size of the auxiliary register subsequently becomes $r=\ell_2+(\ell_1-1)\ell_2/ \inter$. Note that a loop of one vertex is counterproductive, resulting  in a stabilizer $\mathcal{P}^{b}_{jj}=(-1)^{1+b}Z_{j^\prime}$. This only fixes the parity of the auxiliary qubit, which renders it redundant since it is not entangled with the rest of the system. Not just that, it blocks the mode from being part in another loop.\\

Let us now take a look to the  basis of the extended system. As before, the $N$ original modes describing the fermionic Fock space shall make up the data qubit register and the primed auxiliary qubits  be in the register $\aux=(\bigcup_{k\in W}k^\prime)$. An ansatz for a logical basis stabilized by all $\lbrace\mathcal{P}^{b_s}_{\alpha_s \beta_s}\rbrace$ is
\begin{align}
\label{eq:VCbasis}
\ket{\bbs{\omega}}_\data \; \to \; \;  \propto \; \left( \sum_{\bbs{\mu}\in \zetto{r}} \prod_{s=1}^{r} \left[\mathcal{P}^{b_s}_{\alpha_s \beta_s}\right]^{\mu_s} \right) \; \ket{\bbs{\omega}}_{\data}\otimes \ket{\bbs{\chi}}_{\aux} \, ,
\end{align}
where $\ket{\bbs{\chi}}_{\aux}\;=\;(\bigotimes_{k\in W} \ket{\chi_k}_{k^\prime} )$ is a product state on the auxiliary register that can be chosen inside a certain range of parity constraints, which we now want to explain.\\
 These parity constraints are related to a certain freedom in the characterization of the mapping. We have not determined $\bbs{b}$ yet, as up to now the only restrictions we had were on the choice of  $\bbs{\alpha}$ and $\bbs{\beta}$.  In order to understand the role of $\bbs{b}$, let us for a moment assume that the graph spanned by $\bbs{\alpha}$ and $\bbs{\beta}$ is only one loop, which means that $\beta_s=\alpha_{s+1}$ and $\beta_r=\alpha_{1}$. No matter the number of loops, the sum in the basis \eqref{eq:VCbasis} will always contain the product of all stabilizers around a closed loop, here it is  $(\prod_{s=1}^{r} \mathcal{P}^{b_s}_{\alpha_s \beta_{s}})$, met by the summand for which $\bbs{\mu}=(1)^{\otimes r}$. In fact, half of the terms in the sum will differ from the other half only by these operators: (having omitted the normalization factor for that reason) it is alright for some stabilizers to be linearly dependent, as long as they stabilize $\ket{\bbs{\omega}}_{\data}\otimes \ket{\bbs{\chi}}_{\aux}$. Since we are stabilizing a loop, we find by \eqref{eq:VCstabs}, that
 \begin{align}
 \label{eq:VCparity}
 \prod_{s=1}^{r} \mathcal{P}^{b_s}_{\alpha_s \beta_{s}} \; = \; (-1)^{1+\sum_{k=1}^r {b_k}} \, \bigotimes_{j\in W} Z_{j^\prime} \, .
 \end{align}
Since \eqref{eq:VCparity} acts only on $\ket{\bbs{\chi}}_{\aux}$, it becomes apparent that $\bbs{b}$ determines the parity of all auxiliary qubits associated with the loops in the mapping. According to the choice of $\bbs{b}$, we now need to pick a state $\ket{\bbs{\chi}}_\aux$ that meets all parity constraints \eqref{eq:VCparity}. Since we in general have more than one loop in our mapping, we need to fix the parity on several distinct subsets of  $\ket{\bbs{\chi}}_\aux$. For instance if we pick the parity of every loop to be even, we can choose $\ket{\bbs{\chi}}_\aux=\ket{0^r}_\aux$.  \\

We lastly show that $Z$-strings  on the primed qubits come naturally as adjustments to Hamiltonian terms $h_\data$, together with minus signs from the loop parity constraints. The data-qubit substring of the stabilizers \eqref{eq:VCstabs2} is purely a $Z$-string, so we do not need to adjust a string $h_\data \in \lbrace \mathbb{I}, \, Z \rbrace^{\otimes N}$. This means that it is sufficient to consider the changes to be made to a string $(\bigotimes_{j=1}^{k-1} Z_j)\otimes X_k$, in order to describe all fermionic operators $c_k^{(\dagger)}$. This string anticommutes with all stabilizers, that have data qubit substrings $(\bigotimes_{j=s}^t Z_j)$, where $s\leq k$. These stabilizers, $\mathcal{P}^b_{(s-1)t}$ or $\mathcal{P}^b_{t(s-1)}$,  act on the $\aux$-register as
\begin{align}
(-1)^b\; Y_{(s-1)^\prime} \otimes \left( \bigotimes_{\begin{smallmatrix} j\in W\\ t<j<(s-1) \end{smallmatrix}} Z_{j^\prime} \right) \otimes Y_{t^\prime} \qquad \text{or} \qquad (-1)^b\; X_{(s-1)^\prime} \otimes \left( \bigotimes_{\begin{smallmatrix} j\in W\\ t<j<(s-1) \end{smallmatrix}} Z_{j^\prime} \right) \otimes X_{t^\prime} \, ,
\end{align}
which means they change the parity of the subsystem that is spanned by all auxiliary qubits with the labels $j^\prime$, where $j\leq(k-1)$ and $j \in W$. The total parity of all auxiliary qubits is however constant i.e. it does not change with the multiplication of either stabilizer. The total parity is predetermined by $\ket{\bbs{\chi}}_\aux$ and the action of a Majorana-pair operator conserves it.

  If we now multiply $(\bigotimes_{j=1}^{k-1} Z_j)\otimes X_k$ to a basis element \eqref{eq:VCbasis}, we can determine whether it anticommutes with an even or odd number of stabilizers as we move it to the right until it reaches $\ket{\bbs{\omega}}_{\data} \otimes \ket{\bbs{\chi}}_{\aux}$: it anticommutes with an odd number of stabilizers if the parity of  the subsystem, spanned by all auxiliary qubits with labels at most as large as $(k-1)^\prime$, is changed. We therefore extract the parity of said subsystem by the operator $(\bigotimes_{j\in W <k} Z_{j^\prime})$ and add a minus sign in case $(\bigotimes_{j\in W <k} Z_{j^\prime})\ket{\bbs{\chi}}_{\aux}\;=\;(-1)\ket{\bbs{\chi}}_{\aux}$. We hence find
\begin{align}
\left(\bigotimes_{i=1}^{k-1} Z_i\right)\otimes X_k \; \to \;  \pm \left(\bigotimes_{j=1}^{k-1} Z_j\right)\otimes X_k \otimes  \left( \bigotimes_{j \in W <k} Z_{j^\prime} \right) \,
\end{align}
where the sign is determined by $\ket{\bbs{\chi}}$. When we  consider the planar code of the original proposal, we find that string has become
\begin{align}
\pm \,  \left[ \bigotimes_{j=1}^{k-1} (Z_j\otimes Z_{j^\prime}) \right] \otimes X_k  \,
\end{align}
which is the expected string  with perhaps a minus sign, depending on whether we have flipped any stabilizers. Note however that the loop parity constraints have to be fulfilled somewhere, either by minus signs in the logical operators or by flipping stabilizers.

\subsubsection{Adaption to the layout \& Manhattan-distance property}
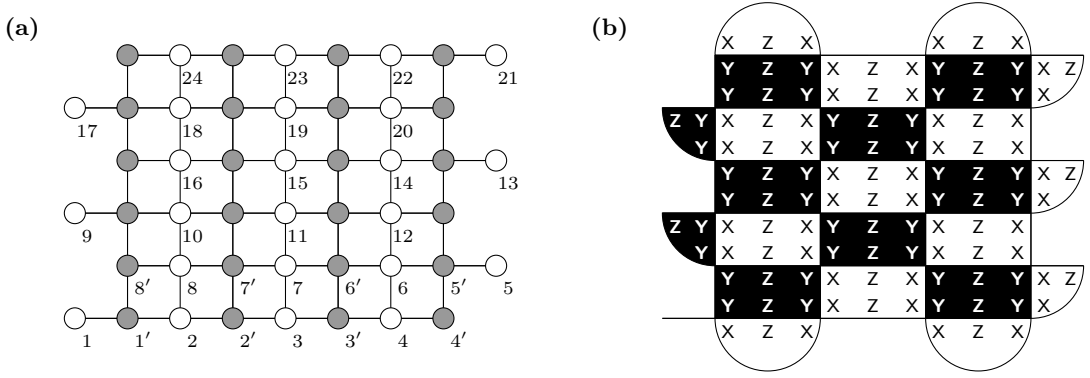
\begin{figure}
\begin{tikzpicture}[scale=0.7,baseline=0]
\node[] at (-1,5.5) {\textbf{(a)}};
\draw[] (0,0)--(7,0);
\draw[] (1,1)--(8,1);
\draw[] (0,2)--(7,2);
\draw[] (1,3)--(8,3);
\draw[] (0,4)--(7,4);
\draw[] (1,5)--(8,5);
\draw[] (1,0)--(1,5);
\draw[] (3,0)--(3,5);
\draw[] (5,0)--(5,5);
\draw[] (7,0)--(7,5);
\draw[] (0,0)--(7,0);
\draw[] (1,1)--(8,1);
\draw[] (0,2)--(7,2);
\draw[] (1,3)--(8,3);
\draw[] (0,4)--(7,4);
\draw[] (1,5)--(8,5);
\draw[] (1,0)--(1,5);
\draw[] (3,0)--(3,5);
\draw[] (5,0)--(5,5);
\draw[] (7,0)--(7,5);
\draw[] (2,0)--(2,5);
\draw[] (4,0)--(4,5);
\draw[] (6,0)--(6,5);

\draw[fill=white](0,0) circle[radius=0.2];
\draw[fill=\clre](1,0) circle[radius=0.2];
\draw[fill=white](2,0) circle[radius=0.2];
\draw[fill=\clre](3,0) circle[radius=0.2];
\draw[fill=white](4,0) circle[radius=0.2];
\draw[fill=\clre](5,0) circle[radius=0.2];
\draw[fill=white](6,0) circle[radius=0.2];
\draw[fill=\clre](7,0) circle[radius=0.2];

\draw[fill=\clre](1,1) circle[radius=0.2];
\draw[fill=white](2,1) circle[radius=0.2];
\draw[fill=\clre](3,1) circle[radius=0.2];
\draw[fill=white](4,1) circle[radius=0.2];
\draw[fill=\clre](5,1) circle[radius=0.2];
\draw[fill=white](6,1) circle[radius=0.2];
\draw[fill=\clre](7,1) circle[radius=0.2];
\draw[fill=white](8,1) circle[radius=0.2];

\draw[fill=white](0,2) circle[radius=0.2];
\draw[fill=\clre](1,2) circle[radius=0.2];
\draw[fill=white](2,2) circle[radius=0.2];
\draw[fill=\clre](3,2) circle[radius=0.2];
\draw[fill=white](4,2) circle[radius=0.2];
\draw[fill=\clre](5,2) circle[radius=0.2];
\draw[fill=white](6,2) circle[radius=0.2];
\draw[fill=\clre](7,2) circle[radius=0.2];

\draw[fill=\clre](1,3) circle[radius=0.2];
\draw[fill=white](2,3) circle[radius=0.2];
\draw[fill=\clre](3,3) circle[radius=0.2];
\draw[fill=white](4,3) circle[radius=0.2];
\draw[fill=\clre](5,3) circle[radius=0.2];
\draw[fill=white](6,3) circle[radius=0.2];
\draw[fill=\clre](7,3) circle[radius=0.2];
\draw[fill=white](8,3) circle[radius=0.2];

\draw[fill=white](0,4) circle[radius=0.2];
\draw[fill=\clre](1,4) circle[radius=0.2];
\draw[fill=white](2,4) circle[radius=0.2];
\draw[fill=\clre](3,4) circle[radius=0.2];
\draw[fill=white](4,4) circle[radius=0.2];
\draw[fill=\clre](5,4) circle[radius=0.2];
\draw[fill=white](6,4) circle[radius=0.2];
\draw[fill=\clre](7,4) circle[radius=0.2];

\draw[fill=\clre](1,5) circle[radius=0.2];
\draw[fill=white](2,5) circle[radius=0.2];
\draw[fill=\clre](3,5) circle[radius=0.2];
\draw[fill=white](4,5) circle[radius=0.2];
\draw[fill=\clre](5,5) circle[radius=0.2];
\draw[fill=white](6,5) circle[radius=0.2];
\draw[fill=\clre](7,5) circle[radius=0.2];
\draw[fill=white](8,5) circle[radius=0.2];

\node[] at (0.3,-.4) {\scriptsize $1^{\phantom{\prime}}$};
\node[] at (1.3,-.4) {\scriptsize $1^\prime$};

\node[] at (2.3,-.4) {\scriptsize $2^{\phantom{\prime}}$};
\node[] at (3.3,-.4) {\scriptsize $2^\prime$};

\node[] at (4.3,-.4) {\scriptsize $3^{\phantom{\prime}}$};
\node[] at (5.3,-.4) {\scriptsize $3^\prime$};

\node[] at (6.3,-.4) {\scriptsize $4^{\phantom{\prime}}$};
\node[] at (7.3,-.4) {\scriptsize $4^\prime$};

\node[] at (2.3,.6) {\scriptsize $8^{\phantom{\prime}}$};
\node[] at (1.3,.6) {\scriptsize $8^\prime$};

\node[] at (4.3,.6) {\scriptsize $7^{\phantom{\prime}}$};
\node[] at (3.3,.6) {\scriptsize $7^\prime$};

\node[] at (6.3,.6) {\scriptsize $6^{\phantom{\prime}}$};
\node[] at (5.3,.6) {\scriptsize $6^\prime$};

\node[] at (8.3,.6) {\scriptsize $5^{\phantom{\prime}}$};
\node[] at (7.3,.6) {\scriptsize $5^\prime$};

\node[] at (0.3,1.6) {\scriptsize $9^{\phantom{\prime}}$};

\node[] at (2.3,1.6) {\scriptsize $10^{\phantom{\prime}}$};

\node[] at (4.3,1.6) {\scriptsize $11^{\phantom{\prime}}$};

\node[] at (6.3,1.6) {\scriptsize $12^{\phantom{\prime}}$};

\node[] at (2.3,2.6) {\scriptsize $16^{\phantom{\prime}}$};

\node[] at (4.3,2.6) {\scriptsize $15^{\phantom{\prime}}$};

\node[] at (6.3,2.6) {\scriptsize $14^{\phantom{\prime}}$};

\node[] at (8.3,2.6) {\scriptsize $13^{\phantom{\prime}}$};

\node[] at (0.3,3.6) {\scriptsize $17^{\phantom{\prime}}$};

\node[] at (2.3,3.6) {\scriptsize $18^{\phantom{\prime}}$};

\node[] at (4.3,3.6) {\scriptsize $19^{\phantom{\prime}}$};

\node[] at (6.3,3.6) {\scriptsize $20^{\phantom{\prime}}$};

\node[] at (2.3,4.6) {\scriptsize $24^{\phantom{\prime}}$};

\node[] at (4.3,4.6) {\scriptsize $23^{\phantom{\prime}}$};

\node[] at (6.3,4.6) {\scriptsize $22^{\phantom{\prime}}$};

\node[] at (8.3,4.6) {\scriptsize $21^{\phantom{\prime}}$};

\end{tikzpicture}$\qquad $\begin{tikzpicture}[scale=0.7,baseline=0]

\node[] at (-1,5.5) {\textbf{(b)}};
\path[fill=black] (1,0)--(3,0)--(3,1)--(1,1)--cycle;
\path[fill=black] (5,0)--(7,0)--(7,1)--(5,1)--cycle;

\path[fill=black] (1,2)--(3,2)--(3,3)--(1,3)--cycle;
\path[fill=black] (5,2)--(7,2)--(7,3)--(5,3)--cycle;

\path[fill=black] (1,4)--(3,4)--(3,5)--(1,5)--cycle;
\path[fill=black] (5,4)--(7,4)--(7,5)--(5,5)--cycle;

\path[fill=black] (3,1)--(5,1)--(5,2)--(3,2)--cycle;
\path[fill=black] (3,3)--(5,3)--(5,4)--(3,4)--cycle;

\draw[] (0,0)--(7,0);
\draw[] (1,1)--(8,1);
\draw[] (0,2)--(7,2);
\draw[] (1,3)--(8,3);
\draw[] (0,4)--(7,4);
\draw[] (1,5)--(8,5);
\draw[] (1,0)--(1,5);
\draw[] (3,0)--(3,5);
\draw[] (5,0)--(5,5);
\draw[] (7,0)--(7,5);
\draw[] (0,0)--(7,0);
\draw[] (1,1)--(8,1);
\draw[] (0,2)--(7,2);
\draw[] (1,3)--(8,3);
\draw[] (0,4)--(7,4);
\draw[] (1,5)--(8,5);
\draw[] (1,0)--(1,5);
\draw[] (3,0)--(3,5);
\draw[] (5,0)--(5,5);
\draw[] (7,0)--(7,5);
\draw[fill=black](1,1) arc (-90:-180:1) -- (1,2) -- cycle;
\draw[fill=black](1,3) arc (-90:-180:1) -- (1,4) -- cycle;

\node[white] at (1.25,.25) {\scriptsize  $\bbs{\mathsf{Y}}$};
\node[white] at (1.25,.75) {\scriptsize  $\bbs{\mathsf{Y}}$};
\node[white] at (2,.25) {\scriptsize  $\bbs{\mathsf{Z}}$};
\node[white] at (2,.75) {\scriptsize  $\bbs{\mathsf{Z}}$};
\node[white] at (2.75,.25) {\scriptsize  $\bbs{\mathsf{Y}}$};
\node[white] at (2.75,.75) {\scriptsize  $\bbs{\mathsf{Y}}$};

\node[] at (3.25,.25) {\scriptsize  $\mathsf{X}$};
\node[] at (3.25,.75) {\scriptsize  $\mathsf{X}$};
\node[] at (4,.25) {\scriptsize  $\mathsf{Z}$};
\node[] at (4,.75) {\scriptsize  $\mathsf{Z}$};
\node[] at (4.75,.25) {\scriptsize  $\mathsf{X}$};
\node[] at (4.75,.75) {\scriptsize  $\mathsf{X}$};

\node[white] at (5.25,.25) {\scriptsize  $\bbs{\mathsf{Y}}$};
\node[white] at (5.25,.75) {\scriptsize  $\bbs{\mathsf{Y}}$};
\node[white] at (6,.25) {\scriptsize  $\bbs{\mathsf{Z}}$};
\node[white] at (6,.75) {\scriptsize  $\bbs{\mathsf{Z}}$};
\node[white] at (6.75,.25) {\scriptsize  $\bbs{\mathsf{Y}}$};
\node[white] at (6.75,.75) {\scriptsize  $\bbs{\mathsf{Y}}$};

\node[] at (1.25,1.25) {\scriptsize  $\mathsf{X}$};
\node[] at (1.25,1.75) {\scriptsize  $\mathsf{X}$};
\node[] at (2,1.25) {\scriptsize  $\mathsf{Z}$};
\node[] at (2,1.75) {\scriptsize  $\mathsf{Z}$};
\node[] at (2.75,1.25) {\scriptsize  $\mathsf{X}$};
\node[] at (2.75,1.75) {\scriptsize  $\mathsf{X}$};

\node[white] at (3.25,1.25) {\scriptsize  $\bbs{\mathsf{Y}}$};
\node[white] at (3.25,1.75) {\scriptsize  $\bbs{\mathsf{Y}}$};
\node[white] at (4,1.25) {\scriptsize  $\bbs{\mathsf{Z}}$};
\node[white] at (4,1.75) {\scriptsize  $\bbs{\mathsf{Z}}$};
\node[white] at (4.75,1.25) {\scriptsize  $\bbs{\mathsf{Y}}$};
\node[white] at (4.75,1.75) {\scriptsize  $\bbs{\mathsf{Y}}$};

\node[] at (5.25,1.25) {\scriptsize  $\mathsf{X}$};
\node[] at (5.25,1.75) {\scriptsize  $\mathsf{X}$};
\node[] at (6,1.25) {\scriptsize  $\mathsf{Z}$};
\node[] at (6,1.75) {\scriptsize  $\mathsf{Z}$};
\node[] at (6.75,1.25) {\scriptsize  $\mathsf{X}$};
\node[] at (6.75,1.75) {\scriptsize  $\mathsf{X}$};

\node[white] at (1.25,2.25) {\scriptsize  $\bbs{\mathsf{Y}}$};
\node[white] at (1.25,2.75) {\scriptsize  $\bbs{\mathsf{Y}}$};
\node[white] at (2,2.25) {\scriptsize  $\bbs{\mathsf{Z}}$};
\node[white] at (2,2.75) {\scriptsize  $\bbs{\mathsf{Z}}$};
\node[white] at (2.75,2.25) {\scriptsize  $\bbs{\mathsf{Y}}$};
\node[white] at (2.75,2.75) {\scriptsize  $\bbs{\mathsf{Y}}$};

\node[] at (3.25,2.25) {\scriptsize  $\mathsf{X}$};
\node[] at (3.25,2.75) {\scriptsize  $\mathsf{X}$};
\node[] at (4,2.25) {\scriptsize  $\mathsf{Z}$};
\node[] at (4,2.75) {\scriptsize  $\mathsf{Z}$};
\node[] at (4.75,2.25) {\scriptsize  $\mathsf{X}$};
\node[] at (4.75,2.75) {\scriptsize  $\mathsf{X}$};

\node[white] at (5.25,2.25) {\scriptsize  $\bbs{\mathsf{Y}}$};
\node[white] at (5.25,2.75) {\scriptsize  $\bbs{\mathsf{Y}}$};
\node[white] at (6,2.25) {\scriptsize  $\bbs{\mathsf{Z}}$};
\node[white] at (6,2.75) {\scriptsize  $\bbs{\mathsf{Z}}$};
\node[white] at (6.75,2.25) {\scriptsize  $\bbs{\mathsf{Y}}$};
\node[white] at (6.75,2.75) {\scriptsize  $\bbs{\mathsf{Y}}$};

\node[] at (1.25,3.25) {\scriptsize  $\mathsf{X}$};
\node[] at (1.25,3.75) {\scriptsize  $\mathsf{X}$};
\node[] at (2,3.25) {\scriptsize  $\mathsf{Z}$};
\node[] at (2,3.75) {\scriptsize  $\mathsf{Z}$};
\node[] at (2.75,3.25) {\scriptsize  $\mathsf{X}$};
\node[] at (2.75,3.75) {\scriptsize  $\mathsf{X}$};

\node[white] at (3.25,3.25) {\scriptsize  $\bbs{\mathsf{Y}}$};
\node[white] at (3.25,3.75) {\scriptsize  $\bbs{\mathsf{Y}}$};
\node[white] at (4,3.25) {\scriptsize  $\bbs{\mathsf{Z}}$};
\node[white] at (4,3.75) {\scriptsize  $\bbs{\mathsf{Z}}$};
\node[white] at (4.75,3.25) {\scriptsize  $\bbs{\mathsf{Y}}$};
\node[white] at (4.75,3.75) {\scriptsize  $\bbs{\mathsf{Y}}$};

\node[] at (5.25,3.25) {\scriptsize  $\mathsf{X}$};
\node[] at (5.25,3.75) {\scriptsize  $\mathsf{X}$};
\node[] at (6,3.25) {\scriptsize  $\mathsf{Z}$};
\node[] at (6,3.75) {\scriptsize  $\mathsf{Z}$};
\node[] at (6.75,3.25) {\scriptsize  $\mathsf{X}$};
\node[] at (6.75,3.75) {\scriptsize  $\mathsf{X}$};

\node[white] at (1.25,4.25) {\scriptsize  $\bbs{\mathsf{Y}}$};
\node[white] at (1.25,4.75) {\scriptsize  $\bbs{\mathsf{Y}}$};
\node[white] at (2,4.25) {\scriptsize  $\bbs{\mathsf{Z}}$};
\node[white] at (2,4.75) {\scriptsize  $\bbs{\mathsf{Z}}$};
\node[white] at (2.75,4.25) {\scriptsize  $\bbs{\mathsf{Y}}$};
\node[white] at (2.75,4.75) {\scriptsize  $\bbs{\mathsf{Y}}$};

\node[] at (3.25,4.25) {\scriptsize  $\mathsf{X}$};
\node[] at (3.25,4.75) {\scriptsize  $\mathsf{X}$};
\node[] at (4,4.25) {\scriptsize  $\mathsf{Z}$};
\node[] at (4,4.75) {\scriptsize  $\mathsf{Z}$};
\node[] at (4.75,4.25) {\scriptsize  $\mathsf{X}$};
\node[] at (4.75,4.75) {\scriptsize  $\mathsf{X}$};

\node[white] at (5.25,4.25) {\scriptsize  $\bbs{\mathsf{Y}}$};
\node[white] at (5.25,4.75) {\scriptsize  $\bbs{\mathsf{Y}}$};
\node[white] at (6,4.25) {\scriptsize  $\bbs{\mathsf{Z}}$};
\node[white] at (6,4.75) {\scriptsize  $\bbs{\mathsf{Z}}$};
\node[white] at (6.75,4.25) {\scriptsize  $\bbs{\mathsf{Y}}$};
\node[white] at (6.75,4.75) {\scriptsize  $\bbs{\mathsf{Y}}$};

\node[] at (7.25,0.75) {\scriptsize  $\mathsf{X}$};
\node[] at (7.75,0.75) {\scriptsize  $\mathsf{Z}$};
\node[] at (7.25,0.25) {\scriptsize  $\mathsf{X}$};

\node[] at (7.25,2.75) {\scriptsize  $\mathsf{X}$};
\node[] at (7.75,2.75) {\scriptsize  $\mathsf{Z}$};
\node[] at (7.25,2.25) {\scriptsize  $\mathsf{X}$};

\node[] at (7.25,4.75) {\scriptsize  $\mathsf{X}$};
\node[] at (7.75,4.75) {\scriptsize  $\mathsf{Z}$};
\node[] at (7.25,4.25) {\scriptsize  $\mathsf{X}$};

\node[white] at (0.75,1.75) {\scriptsize  $\bbs{\mathsf{Y}}$};
\node[white] at (.25,1.75) {\scriptsize  $\bbs{\mathsf{Z}}$};
\node[white] at (0.75,1.25) {\scriptsize  $\bbs{\mathsf{Y}}$};

\node[white] at (0.75,3.75) {\scriptsize  $\bbs{\mathsf{Y}}$};
\node[white] at (.25,3.75) {\scriptsize  $\bbs{\mathsf{Z}}$};
\node[white] at (0.75,3.25) {\scriptsize  $\bbs{\mathsf{Y}}$};

\node[] at (1.25,-0.25) {\scriptsize  ${\mathsf{X}}$};
\node[] at (2,-0.25) {\scriptsize  ${\mathsf{Z}}$};
\node[] at (2.75,-0.25) {\scriptsize  ${\mathsf{X}}$};

\node[] at (5.25,-.25) {\scriptsize  ${\mathsf{X}}$};
\node[] at (6,-.25) {\scriptsize  ${\mathsf{Z}}$};
\node[] at (6.75,-.25) {\scriptsize  ${\mathsf{X}}$};

\node[] at (1.25,5.25) {\scriptsize  ${\mathsf{X}}$};
\node[] at (2,5.25) {\scriptsize  ${\mathsf{Z}}$};
\node[] at (2.75,5.25) {\scriptsize  ${\mathsf{X}}$};

\node[] at (5.25,5.25) {\scriptsize  ${\mathsf{X}}$};
\node[] at (6,5.25) {\scriptsize  ${\mathsf{Z}}$};
\node[] at (6.75,5.25) {\scriptsize  ${\mathsf{X}}$};

\draw[](1,5) arc(180:0:1) -- cycle;
\draw[](5,5) arc(180:0:1) -- cycle;

\draw[](1,0) arc(-180:0:1) -- cycle;
\draw[](5,0) arc(-180:0:1) -- cycle;

\draw[](7,4) arc (-90:0:1);
\draw[](7,2) arc (-90:0:1);
\draw[](7,0) arc (-90:0:1);

\end{tikzpicture}
\caption{VCT as a planar code. \textbf{(a)} Connectivity graph, in which we alternate data (white) and auxiliary qubits (gray), but shift every second row such that the auxiliary qubits align vertically. The labeling of the qubits follows an S-pattern. \textbf{(b)} Stabilizers of the VCT for a graph as in  Figure \ref{fig:vc}(b), the original proposal. We here give the connectivity graph a two-coloring of the stabilizer plaquettes, where the Pauli operators, that make up each stabilizer, are denoted by letters inside the plaquettes close to where their corresponding qubits are. Note that we have not indicated the signs that each stabilizer possibly has attached to it.  }\label{fig:surface}
\end{figure}

We here adapt the Verstraete-Cirac transform to the square lattice connectivity, such that it has the  Manhattan-distance property. In doing so, we will not stray too far from the original proposal, that is built upon the connectivity graph in Figure \ref{fig:vc}(b). The layout is roughly motivated by an S-pattern of the qubits ordered $1 \; 1^\prime \; 2 \; 2^\prime \;\dots N \; N^\prime$. For reasons that become clear later, we need the rows to be connected vertically by the auxiliary qubits, which leads us to shift every second row in order to align the primed qubits. The vertical connections are also placed along the windings of the S-pattern, resulting in a graph that can be studied in  Figure  \ref{fig:surface}(a).  For the initialization of a state, stabilizers that are horizontally adjacent are multiplied pairwise. We fully constrain the auxiliary systems by those localized stabilizers, plus the stabilizers that are local already: the ones along the windings and the horizontal connections in the first and $\ell_2$-th row. The stabilizer tiling to the  layout of Figure \ref{fig:surface}(a) is presented in panel (b) of the same figure.  As  already remarked in \cite{verstraete2005mapping}, the analogy of the stabilizer tilings of this code and the rotated surface code \cite{bombin2007optimal} comes to mind easily. The tiles of the VCT are identical to the surface code on the  primed qubits, but the stabilizers contain some additional $Z$-strings on the data qubits. Also, not all of the stabilizers might have the same sign according to $b$ in the definition $\mathcal{P}^b_{jk}$. Curiously, only the first qubit of the data register is not entangled with the auxiliary system in any way.

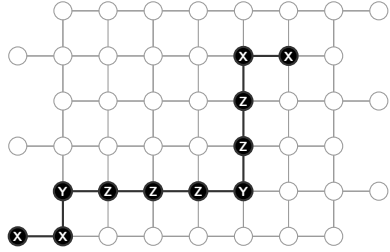
\begin{figure}
\begin{tikzpicture}[scale=0.6,baseline=0]
\draw[\clre] (0,0)--(7,0);
\draw[\clre] (1,1)--(8,1);
\draw[\clre] (0,2)--(7,2);
\draw[\clre] (1,3)--(8,3);
\draw[\clre] (0,4)--(7,4);
\draw[\clre] (1,5)--(8,5);
\draw[\clre] (1,0)--(1,5);
\draw[\clre] (3,0)--(3,5);
\draw[\clre] (5,0)--(5,5);
\draw[\clre] (7,0)--(7,5);
\draw[\clre] (0,0)--(7,0);
\draw[\clre] (1,1)--(8,1);
\draw[\clre] (0,2)--(7,2);
\draw[\clre] (1,3)--(8,3);
\draw[\clre] (0,4)--(7,4);
\draw[\clre] (1,5)--(8,5);
\draw[\clre] (1,0)--(1,5);
\draw[\clre] (3,0)--(3,5);
\draw[\clre] (5,0)--(5,5);
\draw[\clre] (7,0)--(7,5);
\draw[\clre] (2,0)--(2,5);
\draw[\clre] (4,0)--(4,5);
\draw[\clre] (6,0)--(6,5);
\draw[thick, \clrd] (0,0)--(1,0)--(1,1)--(5,1)--(5,4)--(6,4);
\draw[color=\clrb, fill=\clra, thick](0,0) circle[radius=0.2];
\draw[color=\clrb, fill=\clra, thick](1,0) circle[radius=0.2];
\draw[\clre,fill=white](2,0) circle[radius=0.2];
\draw[\clre,fill=white](3,0) circle[radius=0.2];
\draw[\clre,fill=white](4,0) circle[radius=0.2];
\draw[\clre,fill=white](5,0) circle[radius=0.2];
\draw[\clre,fill=white](6,0) circle[radius=0.2];
\draw[\clre,fill=white](7,0) circle[radius=0.2];

\draw[color=\clrb, fill=\clra, thick](1,1) circle[radius=0.2];
\draw[color=\clrb, fill=\clra, thick](2,1) circle[radius=0.2];
\draw[color=\clrb, fill=\clra, thick](3,1) circle[radius=0.2];
\draw[color=\clrb, fill=\clra, thick](4,1) circle[radius=0.2];
\draw[color=\clrb, fill=\clra, thick](5,1) circle[radius=0.2];
\draw[\clre,fill=white](6,1) circle[radius=0.2];
\draw[\clre,fill=white](7,1) circle[radius=0.2];
\draw[\clre,fill=white](8,1) circle[radius=0.2];
\draw[\clre,fill=white](0,2) circle[radius=0.2];
\draw[\clre,fill=white](1,2) circle[radius=0.2];
\draw[\clre,fill=white](2,2) circle[radius=0.2];
\draw[\clre,fill=white](3,2) circle[radius=0.2];
\draw[\clre,fill=white](4,2) circle[radius=0.2];
\draw[color=\clrb, fill=\clra, thick](5,2) circle[radius=0.2];
\draw[\clre,fill=white](6,2) circle[radius=0.2];
\draw[\clre,fill=white](7,2) circle[radius=0.2];

\draw[\clre,fill=white](1,3) circle[radius=0.2];
\draw[\clre,fill=white](2,3) circle[radius=0.2];
\draw[\clre,fill=white](3,3) circle[radius=0.2];
\draw[\clre,fill=white](4,3) circle[radius=0.2];
\draw[color=\clrb, fill=\clra, thick](5,3) circle[radius=0.2];
\draw[\clre,fill=white](6,3) circle[radius=0.2];
\draw[\clre,fill=white](7,3) circle[radius=0.2];
\draw[\clre,fill=white](8,3) circle[radius=0.2];
\draw[\clre,fill=white](0,4) circle[radius=0.2];
\draw[\clre,fill=white](1,4) circle[radius=0.2];
\draw[\clre,fill=white](2,4) circle[radius=0.2];
\draw[\clre,fill=white](3,4) circle[radius=0.2];
\draw[\clre,fill=white](4,4) circle[radius=0.2];
\draw[color=\clrb, fill=\clra, thick](5,4) circle[radius=0.2];
\draw[color=\clrb, fill=\clra, thick](6,4) circle[radius=0.2];
\draw[\clre,fill=white](7,4) circle[radius=0.2];

\draw[\clre,fill=white](1,5) circle[radius=0.2];
\draw[\clre,fill=white](2,5) circle[radius=0.2];
\draw[\clre,fill=white](3,5) circle[radius=0.2];
\draw[\clre,fill=white](4,5) circle[radius=0.2];
\draw[\clre,fill=white](5,5) circle[radius=0.2];
\draw[\clre,fill=white](6,5) circle[radius=0.2];
\draw[\clre,fill=white](7,5) circle[radius=0.2];
\draw[\clre,fill=white](8,5) circle[radius=0.2];

\node[\clrc] at (0,0) {\tiny $\boldsymbol{\mathsf{X}}$};
\node[\clrc] at (1,0) {\tiny $\boldsymbol{\mathsf{X}}$};
\node[\clrc] at (1,1) {\tiny $\boldsymbol{\mathsf{Y}}$};
\node[\clrc] at (2,1) {\tiny $\boldsymbol{\mathsf{Z}}$};
\node[\clrc] at (3,1) {\tiny $\boldsymbol{\mathsf{Z}}$};
\node[\clrc] at (4,1) {\tiny $\boldsymbol{\mathsf{Z}}$};
\node[\clrc] at (5,1) {\tiny $\boldsymbol{\mathsf{Y}}$};
\node[\clrc] at (5,2) {\tiny $\boldsymbol{\mathsf{Z}}$};
\node[\clrc] at (5,3) {\tiny $\boldsymbol{\mathsf{Z}}$};
\node[\clrc] at (5,4) {\tiny $\boldsymbol{\mathsf{X}}$};
\node[\clrc] at (6,4) {\tiny $\boldsymbol{\mathsf{X}}$};

\end{tikzpicture}
\caption{Simulating the term $(i m_{20}\,\obar{m}_1)$ via the VCT, where we have arbitrarily deformed the string by the multiplication of stabilizers. }\label{fig:VCmanhattan}
\end{figure}

Using the interpretations of the stabilizers \eqref{eq:VCstabs}, we can define $\mathcal{O}_{jk} \propto (-1)^{b}\; \mathcal{P}^{b}_{jk} Z_{k^\prime}$  and obtain arbitrary long-range vertical connections over the sequence of vertically aligned stabilizers $\mathcal{P}^{b_s}_{k_{s} k_{s+1}}$, where  $\bbs{k} \in [N]^{\otimes l}$ and $\bbs{b}\in \zetto{l}$,  via  \eqref{eq:manhattanops}:

\begin{align}
\label{eq:manhattanP}
\prod_{t=1}^{l-1} \mathcal{P}^{b_t}_{k_t k_{t+1}}   \; = \;   \mathcal{P}^{a}_{k_1 k_l}\;  \bigotimes_{u=2}^{l-1} Z_{k_u^\prime}\, ,
\end{align}
where $a=(\sum_{s=1}^l b_s)$. Equation \eqref{eq:manhattanP} means that the multiplication of these vertical stabilizers yields a non-local connection $\mathcal{P}^{a}_{k_1 k_l}$, which (is not a stabilizer and) is missing the operators $Z_{k_u^\prime}$ for $1<u<l$. The absence of these $Z$-operators does not cancel them in Pauli strings originating from fermionic terms  like $c_i^{\dagger} c^{\phantom{\dagger}}_j$, where $i\leq k_1<k_l \leq j$. These operators subsequently serve as connection between the qubits labeled $k^\prime_1$ and $k^\prime_l$, as the qubits are vertically aligned by our layout.
With this building block we can  multiply various stabilizers  and so connect the qubits $i$ and $j$ via different paths but with the same number of gates.  In Figure \ref{fig:VCmanhattan}, we present an example of such a term.

 \subsection{Superfast Simulation}
 \subsubsection{Review}
 We here review the original proposal of the Bravyi-Kitaev Superfast simulation,  \cite{bravyi2002fermionic}, which includes the transform of the operators and the structure of the stabilizers.

 In contrast to the other mappings, the Superfast simulation is not defined to  transform fermionic operators, but  pairs of Majoranas. Thus the BKSF only allows us to conveniently consider Hamiltonians that conserve the fermionic parity i.e.  are comprised of operator pairs $c_jc_k$, $c_j^{\dagger}c_k^{\dagger}$ and $c_j^{\dagger}c_k^{\phantom{\dagger}}$. By the relations \eqref{eq:majoran}, these Hamiltonians can then be expressed using only the operators
\begin{align}
\label{eq:aa}
\mathcal{A}_{jk}\;&\hat{=}\; -i\,m_j\, m_k \, , \\
\label{eq:bb}
\mathcal{B}_k\; & \hat{=} \; -i \, m_k \,\obar{m}_k \, ,
\end{align}
where $\mathcal{A}_{jk}$ and $\mathcal{B}_k$ are some Pauli strings. Using these operators, fermionic  Hamiltonians can be transformed via
 \begin{align}
 \label{eq:pairs1}
&c_j^{\phantom{\dagger}}c_k^{\phantom{\dagger}} \;  \hat{=} \; \frac{i}{4} \left( \mathcal{A}_{jk}- \mathcal{A}_{jk} \mathcal{B}_k + \mathcal{B}_j \mathcal{A}_{jk} - \mathcal{B}_j \mathcal{A}_{jk} \mathcal{B}_k\right) \, , \\
 &c_j^{\dagger}c_k^{\dagger} \;  \hat{=} \; \frac{i}{4} \left( \mathcal{A}_{jk}+ \mathcal{A}_{jk} \mathcal{B}_k - \mathcal{B}_j \mathcal{A}_{jk} - \mathcal{B}_j \mathcal{A}_{jk} \mathcal{B}_k\right) \, ,\label{eq:pairs2} \\
\label{eq:pairs3} &c_j^{\dagger}c_k^{\phantom{\dagger}} \;  \hat{=} \; \frac{i}{4} \left( \mathcal{A}_{jk}- \mathcal{A}_{jk} \mathcal{B}_k - \mathcal{B}_j \mathcal{A}_{jk} + \mathcal{B}_j \mathcal{A}_{jk} \mathcal{B}_k\right) \, .
 \end{align}
The BKSF is furthermore not based on the Jordan-Wigner transform, so  $\mathcal{A}_{jk}$ and $\mathcal{B}_k$ are not going to be obtained by transforming the right-hand side of  \eqref{eq:aa} and \eqref{eq:bb} under \eqref{eq:majoran2}. Instead, the $\mathcal{A}$- and $\mathcal{B}$-operators will be defined on a unique qubit layout, that we now introduce.   \\
The Hamiltonian that we want to simulate describes a certain graph of  pairwise interactions between modes, for example there is an edge between vertices $j$, $k$ when it contains at least one of the term \eqref{eq:pairs1}-\eqref{eq:pairs3}. The qubit connectivity graph of the Superfast simulation is then the line graph of this Hamiltonian graph. Here the operators $\mathcal{A}_{jk}$ are associated with  edges in the Hamiltonian graph, i.e. interactions of the Hamiltonian, and the operators $\mathcal{B}_k$ are associated with  vertices, i.e.  fermionic modes. Let  $E$  be the set of undirected edges of the Hamiltonian graph, and $\varepsilon_{jk}$ a number associated to the index pair $jk$, that yields zero if $jk \notin E$. By  means of $\varepsilon_{jk}$  a direction on the graph is fixed by imposing that if $jk \in E$, then $\varepsilon_{jk}=1 $, in case the edge is directed from $j\to k$, and $\varepsilon_{jk}=-1$ when the direction is opposite. With that construction, we will take into account that $\mathcal{A}_{jk}=-\mathcal{A}_{kj}$, which is straightforward to see from \eqref{eq:aa}. Also, on every vertex $k$, we need to  impose an ordering of the edges connected to it. To that end Bravyi and Kitaev introduce the symbolic operator  $\BKsmaller{k}$, such that  two different edges  $jk,\, lk \in E$, $j\neq l$  on vertex $k$ are ordered  by a relation like  $jk\BKsmaller{k}lk$. As we place the qubits on the edges of that graph, both $jk$ and $kj$ shall be identifiers for the same qubit (given $\varepsilon_{jk} \neq 0$). In the original BKSF, the number of qubits equals the number of edges in the graph, so the qubit requirements do not depend on the system size, but on the size of the Hamiltonian. The operators $\mathcal{A}_{jk}$ and $\mathcal{B}_k$ are defined by
\begin{align}
\label{eq:aa2}
\mathcal{B}_k\; & = \; \bigotimes_{a:\, ak\in E}  Z_{ak}\, , \\ \label{eq:bb2}
\mathcal{A}_{jk}\;&=\; \varepsilon_{jk} X_{jk} \left( \bigotimes_{b:\,bk\BKsmaller{k} jk} Z_{bk}\right) \left( \bigotimes_{c:\,jc\BKsmaller{j} jk} Z_{jc}\right) \, .
\end{align}
 As shown in \cite{bravyi2002fermionic}, these operators fulfill all algebraic relations that we  would expect from  representations of \eqref{eq:aa} and \eqref{eq:bb} but one. As it is now, the mapping would allow a  Majorana to unphysically interact with itself via  hopping terms around a closed loop. For a length-$l$ sequence $a_1, \, a_2, \, a_3, \, \dots , a_{l}  $, that describes a closed loop along edges, i.e. $a_j a_{j+1} \in E$   and  $a_1=a_l$, we must impose that
\begin{align}
\label{eq:Superfaststab}
(i)^l\prod_{j=1}^{l-1}\mathcal{A}_{a_j a_{j+1}}
\end{align}
is a stabilizer of the system. As not all closed loops are linearly independent, one needs to stabilize only the smallest closed loops  of the system.

\subsubsection{Adaption to the layout \& Manhattan-distance property}

\begin{figure}
\begin{tikzpicture}[baseline=0]
\node[] at (-.7,3.5) {\textbf{(a)}};
\path[fill=\checker] (-0.5,0)--(0,0.5)--(.5,0)--(0,-0.5)--cycle;
\path[fill=\checker] (0.5,0)--(1,0.5)--(1.5,0)--(1,-0.5)--cycle;
\path[fill=\checker] (1.5,0)--(2,0.5)--(2.5,0)--(2,-0.5)--cycle;
\path[fill=\checker] (2.5,0)--(3,0.5)--(3.5,0)--(3,-0.5)--cycle;
\path[fill=\checker] (3.5,0)--(4,0.5)--(4.5,0)--(4,-0.5)--cycle;

\path[fill=\checker] (-0.5,1)--(0,1.5)--(.5,1)--(0,0.5)--cycle;
\path[fill=\checker] (0.5,1)--(1,1.5)--(1.5,1)--(1,0.5)--cycle;
\path[fill=\checker] (1.5,1)--(2,1.5)--(2.5,1)--(2,0.5)--cycle;
\path[fill=\checker] (2.5,1)--(3,1.5)--(3.5,1)--(3,0.5)--cycle;
\path[fill=\checker] (3.5,1)--(4,1.5)--(4.5,1)--(4,0.5)--cycle;

\path[fill=\checker] (-0.5,2)--(0,2.5)--(.5,2)--(0,01.5)--cycle;
\path[fill=\checker] (0.5,2)--(1,2.5)--(1.5,2)--(1,01.5)--cycle;
\path[fill=\checker] (1.5,2)--(2,2.5)--(2.5,2)--(2,01.5)--cycle;
\path[fill=\checker] (2.5,2)--(3,2.5)--(3.5,2)--(3,01.5)--cycle;
\path[fill=\checker] (3.5,2)--(4,2.5)--(4.5,2)--(4,01.5)--cycle;

\path[fill=\checker] (-0.5,3)--(0,3.5)--(.5,3)--(0,02.5)--cycle;
\path[fill=\checker] (0.5,3)--(1,3.5)--(1.5,3)--(1,02.5)--cycle;
\path[fill=\checker] (1.5,3)--(2,3.5)--(2.5,3)--(2,02.5)--cycle;
\path[fill=\checker] (2.5,3)--(3,3.5)--(3.5,3)--(3,02.5)--cycle;
\path[fill=\checker] (3.5,3)--(4,3.5)--(4.5,3)--(4,02.5)--cycle;

\draw[](0,0)--(4,0);
\draw[](0,1)--(4,1);
\draw[](0,2)--(4,2);
\draw[](0,3)--(4,3);
\draw[](0,0)--(0,3);
\draw[](1,0)--(1,3);
\draw[](2,0)--(2,3);
\draw[](3,0)--(3,3);
\draw[](4,0)--(4,3);
\draw[-latex](0,0)--(0.6,0);
\draw[-latex](1,0)--(1.6,0);
\draw[-latex](2,0)--(2.6,0);
\draw[-latex](3,0)--(3.6,0);
\draw[-latex](0,1)--(0.6,1);
\draw[-latex](1,1)--(1.6,1);
\draw[-latex](2,1)--(2.6,1);
\draw[-latex](3,1)--(3.6,1);
\draw[-latex](0,2)--(0.6,2);
\draw[-latex](1,2)--(1.6,2);
\draw[-latex](2,2)--(2.6,2);
\draw[-latex](3,2)--(3.6,2);
\draw[-latex](0,3)--(0.6,3);
\draw[-latex](1,3)--(1.6,3);
\draw[-latex](2,3)--(2.6,3);
\draw[-latex](3,3)--(3.6,3);

\draw[-latex](0,0)--(0,.6);
\draw[-latex](0,1)--(0,1.6);
\draw[-latex](0,2)--(0,2.6);

\draw[-latex](1,0)--(1,.6);
\draw[-latex](1,1)--(1,1.6);
\draw[-latex](1,2)--(1,2.6);

\draw[-latex](2,0)--(2,.6);
\draw[-latex](2,1)--(2,1.6);
\draw[-latex](2,2)--(2,2.6);

\draw[-latex](3,0)--(3,.6);
\draw[-latex](3,1)--(3,1.6);
\draw[-latex](3,2)--(3,2.6);

\draw[-latex](4,0)--(4,.6);
\draw[-latex](4,1)--(4,1.6);
\draw[-latex](4,2)--(4,2.6);

\draw[color=white,fill=black] (0,0) circle[radius=0.08];
\draw[color=white,fill=black] (1,0) circle[radius=0.08];
\draw[color=white,fill=black] (2,0) circle[radius=0.08];
\draw[color=white,fill=black] (3,0) circle[radius=0.08];
\draw[color=white,fill=black] (4,0) circle[radius=0.08];

\draw[color=white,fill=black] (0,1) circle[radius=0.08];
\draw[color=white,fill=black] (1,1) circle[radius=0.08];
\draw[color=white,fill=black] (2,1) circle[radius=0.08];
\draw[color=white,fill=black] (3,1) circle[radius=0.08];
\draw[color=white,fill=black] (4,1) circle[radius=0.08];

\draw[color=white,fill=black] (0,2) circle[radius=0.08];
\draw[color=white,fill=black] (1,2) circle[radius=0.08];
\draw[color=white,fill=black] (2,2) circle[radius=0.08];
\draw[color=white,fill=black] (3,2) circle[radius=0.08];
\draw[color=white,fill=black] (4,2) circle[radius=0.08];

\draw[color=white,fill=black] (0,3) circle[radius=0.08];
\draw[color=white,fill=black] (1,3) circle[radius=0.08];
\draw[color=white,fill=black] (2,3) circle[radius=0.08];
\draw[color=white,fill=black] (3,3) circle[radius=0.08];
\draw[color=white,fill=black] (4,3) circle[radius=0.08];
\end{tikzpicture} $\qquad$ \begin{tikzpicture}[baseline=0]
\node[] at (-.7,3.5) {\textbf{(b)}};
\path[fill=\checker] (-0.5,0)--(0,0.5)--(.5,0)--(0,-0.5)--cycle;
\path[fill=\checker] (0.5,0)--(1,0.5)--(1.5,0)--(1,-0.5)--cycle;
\path[fill=\checker] (1.5,0)--(2,0.5)--(2.5,0)--(2,-0.5)--cycle;
\path[fill=\checker] (2.5,0)--(3,0.5)--(3.5,0)--(3,-0.5)--cycle;
\path[fill=\checker] (3.5,0)--(4,0.5)--(4.5,0)--(4,-0.5)--cycle;

\path[fill=\checker] (-0.5,1)--(0,1.5)--(.5,1)--(0,0.5)--cycle;
\path[fill=\checker] (0.5,1)--(1,1.5)--(1.5,1)--(1,0.5)--cycle;
\path[fill=\checker] (1.5,1)--(2,1.5)--(2.5,1)--(2,0.5)--cycle;
\path[fill=\checker] (2.5,1)--(3,1.5)--(3.5,1)--(3,0.5)--cycle;
\path[fill=\checker] (3.5,1)--(4,1.5)--(4.5,1)--(4,0.5)--cycle;

\path[fill=\checker] (-0.5,2)--(0,2.5)--(.5,2)--(0,01.5)--cycle;
\path[fill=\checker] (0.5,2)--(1,2.5)--(1.5,2)--(1,01.5)--cycle;
\path[fill=\checker] (1.5,2)--(2,2.5)--(2.5,2)--(2,01.5)--cycle;
\path[fill=\checker] (2.5,2)--(3,2.5)--(3.5,2)--(3,01.5)--cycle;
\path[fill=\checker] (3.5,2)--(4,2.5)--(4.5,2)--(4,01.5)--cycle;

\path[fill=\checker] (-0.5,3)--(0,3.5)--(.5,3)--(0,02.5)--cycle;
\path[fill=\checker] (0.5,3)--(1,3.5)--(1.5,3)--(1,02.5)--cycle;
\path[fill=\checker] (1.5,3)--(2,3.5)--(2.5,3)--(2,02.5)--cycle;
\path[fill=\checker] (2.5,3)--(3,3.5)--(3.5,3)--(3,02.5)--cycle;
\path[fill=\checker] (3.5,3)--(4,3.5)--(4.5,3)--(4,02.5)--cycle;

\draw[] (.5,3)--(0,2.5);
\draw[] (0,1.5)--(1.5,3);
\draw[] (0,0.5)--(2.5,3);
\draw[] (0.5,0.0)--(3.5,3);
\draw[] (1.5,0.0)--(4,2.5);
\draw[] (2.5,0.0)--(4,1.5);
\draw[] (3.5,0.0)--(4,0.5);

\draw[] (0,.5)--(0.5,0);
\draw[] (0,1.5)--(1.5,0);
\draw[] (0,2.5)--(2.5,0);
\draw[] (.5,3)--(3.5,0);
\draw[] (1.5,3)--(4,0.5);
\draw[] (2.5,3)--(4,1.5);
\draw[] (3.5,3)--(4,2.5);

\draw[ fill=white] (0.5,1) circle[radius=0.2];
\draw[ fill=white] (1.5,1) circle[radius=0.2];
\draw[ fill=white] (2.5,1) circle[radius=0.2];
\draw[ fill=white] (3.5,1) circle[radius=0.2];

\draw[ fill=white] (0.5,2) circle[radius=0.2];
\draw[ fill=white] (1.5,2) circle[radius=0.2];
\draw[ fill=white] (2.5,2) circle[radius=0.2];
\draw[ fill=white] (3.5,2) circle[radius=0.2];

\draw[ fill=white] (0.5,3) circle[radius=0.2];
\draw[ fill=white] (1.5,3) circle[radius=0.2];
\draw[ fill=white] (2.5,3) circle[radius=0.2];
\draw[ fill=white] (3.5,3) circle[radius=0.2];

\draw[ fill=white] (0.5,0) circle[radius=0.2];
\draw[ fill=white] (1.5,0) circle[radius=0.2];
\draw[ fill=white] (2.5,0) circle[radius=0.2];
\draw[ fill=white] (3.5,0) circle[radius=0.2];

\draw[ fill=white] (0,0.5) circle[radius=0.2];
\draw[ fill=white] (1,0.5) circle[radius=0.2];
\draw[ fill=white] (2,0.5) circle[radius=0.2];
\draw[ fill=white] (3,0.5) circle[radius=0.2];
\draw[ fill=white] (4,0.5) circle[radius=0.2];

\draw[ fill=white] (0,1.5) circle[radius=0.2];
\draw[ fill=white] (1,1.5) circle[radius=0.2];
\draw[ fill=white] (2,1.5) circle[radius=0.2];
\draw[ fill=white] (3,1.5) circle[radius=0.2];
\draw[ fill=white] (4,1.5) circle[radius=0.2];

\draw[ fill=white] (0,1.5) circle[radius=0.2];
\draw[ fill=white] (1,1.5) circle[radius=0.2];
\draw[ fill=white] (2,1.5) circle[radius=0.2];
\draw[ fill=white] (3,1.5) circle[radius=0.2];
\draw[ fill=white] (4,1.5) circle[radius=0.2];

\draw[ fill=white] (0,2.5) circle[radius=0.2];
\draw[ fill=white] (1,2.5) circle[radius=0.2];
\draw[ fill=white] (2,2.5) circle[radius=0.2];
\draw[ fill=white] (3,2.5) circle[radius=0.2];
\draw[ fill=white] (4,2.5) circle[radius=0.2];
\end{tikzpicture}
\caption{Connectivity graphs for Superfast simulation in limited connectivity. \textbf{(a)} Hamiltonian graph: all vertices correspond to fermionic modes, and in the original setting all edges would indicate the presence of hopping terms between the two modes in the Hamiltonian. We have displayed the direction of every edge in this graph. \textbf{(b)} Qubit connectivity graph: a qubit is placed on each vertex of this rotated square lattice. The underlying checkerboard pattern indicates which qubits are associated with which fermionic modes. Each dark plaquette is associated with an index $k$, such that the qubits on each of its corners have indices $jk \in E$.   }\label{fig:Superfastgraph}
\end{figure}
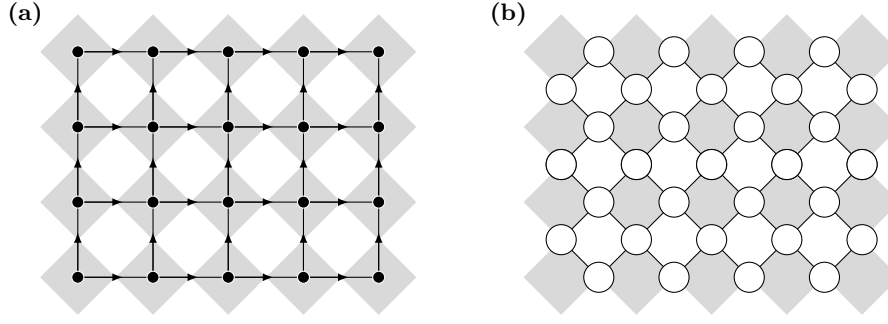

We now adapt the Superfast simulation to the square lattice layout  and give it the Manhattan-distance property. As we are interested in simulating more than square lattice Hamiltonians, we are going to depart a bit from the original concept of the qubit connectivity being related to the Hamiltonian.

\begin{figure}
\begin{tikzpicture}[baseline=0, scale=0.6]
\path[fill=\checker] (-0.5,-1)--(0,-0.5)--(.5,-1)--(0,-1.5)--cycle;
\path[fill=\checker] (0.5,-1)--(1,-0.5)--(1.5,-1)--(1,-1.5)--cycle;
\path[fill=\checker] (1.5,-1)--(2,-0.5)--(2.5,-1)--(2,-1.5)--cycle;
\path[fill=\checker] (2.5,-1)--(3,-0.5)--(3.5,-1)--(3,-1.5)--cycle;
\path[fill=\checker] (3.5,-1)--(4,-0.5)--(4.5,-1)--(4,-1.5)--cycle;
\path[fill=\checker] (4.5,-1)--(5,-0.5)--(5.5,-1)--(5,-1.5)--cycle;
\path[fill=\checker] (5.5,-1)--(6,-0.5)--(6.5,-1)--(6,-1.5)--cycle;
\path[fill=\checker] (6.5,-1)--(7,-0.5)--(7.5,-1)--(7,-1.5)--cycle;

\path[fill=\checker] (-0.5,0)--(0,0.5)--(.5,0)--(0,-0.5)--cycle;
\path[fill=black!60] (0.5,0)--(1,0.5)--(1.5,0)--(1,-0.5)--cycle;
\path[fill=\checker] (1.5,0)--(2,0.5)--(2.5,0)--(2,-0.5)--cycle;
\path[fill=\checker] (2.5,0)--(3,0.5)--(3.5,0)--(3,-0.5)--cycle;
\path[fill=\checker] (3.5,0)--(4,0.5)--(4.5,0)--(4,-0.5)--cycle;

\path[fill=\checker] (4.5,0)--(5,0.5)--(5.5,0)--(5,-0.5)--cycle;
\path[fill=\checker] (5.5,0)--(6,0.5)--(6.5,0)--(6,-0.5)--cycle;
\path[fill=\checker] (6.5,0)--(7,0.5)--(7.5,0)--(7,-0.5)--cycle;

\path[fill=\checker] (-0.5,1)--(0,1.5)--(.5,1)--(0,0.5)--cycle;
\path[fill=black!60] (0.5,1)--(1,1.5)--(1.5,1)--(1,0.5)--cycle;
\path[fill=\checker] (1.5,1)--(2,1.5)--(2.5,1)--(2,0.5)--cycle;
\path[fill=\checker] (2.5,1)--(3,1.5)--(3.5,1)--(3,0.5)--cycle;
\path[fill=\checker] (3.5,1)--(4,1.5)--(4.5,1)--(4,0.5)--cycle;

\path[fill=\checker] (4.5,1)--(5,1.5)--(5.5,1)--(5,0.5)--cycle;
\path[fill=black!60] (5.5,1)--(6,1.5)--(6.5,1)--(6,0.5)--cycle;
\path[fill=\checker] (6.5,1)--(7,1.5)--(7.5,1)--(7,0.5)--cycle;

\path[fill=\checker] (-0.5,2)--(0,2.5)--(.5,2)--(0,01.5)--cycle;
\path[fill=\checker] (0.5,2)--(1,2.5)--(1.5,2)--(1,01.5)--cycle;
\path[fill=\checker] (1.5,2)--(2,2.5)--(2.5,2)--(2,01.5)--cycle;
\path[fill=\checker] (2.5,2)--(3,2.5)--(3.5,2)--(3,01.5)--cycle;
\path[fill=\checker] (3.5,2)--(4,2.5)--(4.5,2)--(4,01.5)--cycle;

\path[fill=\checker] (4.5,2)--(5,2.5)--(5.5,2)--(5,1.5)--cycle;
\path[fill=\checker] (5.5,2)--(6,2.5)--(6.5,2)--(6,1.5)--cycle;
\path[fill=\checker] (6.5,2)--(7,2.5)--(7.5,2)--(7,1.5)--cycle;

\path[fill=\checker] (-0.5,3)--(0,3.5)--(.5,3)--(0,02.5)--cycle;
\path[fill=\checker] (0.5,3)--(1,3.5)--(1.5,3)--(1,02.5)--cycle;
\path[fill=black!60] (1.5,3)--(2,3.5)--(2.5,3)--(2,02.5)--cycle;
\path[fill=\checker] (2.5,3)--(3,3.5)--(3.5,3)--(3,02.5)--cycle;
\path[fill=\checker] (3.5,3)--(4,3.5)--(4.5,3)--(4,02.5)--cycle;

\path[fill=\checker] (4.5,3)--(5,3.5)--(5.5,3)--(5,2.5)--cycle;
\path[fill=\checker] (5.5,3)--(6,3.5)--(6.5,3)--(6,2.5)--cycle;
\path[fill=\checker] (6.5,3)--(7,3.5)--(7.5,3)--(7,2.5)--cycle;

\path[fill=\checker] (-0.5,4)--(0,4.5)--(.5,4)--(0,3.5)--cycle;
\path[fill=\checker] (0.5,4)--(1,4.5)--(1.5,4)--(1,3.5)--cycle;
\path[fill=\checker] (1.5,4)--(2,4.5)--(2.5,4)--(2,3.5)--cycle;
\path[fill=\checker] (2.5,4)--(3,4.5)--(3.5,4)--(3,3.5)--cycle;
\path[fill=\checker] (3.5,4)--(4,4.5)--(4.5,4)--(4,3.5)--cycle;
\path[fill=\checker] (4.5,4)--(5,4.5)--(5.5,4)--(5,3.5)--cycle;
\path[fill=\checker] (5.5,4)--(6,4.5)--(6.5,4)--(6,3.5)--cycle;
\path[fill=\checker] (6.5,4)--(7,4.5)--(7.5,4)--(7,3.5)--cycle;

\draw[] (2.5,3)--(4.5,1.0)--(5.5,2)--(6,1.5);
\draw[] (5,3.5)--(4.5,4)--(4,3.5)--(4.5,3)--cycle;
\draw[white, fill=black] (2.5,3) circle[radius=0.1];
\draw[white, fill=black] (3,2.5) circle[radius=0.1];
\draw[white, fill=black] (3.5,2) circle[radius=0.1];
\draw[white, fill=black] (4,1.5) circle[radius=0.1];
\draw[white, fill=black] (4.5,1) circle[radius=0.1];
\draw[white, fill=black] (5,1.5) circle[radius=0.1];
\draw[white, fill=black] (5.5,2) circle[radius=0.1];
\draw[white, fill=black] (6,1.5) circle[radius=0.1];

\draw[white, fill=black] (1,0.5) circle[radius=0.1];

\draw[white, fill=black] (5,3.5) circle[radius=0.1];
\draw[white, fill=black] (4.5,4) circle[radius=0.1];
\draw[white, fill=black] (4,3.5) circle[radius=0.1];
\draw[white, fill=black] (4.5,3) circle[radius=0.1];

\node[white] at (1,1) { $\boldsymbol{1}$};
\node[white] at (1,0) { $\boldsymbol{1}$};
\node[white] at (2,3) { $\boldsymbol{1}$};
\node[white] at (6,1) { $\boldsymbol{1}$};

\node[] at (0,1.1) {\footnotesize \textbf{(a)}};
\node[] at (2.5,2) {\footnotesize \textbf{(b)}};
\node[] at (5.3,4.1) {\footnotesize \textbf{(c)}};
\end{tikzpicture}
\caption{State preparation in the Superfast simulation. Black dots are flipped qubits and plaquettes with an odd number of flipped qubits are marked with $\boldsymbol{1}$, as a Fermion is created on the corresponding mode. \textbf{(a)} Flipping a qubit with label $(jk)$ creates Fermions on the adjacent modes $j$ and $k$. \textbf{(b)} $X$-strings  (here emphasized by linking the qubits) create non-local pairs of Fermions, as long as we ensure to flip always an even number of qubits on each plaquette, which means  winding around white plaquettes when the string has to change direction.  \textbf{(c)} Flips like this result from stabilizers, and do not excite Fermions, as on all dark plaquettes  an even number of qubits is flipped. } \label{fig:superfastprep}
\end{figure}
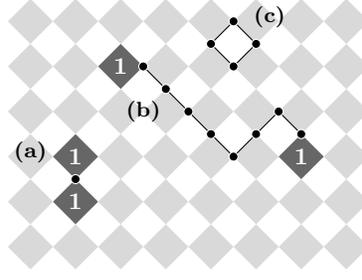

 Instead, we will show  that we can adapt the mapping adequately by pretending that the Hamiltonian graph is a square lattice. On this lattice, modes that are actually subject to hopping interactions in the Hamiltonian, should be locally close. Such a lattice of modes is shown in Figure \ref{fig:Superfastgraph}(a), where  the direction of every edge is indicated. As the direction of every edge $jk$ only determines the factor $\varepsilon_{jk}\in\lbrace +1, \, -1 \rbrace$ in \eqref{eq:aa2},  it has not much influence on the transformation. We will see later that the choice of the order of the edges on every mode is way more relevant for the strings that such a mapping produces. In \ref{fig:Superfastgraph}(a), we have already outlined the tiling of the line graph, to which we now switch. The resulting qubit connectivity graph can be seen in Figure \ref{fig:Superfastgraph}(b), where the plaquettes enclosing a fermionic mode are darkened. Starting from a general set of $\ell_1 \times \ell_2$ modes, we have now ended up with a rotated patch of the square lattice that has $2\ell_1\ell_2-(\ell_1+\ell_2)$ qubits on it. The number of white plaquettes, that are enclosed in the graph, describes the number of smallest possible loops, which means it is the total number of linearly independent stabilizers. We have  $(\ell_1-1)(\ell_2-1)$ of those white plaquettes, which means the system has $2^{\ell_1\ell_2-1}$ degrees of freedom left: since we have mapped only pairs of operators \eqref{eq:pairs1}-\eqref{eq:pairs3}, we are now seemingly stuck in the subspace with an even number of Fermions. This situation is however not terminal: we can simulate the odd-parity subspace separately as well as the entire Fock space.  Let us further illuminate this issue by considering the logical basis of the even-parity subspace first. For that purpose we pick a set $\lbrace S^i \rbrace_i$ of $(\ell_1-1)(\ell_2-1)$  linearly independent  stabilizers from \eqref{eq:Superfaststab}. The set fully constrains the system. Automatically, all stabilizers $S^i$ are  orthogonal in the computational basis, such that the fermionic  vacuum state is encoded as
\begin{align}
\label{eq:superfastvacuum}
\ket{\Theta} \; \hat{=} \; \left[\prod_i\frac{1}{\sqrt{2}}\left( \mathbb{I}+S^i \right) \right] \ket{0^n}\, .
\end{align}
We can then apply operators  $\mathcal{A}_{jk}$ and $\mathcal{B}_j$ in order to prepare other states with an even particle number. While the $\mathcal{A}_{jk}$ are different for every ordering, the operators $\mathcal{B}_k$, are independent of it: an operator $\mathcal{B}_k$ is the string of $Z$-operators around the shaded plaquette associated with mode $k$. If this plaquette is in the interior of the lattice in Figure \ref{fig:Superfastgraph}(b), the string has weight four, three if it is on the boundary edge, and two if in a corner. The one feature that the operators $\mathcal{A}_{jk}$ have in common for every ordering, is that they include an $X$-operator on the qubit $(jk)$. Apart from the  administration of some minus signs, the $\mathcal{A}_{jk}$  has generally the effect to flip  qubit  $(jk)$ in the  all-zero state $\ket{0^n}$ of \eqref{eq:superfastvacuum}. Comparing the encoded operators  \eqref{eq:aa} and \eqref{eq:bb} to the toy picture of the $\mathcal{A}$ and $\mathcal{B}$ operators we have just suggested, we find that a qubit configuration $\ket{\bbs{\xi}}= ( \bigotimes_{jk\in E} \ket{\xi_{jk}}_{jk} )$, with all $\xi_{jk}\in \mathbb{Z}_2$,  has the following correspondence to a fermionic quantum state:
\begin{align}
\label{eq:superfastconfig}
\left[\prod_i\frac{1}{\sqrt{2}}\left( \mathbb{I}+S^i \right) \right] \ket{\bbs{\xi}} \;\; \hat{\propto} \;\; \left[ \prod_{j=1}^{N} \left(c^\dagger_j\right)^{\sum_{i :\, (ij)\in E} \;\xi_{ij} \; \moto} \right] \ket{\Theta} \, .
\end{align}
Note that (as denoted by $\hat{\propto}$)   we have not kept track of any minus signs in  \eqref{eq:superfastconfig}. The relation is however sufficient to show that a fermionic mode $k$ is occupied, if an odd number of qubits  around the plaquette $k$  are in $\ket{1}$. The product of the stabilizers $\prod_i\frac{1}{\sqrt{2}}\left( \mathbb{I}+S^i \right)$ mixes all possible configurations that conserve the common parity of  qubits around a shaded plaquette (as the stabilizers need to commute with $\mathcal{B}_k$, a logical operator), and so the fermionic occupations are conserved as well. In order to prepare a pure fermionic state different from the vacuum, we need to consider a qubit configuration $\ket{\bbs{\xi}}$, in which we flip strings of adjacent qubits in order to create Fermions on the plaquettes at their ends, see Figure \ref{fig:superfastprep}.

So far, we still have not left the even-parity subspace, but we might have systems to solve that are populated by odd numbers of Fermions. In  \cite{setia2017bravyi}, it is suggested to add another mode to the system that is however not coupled to any other term in the Hamiltonian. From the original concept of the BKSF it is however not clear how this mode is brought into the system, since all qubits correspond to couplings of modes in the Hamiltonian, which here do not exist. Let us suggest to couple this mode to exactly one other, without ever using the $\mathcal{A}$-operator  of this link in the Hamiltonian. For state preparation we however can have strings that end at that outer plaquette, creating a mode that does not play a role and so effectively increase the degrees of freedom to $2^{N}$, modeling the entire Fock space. The cost of this increase is the overhead of one qubit. Alternatively there is a way to only map  the odd-parity subspace without using additional quantum resources: the idea is to consider the plaquette $k$ as being switched to `filled', such that the configuration on the right-hand side of \eqref{eq:superfastvacuum} does not correspond to the vacuum state (which is in the even-parity subspace), but to the state $c_k^\dagger\ket{\Theta}$. Flipping  the qubit $(jk)$ will lead to the Fermion on $k$ being annihilated and re-created on $j$, a string of flips that ends at $k$ will in general move the Fermion to the other end. We therefore make the replacement ${\sum_i \xi_{ik}} \to {(1+\sum_i \xi_{ik})}$ in the exponent of the mode-$k$ creation operator $c_k^\dagger$ on the right-hand side of \eqref{eq:superfastconfig}. In order to account for the switched occupation, we also need to update  $\mathcal{B}_k\to (-1) \mathcal{B}_k$ and add minus signs to some $\mathcal{A}$-operators.\\
 After having established an abstract idea of BKSF on the square lattice, we will now consider different versions of this mapping as we delve into detail. As mentioned before, the stabilizers of this mapping roughly flip qubits around white plaquettes. Due to \eqref{eq:Superfaststab},  their exact structure is determined by the operators $\mathcal{A}_{jk}$, which on the other hand depend on the ordering of edges on every vertex in the Hamiltonian graph, Figure \ref{fig:Superfastgraph}(a). In the qubit graph, this means that with every shaded plaquette we associate numbers with the qubits on its edges. The decision for an ordering has to be made consciously, as it influences the weight of strings simulating long-range hoppings. For now let us consider two different versions of this mapping in Table \ref{tab:compass}. For each version we assume that the ordering on every dark plaquette (leaving out missing vertices at the boundaries) is the same. From \eqref{eq:aa2}, we therefore just need to differentiate between  vertical and horizontal version of the operators $\mathcal{A}_{jk}$, i.e. considering the directions of the edges, we need to separate the cases where (1) the plaquette $k$ is the right neighbor of the plaquette $j$  and (2) where the plaquette $j$ is below   $k$. In the Table \ref{tab:compass}, we sketch these operators, along with the  stabilizers that follow from the multiplication of four of those operators to describe a closed loop around a white plaquette.  The first version is  the one already considered in \cite{havlivcek2017operator}, and second one is related to the mapping in \cite{chen2018exact}.
\begin{table}
\begin{tabular}{ccccccc}
Ordering && $\mathcal{A}_{jk}$ (horizontal) && $\mathcal{A}_{jk}$ (vertical)&& Stabilizer  \\ \hline \\
\begin{tikzpicture}[scale=0.5, baseline=0]
\draw[fill=\checker] (-.5,0)--(0,.5)--(0.5,0)--(0,-.5)--cycle;
\node[left] at (-.5,0) {2};
\node[right] at (.5,0) {3};
\node[above] at (0,.5) {1};
\node[below] at (0,-.5) {4};
\end{tikzpicture} &$\qquad$& \begin{tikzpicture}[scale=1, baseline=0]

\draw[color=\clre, fill=\checker] (-.5,0)--(0,.5)--(0.5,0)--(0,-.5)--cycle;
\draw[color=\clre, fill=\checker] (0.5,0)--(1,.5)--(1.5,0)--(1,-.5)--cycle;
\draw[thick] (-0.5,0)--(0,0.5)--(0.5,0)--(1,0.5);
\draw[thick, color=black, fill=white] (-0.5,0) circle[radius=0.2];
\node[] at (-0.5,0) {\footnotesize ${\mathsf{Z}}$};
\draw[thick,color=black, fill=white] (0.5,0) circle[radius=0.2];
\node[] at (0.5,0) {\footnotesize ${\mathsf{X}}$};
\draw[thick,color=black, fill=white](0,0.5) circle[radius=0.2];
\node[] at (0,0.5) {\footnotesize ${\mathsf{Z}}$};
\draw[color=\clre, fill=white] (0,-0.5) circle[radius=0.2];

\draw[color=\clre, fill=white] (1,-0.5) circle[radius=0.2];
\draw[thick,color=black, fill=white] (1,0.5) circle[radius=0.2];
\node[] at (1,0.5) {\footnotesize ${\mathsf{Z}}$};
\draw[color=\clre, fill=white] (1.5,0) circle[radius=0.2];
\node[] at (0,0) {\scriptsize $j$};
\node[] at (1,0) {\scriptsize $k$};
\end{tikzpicture} &$\qquad$& \begin{tikzpicture}[scale=1, baseline=0]

\draw[color=\clre, fill=\checker] (-.5,-1)--(0,-.5)--(0.5,-1)--(0,-1.5)--cycle;
\draw[thick, color=black, fill=\checker] (-.5,0)--(0,.5)--(0.5,0)--(0,-.5)--cycle;

\draw[thick, color=black, fill=white] (-0.5,0) circle[radius=0.2];
\node[] at (-0.5,0) {\footnotesize ${\mathsf{Z}}$};
\draw[thick, color=black, fill=white] (0.5,0) circle[radius=0.2];
\node[] at (0.5,0) {\footnotesize ${\mathsf{Z}}$};
\draw[thick, color=black, fill=white] (0,0.5) circle[radius=0.2];
\node[] at (0,0.5) {\footnotesize ${\mathsf{Z}}$};
\draw[thick, color=black, fill=white] (0,-0.5) circle[radius=0.2];
\node[] at (0, -0.5) {\footnotesize ${\mathsf{X}}$};

\draw[color=\clre, fill=white] (0,-1.5) circle[radius=0.2];
\draw[color=\clre, fill=white] (-0.5,-1) circle[radius=0.2];
\draw[color=\clre, fill=white] (0.5,-1) circle[radius=0.2];
\node[] at (0,0) {\scriptsize $k$};
\node[] at (0,-1) {\scriptsize $j$};

\end{tikzpicture} &$\qquad$& \begin{tikzpicture}[scale=1,  baseline=0]
\draw[color=\clre, fill=\checker] (-.5,0)--(0,.5)--(0.5,0)--(0,-.5)--cycle;
\draw[color=\clre, fill=\checker] (0.5,0)--(1,.5)--(1.5,0)--(1,-.5)--cycle;
\draw[color=\clre, fill=\checker] (-.5,-1)--(0,-.5)--(0.5,-1)--(0,-1.5)--cycle;
\draw[color=\clre, fill=\checker] (0.5,-1)--(1,-.5)--(1.5,-1)--(1,-1.5)--cycle;
\draw[thick] (0,-0.5)--(0.5,-1)--(1,-0.5)--(0.5,0)--cycle;
\draw[thick] (-0.5,-1)--(0,-0.5);
\draw[thick] (1.5,0)--(1,-0.5);
\draw[color=\clre, fill=white] (-0.5,0) circle[radius=0.2];
\draw[thick, color=black, fill=white] (0.5,0) circle[radius=0.2];
\node[] at (0.5,0) {\footnotesize ${\mathsf{X}}$};
\draw[color=\clre, fill=white] (0,0.5) circle[radius=0.2];
\draw[thick, color=black, fill=white] (0,-0.5) circle[radius=0.2];
\node[] at (0,-0.5) {\footnotesize ${\mathsf{Y}}$};

\draw[thick, color=black, fill=white] (1,-0.5) circle[radius=0.2];
\node[] at (1,-0.5) {\footnotesize ${\mathsf{Y}}$};
\draw[color=\clre, fill=white] (1,0.5) circle[radius=0.2];
\draw[thick, color=black, fill=white] (1.5,0) circle[radius=0.2];
\node[] at (1.5,0) {\footnotesize ${\mathsf{Z}}$};

\draw[thick, color=black, fill=white] (-0.5,-1) circle[radius=0.2];
\node[] at (-0.5,-1) {\footnotesize ${\mathsf{Z}}$};
\draw[thick, color=black, fill=white] (0.5,-1) circle[radius=0.2];
\node[] at (0.5,-1) {\footnotesize ${\mathsf{X}}$};

\draw[color=\clre, fill=white] (0,-1.5) circle[radius=0.2];

\draw[color=\clre, fill=white] (1,-1.5) circle[radius=0.2];
\draw[color=\clre, fill=white] (1.5,-1) circle[radius=0.2];

\end{tikzpicture} \\  \\ \hline \\
\begin{tikzpicture}[scale=0.5,baseline=0]
\draw[fill=\checker] (-.5,0)--(0,.5)--(0.5,0)--(0,-.5)--cycle;
\node[left] at (-.5,0) {4};
\node[right] at (.5,0) {2};
\node[above] at (0,.5) {3};
\node[below] at (0,-.5) {1};
\end{tikzpicture}&&\begin{tikzpicture}[scale=1,baseline=0]
\draw[color=\clre, fill=\checker] (-.5,0)--(0,.5)--(0.5,0)--(0,-.5)--cycle;
\draw[thick , fill=\checker] (0.5,0)--(1,.5)--(1.5,0)--(1,-.5)--cycle;
\draw[thick] (0.5,0)--(0,-0.5);
\draw[color=\clre, fill=white] (-0.5,0) circle[radius=0.2];
\draw[thick, fill=white] (0.5,0) circle[radius=0.2];
\node[] at (0.5,0) {\scriptsize $\mathsf{X}$};
\draw[color=\clre, fill=white] (0,0.5) circle[radius=0.2];

\draw[thick, fill=white] (0,-0.5) circle[radius=0.2];
\node[] at (0,-0.5) {\scriptsize $\mathsf{Z}$};
\draw[thick, fill=white] (1,-0.5) circle[radius=0.2];
\node[] at (1,-.5) {\scriptsize $\mathsf{Z}$};
\draw[thick, fill=white](1,0.5) circle[radius=0.2];
\node[] at (1,.5) {\scriptsize $\mathsf{Z}$};
\draw[thick, fill=white] (1.5,0) circle[radius=0.2];
\node[] at (1.5,0) {\scriptsize $\mathsf{Z}$};
\node[] at (0,0) {\scriptsize $j$};
\node[] at (1,0) {\scriptsize $k$};
\end{tikzpicture}&&\begin{tikzpicture}[scale=1,baseline=0]
\draw[color=\clre, fill=\checker] (-.5,0)--(0,.5)--(0.5,0)--(0,-.5)--cycle;
\draw[color=\clre, fill=\checker] (-.5,-1)--(0,-.5)--(0.5,-1)--(0,-1.5)--cycle;
\draw[thick](0,-0.5)--(0.5,-1)--(0,-1.5);
\draw[color=\clre, fill=white] (-0.5,0) circle[radius=0.2];
\draw[color=\clre, fill=white] (0.5,0) circle[radius=0.2];
\draw[color=\clre, fill=white] (0,0.5) circle[radius=0.2];
\draw[thick, fill=white] (0,-0.5) circle[radius=0.2];
\node[] at (0,-0.5) {\scriptsize $\mathsf{X}$};

\draw[thick, fill=white] (0,-1.5) circle[radius=0.2];
\node[] at (0,-1.5) {\scriptsize $\mathsf{Z}$};
\draw[color=\clre, fill=white] (-0.5,-1) circle[radius=0.2];
\draw[thick, fill=white] (0.5,-1) circle[radius=0.2];
\node[] at (.5,-1) {\scriptsize $\mathsf{Z}$};
\node[] at (0,0) {\scriptsize $k$};
\node[] at (0,-1) {\scriptsize $j$};
\end{tikzpicture} && \begin{tikzpicture}[scale=1, baseline=0]
\draw[color=\clre, fill=\checker] (-.5,0)--(0,.5)--(0.5,0)--(0,-.5)--cycle;
\draw[thick, fill=\checker] (0.5,0)--(1,.5)--(1.5,0)--(1,-.5)--cycle;
\draw[color=\clre, fill=\checker] (-.5,-1)--(0,-.5)--(0.5,-1)--(0,-1.5)--cycle;
\draw[color=\clre, fill=\checker] (0.5,-1)--(1,-.5)--(1.5,-1)--(1,-1.5)--cycle;
\draw[thick] (0,-0.5)--(0.5,-1)--(1,-0.5)--(0.5,0)--cycle;

\draw[color=\clre, fill=white] (-0.5,0) circle[radius=0.2];
\draw[thick, color=black, fill=white] (0.5,0) circle[radius=0.2];
\node[] at (0.5,0) {\footnotesize ${\mathsf{X}}$};
\draw[color=\clre, fill=white] (0,0.5) circle[radius=0.2];
\draw[thick, color=black, fill=white] (0,-0.5) circle[radius=0.2];
\node[] at (0,-0.5) {\footnotesize ${\mathsf{Y}}$};

\draw[thick, color=black, fill=white] (1,-0.5) circle[radius=0.2];
\node[] at (1,-0.5) {\footnotesize ${\mathsf{X}}$};
\draw[thick, color=black, fill=white] (1,0.5) circle[radius=0.2];
\node[] at (1,0.5) {\footnotesize ${\mathsf{Z}}$};
\draw[thick, color=black, fill=white] (1.5,0) circle[radius=0.2];
\node[] at (1.5,0) {\footnotesize ${\mathsf{Z}}$};

\draw[color=\clre, fill=white]  (-0.5,-1) circle[radius=0.2];
\draw[thick, color=black, fill=white] (0.5,-1) circle[radius=0.2];
\node[] at (0.5,-1) {\footnotesize ${\mathsf{Y}}$};

\draw[color=\clre, fill=white] (0,-1.5) circle[radius=0.2];

\draw[color=\clre, fill=white] (1,-1.5) circle[radius=0.2];
\draw[color=\clre, fill=white] (1.5,-1) circle[radius=0.2];

\end{tikzpicture} \\  \\ \hline
\end{tabular}
\caption{Different versions of BKSF. The ordering of the edges on each vertex is displayed as well as the operators this ordering entails: horizontal and vertical edge operators $\mathcal{A}_{jk}$ and the stabilizers (signs are omitted). The upper version is the one used in \cite{havlivcek2017operator}, while the lower one is related to the mapping in \cite{chen2018exact}. }\label{tab:compass}
\end{table}

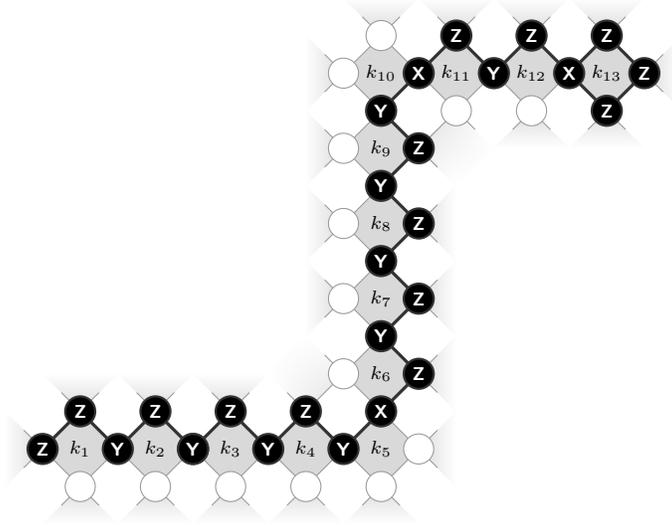
\begin{figure}
\begin{tikzpicture}[scale=1]

\path[top color= white, bottom color = \checker] (2,.5)--(1.5,1)--(2.5,1)--cycle;
\draw[\clre] (2,.5)--(1.75,.75);
\draw[\clre] (2,.5)--(2.25,.75);
\path[top color= white, bottom color = \checker] (3,.5)--(2.5,1)--(3.5,1)--cycle;
\draw[\clre] (3,.5)--(2.75,.75);
\draw[\clre] (3,.5)--(3.25,.75);
\path[bottom color= white, top color = \checker] (2,-.5)--(1.5,-1)--(2.5,-1)--cycle;
\draw[\clre](2,-.5)--(1.75,-.75);
\draw[\clre](2,-.5)--(2.25,-.75);
\path[bottom color= white, top color = \checker] (3,-.5)--(2.5,-1)--(3.5,-1)--cycle;
\draw[\clre](3,-.5)--(2.75,-.75);
\draw[\clre](3,-.5)--(3.25,-.75);
\draw[color=\clre, fill=\checker] (1.5,0)--(2,.5)--(2.5,0)--(2,-.5)--cycle;
\draw[color=\clre, fill=\checker] (2.5,0)--(3,.5)--(3.5,0)--(3,-.5)--cycle;


\path[top color= white, bottom color = \checker] (4,.5)--(3.5,1)--(4.5,1)--cycle;
\draw[\clre] (4,.5)--(3.75,.75);
\draw[\clre] (4,.5)--(4.25,.75);
\shade[lower right=\checker, upper right = white] (5,0.5)--(5.5,1)--(5,1.5)--(4.5,1)--cycle;
\draw[\clre](4.75,.75)--(5,0.5)--(5.5,1)--(5.25,1.25);
\path[bottom color= white, top color = \checker] (4,-.5)--(3.5,-1)--(4.5,-1)--cycle;
\draw[\clre](4,-.5)--(3.75,-.75);
\draw[\clre](4,-.5)--(4.25,-.75);
\path[bottom color= white, top color = \checker] (5,-.5)--(4.5,-1)--(5.5,-1)--cycle;
\draw[\clre](5,-.5)--(4.75,-.75);
\draw[\clre](5,-.5)--(5.25,-.75);
\draw[color=\clre, fill=\checker] (3.5,0)--(4,.5)--(4.5,0)--(4,-.5)--cycle;
\draw[color=\clre, fill=\checker] (4.5,0)--(5,.5)--(5.5,0)--(5,-.5)--cycle;


\path[left color= white, right color = \checker] (1.5,0)--(1,.5)--(1,-.5)--cycle;
\draw[color = \clre] (1.5,0)--(1.25,.25);
\draw[color = \clre] (1.5,0)--(1.25,-.25);
\path[right color= white, left color = \checker] (6.5,0)--(7,.5)--(7,-.5)--cycle;
\draw[color = \clre] (6.5,0)--(6.75,.25);
\draw[color = \clre] (6.5,0)--(6.75,-.25);

\path[bottom color= white, top color = \checker] (6,-.5)--(5.5,-1)--(6.5,-1)--cycle;
\draw[\clre](6,-.5)--(5.75,-.75);
\draw[\clre](6,-.5)--(6.25,-.75);

\draw[color=\clre, fill=\checker] (5.5,0)--(6,.5)--(6.5,0)--(6,-.5)--cycle;

\path[left color= white, right color = \checker] (5.5,2)--(5,2.5)--(5,1.5)--cycle;
\draw[color = \clre] (5.5,2)--(5.25,2.25);
\draw[color = \clre] (5.5,2)--(5.25,1.75);

\path[right color= white, left color = \checker] (6.5,2)--(7,2.5)--(7,1.5)--cycle;
\draw[color = \clre] (6.5,2)--(6.75,2.25);
\draw[color = \clre] (6.5,2)--(6.75,1.75);

\path[right color= white, left color = \checker] (6.5,1)--(7,1.5)--(7,.5)--cycle;
\draw[color = \clre] (6.5,1)--(6.75,1.25);
\draw[color = \clre] (6.5,1)--(6.75,.75);

\draw[color=\clre, fill=\checker] (5.5,2)--(6,2.5)--(6.5,2)--(6,1.5)--cycle;
\draw[color=\clre, fill=\checker] (5.5,1)--(6,1.5)--(6.5,1)--(6,.5)--cycle;


\path[left color= white, right color = \checker] (5.5,4)--(5,4.5)--(5,3.5)--cycle;
\draw[color = \clre] (5.5,4)--(5.25,4.25);
\draw[color = \clre] (5.5,4)--(5.25,3.75);

\path[right color= white, left color = \checker] (6.5,4)--(7,4.5)--(7,3.5)--cycle;
\draw[color = \clre] (6.5,4)--(6.75,4.25);
\draw[color = \clre] (6.5,4)--(6.75,3.75);

\path[left color= white, right color = \checker] (5.5,3)--(5,3.5)--(5,2.5)--cycle;
\draw[color = \clre] (5.5,3)--(5.25,3.25);
\draw[color = \clre] (5.5,3)--(5.25,2.75);

\path[right color= white, left color = \checker] (6.5,3)--(7,3.5)--(7,2.5)--cycle;
\draw[color = \clre] (6.5,3)--(6.75,3.25);
\draw[color = \clre] (6.5,3)--(6.75,2.75);

\draw[color=\clre, fill=\checker] (5.5,4)--(6,4.5)--(6.5,4)--(6,3.5)--cycle;
\draw[color=\clre, fill=\checker] (5.5,3)--(6,3.5)--(6.5,3)--(6,2.5)--cycle;


\path[bottom color=\checker, top color=white] (6,5.5)--(5.5,6)--(6.5,6)--cycle;
\draw[\clre] (5.75,5.75)--(6,5.5)--(6.25,5.75);

\path[left color= white, right color = \checker] (5.5,5)--(5,5.5)--(5,4.5)--cycle;
\draw[color = \clre] (5.5,5)--(5.25,5.25);
\draw[color = \clre] (5.5,5)--(5.25,4.75);


\draw[color=\clre, fill=\checker] (5.5,5)--(6,5.5)--(6.5,5)--(6,4.5)--cycle;


 \shade[ lower right = white, upper left = \checker] (6.5,4)--(7,4.5)--(7.5,4)--(7,3.5)--cycle;
\draw[\clre] (6.75,3.75)--(6.5,4)--(7,4.5)--(7.25,4.25);

\path[top color= white, bottom color = \checker] (7,5.5)--(6.5,6)--(7.5,6)--cycle;
\draw[\clre] (7,5.5)--(6.75,5.75);
\draw[\clre] (7,5.5)--(7.25,5.75);
\path[top color= white, bottom color = \checker] (8,5.5)--(7.5,6)--(8.5,6)--cycle;
\draw[\clre] (8,5.5)--(7.75,5.75);
\draw[\clre] (8,5.5)--(8.25,5.75);

\path[bottom color= white, top color = \checker] (8,4.5)--(7.5,4)--(8.5,4)--cycle;
\draw[\clre](8,4.5)--(7.75,4.25);
\draw[\clre](8,4.5)--(8.25,4.25);
\draw[color=\clre, fill=\checker] (6.5,5)--(7,5.5)--(7.5,5)--(7,4.5)--cycle;
\draw[color=\clre, fill=\checker] (7.5,5)--(8,5.5)--(8.5,5)--(8,4.5)--cycle;
\draw[color=\clre, fill=\checker] (8.5,5)--(9,5.5)--(9.5,5)--(9,4.5)--cycle;

\path[bottom color= white, top color = \checker] (9,4.5)--(8.5,4)--(9.5,4)--cycle;
\draw[\clre](9,4.5)--(8.75,4.25);
\draw[\clre](9,4.5)--(9.25,4.25);

\path[top color= white, bottom color = \checker] (9,5.5)--(8.5,6)--(9.5,6)--cycle;
\draw[\clre] (9,5.5)--(8.75,5.75);
\draw[\clre] (9,5.5)--(9.25,5.75);

\path[right color= white, left color = \checker] (9.5,5)--(10,5.5)--(10,4.5)--cycle;
\draw[color = \clre] (9.5,5)--(9.75,5.25);
\draw[color = \clre] (9.5,5)--(9.75,4.75);
\draw[\clrd, very thick] (1.5,0)--(2,0.5)--(2.5,0)--(3,0.5)--(3.5,0)--(4,0.5)--(4.5,0)--(5,.5)--(5.5,0)--(6.5,1)--(6,1.5)--(6.5,2)--(6,2.5)--(6.5,3)--(6,3.5)--(6.5,4)--(6,4.5)--(7,5.5)--(7.5,5)--(8,5.5)--(8.5,5)--(9,5.5)--(9.5,5)--(9,4.5)--(8.5,5);
\draw[color=\clrb, fill=\clra, thick] (1.5,0) circle[radius=0.2];
\draw[color=\clrb, fill=\clra, thick] (2.5,0) circle[radius=0.2];
\draw[color=\clrb, fill=\clra, thick] (2,0.5) circle[radius=0.2];
\draw[color=\clre, fill=white] (2,-0.5) circle[radius=0.2];

\draw[color=\clre, fill=white] (3,-0.5) circle[radius=0.2];
\draw[color=\clrb, fill=\clra, thick] (3,0.5) circle[radius=0.2];
\draw[color=\clrb, fill=\clra, thick] (3.5,0) circle[radius=0.2];

\draw[color=\clre, fill=white] (6,-0.5) circle[radius=0.2];
\draw[color=\clrb, fill=\clra, thick] (6,0.5) circle[radius=0.2];
\draw[color=\clre, fill=white] (6.5,0) circle[radius=0.2];

\draw[color=\clrb, fill=\clra, thick] (3.5,0) circle[radius=0.2];
\draw[color=\clrb, fill=\clra, thick] (4.5,0) circle[radius=0.2];
\draw[color=\clrb, fill=\clra, thick] (4,0.5) circle[radius=0.2];
\draw[color=\clre, fill=white] (4,-0.5) circle[radius=0.2];

\draw[color=\clre, fill=white] (5,-0.5) circle[radius=0.2];
\draw[color=\clrb, fill=\clra, thick] (5,0.5) circle[radius=0.2];
\draw[color=\clrb, fill=\clra, thick] (5.5,0) circle[radius=0.2];

\draw[color=\clre, fill=white] (5.5,2) circle[radius=0.2];
\draw[color=\clrb, fill=\clra, thick] (6.5,2) circle[radius=0.2];
\draw[color=\clrb, fill=\clra, thick] (6,2.5) circle[radius=0.2];
\draw[color=\clrb, fill=\clra, thick] (6,1.5) circle[radius=0.2];

\draw[color=\clre, fill=white] (5.5,1) circle[radius=0.2];
\draw[color=\clrb, fill=\clra, thick] (6.5,1) circle[radius=0.2];

\draw[color=\clre, fill=white] (5.5,4) circle[radius=0.2];
\draw[color=\clrb, fill=\clra, thick] (6.5,4) circle[radius=0.2];
\draw[color=\clrb, fill=\clra, thick] (6,4.5) circle[radius=0.2];
\draw[color=\clrb, fill=\clra, thick] (6,3.5) circle[radius=0.2];

\draw[color=\clre, fill=white] (5.5,3) circle[radius=0.2];
\draw[color=\clrb, fill=\clra, thick] (6.5,3) circle[radius=0.2];

\draw[color=\clre, fill=white] (6,5.5) circle[radius=0.2];

\draw[color=\clre, fill=white] (5.5,5) circle[radius=0.2];

\draw[color=\clrb, fill=\clra, thick] (6.5,5) circle[radius=0.2];
\draw[color=\clrb, fill=\clra, thick] (7.5,5) circle[radius=0.2];
\draw[color=\clre, fill=white] (7,4.5) circle[radius=0.2];

\draw[color=\clre, fill=white] (8,4.5) circle[radius=0.2];
\draw[color=\clrb, fill=\clra, thick] (8,5.5) circle[radius=0.2];
\draw[color=\clrb, fill=\clra, thick] (8.5,5) circle[radius=0.2];

\draw[color=\clrb, fill=\clra, thick] (9,4.5) circle[radius=0.2];
\draw[color=\clrb, fill=\clra, thick] (9,5.5) circle[radius=0.2];
\draw[color=\clrb, fill=\clra, thick] (9.5,5) circle[radius=0.2];
\draw[color=\clrb, fill=\clra, thick] (7,5.5) circle[radius=0.2];

\node[\clrc] at (1.5,0) {\scriptsize $\boldsymbol{\mathsf{Z}}$};
\node[\clrc] at (2,0.5) {\scriptsize $\boldsymbol{\mathsf{Z}}$};
\node[\clrc] at (2.5,0) {\scriptsize $\boldsymbol{\mathsf{Y}}$};
\node[\clrc] at (3,0.5) {\scriptsize $\boldsymbol{\mathsf{Z}}$};
\node[\clrc] at (3.5,0) {\scriptsize $\boldsymbol{\mathsf{Y}}$};
\node[\clrc] at (4,0.5) {\scriptsize $\boldsymbol{\mathsf{Z}}$};
\node[\clrc] at (4.5,0) {\scriptsize $\boldsymbol{\mathsf{Y}}$};
\node[\clrc] at (5,.5)  {\scriptsize $\boldsymbol{\mathsf{Z}}$};
\node[\clrc] at (5.5,0) {\scriptsize $\boldsymbol{\mathsf{Y}}$};
\node[\clrc] at (6,0.5) {\scriptsize $\boldsymbol{\mathsf{X}}$};
\node[\clrc] at (6.5,1) {\scriptsize $\boldsymbol{\mathsf{Z}}$};
\node[\clrc] at (6,1.5) {\scriptsize $\boldsymbol{\mathsf{Y}}$};
\node[\clrc] at (6.5,2) {\scriptsize $\boldsymbol{\mathsf{Z}}$};
\node[\clrc] at (6,2.5) {\scriptsize $\boldsymbol{\mathsf{Y}}$};
\node[\clrc] at (6.5,3) {\scriptsize $\boldsymbol{\mathsf{Z}}$};
\node[\clrc] at (6,3.5) {\scriptsize $\boldsymbol{\mathsf{Y}}$};
\node[\clrc] at (6.5,4) {\scriptsize $\boldsymbol{\mathsf{Z}}$};
\node[\clrc] at (6,4.5) {\scriptsize $\boldsymbol{\mathsf{Y}}$};
\node[\clrc] at (6.5,5) {\scriptsize $\boldsymbol{\mathsf{X}}$};
\node[\clrc] at (7,5.5) {\scriptsize $\boldsymbol{\mathsf{Z}}$};
\node[\clrc] at (7.5,5) {\scriptsize $\boldsymbol{\mathsf{Y}}$};
\node[\clrc] at (8,5.5) {\scriptsize $\boldsymbol{\mathsf{Z}}$};
\node[\clrc] at (8.5,5) {\scriptsize $\boldsymbol{\mathsf{X}}$};
\node[\clrc] at (9,5.5) {\scriptsize $\boldsymbol{\mathsf{Z}}$};
\node[\clrc] at (9.5,5) {\scriptsize $\boldsymbol{\mathsf{Z}}$};
\node[\clrc] at (9,4.5) {\scriptsize $\boldsymbol{\mathsf{Z}}$};
\node[black] at (2,0) {\scriptsize $k_1$};
\node[black] at (3,0) {\scriptsize $k_2$};
\node[black] at (4,0) {\scriptsize $k_3$};
\node[black] at (5,0) {\scriptsize $k_4$};
\node[black] at (6,0) {\scriptsize $k_5$};
\node[black] at (6,1) {\scriptsize $k_6$};
\node[black] at (6,2) {\scriptsize $k_7$};
\node[black] at (6,3) {\scriptsize $k_8$};
\node[black] at (6,4) {\scriptsize $k_9$};
\node[black] at (6,5) {\scriptsize $k_{10}$};
\node[black] at (7,5) {\scriptsize $k_{11}$};
\node[black] at (8,5) {\scriptsize $k_{12}$};
\node[black] at (9,5) {\scriptsize $k_{13}$};
\end{tikzpicture}
\caption{Superfast simulation of a hopping operator in the between modes $k_1$ and $k_{13}$, coupling the respective shaded plaquettes in a string of length scaling with their Manhattan distance, where the path taken is defined by the locally connected chain of modes $k_2$ to $k_{12}$. The string simulated is  $(-i\mathcal{B}_{k_1} \mathcal{A}_{k_1k_{13}})$, which in Jordan-Wigner transform would be  $h_\data=(X_{k_1}  \otimes Z_{k_1+1} \otimes \dots \otimes Z_{k_{13}-1} \otimes X_{k_{13}})$. The plaquettes $(k_1, \, ... \, , \, k_{13})$ are labeled on this lattice. } \label{fig:Superfastmanhattan}
\end{figure}

  We can now describe Fermion-operator-pairs via Table \ref{tab:compass} with \eqref{eq:pairs1}-\eqref{eq:pairs3}. The latter equations hold for operators $\mathcal{A}_{jk}$ of every link, whereas the table only provides us with operators in which $j$ and $k$ are adjacent plaquettes. We will now cease to pretend that the Hamiltonian is just composed of nearest-neighbor interactions, and derive non-local operators $\mathcal{A}_{jk}$. By \eqref{eq:aa} we set $\mathcal{A}_{jk}\propto \mathcal{O}_{jk}$ and using \eqref{eq:manhattanops}  we find

\begin{align}
\label{eq:manhattanA}
\mathcal{A}_{k_1 k_l}  = (i)^{l-1} \prod_{s=1}^{l-1} \mathcal{A}_{k_s k_{s+1}}
\end{align}
for any sequence $k_1, \, k_2, \, \dots, \, k_l$, where for all $s \in [l-1]$: $k_sk_{s+1} \in E$. This means we can multiply several of the nearest-neighbor $\mathcal{A}$-operators from Table \ref{tab:compass}. The choice of the ordering turns out to be crucial, as for various orderings, the resulting mapping is not a good one according to the  criteria of Section \ref{sec:one}.  The first mapping in  Table \ref{tab:compass} for instance does not produce a continuous Pauli string \eqref{eq:manhattanA} when making a chain of several horizontal $\mathcal{A}_{jk}$. For a vertical chain, we have a maximal operator weight. The second mapping on the other hand is better behaved:  horizontal and vertical $\mathcal{A}$-operators are connected  and their weight is minimal.
In Figure \ref{fig:Superfastmanhattan}, we present an example of the simulation of the Pauli string $(-i\mathcal{B}_{k_1} \mathcal{A}_{k_1k_{l}})$, where  $\mathcal{A}_{k_1k_l}$ is non-local as in \eqref{eq:manhattanA}, with  $l=13$. The string here extends on  a zig zag line along the edges of the plaquettes involved, $\lbrace k_s \rbrace_s$ connecting the plaquettes $k_1$ and $k_{13}$. The weight of this string can perhaps be optimized in cutting more corners like at plaquette $k_5$. In any case, we have  adapted the BKSF  as a two-dimensional Fermion-to-qubit mapping on the square lattice.

\subsection{Fermi-Hubbard model}
In this Section we test the proposed square lattice implementations of the Superfast simulation and the Verstraete-Cirac transform on the Fermi-Hubbard model. \\
For both mappings, we have to decide where to place spin-up and -down modes of the same spatial site. On the one hand should the qubits representing these modes  be locally close, perhaps even horizontally or vertically adjacent, but on the other hand they will increase the weight of the strings simulating hopping terms, as they are `in the way'. For the BKSF, it is almost inconsequential whether the spin pairs are vertically or horizontally stacked, so we decide for the latter. For the VCT, the situation is different as it produces shorter hopping strings in the vertical direction, which leads us to make the spin pairs vertical neighbors on the grid. In order to do that, we need to compensate for the shift that has emerged aligning the primed qubits: in Figure \ref{fig:surface}(a),  qubit $4$ is for instance below qubit $6$, not qubit $5$. Without this shift, there would be additional costs for horizontal or vertical hoppings, but with the shift, additional costs emerge for the Hubbard terms. As a fix, we simulate the model with $\ell_2$ additional modes, that remain empty. The qubits corresponding to those modes are the ones  at the horizontal perimeter of the qubit lattice,  i.e.~the qubits labeled 1, 5, 9, 13, 17 and 21 in Figure \ref{fig:surface}(a). Those data qubits, fixed to $\ket{0}$, can as well be removed, but their primed counterparts must remain and be part of the code. The spin-partners can now be placed vertically adjacent on the grid.  The Hubbard model with $L\times L$ spatial sites is thus simulated with $4L^2+2L$ qubits in the VCT, and with $4L^2-3L$ qubits in the BKSF. The resulting Pauli strings can be found in Table \ref{tab:hubbardx}.

\begin{table} \begin{tabular}{ccc}
Verstraete-Cirac transform &$\qquad$&Superfast simulation  \\ \hline
\begin{tikzpicture} [scale=0.7, baseline=0]
\draw[\clre] (3.5,0)--(0.5,0);
\draw[\clre] (1,-0.5)--(1,2.5);
\draw[\clre] (2,-0.5)--(2,2.5);
\draw[\clre] (3,-0.5)--(3,2.5);
\draw[\clre] (3.5,1)--(0.5,1);
\draw[\clre] (3.5,2)--(0.5,2);
\draw[very thick, \clrd] (2,0)--(3,0)--(3,2)--(2,2);
\draw[color=\clrb, fill=\clra, thick] (3,0) circle[radius=0.2];
\draw[color=\clre, fill=white] (1,0) circle[radius=0.2];
\draw[color=\clrb, fill=\clra, thick] (2,0) circle[radius=0.2];
\draw[color=\clrb, fill=\clra, thick] (3,1) circle[radius=0.2];
\draw[color=\clre, fill=white](1,1) circle[radius=0.2];
\draw[color=\clre, fill=white](2,1) circle[radius=0.2];
\draw[color=\clrb, fill=\clra, thick] (3,2) circle[radius=0.2];
\draw[color=\clre, fill=white](1,2) circle[radius=0.2];
\draw[color=\clrb, fill=\clra, thick](2,2) circle[radius=0.2];

\node[\clrc] at (2,0) {\tiny $\boldsymbol{\mathsf{X}}$};
\node[\clrc] at (3,0) {\tiny $\boldsymbol{\mathsf{Y}}$};
\node[\clrc] at (3,1) {\tiny $\boldsymbol{\mathsf{Z}}$};
\node[\clrc] at (3,2) {\tiny $\boldsymbol{\mathsf{X}}$};
\node[\clrc] at (2,2) {\tiny $\boldsymbol{\mathsf{Y}}$};
\end{tikzpicture}$\quad$ \begin{tikzpicture} [scale=0.7, baseline=0]
\draw[\clre] (3.5,0)--(0.5,0);
\draw[\clre] (1,-0.5)--(1,2.5);
\draw[\clre] (2,-0.5)--(2,2.5);
\draw[\clre] (3,-0.5)--(3,2.5);
\draw[\clre] (3.5,1)--(0.5,1);
\draw[\clre] (3.5,2)--(0.5,2);
\draw[very thick, \clrd] (2,0)--(3,0)--(3,2)--(2,2);
\draw[color=\clrb, fill=\clra, thick] (3,0) circle[radius=0.2];
\draw[color=\clre, fill=white] (1,0) circle[radius=0.2];
\draw[color=\clrb, fill=\clra, thick] (2,0) circle[radius=0.2];
\draw[color=\clrb, fill=\clra, thick] (3,1) circle[radius=0.2];
\draw[color=\clre, fill=white](1,1) circle[radius=0.2];
\draw[color=\clre, fill=white](2,1) circle[radius=0.2];
\draw[color=\clrb, fill=\clra, thick] (3,2) circle[radius=0.2];
\draw[color=\clre, fill=white](1,2) circle[radius=0.2];
\draw[color=\clrb, fill=\clra, thick](2,2) circle[radius=0.2];

\node[\clrc] at (2,0) {\tiny $\boldsymbol{\mathsf{Y}}$};
\node[\clrc] at (3,0) {\tiny $\boldsymbol{\mathsf{Y}}$};
\node[\clrc] at (3,1) {\tiny $\boldsymbol{\mathsf{Z}}$};
\node[\clrc] at (3,2) {\tiny $\boldsymbol{\mathsf{X}}$};
\node[\clrc] at (2,2) {\tiny $\boldsymbol{\mathsf{X}}$};
\end{tikzpicture} & \begin{tabular}{c}  Vertical hoppings \\ $\,$ \\ $\,$ \\  $\,$ \end{tabular}& \begin{tikzpicture}[scale=0.8,baseline=-30]
\path[left color= white, right color = \checker] (-.5,0)--(-1,.5)--(-1,-.5)--cycle;
\draw[color = \clre] (-.5,0)--(-.75,.25);
\draw[color = \clre] (-.5,0)--(-.75,-.25);

\path[right color= white, left color = \checker] (.5,0)--(1,.5)--(1,-.5)--cycle;
\draw[color = \clre] (.5,0)--(.75,.25);
\draw[color = \clre] (.5,0)--(.75,-.25);

\path[left color= white, right color = \checker] (-.5,-1)--(-1,-.5)--(-1,-1.5)--cycle;
\draw[color = \clre] (-.5,-1)--(-.75,-.75);
\draw[color = \clre] (-.5,-1)--(-.75,-1.25);

\path[right color= white, left color = \checker] (.5,-1)--(1,-.5)--(1,-1.5)--cycle;
\draw[color = \clre] (.5,-1)--(.75,-.75);
\draw[color = \clre] (.5,-1)--(.75,-1.25);

\path[top color= white, bottom color = \checker] (0,.5)--(-0.5,1)--(0.5,1)--cycle;
\draw[\clre] (0,.5)--(-.25,.75);
\draw[\clre] (0,.5)--(.25,.75);

\path[bottom color= white, top color = \checker] (0,-1.5)--(-0.5,-2)--(0.5,-2)--cycle;
\draw[\clre](0,-1.5)--(-0.25,-1.75);
\draw[\clre](0,-1.5)--(0.25,-1.75);

\draw[color=\clre, fill=\checker] (-.5,0)--(0,.5)--(0.5,0)--(0,-.5)--cycle;
\draw[color=\clre, fill=\checker] (-.5,-1)--(0,-.5)--(0.5,-1)--(0,-1.5)--cycle;
\draw[very thick, color=\clrd] (0,-0.5)--(-0.5,-1);
\draw[color=\clre, fill=white] (-0.5,0) circle[radius=0.2];
\draw[color=\clre, fill=white] (0.5,0) circle[radius=0.2];
\draw[color=\clre, fill=white] (0,0.5) circle[radius=0.2];
\draw[color=\clrb, fill=\clra, thick] (0,-0.5) circle[radius=0.2];
\node[\clrc] at (0,-0.5) {\tiny $\boldsymbol{\mathsf{Y}}$};
\draw[color=\clre, fill=white] (0,-1.5) circle[radius=0.2];
\draw[color=\clrb, fill=\clra, thick] (-0.5,-1) circle[radius=0.2];
\node[\clrc] at (-0.5,-1) {\tiny $\boldsymbol{\mathsf{Z}}$};
\draw[color=\clre, fill=white] (0.5,-1) circle[radius=0.2];

\end{tikzpicture} $\quad$\begin{tikzpicture}[scale=0.8, baseline=-30]
\path[left color= white, right color = \checker] (-.5,0)--(-1,.5)--(-1,-.5)--cycle;
\draw[color = \clre] (-.5,0)--(-.75,.25);
\draw[color = \clre] (-.5,0)--(-.75,-.25);

\path[right color= white, left color = \checker] (.5,0)--(1,.5)--(1,-.5)--cycle;
\draw[color = \clre] (.5,0)--(.75,.25);
\draw[color = \clre] (.5,0)--(.75,-.25);

\path[left color= white, right color = \checker] (-.5,-1)--(-1,-.5)--(-1,-1.5)--cycle;
\draw[color = \clre] (-.5,-1)--(-.75,-.75);
\draw[color = \clre] (-.5,-1)--(-.75,-1.25);

\path[right color= white, left color = \checker] (.5,-1)--(1,-.5)--(1,-1.5)--cycle;
\draw[color = \clre] (.5,-1)--(.75,-.75);
\draw[color = \clre] (.5,-1)--(.75,-1.25);

\path[top color= white, bottom color = \checker] (0,.5)--(-0.5,1)--(0.5,1)--cycle;
\draw[\clre] (0,.5)--(-.25,.75);
\draw[\clre] (0,.5)--(.25,.75);

\path[bottom color= white, top color = \checker] (0,-1.5)--(-0.5,-2)--(0.5,-2)--cycle;
\draw[\clre](0,-1.5)--(-0.25,-1.75);
\draw[\clre](0,-1.5)--(0.25,-1.75);

\draw[very thick, color=\clrd, fill=\checker] (-.5,0)--(0,.5)--(0.5,0)--(0,-.5)--cycle;
\draw[color=\clre, fill=\checker] (-.5,-1)--(0,-.5)--(0.5,-1)--(0,-1.5)--cycle;
\draw[very thick, color=\clrd] (0,-.5)--(0.5,-1)--(0,-1.5);
\draw[color=\clrb, fill=\clra, thick] (-0.5,0) circle[radius=0.2];
\node[\clrc] at (-0.5,0) {\tiny $\boldsymbol{\mathsf{Z}}$};
\draw[color=\clrb, fill=\clra, thick] (0.5,0) circle[radius=0.2];
\node[\clrc] at (0.5,0) {\tiny $\boldsymbol{\mathsf{Z}}$};
\draw[color=\clrb, fill=\clra, thick] (0,0.5) circle[radius=0.2];
\node[\clrc] at (0,0.5) {\tiny $\boldsymbol{\mathsf{Z}}$};
\draw[color=\clrb, fill=\clra, thick] (0,-0.5) circle[radius=0.2];
\node[\clrc] at (0,-0.5) {\tiny $\boldsymbol{\mathsf{Y}}$};

\draw[color=\clrb, fill=\clra, thick] (0,-1.5) circle[radius=0.2];
\node[\clrc] at (0,-1.5) {\tiny $\boldsymbol{\mathsf{Z}}$};
\draw[color=\clre, fill=white](-0.5,-1) circle[radius=0.2];

\draw[color=\clrb, fill=\clra, thick] (0.5,-1) circle[radius=0.2];
\node[\clrc] at (0.5,-1) {\tiny $\boldsymbol{\mathsf{Z}}$};

\end{tikzpicture}\\
\begin{tikzpicture} [scale=0.7, baseline=0]
\draw[\clre] (2.5,0)--(-1.5,0);
\draw[\clre] (0,-0.5)--(0,1.5);
\draw[\clre] (2.5,1)--(-1.5,1);
\draw[\clre] (2,-0.5)--(2,1.5);
\draw[\clre] (1,-0.5)--(1,1.5);
\draw[\clre] (-1,-0.5)--(-1,1.5);
\draw[very thick, \clrd] (-1,0)--(1,0);
\draw[color=\clre, fill=white] (0,0) circle[radius=0.2];
\draw[color=\clre, fill=white] (1,0) circle[radius=0.2];
\draw[color=\clre, fill=white] (0,1) circle[radius=0.2];
\draw[color=\clre, fill=white] (1,1) circle[radius=0.2];
\draw[color=\clre, fill=white] (2,1) circle[radius=0.2];
\draw[color=\clre, fill=white] (2,0) circle[radius=0.2];
\draw[color=\clrb, fill=\clra, thick] (-1,0) circle[radius=0.2];

\draw[color=\clre, fill=white] (-1,1) circle[radius=0.2];

\node[\clrc] at (-1,0) {\tiny $\boldsymbol{\mathsf{X}}$};
\draw[color=\clrb, fill=\clra, thick] (0,0) circle[radius=0.2];
\node[\clrc] at (0,0) {\tiny $\boldsymbol{\mathsf{Z}}$};
\draw[color=\clrb, fill=\clra, thick] (1,0) circle[radius=0.2];
\node[\clrc] at (1,0) {\tiny $\boldsymbol{\mathsf{X}}$};
\end{tikzpicture} & Horizontal hoppings&
\begin{tikzpicture}[scale=0.8,baseline=-10]
\path[left color= white, right color = \checker] (-.5,0)--(-1,.5)--(-1,-.5)--cycle;
\draw[color = \clre] (-.5,0)--(-.75,.25);
\draw[color = \clre] (-.5,0)--(-.75,-.25);
\path[right color= white, left color = \checker] (2.5,0)--(3,.5)--(3,-.5)--cycle;
\draw[color = \clre] (2.5,0)--(2.75,.25);
\draw[color = \clre] (2.5,0)--(2.75,-.25);
\path[top color= white, bottom color = \checker] (0,.5)--(-0.5,1)--(0.5,1)--cycle;
\draw[\clre] (0,.5)--(-.25,.75);
\draw[\clre] (0,.5)--(.25,.75);
\path[top color= white, bottom color = \checker] (2,.5)--(1.5,1)--(2.5,1)--cycle;
\draw[\clre] (2,.5)--(1.75,.75);
\draw[\clre] (2,.5)--(2.25,.75);
\path[bottom color= white, top color = \checker] (0,-.5)--(-0.5,-1)--(0.5,-1)--cycle;
\draw[\clre](0,-.5)--(-0.25,-.75);
\draw[\clre](0,-.5)--(0.25,-.75);
\path[bottom color= white, top color = \checker] (2,-.5)--(1.5,-1)--(2.5,-1)--cycle;
\draw[\clre](2,-.5)--(1.75,-.75);
\draw[\clre](2,-.5)--(2.25,-.75);

\path[bottom color= white, top color = \checker] (1,-.5)--(0.5,-1)--(1.5,-1)--cycle;
\draw[\clre](1,-.5)--(0.75,-.75);
\draw[\clre](1,-.5)--(1.25,-.75);

\path[top color= white, bottom color = \checker] (1,.5)--(0.5,1)--(1.5,1)--cycle;
\draw[\clre] (1,.5)--(.75,.75);
\draw[\clre] (1,.5)--(1.25,.75);

\draw[color=\clre, fill=\checker] (-.5,0)--(0,.5)--(0.5,0)--(0,-.5)--cycle;
\draw[color=\clre, fill=\checker] (1.5,0)--(2,.5)--(2.5,0)--(2,-.5)--cycle;
\draw[color=\clre, fill=\checker] (0.5,0)--(1,.5)--(1.5,0)--(1,-.5)--cycle;
\draw[very thick,color=\clrd ] (-0.5,0)--(0,0.5)--(0.5,0)--(1,0.5)--(1.5,0)--(2,0.5)--(2.5,0)--(2,-0.5)--(1.5,0);

\draw[color=\clrb, fill=\clra, thick] (-0.5,0) circle[radius=0.2];
\node[\clrc] at (-0.5,0) {\tiny $\boldsymbol{\mathsf{Z}}$};
\draw[color=\clrb, fill=\clra, thick] (0.5,0) circle[radius=0.2];
\node[\clrc] at (0.5,0) {\tiny $\boldsymbol{\mathsf{Y}}$};
\draw[color=\clrb, fill=\clra, thick] (0,0.5) circle[radius=0.2];
\node[\clrc] at (0,0.5) {\tiny $\boldsymbol{\mathsf{Z}}$};
\draw[color=\clre, fill=white] (0,-0.5) circle[radius=0.2];

\draw[color=\clrb, fill=\clra, thick] (2,-0.5) circle[radius=0.2];
\node[\clrc] at (2,-0.5) {\tiny $\boldsymbol{\mathsf{Z}}$};
\draw[color=\clrb, fill=\clra, thick] (2,0.5) circle[radius=0.2];
\node[\clrc] at (2,0.5) {\tiny $\boldsymbol{\mathsf{Z}}$};
\draw[color=\clrb, fill=\clra, thick] (2.5,0) circle[radius=0.2];
\node[\clrc] at (2.5,0) {\tiny $\boldsymbol{\mathsf{Z}}$};
\draw[color=\clre, fill=white] (1,-0.5) circle[radius=0.2];
\draw[color=\clrb, fill=\clra, thick] (1,0.5) circle[radius=0.2];
\node[\clrc] at (1,0.5) {\tiny $\boldsymbol{\mathsf{Z}}$};
\draw[color=\clrb, fill=\clra, thick] (1.5,0) circle[radius=0.2];
\node[\clrc] at (1.5,0) {\tiny $\boldsymbol{\mathsf{Y}}$};

\end{tikzpicture}\\\begin{tikzpicture} [scale=0.7, baseline=0]
\draw[\clre] (2.5,0)--(-1.5,0);
\draw[\clre] (0,-0.5)--(0,1.5);
\draw[\clre] (2.5,1)--(-1.5,1);
\draw[\clre] (2,-0.5)--(2,1.5);
\draw[\clre] (1,-0.5)--(1,1.5);
\draw[\clre] (-1,-0.5)--(-1,1.5);
\draw[very thick, \clrd] (-1,0)--(1,0);
\draw[color=\clre, fill=white] (0,0) circle[radius=0.2];
\draw[color=\clre, fill=white] (1,0) circle[radius=0.2];
\draw[color=\clre, fill=white] (0,1) circle[radius=0.2];
\draw[color=\clre, fill=white] (1,1) circle[radius=0.2];
\draw[color=\clre, fill=white] (2,1) circle[radius=0.2];
\draw[color=\clre, fill=white] (2,0) circle[radius=0.2];
\draw[color=\clrb, fill=\clra, thick] (-1,0) circle[radius=0.2];

\draw[color=\clre, fill=white] (-1,1) circle[radius=0.2];

\node[\clrc] at (-1,0) {\tiny $\boldsymbol{\mathsf{Y}}$};
\draw[color=\clrb, fill=\clra, thick] (0,0) circle[radius=0.2];
\node[\clrc] at (0,0) {\tiny $\boldsymbol{\mathsf{Z}}$};
\draw[color=\clrb, fill=\clra, thick] (1,0) circle[radius=0.2];
\node[\clrc] at (1,0) {\tiny $\boldsymbol{\mathsf{Y}}$};
\end{tikzpicture} & \begin{tabular}{c}
 \\ \\ \\
\end{tabular} &\begin{tikzpicture}[scale=0.8, baseline=-10]
\path[left color= white, right color = \checker] (-.5,0)--(-1,.5)--(-1,-.5)--cycle;
\draw[color = \clre] (-.5,0)--(-.75,.25);
\draw[color = \clre] (-.5,0)--(-.75,-.25);
\path[right color= white, left color = \checker] (2.5,0)--(3,.5)--(3,-.5)--cycle;
\draw[color = \clre] (2.5,0)--(2.75,.25);
\draw[color = \clre] (2.5,0)--(2.75,-.25);
\path[top color= white, bottom color = \checker] (0,.5)--(-0.5,1)--(0.5,1)--cycle;
\draw[\clre] (0,.5)--(-.25,.75);
\draw[\clre] (0,.5)--(.25,.75);
\path[top color= white, bottom color = \checker] (2,.5)--(1.5,1)--(2.5,1)--cycle;
\draw[\clre] (2,.5)--(1.75,.75);
\draw[\clre] (2,.5)--(2.25,.75);
\path[bottom color= white, top color = \checker] (0,-.5)--(-0.5,-1)--(0.5,-1)--cycle;
\draw[\clre](0,-.5)--(-0.25,-.75);
\draw[\clre](0,-.5)--(0.25,-.75);
\path[bottom color= white, top color = \checker] (2,-.5)--(1.5,-1)--(2.5,-1)--cycle;
\draw[\clre](2,-.5)--(1.75,-.75);
\draw[\clre](2,-.5)--(2.25,-.75);

\path[bottom color= white, top color = \checker] (1,-.5)--(0.5,-1)--(1.5,-1)--cycle;
\draw[\clre](1,-.5)--(0.75,-.75);
\draw[\clre](1,-.5)--(1.25,-.75);

\path[top color= white, bottom color = \checker] (1,.5)--(0.5,1)--(1.5,1)--cycle;
\draw[\clre] (1,.5)--(.75,.75);
\draw[\clre] (1,.5)--(1.25,.75);

\draw[color=\clre, fill=\checker] (-.5,0)--(0,.5)--(0.5,0)--(0,-.5)--cycle;
\draw[color=\clre, fill=\checker] (1.5,0)--(2,.5)--(2.5,0)--(2,-.5)--cycle;
\draw[color=\clre, fill=\checker] (0.5,0)--(1,.5)--(1.5,0)--(1,-.5)--cycle;
\draw[very thick, color=\clrd] (0,-0.5)--(1,0.5)--(1.5,0);

\draw[color=\clre, fill=white] (-0.5,0) circle[radius=0.2];
\draw[color=\clrb, fill=\clra, thick] (0.5,0) circle[radius=0.2];
\node[\clrc] at (0.5,0) {\tiny $\boldsymbol{\mathsf{X}}$};
\draw[color=\clre, fill=white] (0,0.5) circle[radius=0.2];
\draw[color=\clrb, fill=\clra, thick] (0,-0.5) circle[radius=0.2];
\node[\clrc] at (0,-0.5) {\tiny $\boldsymbol{\mathsf{Z}}$};

\draw[color=\clre, fill=white] (2,-0.5) circle[radius=0.2];
\draw[color=\clre, fill=white] (2,0.5) circle[radius=0.2];
\draw[color=\clre, fill=white] (2.5,0) circle[radius=0.2];
\draw[color=\clre, fill=white] (1,-0.5) circle[radius=0.2];
\draw[color=\clrb, fill=\clra, thick] (1,0.5) circle[radius=0.2];
\node[\clrc] at (1,0.5) {\tiny $\boldsymbol{\mathsf{Z}}$};
\draw[color=\clrb, fill=\clra, thick] (1.5,0) circle[radius=0.2];
\node[\clrc] at (1.5,0) {\tiny $\boldsymbol{\mathsf{X}}$};

\end{tikzpicture} \\ \begin{tikzpicture} [scale=0.7, baseline=0]
\draw[\clre] (2.5,0)--(-0.5,0);

\draw[\clre] (0,-0.5)--(0,1.5);
\draw[\clre] (2.5,1)--(-0.5,1);
\draw[\clre] (2,-0.5)--(2,1.5);
\draw[\clre] (1,-0.5)--(1,0);
\draw[\clre] (1,1.5)--(1,1);
\draw[color=\clrb, fill=\clra, very thick] (1,0)--(1,1);
\draw[color=\clre, fill=white] (0,0) circle[radius=0.2];
\draw[color=\clrb, fill=\clra, thick] (1,0) circle[radius=0.2];
\node[\clrc] at (1,0) {\tiny $\boldsymbol{\mathsf{Z}}$};
\draw[color=\clre, fill=white] (0,1) circle[radius=0.2];
\draw[color=\clrb, fill=\clra, thick] (1,1) circle[radius=0.2];
\node[\clrc] at (1,1) {\tiny $\boldsymbol{\mathsf{Z}}$};
\draw[color=\clre, fill=white] (2,1) circle[radius=0.2];
\draw[color=\clre, fill=white] (2,0) circle[radius=0.2];
\end{tikzpicture} $\quad$\begin{tikzpicture} [scale=0.7, baseline=0]
\draw[\clre] (2.5,0)--(-0.5,0);
\draw[\clre] (1,-0.5)--(1,1.5);
\draw[\clre] (0,-0.5)--(0,1.5);
\draw[\clre] (2.5,1)--(-0.5,1);
\draw[\clre] (2,-0.5)--(2,1.5);
\draw[color=\clre, fill=white] (0,0) circle[radius=0.2];
\draw[color=\clrb, fill=\clra, thick] (1,0) circle[radius=0.2];
\node[\clrc] at (1,0) {\tiny $\boldsymbol{\mathsf{Z}}$};
\draw[color=\clre, fill=white] (0,1) circle[radius=0.2];
\draw[color=\clre, fill=white] (1,1) circle[radius=0.2];
\draw[color=\clre, fill=white] (2,1) circle[radius=0.2];
\draw[color=\clre, fill=white] (2,0) circle[radius=0.2];
\end{tikzpicture}&\begin{tabular}{c}Hubbard terms \\ $\,$ \end{tabular} &  \begin{tikzpicture}[scale=0.8,baseline=-10]
\path[left color= white, right color = \checker] (-.5,0)--(-1,.5)--(-1,-.5)--cycle;
\draw[color = \clre] (-.5,0)--(-.75,.25);
\draw[color = \clre] (-.5,0)--(-.75,-.25);
\path[right color= white, left color = \checker] (1.5,0)--(2,.5)--(2,-.5)--cycle;
\draw[color = \clre] (1.5,0)--(1.75,.25);
\draw[color = \clre] (1.5,0)--(1.75,-.25);
\path[top color= white, bottom color = \checker] (0,.5)--(-0.5,1)--(0.5,1)--cycle;
\draw[\clre] (0,.5)--(-.25,.75);
\draw[\clre] (0,.5)--(.25,.75);
\path[top color= white, bottom color = \checker] (1,.5)--(0.5,1)--(1.5,1)--cycle;
\draw[\clre] (1,.5)--(.75,.75);
\draw[\clre] (1,.5)--(1.25,.75);
\path[bottom color= white, top color = \checker] (0,-.5)--(-0.5,-1)--(0.5,-1)--cycle;
\draw[\clre](0,-.5)--(-0.25,-.75);
\draw[\clre](0,-.5)--(0.25,-.75);
\path[bottom color= white, top color = \checker] (1,-.5)--(0.5,-1)--(1.5,-1)--cycle;
\draw[\clre](1,-.5)--(0.75,-.75);
\draw[\clre](1,-.5)--(1.25,-.75);
\draw[color=\clre, fill=\checker] (-.5,0)--(0,.5)--(0.5,0)--(0,-.5)--cycle;
\draw[color=\clre, fill=\checker] (0.5,0)--(1,.5)--(1.5,0)--(1,-.5)--cycle;
\draw[very thick, color=\clrd](0,-.5)--(-.5,0)--(0,.5);
\draw[very thick, color=\clrd](1,-.5)--(1.5,0)--(1,.5);
\draw[very thick,  color=\clrd] (0,-0.5)--(1,0.5);
\draw[color=\clrb, fill=\clra, thick] (-0.5,0) circle[radius=0.2];
\node[\clrc] at (-0.5,0) {\tiny $\boldsymbol{\mathsf{Z}}$};
\draw[thick, color=\clrb, fill=\clra] (0.5,0) circle[radius=0.2];
\draw[color=\clrb, fill=\clra, thick] (0,0.5) circle[radius=0.2];
\node[\clrc] at (0,0.5) {\tiny $\boldsymbol{\mathsf{Z}}$};
\draw[color=\clrb, fill=\clra, thick] (0,-0.5) circle[radius=0.2];
\node[\clrc] at (0,-0.5) {\tiny $\boldsymbol{\mathsf{Z}}$};
\draw[color=\clrb, fill=\clra, thick] (1,-0.5) circle[radius=0.2];
\node[\clrc] at (1,-0.5) {\tiny $\boldsymbol{\mathsf{Z}}$};
\draw[color=\clrb, fill=\clra, thick] (1,0.5) circle[radius=0.2];
\node[white] at (1,0.5) {\tiny $\boldsymbol{\mathsf{Z}}$};
\draw[color=\clrb, fill=\clra, thick] (1.5,0) circle[radius=0.2];
\node[\clrc] at (1.5,0) {\tiny $\boldsymbol{\mathsf{Z}}$};

\end{tikzpicture} $\quad$ \begin{tikzpicture}[scale=0.8, baseline=-10]
\path[left color= white, right color = \checker] (-.5,0)--(-1,.5)--(-1,-.5)--cycle;
\draw[color = \clre] (-.5,0)--(-.75,.25);
\draw[color = \clre] (-.5,0)--(-.75,-.25);
\path[right color= white, left color = \checker] (0.5,0)--(1,.5)--(1,-.5)--cycle;
\draw[color = \clre] (0.5,0)--(0.75,.25);
\draw[color = \clre] (0.5,0)--(0.75,-.25);
\path[top color= white, bottom color = \checker] (0,.5)--(-0.5,1)--(0.5,1)--cycle;
\draw[\clre] (0,.5)--(-.25,.75);
\draw[\clre] (0,.5)--(.25,.75);

\path[bottom color= white, top color = \checker] (0,-.5)--(-0.5,-1)--(0.5,-1)--cycle;
\draw[\clre](0,-.5)--(-0.25,-.75);
\draw[\clre](0,-.5)--(0.25,-.75);

\draw[very thick, color=\clrd, fill=\checker] (-.5,0)--(0,.5)--(0.5,0)--(0,-.5)--cycle;
\draw[color=\clrb, fill=\clra, thick] (-0.5,0) circle[radius=0.2];
\node[\clrc] at (-.5,0) {\tiny $\boldsymbol{\mathsf{Z}}$};
\draw[color=\clrb, fill=\clra, thick] (0.5,0) circle[radius=0.2];
\node[\clrc] at (.5,0) {\tiny $\boldsymbol{\mathsf{Z}}$};
\draw[color=\clrb, fill=\clra, thick] (0,0.5) circle[radius=0.2];
\node[\clrc] at (0,.5) {\tiny $\boldsymbol{\mathsf{Z}}$};
\draw[color=\clrb, fill=\clra, thick] (0,-0.5) circle[radius=0.2];
\node[\clrc] at (0,-.5) {\tiny $\boldsymbol{\mathsf{Z}}$};

\end{tikzpicture}
\\ \hline

\end{tabular}
\caption{Transforming terms of the Hubbard model according to the Verstraete-Cirac and Superfast simulation mapping. For the hoppings, we consider the real hopping terms, i.e. transforms of  $(i m_j \obar{m}_k)$ and $(i m_k \,\obar{m}_j)$ for $j<k$.  Note that for the Verstraete-Cirac transform, the vertical hopping terms are different for even/odd rows and columns.  Here the south east qubit is in an even column and odd row. The  qubit marked, but  not labeled with  $\mathsf{X}$, $\mathsf{Y}$ or $\mathsf{Z}$,  is skipped.  }\label{tab:hubbardx}
\end{table}

\section{Notation}
\label{sec:notations}
\newcommand{\colw}{0.5\textwidth}
\begin{tabular}{cc|l}
\textbf{Notation} & \textbf{(Value)} &  \textbf{Definition}  \\
\hline $[... ]$ && \begin{minipage}{\colw}Set of integers from $1$ to argument. \end{minipage}\\
\hline $(i ,\, j)$ && \begin{minipage}{\colw}Spacial coordinates replacing qubit labels in Section \ref{sec:three}.\end{minipage} \\
\hline $\hat{=}$ && \begin{minipage}{\colw} $\,$\\ Equivalence  between fermionic operators/states to qubit counterparts. \\  \end{minipage}\\
\hline $A^{\vphantom{-1}}, \, A^{-1}$ &&  \begin{minipage}{\colw} $\,$ \\ Binary $(N \times N)$ matrix defining linear Fermion-to-qubit mappings, see \eqref{eq:lineartransform}. \\ \end{minipage}\\
\hline $\aux$ &  $ \bigcup_{m \in [r]} \; \lbrace N+m \rbrace$ & \begin{minipage}{\colw} Auxiliary register labels, $(N+1)$ to $(N+r)$. \end{minipage} \\
\hline $ c^{\dagger}_j \, , c^{\phantom{\dagger}}_j$ &$\phantom{\begin{matrix} 1 \\ & 1 \end{matrix}}$ &  \begin{minipage}{\colw} Fermionic annihilation, creation operator on mode $j$, see \eqref{eq:aannttii} and \eqref{eq:singleop}. \end{minipage}\\
\hline $\textsc{CNot}(i \to j)$ &$\ket{0}\!\!\bra{0}_i + \ket{1}\!\!\bra{1}_i \otimes X_j $& \begin{minipage}{\colw} Controlled-Not gate. $\vphantom{\begin{matrix} 1 \\ 1\end{matrix} }$ \end{minipage}\\
\hline $\data$ &$[N]$&  \begin{minipage}{\colw} Data register labels, $1$ to $N$. \end{minipage}\\
\hline $F(j)$ &&\begin{minipage}{\colw}  Flip set of mode $j$.\end{minipage} \\
\hline $h_\data, \, \widetilde{h}_{\auxdata}$ && \begin{minipage}{\colw} $\,$ \\ $N$-qubit Pauli strings occurring in the Hamiltonian \\ and their logical equivalents on $N+r$ qubits. Note that $\widetilde{h}_\auxdata$ can differ from $(h_\data \otimes \kappa^h_{\aux})$ by the multiplication with stabilizers.\\ \end{minipage} \\
\hline $H_\data, \, \widetilde{H}_{\auxdata}$ && \begin{minipage}{\colw} $\,$ \\ $N$-qubit Hamiltonian \eqref{eq:hamil}, as for instance obtained by Jordan-Wigner transform, and  ($N+r$)-qubit logical Hamiltonian \eqref{eq:stabhamil2}.\\  \end{minipage} \\
\hline $\mathrm{H}_j$&$\frac{1}{\sqrt{2}}\left[\begin{smallmatrix} 1 & 1 \\ 1 & -1 \end{smallmatrix}\right]$& \begin{minipage}{\colw} Hadamard gate on qubit $j$. \end{minipage} \\
\hline $\mathbb{I}$ && \begin{minipage}{\colw} Identity matrix/operation.\end{minipage} \\
\hline $\inter$ && \begin{minipage}{\colw} Periodicity in the  sparse AQM, see Table \ref{tab:results}.\end{minipage}\\
\hline $\kappa^h_\aux$ && \begin{minipage}{\colw} Adjustments to $h_\data$, see \eqref{eq:stabhamil}. \end{minipage} \\
\hline $\ell_1, \, \ell_2 $ && \begin{minipage}{\colw} $\,$\\ $\ell_1 \times \ell_2$ is the dimension of the fermionic lattice depicted in Figure \ref{fig:layers}(a).\\ \end{minipage}\\
\hline $L$ && \begin{minipage}{\colw}$\,$\\  $2L\times L$ is the dimension of the Fermi-Hubbard-lattice in Section \ref{sec:four}. \\ \end{minipage}\\
\hline $m_j, \, \obar{m}_j$ && \begin{minipage}{\colw} Majorana operators, see \eqref{eq:anticommix}-\eqref{eq:majoran2}.\end{minipage}\\
\hline $n$ &$N+r$& \begin{minipage}{\colw} Number of qubits. \end{minipage}\\
\hline $N$ & $\ell_1 \times \ell_2$& \begin{minipage}{\colw} Number of fermionic modes. \end{minipage}\\
\hline $P(j)$ && \begin{minipage}{\colw} Parity set of mode $j$, see \eqref{eq:singleop}. \end{minipage} \\
\hline $p^i_\data$ && \begin{minipage}{\colw} Data-qubit part of the stabilizers $(p^i_\data \otimes \sigma^i_{N+i})$, see \eqref{eq:stabsystem}. \end{minipage}\\
\hline $r$ && \begin{minipage}{\colw} Number of auxiliary qubits. \end{minipage}\\
\hline $R$ && \begin{minipage}{\colw} Lower triangular matrix, see \eqref{eq:triangular}. \end{minipage}\\
\hline $\sigma^i_{N+i}$ && \begin{minipage}{\colw}Part of the stabilizers $(p^i_\data \otimes \sigma^i_{N+i})$, see \eqref{eq:stabsystem}.\end{minipage} \\
\hline $U(j)$ && \begin{minipage}{\colw} Update set of mode $j$, see \eqref{eq:singleop}.  \end{minipage}\\
\hline $V_\auxdata$ && \begin{minipage}{\colw} Initialization circuit for the code space, see \eqref{eq:stabsystem}.\end{minipage}\\
\hline $X_j, \, Y_j, \, Z_j $ & $ \phantom{\begin{matrix} 1  \\ & -1 \end{matrix}}\left[\begin{smallmatrix} & 1  \\ 1 \end{smallmatrix} \right] , \, \left[\begin{smallmatrix}  & -i  \\ i \end{smallmatrix} \right], \, \left[\begin{smallmatrix} 1  \\ & -1 \end{smallmatrix} \right]  \phantom{\begin{matrix} 1  \\ & -1 \end{matrix}} $ & \begin{minipage}{\colw} Pauli operators acting on qubit $j$. \end{minipage} \\

\hline $\mathbb{Z}_2$ & $\lbrace 0,\, 1\rbrace $ & \begin{minipage}{\colw} Binary digits. \end{minipage}
\end{tabular}
 \end{document}